\documentclass[preprint,prd,aps,showpacs,showkeys,nofootinbib]{revtex4}
\usepackage{amsmath}
\usepackage{graphicx}
\allowdisplaybreaks
\begin{document}


\title{Gauss Relations in Feynman Integrals}

\author{Tai-Fu Feng\footnote{email:fengtf@hbu.edu.cn}$^{a,b,c,d}$,
Yang Zhou$^{a,b,c}$,
Hai-Bin Zhang\footnote{email:hbzhang@hbu.edu.cn}$^{a,b,c}$}

\affiliation{$^a$Department of Physics, Hebei University, Baoding, 071002, China}
\affiliation{$^b$Hebei Key Laboratory of High-precision Computation and Application of Quantum Field Theory, Baoding, 071002, China}
\affiliation{$^c$Hebei Research Center of the Basic Discipline for Computational Physics, Baoding, 071002, China}
\affiliation{$^d$Department of Physics, Chongqing University, Chongqing, 401331, China}

\begin{abstract}
Embedding Feynman integrals in Grassmannians, we can write Feynman integrals as some finite linear combinations
of generalized hypergeometric functions. In this paper we present a general method
to obtain Gauss relations among those generalized hypergeometric functions. The hypergeometric expressions of
Feynman integral are continued from a connected component to
another by the inverse Gauss relations, then continued to the whole domain
of definition by the Gauss-Kummer relations. The Laurent series of the Feynman integral around the space-time
dimension $D=4$ is obtained by the Gauss adjacent relations where the coefficient of the term with power of $D-4$ is
given as the linear combinations of hypergeometric functions with integer parameters.
As examples, we illustrate how to obtain the expressions for the Feynman integrals of the 1-loop self-energy and
a 2-loop massless triangle diagram in the domains of definition.
\end{abstract}


\maketitle

\section{Introduction\label{sec1}}
\indent\indent
Precision tests of the standard model (SM) require the calculation of the multi-loop Feynman integrals involving several energy scales~\cite{Heinrich2021}.
Nevertheless the popular approaches in the literature are either only applicable to some special cases, or are not efficient
since the corresponding numerical programs occupy computer memory and consume a large amount of machine time~\cite{Smirnov2012}.
Even if the development of computational technology does not impede numerical evaluations of Feynman integrals efficiently,
analytic expressions are still important because they provide the cross checks for relevant numerical programs. Additionally
the analytic expression of Feynman integral is convenient to reveal the threshold effects through the asymptotic
expansion~\cite{Smirnov1990}.

Any commonly used function of one variable of analysis can be expressed as the Gauss function with three parameters
$a,\;b,\;c$:
\begin{eqnarray}
&&\;_2F_1\left(\left.\begin{array}{c}a,b\\c\end{array}\right| x\right)=\sum\limits_{n=0}^\infty{(a)_{_n}(b)_{_n}\over n!\;(c)_{_n}}\;x^n\;,
\label{Intro1}
\end{eqnarray}
where $(a)_{_n}=\Gamma(a+n)/\Gamma(a)$ is the Pochhammer notation~\cite{Slater1966}.
Similarly any Feynman integral involving several energy scales can be given by some finite linear combination of
generalized hypergeometric functions~\cite{Kashiwara1976,Davydychev1987,Davydychev2000,
Davydychev1993NPB,Davydychev3,Davydychev1992JPA,Davydychev2006,Davydychev1992JMP,Davydychev1991JMP,
Davydychev1993,Tarasov2000,Tarasov2003,Kalmykov2009,Bytev2010,Kalmykov2011,Bytev2013,Bytev2015,Bytev2016,
Kalmykov2012,Kalmykov2017,Davydychev1998,Davydychev2000a,Davydychev2001a,Davydychev2017,Suzuki2002,
Suzuki2003a,Suzuki2003b,Davydychev2004,Jegerlehner2004}, where the variables are dimensionless ratios
among the external squared momenta and the virtual squared masses, and the parameters depend on
the space-time dimension in dimensional regularization.

There are three types of Gauss relations among those 24 hypergeometric
solutions of the partial differential equation (PDE) which can be written as the GKZ-system on the Grassmannians $G_{_{2,4}}$ ~\cite{Gelfand1986-1}.
The inverse Gauss relations include the following analytic continuation and its various variants
\begin{eqnarray}
&&\;_2F_1\left(\left.\begin{array}{c}a,b\\c\end{array}\right| x\right)
={\Gamma(c)\Gamma(b-a)\over\Gamma(b)\Gamma(c-a)}(-x)^{-a}\;_2F_1\left(\left.\begin{array}{c}a,1+a-c\\1+a-b\end{array}\right|{1\over x}\right)
\nonumber\\
&&\hspace{3.2cm}
+{\Gamma(c)\Gamma(a-b)\over\Gamma(a)\Gamma(c-b)}(-x)^{-b}\;_2F_1\left(\left.\begin{array}{c}b,1+b-c\\1-a+b\end{array}\right|{1\over x}\right)\;.
\label{Intro2}
\end{eqnarray}
This transformation satisfies the idempotent property. Performing the inverse transformation on the terms of
the right-hand side, we find that the sum of the images is exactly the term on the left-hand side.
The independent Gauss adjacent relations include the following two equations
\begin{eqnarray}
&&c\;_2F_1\left(\left.\begin{array}{c}a,b\\c\end{array}\right| x\right)=a\;x
\;_2F_1\left(\left.\begin{array}{c}a+1,b\\c+1\end{array}\right| x\right)
+c\;_2F_1\left(\left.\begin{array}{c}a,b-1\\c\end{array}\right| x\right)
\;,\nonumber\\
&&(a-c+1)\;_2F_1\left(\left.\begin{array}{c}a,b\\c\end{array}\right| x\right)
=a\;_2F_1\left(\left.\begin{array}{c}a+1,b\\c\end{array}\right| x\right)
-(c-1)\;_2F_1\left(\left.\begin{array}{c}a,b\\c-1\end{array}\right| x\right)\;,
\label{Intro3}
\end{eqnarray}
and there are other two equations which are obtained by interchanging $a\leftrightarrow b$ in Eq.(\ref{Intro3}).
It is easy to verify that all of the Gauss adjacent relations presented in (1.4.1)-(1.4.15) in the literature~\cite{Slater1966} are
linear combinations of these four relations. The third type Gauss relations contain
\begin{eqnarray}
&&\;_{_2}F_{_1}\left(\left.\begin{array}{c}a,\;b\\ c\end{array}\right|\;x\right)=(1-x)^{c-a-b}\;_{_2}F_{_1}\left(\left.\begin{array}{c}c-a,\;c-b\\ c\end{array}\right|\;x\right)
\nonumber\\
&&\hspace{0.0cm}=
(1-x)^{-a}\;_{_2}F_{_1}\left(\left.\begin{array}{c}a,\;c-b\\ c\end{array}\right|\;{x\over x-1}\right)
=(1-x)^{-b}\;_{_2}F_{_1}\left(\left.\begin{array}{c}c-a,\;b\\ c\end{array}\right|\;{x\over x-1}\right)
\label{Intro4}
\end{eqnarray}
and its various variants which are derived through Kummer's classification.

Applying the GKZ-system satisfied by the concerned Feynman integral, we obtain the fundamental solution systems which are consisted of
generalized hypergeometric functions in proper subsets of some connected components of its domain of definition.
In order to continue the expressions properly in the domain of definition,
we should also present the generalized Gaussian relations among the hypergeometric functions
constituting the fundamental solution systems. The Feynman integral has the analytic expressions
at some regular singularities. Taking those analytic expressions as boundary conditions
and applying the generalized inverse Gauss relations, we obtain the unique hypergeometric expression of the
Feynman integral in each proper subset of some connected components of the domain of definition. The hypergeometric expressions
can be continued to the whole domain of definition of the Feynman integral through the Gauss-Kummer
relations. The Gauss adjacent relations are adopted to explain the equivalence between the different fundamental solution
systems of a Feynman integral. Moreover, the adjacent relations can also be used to formulate the coefficient of the
term with power of $\varepsilon=2-D/2$ in Feynman integral expansions as a linear combination
of generalized hypergeometric functions with integer parameters.

When the Feynman integral is regarded as a function of Grassmannians~\cite{Feng2022},
the inverse Gauss relations among the hypergeometric functions contained in the fundamental solution systems are derived by
the Mellin-Barnes's contour~\cite{Barnes1907}, and the Gauss adjacent relations depend on the matroid representing the Grassmannian
and its dual~\cite{Oxley2011,Gelfand1987}, respectively. Requiring equality of the principal branch of those hypergeometric
functions which correspond to a same geometric representation,
we generalize the relations of Gauss functions originating from Kummer's classification.

Certainly the Feynman integral can also be regarded as a function on a projective space~\cite{Feng2018,Feng2019,Feng2020,Feng2023,Zhang2023,Zhang2024,
Cruz2019,Klausen2019,Ananthanarayan2019,Ananthanarayan2022,Ananthanarayan2022a,Bera2023,Bera2024,Chestnov2022,Chestnov2023,Munch2022,Munch2024,Klemm2020,Bonisch2021},
and the fundamental solution systems are obtained through the corresponding GKZ-system.
The inverse Gauss relations of the hypergeometric functions are obtained by the Mellin-Barnes's contour,
and the adjacent relations of the hypergeometric functions on the $n-1$
dimensional projective space are determined by the quotient module of the free module ${\bf C}^n$ by the submodule which is generated
by the coefficient vectors of the corresponding GKZ-system~\cite{Gelfand1992}. Because the GKZ-system
does not contain sufficiently combinatorial information, we cannot obtain the generalized Gauss-Kummer relations
of the hypergeometric functions~\cite{Gelfand1986-2}.

It is well-known that the Gauss functions can be regarded as the functions on the Grassmannian $G_{_{2,4}}$~\cite{Gelfand1986-1}.
Similarly the Feynman integrals of the 1-loop self-energy, the 1-loop massless triangle diagram,
the 2-loop vacuum, and a 2-loop massless triangle diagram investigated below all can be regarded as some linear combinations of hypergeometric functions
on the Grassmannian $G_{_{3,5}}$ with different exponent vectors. The splitting coordinates of the Grassmannian $G_{_{3,5}}$~\cite{Feng2022}
are written as
\begin{eqnarray}
&&\boldsymbol{{\xi}}=\left(\begin{array}{ccccc}\;1\;&\;0\;&\;0\;&\;1\;&\;r_{_1}\;\\\;0\;&\;1\;&\;0\;&\;1\;&\;r_{_2}\;\\
\;0\;&\;0\;&\;1\;&\;1\;&\;r_{_3}\;\end{array}\right)\;,
\label{Intro5}
\end{eqnarray}
where $r_{_{1,2}}=m_{_{1,2}}^2,\;r_{_3}=p^2$ for the Feynman integral of the 1-loop self energy,
$r_{_{1,2}}=p_{_{1,2}}^2,\;r_{_3}=p_{_3}^2=(p_{_1}+p_{_2})^2$ for that of the 1-loop massless triangle diagram
and the 2-loop massless triangle diagram below, and $r_{_i}=m_{_i}^2,\;i=1,2,3$ for that of the 2-loop vacuum, respectively.
Correspondingly the fundamental solution systems are consisted of the first type Appell functions
and some Horn functions in two variables. After giving the hypergeometric functions in all different affine spanning, we generalize all Gauss relations
among the hypergeometric functions on the Grassmannian $G_{_{3,5}}$ here.
Actually the methods presented here can be adopted to derive the Gauss relations among the generalized
hypergeometric functions on any Grassmannians $G_{_{k,n}},\;k<n$. As examples, we illustrate how to obtain the expressions of
the Feynman integrals of the 1-loop self energy and a 2-loop massless triangle diagram in the approach.
The GKZ-system on the Grassmannian $G_{_{2,4}}$ has 24 Gauss function solutions
which are divided into 6 groups, where the Gauss functions in each group are equal to
each other in the intersection of their convergent regions, respectively. Similarly
there are 120 generalized hypergeometric function solutions
totally for the GKZ-hypergeometric system on the Grassmannian $G_{_{3,5}}$, where 60 solutions
are the first type Appell functions, and another 60 solutions are the Horn series of two variables.
Those Appell functions are divided into 10 groups, where all the 6 hypergeometric series in each group are equal to
each other in the intersection of their convergent regions, respectively.
Accordingly those Horn series are divided into 15 groups, where all the 4 hypergeometric series in each group are equal to
each other in the intersection of their convergent regions, respectively.

Our presentation is as follows. After giving the hypergeometric functions
in different affine spanning of the Grassmannian $G_{_{3,5}}$ in section \ref{sec2},
we generalize the inverse Gauss relations among the hypergeometric functions
through the Mellin-Barnes's contour in section \ref{sec3}, and we also generalize the Gauss adjacent relations
in section \ref{sec4}, respectively. Requiring equality on the principal branch of those hypergeometric
functions with a same geometric representation, we present the Gauss-Kummer relations
in section \ref{sec5}. For illustration, we show  how to obtain the expressions of
the Feynman integral of 1-loop self energy in section \ref{sec6}, and how to obtain the expressions of
the Feynman integral of a 2-loop massless triangle diagram in section \ref{sec7}, respectively. Some conclusions are summarized
in section \ref{sec8}. In order to shorten the length of text, we collect most of formulas in the appendices.

\section{The hypergeometric functions\label{sec2}}
\indent\indent
A point of the Grassmannian $G_{_{3,5}}$ is a projective plane in the projective space $P^4$,
or equivalently is regarded as a 3-dimensional vector subspace ${\bf C}^3$ of the vector space ${\bf C}^5$.
For convenience, we denote the set of indices of columns in Eq.(\ref{Intro5}) by ${\mathcal N}=\{1,\cdots,5\}$. For the general vector
$(r_{_1},\;r_{_2},\;r_{_3})\in{\bf C}^3$, any three column vectors of the matrix $\boldsymbol{{\xi}}$ are linearly independent,
and therefore can be chosen as an affine spanning of the vector subspace ${\bf C}^3$. Denoting by ${\cal B}$ the subset of indices of
three columns vectors to span ${\bf C}^3$, one finds that the hypergeometric functions also depend on
the $3\times5$ integer lattice matrix whose entries of the columns with indices from the subset ${\mathcal N}\backslash{\cal B}$
are formulated as $\pm n_{_1}E_{_3}^{(i)}\pm n_{_2}E_{_3}^{(j)}$, where $n_{_{1,2}}\ge0$, $(i,j)\in\{(1,2),(1,3),(2,3)\}$,
and other entries are all zero. Here
\begin{eqnarray}
&&E_{_3}^{(1)}=\left(\begin{array}{cc}\;\;0\;\;&\;\;0\;\;\\\;\;1\;\;&\;\;-1\;\;\\
\;\;-1\;\;&\;\;1\;\;\end{array}\right)\;,\;
E_{_3}^{(2)}=\left(\begin{array}{cc}\;\;1\;\;&\;\;-1\;\;\\ \;\;0\;\;&\;\;0\;\;\\
\;\;-1\;\;&\;\;1\;\;\end{array}\right)\;,
\nonumber\\
&&E_{_3}^{(3)}=\left(\begin{array}{cc}\;\;1\;\;&\;\;-1\;\;\\
\;\;-1\;\;&\;\;1\;\; \\ \;\;0\;\;&\;\;0\;\;\end{array}\right)\;.
\label{sec2-1}
\end{eqnarray}
In addition, the hypergeometric function on the Grassmannian $G_{_{3,5}}$ with the splitting coordinates in Eq.(\ref{Intro5})
satisfies the GKZ-system
\begin{eqnarray}
&&\Big\{\vartheta_{_{1,4}}+\vartheta_{_{1,5}}\Big\}\Phi(\boldsymbol{{\beta}},\;\boldsymbol{{\xi}})=
-\beta_{_1}\Phi(\boldsymbol{{\beta}},\;\boldsymbol{{\xi}})
\;,\nonumber\\
&&\Big\{\vartheta_{_{2,4}}+\vartheta_{_{2,5}}\Big\}\Phi(\boldsymbol{{\beta}},\;\boldsymbol{{\xi}})=
-\beta_{_2}\Phi(\boldsymbol{{\beta}},\;\boldsymbol{{\xi}})
\;,\nonumber\\
&&\Big\{\vartheta_{_{3,4}}+\vartheta_{_{3,5}}\Big\}\Phi(\boldsymbol{{\beta}},\;\boldsymbol{{\xi}})=
-\beta_{_3}\Phi(\boldsymbol{{\beta}},\;\boldsymbol{{\xi}})
\;,\nonumber\\
&&\Big\{\vartheta_{_{1,4}}+\vartheta_{_{2,4}}+\vartheta_{_{3,4}}\Big\}\Phi(\boldsymbol{{\beta}},\;\boldsymbol{{\xi}})=
(\beta_{_4}-1)\Phi(\boldsymbol{{\beta}},\;\boldsymbol{{\xi}})
\;,\nonumber\\
&&\Big\{\vartheta_{_{1,5}}+\vartheta_{_{2,5}}+\vartheta_{_{3,5}}\Big\}\Phi(\boldsymbol{{\beta}},\;\boldsymbol{{\xi}})=
(\beta_{_5}-1)\Phi(\boldsymbol{{\beta}},\;\boldsymbol{{\xi}})\;,
\label{sec2-2}
\end{eqnarray}
where the Euler operators $\vartheta_{_{i,j}}=\xi_{_{i,j}}\partial/\partial \xi_{_{i,j}}$, and the exponent vector $\boldsymbol{{\beta}}=(\beta_{_1},\cdots,\beta_{_5})\in{\bf C}^5$
satisfying $\sum\beta_{_i}=2$.
\begin{figure}[ht]
\setlength{\unitlength}{1cm}
\centering
\vspace{0.0cm}\hspace{-1.5cm}
\includegraphics[height=8cm,width=15.0cm]{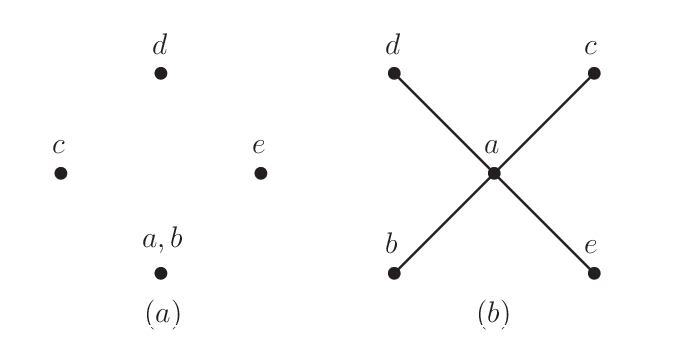}
\vspace{0cm}
\caption[]{The geometric configurations of the hypergeometric functions on the projective plane $P^{2}$, where the points $a,\cdots,e$
denote the indices of columns of the $3\times5$ matrix.}
\label{fig1}
\end{figure}
Choosing the spanning subset ${\cal B}$ and the integer lattice on the complement ${\mathcal N}\backslash{\cal B}$,
one gets the hypergeometric function accordingly. As ${\cal B}=\{1,2,3\}$ and the integer lattice
on the complement ${\mathcal N}\backslash{\cal B}=\{4,5\}$ is given by $n_{_1}E_{_3}^{(1)}+ n_{_2}E_{_3}^{(2)}$,
the hypergeometric function is written as
\begin{eqnarray}
&&\Phi_{_{\{1,2,3\}}}^{(1)}({\boldsymbol{\beta}},\;\boldsymbol{{\xi}})=A_{_{\{1,2,3\}}}^{(1)}({\boldsymbol{\beta}})(r_{_1})^{-\beta_{_1}}(r_{_2})^{-\beta_{_2}}(r_{_3})^{1-\beta_{_3}-\beta_{_4}}
\varphi_{_{\{1,2,3\}}}^{(1)}({\boldsymbol{\beta}},\;{r_{_3}\over r_{_2}},\;{r_{_3}\over r_{_1}})\;,\nonumber\\
&&\varphi_{_{\{1,2,3\}}}^{(1)}({\boldsymbol{\beta}},\;x_{_1},\;x_{_2})=\sum\limits_{n_{_1},n_{_2}}c_{_{\{1,2,3\}}}^{(1)}({\boldsymbol{\beta}},\;n_{_1},n_{_2})
x_{_1}^{n_{_1}}x_{_2}^{n_{_2}}
\nonumber\\
&&\hspace{3.1cm}=
F_{_1}\left(\left.\begin{array}{c}1-\beta_{_4},\;\beta_{_2},\;\beta_{_1}\\ 2-\beta_{_3}-\beta_{_4}\end{array}\right|\;x_{_1},x_{_2}\right)\;,
\label{Interger-lattice1-2}
\end{eqnarray}
where
\begin{eqnarray}
&&A_{_{\{1,2,3\}}}^{(1)}({\boldsymbol{\beta}})={\Gamma(\beta_{_5})\over\Gamma(1-\beta_{_1})\Gamma(1-\beta_{_2})\Gamma(2-\beta_{_3}-\beta_{_4})}\;,\nonumber\\
&&c_{_{\{1,2,3\}}}^{(1)}({\boldsymbol{\beta}},\;n_{_1},n_{_2})={(\beta_{_2})_{_{n_{_1}}}(\beta_{_1})_{_{n_{_2}}}(1-\beta_{_4})_{_{n_{_1}+n_{_2}}}\over n_{_1}!n_{_2}!
(2-\beta_{_3}-\beta_{_4})_{_{n_{_1}+n_{_2}}}}\;,
\label{Interger-lattice1-3}
\end{eqnarray}
and $F_{_1}$ is the first type Appell function. The geometric representation of the function in
Eq.(\ref{Interger-lattice1-2}) is drawn in Fig.\ref{fig1}(a) where $\{a,b\}=\{3,4\}$ and $\{c,d,e\}=\{1,2,5\}$,
respectively. In the hypergeometric function whose geometric representation
corresponds to the first geometric configuration presented in Fig.\ref{fig1}(a), its dependence on the
parameter $\beta_{_i}$ can be obtained from the function $\varphi_{_{\{1,2,3\}}}^{(1)}$
by some permutation on components of the exponent vector ${\boldsymbol{\beta}}$.
In order to avoid the cluster of indices, we adopt the notation of the literature~\cite{Gelfand1992}.

As the lattice matrix is given by $n_{_1}E_{_3}^{(1)}+ n_{_2}E_{_3}^{(3)}$,
the generalized hypergeometric function is formulated as
\begin{eqnarray}
&&\Phi_{_{\{1,2,3\}}}^{(2)}({\boldsymbol{\beta}},\;\boldsymbol{{\xi}})=A_{_{\{1,2,3\}}}^{(2)}({\boldsymbol{\beta}})(r_{_1})^{-\beta_{_1}}(r_{_2})^{\beta_{_1}+\beta_{_5}-1}
\varphi_{_{\{1,2,3\}}}^{(2)}({\boldsymbol{\beta}},\;{r_{_3}\over r_{_2}},\;{r_{_2}\over r_{_1}})\;,\nonumber\\
&&\varphi_{_{\{1,2,3\}}}^{(2)}({\boldsymbol{\beta}},\;x_{_1},\;x_{_2})=\sum\limits_{n_{_1},n_{_2}}c_{_{\{1,2,3\}}}^{(2)}({\boldsymbol{\beta}},\;n_{_1},n_{_2})
x_{_1}^{n_{_1}}x_{_2}^{n_{_2}}
\nonumber\\
&&\hspace{3.1cm}=
H_{_1}\left(\left.\begin{array}{c}\beta_{_3},\;\beta_{_1}\\ \beta_{_3}+\beta_{_4},\;\beta_{_1}+\beta_{_5}\end{array}\right|\;x_{_1},x_{_2}\right)\;,
\label{Interger-lattice2-2}
\end{eqnarray}
where
\begin{eqnarray}
&&A_{_{\{1,2,3\}}}^{(2)}({\boldsymbol{\beta}})={\Gamma(\beta_{_4})\Gamma(\beta_{_5})\over\Gamma(1-\beta_{_1})
\Gamma(1-\beta_{_3})\Gamma(\beta_{_1}+\beta_{_5})\Gamma(\beta_{_3}+\beta_{_4})}\;,\nonumber\\
&&c_{_{\{1,2,3\}}}^{(2)}({\boldsymbol{\beta}},\;n_{_1},n_{_2})={(-1)^{n_{_1}+n_{_2}}(\beta_{_3})_{_{n_{_1}}}(\beta_{_1})_{_{n_{_2}}}\over n_{_1}!n_{_2}!
(\beta_{_1}+\beta_{_5})_{_{-n_{_1}+n_{_2}}}(\beta_{_3}+\beta_{_4})_{_{n_{_1}-n_{_2}}}}\;,
\nonumber\\
&&H_{_1}\left(\left.\begin{array}{c}a,\;b\\ c,\;d\end{array}\right|\;x,y\right)
=\sum\limits_{m,n}^\infty{(-1)^{m+n}(a)_{_m}(b)_{_n}\over m!n!(c)_{_{m-n}}(d)_{_{-m+n}}}x^my^n\;.
\label{Interger-lattice2-3}
\end{eqnarray}
The geometric representation of the function in Eq.(\ref{Interger-lattice2-2})
is drawn in Fig.\ref{fig1}(b) with $a=2,\;\{b,c\}=\{1,5\}$ and $\{d,e\}=\{3,4\}$, respectively.
Note that the Horn function $H_{_1}$ is transformed as the function $G_{_2}$ in
Ref.\cite{Erdelyi1950} through
\begin{eqnarray}
&&H_{_1}\left(\left.\begin{array}{c}a,\;b\\ c,\;d\end{array}\right|\;x,y\right)
=G_{_2}(a,b,1-c,1-d;\;-x,-y)\;.
\label{Interger-lattice2-3a}
\end{eqnarray}
In the hypergeometric function whose geometric representation
corresponds to the second geometric configuration presented in Fig.\ref{fig1}(b), its dependence on the
parameter $\beta_{_i}$ can be obtained from the function $\varphi_{_{\{1,2,3\}}}^{(2)}$
by some permutation on components of the exponent vector ${\boldsymbol{\beta}}$.
To shorten the length of text, we present the hypergeometric functions corresponding
to other integer lattices in the appendix \ref{app1}.
Where $\varphi_{_{\{1,2,3\}}}^{(i)}$, $i=1,3,5,8,10,12$ are the first Appell functions whose geometric configuration
is plotted in Fig.\ref{fig1}(a), while $\varphi_{_{\{1,2,3\}}}^{(j)}$, $j=2,4,6,7,9,11$ are the Horn series whose geometric configuration
is plotted in Fig.\ref{fig1}(b), respectively.

The dimension of the solution space of generalized hypergeometric functions on $G_{_{3,5}}$ is $C_{n-2}^{k-1}=C_3^2=3$ for the general exponent
vector $\boldsymbol{{\beta}}\in{\bf C}^5$~\cite{Gelfand1986-2,M.Saito2000}. It is easy to find that the convergent regions of $\varphi_{_{\{1,2,3\}}}^{(1)}$,
$\varphi_{_{\{1,2,3\}}}^{(2)}$, and $\varphi_{_{\{1,2,3\}}}^{(3)}$  have nonempty intersections in a connected component of the domain of definition,
and therefore they constitute a fundamental solution system in the nonempty proper subset of the parameter space.
Similarly, we obtain the fundamental solution systems explicitly on other nonempty proper subsets in the parameter space.
The Feynman integral is some linear combinations of the hypergeometric functions from the fundamental solution systems
in these nonempty proper subsets of the domain of definition.
The linear combinations of hypergeometric functions on the different nonempty proper subsets of the parameter space are regarded
as analytic continuations of each other, thus
\begin{eqnarray}
&&\Psi(\boldsymbol{{\beta}},\;\boldsymbol{{\xi}})=
\sum\limits_{i=\{1,2,3\}}C^{(i)}(\boldsymbol{{\beta}})\Phi_{_{\{1,2,3\}}}^{(i)}({\boldsymbol{\beta}},\;\boldsymbol{{\xi}})
\nonumber\\
&&\hspace{1.5cm}=
\sum\limits_{i=\{1,5,6\}}C^{(i)}(\boldsymbol{{\beta}})\Phi_{_{\{1,2,3\}}}^{(i)}({\boldsymbol{\beta}},\;\boldsymbol{{\xi}})
\nonumber\\
&&\hspace{1.5cm}=
\sum\limits_{i=\{3,7,8\}}C^{(i)}(\boldsymbol{{\beta}})\Phi_{_{\{1,2,3\}}}^{(i)}({\boldsymbol{\beta}},\;\boldsymbol{{\xi}})
\nonumber\\
&&\hspace{1.5cm}=
\sum\limits_{i=\{4,5,12\}}C^{(i)}(\boldsymbol{{\beta}})\Phi_{_{\{1,2,3\}}}^{(i)}({\boldsymbol{\beta}},\;\boldsymbol{{\xi}})
\nonumber\\
&&\hspace{1.5cm}=
\sum\limits_{i=\{8,9,10\}}C^{(i)}(\boldsymbol{{\beta}})\Phi_{_{\{1,2,3\}}}^{(i)}({\boldsymbol{\beta}},\;\boldsymbol{{\xi}})
\nonumber\\
&&\hspace{1.5cm}=
\sum\limits_{i=\{10,11,12\}}C^{(i)}(\boldsymbol{{\beta}})\Phi_{_{\{1,2,3\}}}^{(i)}({\boldsymbol{\beta}},\;\boldsymbol{{\xi}})\;.
\label{sec2-3}
\end{eqnarray}

In order to obtain the expressions in the whole domain of definition, we present the fundamental solution systems
under all possible affine spanning ${\cal B}$. We put those concrete expressions for $\Phi_{_{\cal B}}^{(i)},\;i\in\{1,\cdots,12\}$
for concision in the appendix \ref{app1}. Except for the constant factors and power functions, the corresponding
hypergeometric functions are written as
\begin{eqnarray}
&&\varphi_{_{\{1,2,4\}}}^{(i)}({\boldsymbol{\beta}},\;x_{_1},\;x_{_2})=\varphi_{_{\{1,2,3\}}}^{(i)}(\widehat{(34)}{\boldsymbol{\beta}},\;x_{_1},\;x_{_2})
\;,\nonumber\\
&&\varphi_{_{\{1,2,5\}}}^{(i)}({\boldsymbol{\beta}},\;x_{_1},\;x_{_2})=\varphi_{_{\{1,2,3\}}}^{(i)}(\widehat{(354)}{\boldsymbol{\beta}},\;x_{_1},\;x_{_2})
\;,\nonumber\\
&&\varphi_{_{\{1,3,4\}}}^{(i)}({\boldsymbol{\beta}},\;x_{_1},\;x_{_2})=\varphi_{_{\{1,2,3\}}}^{(i)}(\widehat{(234)}{\boldsymbol{\beta}},\;x_{_1},\;x_{_2})
\;,\nonumber\\
&&\varphi_{_{\{1,3,5\}}}^{(i)}({\boldsymbol{\beta}},\;x_{_1},\;x_{_2})=\varphi_{_{\{1,2,3\}}}^{(i)}(\widehat{(2354)}{\boldsymbol{\beta}},\;x_{_1},\;x_{_2})
\;,\nonumber\\
&&\varphi_{_{\{2,3,4\}}}^{(i)}({\boldsymbol{\beta}},\;x_{_1},\;x_{_2})=\varphi_{_{\{1,2,3\}}}^{(i)}(\widehat{(1234)}{\boldsymbol{\beta}},\;x_{_1},\;x_{_2})
\;,\nonumber\\
&&\varphi_{_{\{2,3,5\}}}^{(i)}({\boldsymbol{\beta}},\;x_{_1},\;x_{_2})=\varphi_{_{\{1,2,3\}}}^{(i)}(\widehat{(12354)}{\boldsymbol{\beta}},\;x_{_1},\;x_{_2})
\;,\nonumber\\
&&\varphi_{_{\{1,4,5\}}}^{(i)}({\boldsymbol{\beta}},\;x_{_1},\;x_{_2})=\varphi_{_{\{1,2,3\}}}^{(i)}(\widehat{(24)}\widehat{(35)}{\boldsymbol{\beta}},\;x_{_1},\;x_{_2})
\;,\nonumber\\
&&\varphi_{_{\{2,4,5\}}}^{(i)}({\boldsymbol{\beta}},\;x_{_1},\;x_{_2})=\varphi_{_{\{1,2,3\}}}^{(i)}(\widehat{(124)}\widehat{(35)}{\boldsymbol{\beta}},\;x_{_1},\;x_{_2})
\;,\nonumber\\
&&\varphi_{_{\{3,4,5\}}}^{(i)}({\boldsymbol{\beta}},\;x_{_1},\;x_{_2})=\varphi_{_{\{1,2,3\}}}^{(i)}(\widehat{(13524)}{\boldsymbol{\beta}},\;x_{_1},\;x_{_2})\;,
\label{sec2-4}
\end{eqnarray}
where the symbols $\widehat{(34)},\;\widehat{(354)},\;\cdots$ are the elements of the permutation
group $S_{_5}$ acting on components of the exponent vector ${\boldsymbol{\beta}}\in{\bf C}^5$,
and $\varphi_{_{\cal B}}^{(i)}$, $i=1,3,5,8,10,12$ are the first Appell functions whose geometric configuration
is presented in Fig.\ref{fig1}(a), while $\varphi_{_{\cal B}}^{(j)}$, $j=2,4,6,7,9,11$ are the Horn functions whose geometric configuration
is presented in Fig.\ref{fig1}(b), respectively.

\section{The Inverse Gauss relations\label{sec3}}
\indent\indent
The inverse Gauss relations are obtained through the Mellin-Barnes's contour on the corresponding complex plane.
The Mellin-Barnes representation of the hypergeometric function $\varphi_{_{\{1,2,3\}}}^{(1)}$ is
\begin{eqnarray}
&&{\Gamma(\beta_{_2})\Gamma(\beta_{_1})\Gamma(1-\beta_{_4})\over
\Gamma(2-\beta_{_3}-\beta_{_4})}
\varphi_{_{\{1,2,3\}}}^{(1)}({\boldsymbol{\beta}},\;x_{_1},\;x_{_2})
\nonumber\\
&&\hspace{-0.5cm}=
{1\over(2\pi i)^2}\int_{-i\infty}^{i\infty}
{\Gamma(\beta_{_2}+s_{_1})\Gamma(\beta_{_1}+s_{_2})\Gamma(1-\beta_{_4}+s_{_1}+s_{_2})\over
\Gamma(2-\beta_{_3}-\beta_{_4}+s_{_1}+s_{_2})}
\nonumber\\
&&\hspace{0.0cm}\times
\Gamma(-s_{_1})\Gamma(-s_{_2})(-x_{_1})^{s_{_1}}(-x_{_2})^{s_{_2}}ds_{_1}\bigwedge ds_{_2}\;.
\label{sec3-1}
\end{eqnarray}
Performing the transformation $\beta_{_1}+s_{_2}=-s_{_2}^\prime$ on the complex plane $s_{_2}$, we rewrite the Barnes's contour
integral on the right-hand side of above equation as
\begin{eqnarray}
&&{(-x_{_2})^{-\beta_{_1}}\over(2\pi i)^2}\int_{-i\infty}^{i\infty}
{\Gamma(\beta_{_2}+s_{_1})\Gamma(\beta_{_1}+s_{_2}^\prime)\Gamma(1-\beta_{_1}-\beta_{_4}+s_{_1}-s_{_2}^\prime)\over
\Gamma(\beta_{_2}+\beta_{_5}+s_{_1}-s_{_2}^\prime)}
\nonumber\\
&&\hspace{0.0cm}\times
\Gamma(-s_{_1})\Gamma(-s_{_2}^\prime)(-x_{_1})^{s_{_1}}(-x_{_2})^{-s_{_2}^\prime}ds_{_1}\bigwedge ds_{_2}^\prime
\nonumber\\
&&\hspace{-0.5cm}=
{\Gamma(\beta_{_1})\Gamma(\beta_{_2})\Gamma(1-\beta_{_1}-\beta_{_4})\over\Gamma(\beta_{_2}+\beta_{_5})}
(-x_{_2})^{-\beta_{_1}}\varphi_{_{\{1,2,3\}}}^{(4)}({\boldsymbol{\beta}},\;x_{_1},\;{1\over x_{_2}})\;.
\label{sec3-2}
\end{eqnarray}
Under the affine transformation $1-\beta_{_4}+s_{_1}+s_{_2}=-s_{_2}^\prime$ on the complex plane $s_{_2}$, the Barnes's contour integral on
the right-hand side of Eq.(\ref{sec3-1}) can be written as
\begin{eqnarray}
&&{(-x_{_2})^{\beta_{_4}-1}\over(2\pi i)^2}\int_{-i\infty}^{i\infty}
{\Gamma(\beta_{_2}+s_{_1})\Gamma(\beta_{_1}+\beta_{_4}-1-s_{_1}-s_{_2}^\prime)\Gamma(1-\beta_{_4}+s_{_1}+s_{_2}^\prime)\over
\Gamma(1-\beta_{_3}-s_{_2}^\prime)}
\nonumber\\
&&\hspace{0.0cm}\times
\Gamma(-s_{_1})\Gamma(-s_{_2}^\prime)(-x_{_1})^{s_{_1}}(-x_{_2})^{-s_{_1}-s_{_2}^\prime}ds_{_1}\bigwedge ds_{_2}^\prime
\nonumber\\
&&\hspace{-0.5cm}=
{\Gamma(\beta_{_1}+\beta_{_4}-1)\Gamma(\beta_{_2})\Gamma(1-\beta_{_4})\over\Gamma(1-\beta_{_3})}(-x_{_2})^{\beta_{_4}-1}
\varphi_{_{\{1,2,3\}}}^{(12)}({\boldsymbol{\beta}},\;{1\over x_{_2}},\;{x_{_1}\over x_{_2}})\;.
\label{sec3-3}
\end{eqnarray}
Then the residue theorem implies the following equation
\begin{eqnarray}
&&\varphi_{_{\{1,2,3\}}}^{(1)}({\boldsymbol{\beta}},\;x_{_1},\;x_{_2})
\nonumber\\
&&\hspace{-0.5cm}=
{\Gamma(1-\beta_{_1}-\beta_{_4})\Gamma(2-\beta_{_3}-\beta_{_4})
\over\Gamma(\beta_{_2}+\beta_{_5})\Gamma(1-\beta_{_4})}(-x_{_2})^{-\beta_{_1}}
\varphi_{_{\{1,2,3\}}}^{(4)}({\boldsymbol{\beta}},\;x_{_1},\;{1\over x_{_2}})
\nonumber\\
&&\hspace{0.0cm}
+{\Gamma(\beta_{_1}+\beta_{_4}-1)\Gamma(2-\beta_{_3}-\beta_{_4})\over\Gamma(\beta_{_1})\Gamma(1-\beta_{_3})}(-x_{_2})^{\beta_{_4}-1}
\varphi_{_{\{1,2,3\}}}^{(12)}({\boldsymbol{\beta}},\;{1\over x_{_2}},\;{x_{_1}\over x_{_2}})\;.
\label{sec3-3a}
\end{eqnarray}
Taking the affine transformation $\beta_{_2}+s_{_1}=-s_{_1}^\prime$ on the complex plane $s_{_1}$,
we formulate accordingly the Barnes's contour integral on the right-hand side of Eq.(\ref{sec3-1}) as
\begin{eqnarray}
&&{(-x_{_1})^{-\beta_{_2}}\over(2\pi i)^2}\int_{-i\infty}^{i\infty}
{\Gamma(\beta_{_2}+s_{_1}^\prime)\Gamma(\beta_{_1}+s_{_2})\Gamma(1-\beta_{_2}-\beta_{_4}-s_{_1}^\prime+s_{_2})\over
\Gamma(\beta_{_1}+\beta_{_5}-s_{_1}^\prime+s_{_2})}
\nonumber\\
&&\hspace{0.0cm}\times
\Gamma(-s_{_1}^\prime)\Gamma(-s_{_2})(-x_{_1})^{-s_{_1}^\prime}(-x_{_2})^{s_{_2}}ds_{_1}^\prime\bigwedge ds_{_2}
\nonumber\\
&&\hspace{-0.5cm}=
{\Gamma(\beta_{_1})\Gamma(\beta_{_2})\Gamma(1-\beta_{_2}-\beta_{_4})\over\Gamma(\beta_{_1}+\beta_{_5})}
(-x_{_1})^{-\beta_{_2}}\varphi_{_{\{1,2,3\}}}^{(7)}({\boldsymbol{\beta}},\;{1\over x_{_1}},\;x_{_2})\;.
\label{sec3-4}
\end{eqnarray}
Under the affine transformation $1-\beta_{_4}+s_{_1}+s_{_2}=-s_{_1}^\prime$ on the complex plane $s_{_1}$,
the Barnes's contour integral on the right-hand side of Eq.(\ref{sec3-1}) is rewritten accordingly as
\begin{eqnarray}
&&{(-x_{_1})^{\beta_{_4}-1}\over(2\pi i)^2}\int_{-i\infty}^{i\infty}
{\Gamma(\beta_{_1}+s_{_2})\Gamma(\beta_{_2}+\beta_{_4}-1-s_{_1}^\prime-s_{_2})\Gamma(1-\beta_{_4}+s_{_1}^\prime+s_{_2})\over
\Gamma(1-\beta_{_3}-s_{_1}^\prime)}
\nonumber\\
&&\hspace{0.0cm}\times
\Gamma(-s_{_1}^\prime)\Gamma(-s_{_2})(-x_{_1})^{-s_{_1}^\prime-s_{_2}}(-x_{_2})^{s_{_2}}ds_{_1}^\prime\bigwedge ds_{_2}
\nonumber\\
&&\hspace{-0.5cm}=
{\Gamma(\beta_{_1})\Gamma(\beta_{_2}+\beta_{_4}-1)\Gamma(1-\beta_{_4})\over\Gamma(1-\beta_{_3})}(-x_{_1})^{\beta_{_4}-1}
\varphi_{_{\{1,2,3\}}}^{(8)}({\boldsymbol{\beta}},\;{1\over x_{_1}},\;{x_{_2}\over x_{_1}})\;.
\label{sec3-5}
\end{eqnarray}
Similarly the residue theorem indicates the following equation
\begin{eqnarray}
&&\varphi_{_{\{1,2,3\}}}^{(1)}({\boldsymbol{\beta}},\;x_{_1},\;x_{_2})
\nonumber\\
&&\hspace{-0.5cm}=
{\Gamma(1-\beta_{_2}-\beta_{_4})\Gamma(2-\beta_{_3}-\beta_{_4})
\over\Gamma(\beta_{_1}+\beta_{_5})\Gamma(1-\beta_{_4})}
(-x_{_1})^{-\beta_{_2}}\varphi_{_{\{1,2,3\}}}^{(7)}({\boldsymbol{\beta}},\;{1\over x_{_1}},\;x_{_2})
\nonumber\\
&&\hspace{0.0cm}
+{\Gamma(\beta_{_2}+\beta_{_4}-1)\Gamma(2-\beta_{_3}-\beta_{_4})\over\Gamma(\beta_{_2})\Gamma(1-\beta_{_3})}(-x_{_1})^{\beta_{_4}-1}
\varphi_{_{\{1,2,3\}}}^{(8)}({\boldsymbol{\beta}},\;{1\over x_{_1}},\;{x_{_2}\over x_{_1}})\;.
\label{sec3-6}
\end{eqnarray}
The equations in (\ref{sec3-3a}) and (\ref{sec3-6}) constitute the inverse Gauss relations originating from the inverse
transformations of the variables in the function $\varphi_{_{\{1,2,3\}}}^{(1)}$. The results presented here coincide with
the analytic continuations in the Refs.~\cite{Ananthanarayan2022,Ananthanarayan2019,Bera2023,Bera2024,Olsson1964} which are
derived from the transformation theory of low variable hypergeometric functions to find the transformation formulae of
higher variable hypergeometric functions. The method presented here generalizes the approach adopted in the work~\cite{Barnes1907},
and can be used to analyze the continuations of any generalized hypergeometric functions.

Similarly we obtain the analytic continuations of other generalized hypergeometric functions $\varphi_{_{\{1,2,3\}}}^{(i)}$,
$i=2,\cdots,12$. To shorten the length of text, we put the expressions in the appendix~\ref{app2}.
For the generalized hypergeometric functions under other affine spanning, the corresponding inverse Gauss relations are obtained from that
of the affine spanning ${\cal B}=\{1,2,3\}$ through some permutations on components of the exponent vector
${\boldsymbol{\beta}}$.
Using the well-known equation $1/(a)_{_{-n}}=(-1)^n(1-a)_{_n}$ and $\sum\beta_{_i}=2$, we have
\begin{eqnarray}
&&\varphi_{_{\{1,2,3\}}}^{(1)}({\boldsymbol{\beta}},\;0,\;x)=
\;_2F_1\left(\left.\begin{array}{c}\beta_{_1},1-\beta_{_4}\\2-\beta_{_3}-\beta_{_4}\end{array}\right| x\right)
\;,\nonumber\\
&&\varphi_{_{\{1,2,3\}}}^{(4)}({\boldsymbol{\beta}},\;0,\;{1\over x})=
\;_2F_1\left(\left.\begin{array}{c}\beta_{_1},1-\beta_{_5}\\ \beta_{_1}+\beta_{_4}\end{array}\right| {1\over x}\right)
\;,\nonumber\\
&&\varphi_{_{\{1,2,3\}}}^{(12)}({\boldsymbol{\beta}},\;0,\;{1\over x})=
\;_2F_1\left(\left.\begin{array}{c}\beta_{_3},1-\beta_{_4}\\ 2-\beta_{_1}-\beta_{_4}\end{array}\right| {1\over x}\right)\;.
\label{sec3-7}
\end{eqnarray}
Thus the analytic continuation of Eq.(\ref{sec3-3a}) recovers the equation (\ref{Intro2}).

The inverse Gauss relations satisfy the idempotent property, i.e. the image of a hypergeometric function under
the composed map of two consecutive inverse transformations of certain variable is itself. Using the analytic continuation in Eq.(\ref{sec3-3a})
and the relevant analytic continuations in the appendix \ref{app2}
\begin{eqnarray}
&&\varphi_{_{\{1,2,3\}}}^{(4)}({\boldsymbol{\beta}},\;x_{_1},\;x_{_2})
\nonumber\\
&&\hspace{-0.5cm}=
{\Gamma(\beta_{_3}+\beta_{_4}-1)\Gamma(\beta_{_1}+\beta_{_4})\over\Gamma(1-\beta_{_2}-\beta_{_5})\Gamma(\beta_{_4})}
(-x_{_2})^{-\beta_{_1}}\varphi_{_{\{1,2,3\}}}^{(1)}({\boldsymbol{\beta}},\;x_{_1},\;{1\over x_{_2}})
\nonumber\\
&&\hspace{0.0cm}
+{\Gamma(\beta_{_1}+\beta_{_4})\Gamma(1-\beta_{_3}-\beta_{_4})\over\Gamma(\beta_{_1})\Gamma(1-\beta_{_3})}(-x_{_2})^{\beta_{_2}+\beta_{_5}-1}
\varphi_{_{\{1,2,3\}}}^{(6)}({\boldsymbol{\beta}},\;{1\over x_{_2}},\;x_{_1}x_{_2})
\;,\nonumber\\
&&\varphi_{_{\{1,2,3\}}}^{(12)}({\boldsymbol{\beta}},\;x_{_1},\;x_{_2})
\nonumber\\
&&\hspace{-0.5cm}=
{\Gamma(2-\beta_{_1}-\beta_{_4})\Gamma(1-\beta_{_3}-\beta_{_4})\over\Gamma(\beta_{_2}+\beta_{_5})\Gamma(1-\beta_{_4})}
(-x_{_1})^{-\beta_{_3}}\varphi_{_{\{1,2,3\}}}^{(6)}({\boldsymbol{\beta}},\;{1\over x_{_1}},\;x_{_2})
\nonumber\\
&&\hspace{0.0cm}
+{\Gamma(\beta_{_3}+\beta_{_4}-1)\Gamma(2-\beta_{_1}-\beta_{_4})\over\Gamma(\beta_{_3})\Gamma(1-\beta_{_1})}(-x_{_1})^{\beta_{_4}-1}
\varphi_{_{\{1,2,3\}}}^{(1)}({\boldsymbol{\beta}},\;{x_{_2}\over x_{_1}},\;{1\over x_{_1}})\;,
\label{Idempotent-1-1}
\end{eqnarray}
we have
\begin{eqnarray}
&&\varphi_{_{\{1,2,3\}}}^{(1)}({\boldsymbol{\beta}},\;x_{_1},\;x_{_2})
\nonumber\\
&&\hspace{-0.5cm}=
\Big\{{\Gamma(1-\beta_{_1}-\beta_{_4})\Gamma(\beta_{_1}+\beta_{_4})\Gamma(\beta_{_3}+\beta_{_4}-1)\Gamma(2-\beta_{_3}-\beta_{_4})\over
\Gamma(\beta_{_2}+\beta_{_5})\Gamma(1-\beta_{_2}-\beta_{_5})\Gamma(\beta_{_4})\Gamma(1-\beta_{_4})}
\nonumber\\
&&\hspace{0.0cm}
+{\Gamma(\beta_{_1}+\beta_{_4}-1)\Gamma(2-\beta_{_1}-\beta_{_4})\Gamma(\beta_{_3}+\beta_{_4}-1)\Gamma(2-\beta_{_3}-\beta_{_4})\over
\Gamma(\beta_{_1})\Gamma(1-\beta_{_1})\Gamma(\beta_{_3})\Gamma(1-\beta_{_3})}\Big\}
\nonumber\\
&&\hspace{0.0cm}\times
\varphi_{_{\{1,2,3\}}}^{(1)}({\boldsymbol{\beta}},\;x_{_1},\;x_{_2})
\nonumber\\
&&\hspace{0.0cm}
+\Big\{{\Gamma(1-\beta_{_1}-\beta_{_4})\Gamma(\beta_{_1}+\beta_{_4})\Gamma(\beta_{_3}+\beta_{_4})\Gamma(1-\beta_{_3}-\beta_{_4})\over\Gamma(\beta_{_3})\Gamma(1-\beta_{_3})}
\nonumber\\
&&\hspace{0.0cm}
+{\Gamma(\beta_{_1}+\beta_{_4}-1)\Gamma(2-\beta_{_1}-\beta_{_4})\Gamma(1-\beta_{_3}-\beta_{_4})\Gamma(\beta_{_3}+\beta_{_4})\over\Gamma(\beta_{_3})\Gamma(1-\beta_{_3})}\Big\}
\nonumber\\
&&\hspace{0.0cm}\times
(-x_{_2})^{\beta_{_3}+\beta_{_4}-1}\varphi_{_{\{1,2,3\}}}^{(6)}({\boldsymbol{\beta}},\;x_{_2},\;{x_{_1}\over x_{_2}})
\nonumber\\
&&\hspace{-0.5cm}=\varphi_{_{\{1,2,3\}}}^{(1)}({\boldsymbol{\beta}},\;x_{_1},\;x_{_2})\;.
\label{Idempotent-1-2}
\end{eqnarray}
The combinations of the inverse Gauss relations and the Gauss-Kummer relations presented in section~\ref{sec5}
generalize the analytic continuations of the Gauss functions presented in (1.8.1.11)-(1.8.1.19) of Ref.~\cite{Slater1966}.
Furthermore, the images of the generalized hypergeometric functions under the inverse transformation of certain variable
are the linear combinations of the generalized hypergeometric solutions of the GKZ-system in the same affine spanning.

\section{The adjacent relations\label{sec4}}
\indent\indent
The adjacent relations of the hypergeometric functions in the (n-1)-dimensional projective space are determined from the quotient module of the free module
${\bf C}^n$ by the submodule which is generated by the coefficient vectors of the corresponding GKZ-system~\cite{Gelfand1987,Gelfand1992}. For the GKZ-system on the Grassmannian,
the adjacent relations of the hypergeometric functions are determined by $G_{_{k,n}}$ and its dual $G_{_{k,n}}^{\bot}$~\cite{Gelfand1987,Gelfand1986-2}.
More precisely, if the exponent vector ${\boldsymbol{\beta}}\in{\bf C}^n$ satisfies $|{\boldsymbol{\beta}}|=\sum\limits_{i=1}^n\beta_{_i}=n-k$,
the adjacent relations can be written respectively as followings.
\begin{itemize}
\item If ${\boldsymbol{\gamma}}={\boldsymbol{\beta}}+{\bf e}_{_i}\in{\bf C}^n,\;i\in\{1,\cdots,n\}$,
and $\sum\limits_{i=1}^na_{_i}{\bf e}_{_i}\in G_{_{k,n}}$, then we have
\begin{eqnarray}
&&\sum\limits_{j=1}^n a_{_j}(\gamma_{_j}-1)\Phi_{_{\bf{\cal B}}}({\boldsymbol{\gamma}}-{\bf e}_{_j},\;{\boldsymbol{\xi}})=0\;.
\label{Gauss-a}
\end{eqnarray}
Where ${\bf{\cal B}}\subset\{1,\cdots,n\}$ is the proper subset of the columns which spans the k-dimensional vector subspace in the ${\bf C}^n$,
and $\{{\bf e}_{_i},\;i=1,\cdots,n\}$ denotes the standard basis of ${\bf C}^n$.
\item If ${\boldsymbol{\gamma}}={\boldsymbol{\beta}}-{\bf e}_{_i}\in{\bf C}^n,\;i\in\{1,\cdots,n\}$,
and $\sum\limits_{i=1}^na_{_i}{\bf e}_{_i}\in G_{_{k,n}}^\bot$, then we have
\begin{eqnarray}
&&\sum\limits_{j=1}^n a_{_j}\Phi_{_{\bf{\cal B}}}({\boldsymbol{\gamma}}+{\bf e}_{_j},\;{\boldsymbol{\xi}})=0\;.
\label{Gauss-b}
\end{eqnarray}
\end{itemize}
There are totally $n$ independent adjacent relations from above two equations. Since the Gauss functions can be regarded as functions
on the Grassmannian $G_{_{2,4}}$, there are four independent adjacent relations in Eq.(\ref{Intro3}) for the Gauss functions.

In order to continue our discussion, we assume that the exponent vector
${\boldsymbol{\beta}}=(\beta_{_1},\cdots,\beta_{_5})\in{\bf C}^5$ satisfies $\sum\limits_{i=1}^5\beta_{_i}=2$,
and ${\bf{\cal B}}=\{1,\;2,\;3\}$.
The dual variety of the Grassmannian $\boldsymbol{{\xi}}$ in Eq.(\ref{Intro5}) is given by the matroid~\cite{Oxley2011}
\begin{eqnarray}
&&\boldsymbol{{\xi}}_{\bot}=\left(\begin{array}{ccccc}\;-1\;&\;-1\;&\;-1\;&\;1\;&\;0\;\\
\;-r_{_1}\;&\;-r_{_2}\;&\;-r_{_3}\;&\;0\;&\;1\;\end{array}\right)\;.
\label{Gauss-c}
\end{eqnarray}
Corresponding to ${\boldsymbol{\beta}}+{\bf e}_{_1}=(1+\beta_{_1},\;\beta_{_2},\;\cdots,\beta_{_5})$, we obtain three independent
relations among $\Phi_{_{\{1,2,3\}}}^{(i)},\;i\in\{1,\cdots,12\}$ with contiguous parameters from Eq.(\ref{Gauss-a}) as
\begin{eqnarray}
&&\beta_{_1}\Phi_{_{\{1,2,3\}}}^{(i)}({\boldsymbol{\beta}},\;\boldsymbol{{\xi}})
+(\beta_{_4}-1)\Phi_{_{\{1,2,3\}}}^{(i)}({\boldsymbol{\beta}}+{\bf e}_{_1}-{\bf e}_{_4},\;\boldsymbol{{\xi}})
\nonumber\\
&&+(\beta_{_5}-1)r_{_1}\Phi_{_{\{1,2,3\}}}^{(i)}({\boldsymbol{\beta}}+{\bf e}_{_1}-{\bf e}_{_5},\;\boldsymbol{{\xi}})\equiv0
\;,\nonumber\\
&&\beta_{_2}\Phi_{_{\{1,2,3\}}}^{(i)}({\boldsymbol{\beta}},\;\boldsymbol{{\xi}})
+(\beta_{_4}-1)\Phi_{_{\{1,2,3\}}}^{(i)}({\boldsymbol{\beta}}+{\bf e}_{_2}-{\bf e}_{_4},\;\boldsymbol{{\xi}})
\nonumber\\
&&+(\beta_{_5}-1)r_{_2}\Phi_{_{\{1,2,3\}}}^{(i)}({\boldsymbol{\beta}}+{\bf e}_{_2}-{\bf e}_{_5},\;\boldsymbol{{\xi}})\equiv0
\;,\nonumber\\
&&\beta_{_3}\Phi_{_{\{1,2,3\}}}^{(i)}({\boldsymbol{\beta}},\;\boldsymbol{{\xi}})
+(\beta_{_4}-1)\Phi_{_{\{1,2,3\}}}^{(i)}({\boldsymbol{\beta}}+{\bf e}_{_3}-{\bf e}_{_4},\;\boldsymbol{{\xi}})
\nonumber\\
&&+(\beta_{_5}-1)r_{_3}\Phi_{_{\{1,2,3\}}}^{(i)}({\boldsymbol{\beta}}+{\bf e}_{_3}-{\bf e}_{_5},\;\boldsymbol{{\xi}})\equiv0\;.
\label{Gauss-d}
\end{eqnarray}
Corresponding to ${\boldsymbol{\beta}}-{\bf e}_{_1}=(\beta_{_1}-1,\;\beta_{_2},\;\cdots,\beta_{_5})$, we obtain two independent
relations among $\Phi_{_{\{1,2,3\}}}^{(i)}$ with contiguous parameters from Eq.(\ref{Gauss-b}) as
\begin{eqnarray}
&&\Phi_{_{\{1,2,3\}}}^{(i)}({\boldsymbol{\beta}},\;\boldsymbol{{\xi}})
+\Phi_{_{\{1,2,3\}}}^{(i)}({\boldsymbol{\beta}}-{\bf e}_{_1}+{\bf e}_{_2},\;\boldsymbol{{\xi}})
\nonumber\\
&&+\Phi_{_{\{1,2,3\}}}^{(i)}({\boldsymbol{\beta}}-{\bf e}_{_1}+{\bf e}_{_3},\;\boldsymbol{{\xi}})
-\Phi_{_{\{1,2,3\}}}^{(i)}({\boldsymbol{\beta}}-{\bf e}_{_1}+{\bf e}_{_4},\;\boldsymbol{{\xi}})\equiv0
\;,\nonumber\\
&&r_{_1}\Phi_{_{\{1,2,3\}}}^{(i)}({\boldsymbol{\beta}},\;\boldsymbol{{\xi}})
+r_{_2}\Phi_{_{\{1,2,3\}}}^{(i)}({\boldsymbol{\beta}}-{\bf e}_{_1}+{\bf e}_{_2},\;\boldsymbol{{\xi}})
\nonumber\\
&&+r_{_3}\Phi_{_{\{1,2,3\}}}^{(i)}({\boldsymbol{\beta}}-{\bf e}_{_1}+{\bf e}_{_3},\;\boldsymbol{{\xi}})
-\Phi_{_{\{1,2,3\}}}^{(i)}({\boldsymbol{\beta}}-{\bf e}_{_1}+{\bf e}_{_5},\;\boldsymbol{{\xi}})\equiv0\;.
\label{Gauss-e}
\end{eqnarray}

For $\Phi_{_{\{1,2,3\}}}^{(1)}$, the adjacent relations of the generalized hypergeometric function
$\varphi_{_{\{1,2,3\}}}^{(1)}$ can be derived from equations (\ref{Gauss-d}), (\ref{Gauss-e})
\begin{eqnarray}
&&(2-\beta_{_3}-\beta_{_4})\Big[\varphi_{_{\{1,2,3\}}}^{(1)}({\boldsymbol{\beta}})
-\varphi_{_{\{1,2,3\}}}^{(1)}({\boldsymbol{\beta}}+{\bf e}_{_1}-{\bf e}_{_5})\Big]
\nonumber\\
&&\hspace{0.0cm}
+(1-\beta_{_4})x_{_2}\varphi_{_{\{1,2,3\}}}^{(1)}({\boldsymbol{\beta}}+{\bf e}_{_1}-{\bf e}_{_4})
\equiv0
\;,\nonumber\\
&&(2-\beta_{_3}-\beta_{_4})\Big[\varphi_{_{\{1,2,3\}}}^{(1)}({\boldsymbol{\beta}})
-\varphi_{_{\{1,2,3\}}}^{(1)}({\boldsymbol{\beta}}+{\bf e}_{_2}-{\bf e}_{_5})\Big]
\nonumber\\
&&\hspace{0.0cm}
+(1-\beta_{_4})x_{_1}\varphi_{_{\{1,2,3\}}}^{(1)}({\boldsymbol{\beta}}+{\bf e}_{_2}-{\bf e}_{_4})\equiv0
\;,\nonumber\\
&&\beta_{_3}\varphi_{_{\{1,2,3\}}}^{(1)}({\boldsymbol{\beta}})
+(\beta_{_4}-1)\varphi_{_{\{1,2,3\}}}^{(1)}({\boldsymbol{\beta}}+{\bf e}_{_3}-{\bf e}_{_4})
\nonumber\\
&&\hspace{0.0cm}
+(1-\beta_{_3}-\beta_{_4})\varphi_{_{\{1,2,3\}}}^{(1)}({\boldsymbol{\beta}}+{\bf e}_{_3}-{\bf e}_{_5})\equiv0
\;,\nonumber\\
&&(\beta_{_1}-1)x_{_2}\varphi_{_{\{1,2,3\}}}^{(1)}({\boldsymbol{\beta}})
+\beta_{_2}x_{_1}\varphi_{_{\{1,2,3\}}}^{(1)}({\boldsymbol{\beta}}-{\bf e}_{_1}+{\bf e}_{_2})
\nonumber\\
&&+(\beta_{_3}+\beta_{_4}-1)\Big[\varphi_{_{\{1,2,3\}}}^{(1)}({\boldsymbol{\beta}}-{\bf e}_{_1}+{\bf e}_{_3})
-\varphi_{_{\{1,2,3\}}}^{(1)}({\boldsymbol{\beta}}-{\bf e}_{_1}+{\bf e}_{_4})\Big]\equiv0
\;,\nonumber\\
&&(\beta_{_1}-1)\varphi_{_{\{1,2,3\}}}^{(1)}({\boldsymbol{\beta}})
+\beta_{_2}\varphi_{_{\{1,2,3\}}}^{(1)}({\boldsymbol{\beta}}-{\bf e}_{_1}+{\bf e}_{_2})
\nonumber\\
&&+(\beta_{_3}+\beta_{_4}-1)\varphi_{_{\{1,2,3\}}}^{(1)}({\boldsymbol{\beta}}-{\bf e}_{_1}+{\bf e}_{_3})
+\beta_{_5}\varphi_{_{\{1,2,3\}}}^{(1)}({\boldsymbol{\beta}}-{\bf e}_{_1}+{\bf e}_{_5})\equiv0\;,
\label{Gauss-adjacent1-1}
\end{eqnarray}
where the variables $x_{_{1,2}}$ in the hypergeometric function $\varphi_{_{\{1,2,3\}}}^{(1)}$ have been omitted for simplicity.
Thus the first adjacent relation is reduced as
\begin{eqnarray}
&&F_{_1}\left(\begin{array}{c}\;1-\beta_{_4},\;\beta_{_2},\;\beta_{_1}\;\\2-\beta_{_3}-\beta_{_4}\end{array}\right)
+{(1-\beta_{_4})x_{_2}\over 2-\beta_{_3}-\beta_{_4}}F_{_1}\left(\begin{array}{c}\;1-\beta_{_4},\;\beta_{_2},\;1+\beta_{_1}\;\\3-\beta_{_3}-\beta_{_4}\end{array}\right)
\nonumber\\
&&-F_{_1}\left(\begin{array}{c}\;1-\beta_{_4},\;\beta_{_2},\;1+\beta_{_1}\;\\2-\beta_{_3}-\beta_{_4}\end{array}\right)\equiv0\;.
\label{Gauss-adjacent1-3}
\end{eqnarray}
When $x_{_2}=0$, this relation is simplified to the trivial equation
\begin{eqnarray}
&&-\;_{_2}F_{_1}\left(\left.\begin{array}{c}1-\beta_{_4},\;\beta_{_2}\\ 2-\beta_{_3}-\beta_{_4}\end{array}\right|\;x_{_1}\right)
+\;_{_2}F_{_1}\left(\left.\begin{array}{c}1-\beta_{_4},\;\beta_{_2}\\ 2-\beta_{_3}-\beta_{_4}\end{array}\right|\;x_{_1}\right)=0\;.
\label{Gauss-adjacent1-4}
\end{eqnarray}
When $x_{_1}=0$, the adjacent relation in Eq.(\ref{Gauss-adjacent1-3}) is simplified as
\begin{eqnarray}
&&\;_{_2}F_{_1}\left(\left.\begin{array}{c}1-\beta_{_4},\;\beta_{_1}\\ 2-\beta_{_3}-\beta_{_4}\end{array}\right|\;x_{_2}\right)
+{(1-\beta_{_4})x_{_2}\over2-\beta_{_3}-\beta_{_4}}\;_{_2}F_{_1}\left(\left.\begin{array}{c}1-\beta_{_4},\;1+\beta_{_1}\\
3-\beta_{_3}-\beta_{_4}\end{array}\right|\;x_{_2}\right)
\nonumber\\
&&-\;_{_2}F_{_1}\left(\left.\begin{array}{c}1-\beta_{_4},\;1+\beta_{_1}\\ 2-\beta_{_3}-\beta_{_4}\end{array}\right|\;x_{_2}\right)\equiv0\;,
\label{Gauss-adjacent1-5}
\end{eqnarray}
which coincides with the first equation in Eq.(\ref{Intro3}).

Similarly we can verify that other adjacent relations in Eq.(\ref{Gauss-adjacent1-1}) recover the well-known adjacent relations
of the Gauss functions as $x_{_1}=0$ or $x_{_2}=0$, respectively.
To shorten the length of text, we put the adjacent relations of the generalized hypergeometric functions
$\varphi_{_{\{1,2,3\}}}^{(i)}({\boldsymbol{\beta}})$, $i=2,\cdots,12$ in the appendix \ref{app3}.
The adjacent relations of other affine spanning ${\cal B}$ are obtained from the adjacent relations
in $\varphi_{_{\{1,2,3\}}}^{(i)}({\boldsymbol{\beta}})$ through some permutation on components of the
exponent vector ${\boldsymbol{\beta}}\in{\bf C}^5$. As mentioned before, the adjacent relations are adopted to
formulate the coefficient of the term with power of $\varepsilon=2-D/2$ in Feynman integral expansion
as a linear combination of generalized hypergeometric functions with integer parameters~\cite{Davydychev1993NPB}.
In addition, those adjacent relations account for the equivalence between the different fundamental solution
systems of a Feynman integral.

\section{Gauss-Kummer relations\label{sec5}}
\indent\indent
When the hypergeometric functions with the same geometric representation are equal to each other on
the principal branch, we can obtain the generalized Gauss relations originating from Kummer's classification.
Corresponding to the geometric representation shown in Fig.\ref{fig1}(a) with $\{a,b\}=\{3,5\}$, $\{c,d,e\}=\{1,2,4\}$,
we derive the following six solutions of the GKZ-system presented in Eq.(\ref{sec2-2}) which
are proportional to each other in the intersection of their convergent regions,
\begin{eqnarray}
&&\Phi_{_{\{1,2,3\}}}^{(10)}\sim\Phi_{_{\{1,2,5\}}}^{(1)}\sim\Phi_{_{\{1,3,4\}}}^{(5)}
\sim\Phi_{_{\{1,4,5\}}}^{(10)}\sim\Phi_{_{\{2,4,5\}}}^{(10)}\sim\Phi_{_{\{2,3,4\}}}^{(5)}\;,
\label{sec4-1}
\end{eqnarray}
where the exponent vector ${\boldsymbol{\beta}}$ and the local splitting coordinates
${\boldsymbol{\xi}}$ have been suppressed.
When the products of power factors and the hypergeometric functions in these solutions are equal to each other on the concrete
principal branch, we obtain the generalized Gauss-Kummer relations as
\begin{eqnarray}
&&\varphi_{_{\{1,2,3\}}}^{(10)}({\boldsymbol{\beta}},\;x,\;y)
\nonumber\\
&&\hspace{-0.5cm}=
(1-x)^{-\beta_{_1}}(1-y)^{-\beta_{_2}}
\varphi_{_{\{1,2,5\}}}^{(1)}({\boldsymbol{\beta}},\;{y\over y-1},\;{x\over x-1})
\nonumber\\
&&\hspace{-0.5cm}=
(1-y)^{\beta_{_5}-1}
\varphi_{_{\{1,3,4\}}}^{(5)}({\boldsymbol{\beta}},\;{y\over y-1},\;{y-x\over y-1})
\nonumber\\
&&\hspace{-0.5cm}=
(1-x)^{\beta_{_5}-1}
\varphi_{_{\{2,3,4\}}}^{(5)}({\boldsymbol{\beta}},\;{x\over x-1},\;{x-y\over x-1})
\nonumber\\
&&\hspace{-0.5cm}=
(1-x)^{-\beta_{_1}}(1-y)^{1-\beta_{_2}-\beta_{_3}}
\varphi_{_{\{1,4,5\}}}^{(10)}({\boldsymbol{\beta}},\;y,\;{y-x\over 1-x})
\nonumber\\
&&\hspace{-0.5cm}=
(1-x)^{1-\beta_{_1}-\beta_{_3}}(1-y)^{-\beta_{_2}}
\varphi_{_{\{2,4,5\}}}^{(10)}({\boldsymbol{\beta}},\;x,\;{x-y\over 1-y})\;,
\label{sec4-2}
\end{eqnarray}
with $x=r_{_1}/r_{_3},\;y=r_{_2}/r_{_3}$. In order to compare the above equations with the known identities in the literature,
we rewrite those $\varphi$ functions in terms of the first type Appell function $F_{_1}$ as
\begin{eqnarray}
&&\varphi_{_{\{1,2,3\}}}^{(10)}({\boldsymbol{\beta}},\;x,\;y)=F_{_1}\left(\left.\begin{array}{c}1-\beta_{_5},\;\beta_{_1},\;\beta_{_2}\\ 2-\beta_{_3}-\beta_{_5}\end{array}\right|\;x,y\right)
\;,\nonumber\\
&&(1-x)^{-\beta_{_1}}(1-y)^{-\beta_{_2}}\varphi_{_{\{1,2,5\}}}^{(1)}({\boldsymbol{\beta}},\;{y\over y-1},\;{x\over x-1})
\nonumber\\
&&\hspace{-0.5cm}=
(1-x)^{-\beta_{_1}}(1-y)^{-\beta_{_2}}F_{_1}\left(\left.\begin{array}{c}1-\beta_{_3},\;\beta_{_1},\;\beta_{_2}\\2-\beta_{_3}-\beta_{_5}\end{array}\right|\;{x\over x-1},\;{y\over y-1}\right)
\nonumber\\
&&(1-y)^{\beta_{_5}-1}\varphi_{_{\{1,3,4\}}}^{(5)}({\boldsymbol{\beta}},\;{y\over y-1},\;{y-x\over y-1})
\nonumber\\
&&\hspace{-0.5cm}=
(1-y)^{\beta_{_5}-1}F_{_1}\left(\left.\begin{array}{c}1-\beta_{_5},\;\beta_{_1},\;\beta_{_4}\\2-\beta_{_3}-\beta_{_5}
\end{array}\right|\;{y-x\over y-1},\;{y\over y-1}\right)
\;,\nonumber\\
&&(1-x)^{\beta_{_5}-1}
\varphi_{_{\{2,3,4\}}}^{(5)}({\boldsymbol{\beta}},\;{x\over x-1},\;{x-y\over x-1})
\nonumber\\
&&\hspace{-0.5cm}=
(1-x)^{\beta_{_5}-1}F_{_1}\left(\left.\begin{array}{c}1-\beta_{_5},\;\beta_{_4},\;\beta_{_2}\\2-\beta_{_3}-\beta_{_5}
\end{array}\right|\;{x\over x-1},\;{x-y\over x-1}\right)
\;,\nonumber\\
&&(1-x)^{-\beta_{_1}}(1-y)^{1-\beta_{_2}-\beta_{_3}}
\varphi_{_{\{1,4,5\}}}^{(10)}({\boldsymbol{\beta}},\;y,\;{y-x\over 1-x})
\nonumber\\
&&\hspace{-0.5cm}=
(1-x)^{-\beta_{_1}}(1-y)^{1-\beta_{_2}-\beta_{_3}}F_{_1}\left(\left.\begin{array}{c}1-\beta_{_3},\;\beta_{_4},\;\beta_{_1}\\2-\beta_{_3}-\beta_{_5}
\end{array}\right|\;y,\;{y-x\over 1-x}\right)
\;,\nonumber\\
&&(1-x)^{1-\beta_{_1}-\beta_{_3}}(1-y)^{-\beta_{_2}}
\varphi_{_{\{2,4,5\}}}^{(10)}({\boldsymbol{\beta}},\;x,\;{x-y\over 1-y})
\nonumber\\
&&\hspace{-0.5cm}=
(1-x)^{1-\beta_{_1}-\beta_{_3}}(1-y)^{-\beta_{_2}}F_{_1}\left(\left.\begin{array}{c}1-\beta_{_3},\;\beta_{_4},\;\beta_{_2}\\2-\beta_{_3}-\beta_{_5}
\end{array}\right|\;x,\;{x-y\over 1-y}\right)\;.
\label{sec4-2a}
\end{eqnarray}
Inserting the above equations into Eq.(\ref{sec4-2}) and using $\sum\beta_{_i}=2$, we obtain
\begin{eqnarray}
&&F_{_1}\left(\left.\begin{array}{c}\alpha,\;\beta,\;\beta^\prime\\ \gamma\end{array}\right|\;x,y\right)
\nonumber\\
&&\hspace{-0.5cm}=
(1-x)^{-\beta}(1-y)^{-\beta^\prime}F_{_1}\left(\left.\begin{array}{c}\gamma-\alpha,\;\beta,\;\beta^\prime\\ \gamma
\end{array}\right|\;{x\over x-1},\;{y\over y-1}\right)
\nonumber\\
&&\hspace{-0.5cm}=
(1-y)^{-\alpha}F_{_1}\left(\left.\begin{array}{c}\alpha,\;\beta,\;\gamma-\beta-\beta^\prime\\ \gamma
\end{array}\right|\;{y-x\over y-1},\;{y\over y-1}\right)
\nonumber\\
&&\hspace{-0.5cm}=
(1-x)^{-\alpha}F_{_1}\left(\left.\begin{array}{c}\alpha,\;\gamma-\beta-\beta^\prime,\;\beta^\prime\\ \gamma
\end{array}\right|\;{x\over x-1},\;{x-y\over x-1}\right)
\nonumber\\
&&\hspace{-0.5cm}=
(1-x)^{-\beta}(1-y)^{\gamma-\alpha-\beta^\prime}F_{_1}\left(\left.\begin{array}{c}\gamma-\alpha,\;\beta,\;\gamma-\beta-\beta^\prime\\ \gamma
\end{array}\right|\;{y-x\over 1-x},\;y\right)
\nonumber\\
&&\hspace{-0.5cm}=
(1-x)^{\gamma-\alpha-\beta}(1-y)^{-\beta^\prime}F_{_1}\left(\left.\begin{array}{c}\gamma-\alpha,\;\gamma-\beta-\beta^\prime,\;\beta^\prime\\ \gamma
\end{array}\right|\;x,\;{x-y\over 1-y}\right)\;,
\label{sec4-2b}
\end{eqnarray}
where $\alpha=1-\beta_{_5}$, $\beta=\beta_{_1}$, $\beta^\prime=\beta_{_2}$, $\gamma=2-\beta_{_3}-\beta_{_5}$.
The relations coincide with those equations presented in Eq.(1)$\sim$Eq.(5)
in the section 9.4 of Ref.~\cite{Bailey1964}, which are derived from the Euler integral expression of the first type Appell
function. Here we provide some reasonable explanations on those relations from the point of view of combinatorial geometry~\cite{Oxley2011}.

There are totally 60 solutions $\Phi_{_{\cal B}}^{(i)},\;i=1,3,5,8,10,12$ which are
formulated as the first type Appell functions with different parameters. Here ${\cal B}$ denotes an affine
spanning of the vector subspace ${\bf C}^3$. In another group of the first type Appell functions which correspond to
a same geometric representation, the obtained Gauss-Kummer relations are various variants of the equations presented in Eq.(\ref{sec4-2b}).
In order to shorten the length of text, we collect them in the appendix~\ref{app4}.

Corresponding to the geometric representation shown in Fig.\ref{fig1}(b) with $a=2$, $\{b,c\}=\{1,4\}$, and $\{d,e\}=\{3,5\}$,
we have the following four solutions of the GKZ-system presented in Eq.(\ref{sec2-2})  which
are proportional to each other in the intersection of their convergent regions,
\begin{eqnarray}
&&\Phi_{_{\{1,2,3\}}}^{(11)}\sim\Phi_{_{\{1,2,5\}}}^{(2)}\sim\Phi_{_{\{2,3,4\}}}^{(6)}
\sim\Phi_{_{\{2,4,5\}}}^{(9)}\;.
\label{sec4-3}
\end{eqnarray}
If the products of power factors and the hypergeometric functions in these solutions are equal to each other on the
principal branch, we have the generalized Gauss-Kummer relations as
\begin{eqnarray}
&&y^{\beta_{_3}+\beta_{_5}-1}\varphi_{_{\{1,2,3\}}}^{(11)}({\boldsymbol{\beta}},\;y,\;{x\over y})
\nonumber\\
&&\hspace{-0.5cm}=
y^{\beta_{_3}+\beta_{_5}-1}(1-x)^{-\beta_{_1}}(1-y)^{\beta_{_1}+\beta_{_4}-1}
\varphi_{_{\{1,2,5\}}}^{(2)}({\boldsymbol{\beta}},\;{y\over y-1},\;{x(y-1)\over y(x-1)})
\nonumber\\
&&\hspace{-0.5cm}=
(1-x)^{-\beta_{_3}}(y-x)^{\beta_{_3}+\beta_{_5}-1}
\varphi_{_{\{2,3,4\}}}^{(6)}({\boldsymbol{\beta}},\;{x\over x-y},\;{y-x\over 1-x})
\nonumber\\
&&\hspace{-0.5cm}=
(1-x)^{1-\beta_{_1}-\beta_{_3}}(1-y)^{\beta_{_1}+\beta_{_4}-1}(y-x)^{\beta_{_3}+\beta_{_5}-1}
\varphi_{_{\{2,4,5\}}}^{(9)}({\boldsymbol{\beta}},\;{x-y\over 1-y},\;{x(1-y)\over x-y})\;.
\label{sec4-4}
\end{eqnarray}
In order to compare the above equations with the known identities in the literature,
we rewrite the $\varphi$ functions in terms of the Horn function $H_{_1}$ as
\begin{eqnarray}
&&y^{\beta_{_3}+\beta_{_5}-1}\varphi_{_{\{1,2,3\}}}^{(11)}({\boldsymbol{\beta}},\;y,\;{x\over y})
=y^{\beta_{_3}+\beta_{_5}-1}H_{_1}\left(\left.\begin{array}{c}\beta_{_1},\;\beta_{_3}\\
\beta_{_1}+\beta_{_4},\;\beta_{_3}+\beta_{_5}\end{array}\right|\;{x\over y},y\right)
\;,\nonumber\\
&&y^{\beta_{_3}+\beta_{_5}-1}(1-x)^{-\beta_{_1}}(1-y)^{\beta_{_1}+\beta_{_4}-1}
\varphi_{_{\{1,2,5\}}}^{(2)}({\boldsymbol{\beta}},\;{y\over y-1},\;{x(y-1)\over y(x-1)})
\nonumber\\
&&\hspace{-0.5cm}=
y^{\beta_{_3}+\beta_{_5}-1}(1-x)^{-\beta_{_1}}(1-y)^{\beta_{_1}+\beta_{_4}-1}
H_{_1}\left(\left.\begin{array}{c}\beta_{_1},\;\beta_{_5}\\
\beta_{_1}+\beta_{_4},\;\beta_{_3}+\beta_{_5}\end{array}\right|\;{x(y-1)\over y(x-1)},{y\over y-1}\right)
\;,\nonumber\\
&&(1-x)^{-\beta_{_3}}(y-x)^{\beta_{_3}+\beta_{_5}-1}
\varphi_{_{\{2,3,4\}}}^{(6)}({\boldsymbol{\beta}},\;{x\over x-y},\;{y-x\over 1-x})
\nonumber\\
&&\hspace{-0.5cm}=
(1-x)^{-\beta_{_3}}(y-x)^{\beta_{_3}+\beta_{_5}-1}H_{_1}\left(\left.\begin{array}{c}\beta_{_4},\;\beta_{_3}\\
\beta_{_3}+\beta_{_5},\;\beta_{_1}+\beta_{_4}\end{array}\right|\;{x\over x-y},{y-x\over 1-x}\right)
\;,\nonumber\\
&&(1-x)^{1-\beta_{_1}-\beta_{_3}}(1-y)^{\beta_{_1}+\beta_{_4}-1}(y-x)^{\beta_{_3}+\beta_{_5}-1}
\varphi_{_{\{2,4,5\}}}^{(9)}({\boldsymbol{\beta}},\;{x-y\over 1-y},\;{x(1-y)\over x-y})
\nonumber\\
&&\hspace{-0.5cm}=
(1-x)^{1-\beta_{_1}-\beta_{_3}}(1-y)^{\beta_{_1}+\beta_{_4}-1}(y-x)^{\beta_{_3}+\beta_{_5}-1}
H_{_1}\left(\left.\begin{array}{c}\beta_{_5},\;\beta_{_4}\\
\beta_{_3}+\beta_{_5},\;\beta_{_1}+\beta_{_4}\end{array}\right|\;{x-y\over 1-y},{x(1-y)\over x-y}\right)\;.
\label{sec4-4a}
\end{eqnarray}
Replacing $x\rightarrow xy$ and denoting $\alpha=\beta_{_1}$, $\beta=\beta_{_1}+\beta_{_4}$, $\alpha^\prime=\beta_{_3}$,
$\beta^\prime=\beta_{_3}+\beta_{_5}$, we have the following relations through Eq.(\ref{sec4-4})
\begin{eqnarray}
&&H_{_1}\left(\left.\begin{array}{c}\alpha,\;\alpha^\prime\\
\beta,\;\beta^\prime\end{array}\right|\;x,y\right)
\nonumber\\
&&\hspace{-0.5cm}=
(1-xy)^{-\alpha}(1-y)^{\beta-1}
H_{_1}\left(\left.\begin{array}{c}\alpha,\;\beta^\prime-\alpha^\prime\\
\beta,\;\beta^\prime\end{array}\right|\;{x(y-1)\over xy-1},{y\over y-1}\right)
\nonumber\\
&&\hspace{-0.5cm}=
(1-x)^{\beta^\prime-1}(1-xy)^{-\alpha^\prime}
H_{_1}\left(\left.\begin{array}{c}\beta-\alpha,\;\alpha^\prime\\
\beta,\;\beta^\prime\end{array}\right|\;{x\over x-1},{y(x-1)\over xy-1}\right)
\nonumber\\
&&\hspace{-0.5cm}=
(1-x)^{\beta^\prime-1}(1-y)^{\beta-1}(1-xy)^{1-\alpha-\alpha^\prime}
H_{_1}\left(\left.\begin{array}{c}\beta-\alpha,\;\beta^\prime-\alpha^\prime\\
\beta,\;\beta^\prime\end{array}\right|\;{x(y-1)\over 1-x},{y(x-1)\over y-1}\right)\;.
\label{sec4-4b}
\end{eqnarray}
Some relations among the hypergeometric series of two variables are derived through the corresponding
Euler integral expression in Ref.\cite{Erdelyi1950}. The first relation coincides with the equation presented
in Eq.(19), the second relation coincides with the equation presented in Eq.(18), and the third relation coincides with the
equation presented in Eq.(20) of Ref.\cite{Erdelyi1950}, respectively.

There are totally 60 solutions $\Phi_{_{\cal B}}^{(i)},\;i=2,4,6,7,9,11$ which are written in terms of
the Horn series with different parameters. Here ${\cal B}$ denotes an affine
spanning of the vector subspace ${\bf C}^3$. In another group of the Horn series which correspond to
a same geometric representation, the obtained Gauss-Kummer relations are various variants of the equations presented in Eq.(\ref{sec4-4b}).
In order to shorten the length of text, we collect them in the appendix~\ref{app4}.

\section{The expressions of the 1-loop self-energy\label{sec6}}
\indent\indent
In $\alpha$-parametrization, the Feynman integral of one-loop self-energy is
\begin{eqnarray}
&&iA_{_{1SE}}(p^2,m_{_1}^2,m_{_2}^2)=
-\Big(\Lambda_{_{\rm RE}}^2\Big)^{2-D/2}\int_0^\infty d\alpha_{_1}d\alpha_{_2}\int{d^Dq\over(2\pi)^D}
\exp\Big\{i\Big[\alpha_{_1}(q^2-m_{_1}^2)
\nonumber\\
&&\hspace{3.5cm}
+\alpha_{_2}((q+p)^2-m_{_2}^2)\Big]\Big\}
\nonumber\\
&&\hspace{3.0cm}=
{i^{2-D/2}\exp\Big\{{i\pi(2-D)\over4}\Big\}\Gamma(2-D/2)
\Big(\Lambda_{_{\rm RE}}^2\Big)^{2-D/2}\over(4\pi)^{D/2}}
\nonumber\\
&&\hspace{3.5cm}\times
\int_S \omega_3(t)\delta(t_{_1}t_{_2}+t_{_1}t_{_3}+t_{_2}t_{_3})
(t_{_1}t_{_2})^{1-D/2}t_{_3}^{D/2-1}
\nonumber\\
&&\hspace{3.5cm}\times
\Big[t_{_1}m_{_1}^2+t_{_2}m_{_2}^2+t_{_3}p^2\Big]^{D/2-2}\;,
\label{1SE-2-1}
\end{eqnarray}
where the hyperplane $S$ is given by the equation $t_{_3}+1=0$, and $\omega_3(t)$ is the volume element in the projective plane $P^2$, respectively.
The integral can be embedded in the subvariety of the Grassmannian $G_{_{3,6}}$
\begin{eqnarray}
&&\boldsymbol{{\xi}}^{\prime 1S}=\left(\begin{array}{cccccc}1&0&0&1&1&m_{_1}^2\\0&1&0&1&1&m_{_2}^2\\
0&0&1&1&1&p^2\end{array}\right)\;.
\label{1SE-2-2}
\end{eqnarray}
Because the fourth and fifth columns in the matroid Eq.(\ref{1SE-2-2}) coalesce into a same point in projective space
$P^2$, $\boldsymbol{{\xi}}^{\prime 1S}$ is reduced to the matroid $\boldsymbol{{\xi}}^{1S}$ of size $3\times 5$~\cite{Oxley2011}
\begin{eqnarray}
&&\boldsymbol{{\xi}}^{1S}=\left(\begin{array}{ccccc}1&0&0&1&m_{_1}^2\\0&1&0&1&m_{_2}^2\\
0&0&1&1&p^2\end{array}\right)\;.
\label{1SE-2-3}
\end{eqnarray}
Where the splitting coordinates $r_{_1}=m_{_1}^2$, $r_{_2}=m_{_2}^2$, and $r_{_3}=p^2$, respectively.

For convenience we take the exponent vector ${\boldsymbol{\beta}}={\boldsymbol{\beta}}_{_{(1S)}}
-\Delta\;{\bf e}_{_1}-\Delta\;{\bf e}_{_2}+2\Delta\;{\bf e}_{_3}$ in Eq.(\ref{sec2-2}), where
${\boldsymbol{\beta}}_{_{(1S)}}=(2-{D\over2},\;2-{D\over2},\;{D\over2},\;-1,\;{D\over2}-1)\in{\bf C}^5$,
and $\Delta$ is a nonzero c-number. Certainly at the end of calculation, we take the limit $\Delta\rightarrow0$. It is easy to get the boundary
values of the Feynman integral of the 1-loop self-energy as
\begin{eqnarray}
&&iA_{_{1SE}}(p^2,0,0)={i\Gamma(2-{D\over2})\Gamma^2({D\over2}-1)\over(4\pi)^{D/2}\Gamma(D-2)}
\Big({-p^2\over\Lambda_{_{\rm RE}}^2}\Big)^{{D\over2}-2}
\;,\nonumber\\
&&iA_{_{1SE}}(0,m^2,0)=iA_{_{1SE}}(0,0,m^2)={i\Gamma(2-{D\over2})\Gamma({D\over2}-1)\over(4\pi)^{D/2}\Gamma({D\over2})}
\Big({m^2\over\Lambda_{_{\rm RE}}^2}\Big)^{{D\over2}-2}\;,
\label{boundary-conditions}
\end{eqnarray}
which are used to obtain the combinatorial coefficients.

In a proper subset of the parameter space $|p^2|<m_{_2}^2<m_{_1}^2$, generally the Feynman integral can be written as
\begin{eqnarray}
&&\Big(\Lambda_{_{\rm RE}}^2\Big)^{D/2-2}A_{_{1SE}}(p^2,m_{_1}^2,m_{_2}^2)
\nonumber\\
&&\hspace{-0.5cm}=
C_{_{\{1,2,3\}}}^{(1)}({\boldsymbol{\beta}})(m_{_1}^2)^{-\beta_{_1}}(m_{_2}^2)^{-\beta_{_2}}(p^2)^{1-\beta_{_3}-\beta_{_4}}
\varphi_{_{\{1,2,3\}}}^{(1)}({\boldsymbol{\beta}},\;{p^2\over m_{_2}^2},\;{p^2\over m_{_1}^2})
\nonumber\\
&&\hspace{0.0cm}
+C_{_{\{1,2,3\}}}^{(2)}({\boldsymbol{\beta}})(m_{_1}^2)^{-\beta_{_1}}(m_{_2}^2)^{\beta_{_1}+\beta_{_5}-1}
\varphi_{_{\{1,2,3\}}}^{(2)}({\boldsymbol{\beta}},\;{p^2\over m_{_2}^2},\;{m_{_2}^2\over m_{_1}^2})
\nonumber\\
&&\hspace{0.0cm}
+C_{_{\{1,2,3\}}}^{(3)}({\boldsymbol{\beta}})(m_{_1}^2)^{\beta_{_5}-1}
\varphi_{_{\{1,2,3\}}}^{(3)}({\boldsymbol{\beta}},\;{p^2\over m_{_1}^2},\;{m_{_2}^2\over m_{_1}^2})\;.
\label{self-energy1}
\end{eqnarray}

In a proper subset of the parameter space $|p^2|<m_{_1}^2<m_{_2}^2$, similarly the Feynman integral can be written as
\begin{eqnarray}
&&\Big(\Lambda_{_{\rm RE}}^2\Big)^{D/2-2}A_{_{1SE}}(p^2,m_{_1}^2,m_{_2}^2)
\nonumber\\
&&\hspace{-0.5cm}=
C_{_{\{1,2,3\}}}^{(1)}({\boldsymbol{\beta}})(m_{_1}^2)^{-\beta_{_1}}(m_{_2}^2)^{-\beta_{_2}}(p^2)^{1-\beta_{_3}-\beta_{_4}}
\varphi_{_{\{1,2,3\}}}^{(1)}({\boldsymbol{\beta}},\;{p^2\over m_{_2}^2},\;{p^2\over m_{_1}^2})
\nonumber\\
&&\hspace{0.0cm}
+C_{_{\{1,2,3\}}}^{(5)}({\boldsymbol{\beta}})(m_{_2}^2)^{\beta_{_5}-1}
\varphi_{_{\{1,2,3\}}}^{(5)}({\boldsymbol{\beta}},\;{p^2\over m_{_2}^2},\;{m_{_1}^2\over m_{_2}^2})
\nonumber\\
&&\hspace{0.0cm}
+C_{_{\{1,2,3\}}}^{(6)}({\boldsymbol{\beta}})(m_{_1}^2)^{\beta_{_2}+\beta_{_5}-1}(m_{_2}^2)^{-\beta_{_2}}
\varphi_{_{\{1,2,3\}}}^{(6)}({\boldsymbol{\beta}},\;{p^2\over m_{_1}^2},\;{m_{_1}^2\over m_{_2}^2})\;.
\label{self-energy2}
\end{eqnarray}

In a proper subset of the parameter space $m_{_2}^2<|p^2|<m_{_1}^2$, the Feynman integral can be written as
\begin{eqnarray}
&&\Big(\Lambda_{_{\rm RE}}^2\Big)^{D/2-2}A_{_{1SE}}(p^2,m_{_1}^2,m_{_2}^2)
\nonumber\\
&&\hspace{-0.5cm}=
C_{_{\{1,2,3\}}}^{(3)}({\boldsymbol{\beta}})(m_{_1}^2)^{\beta_{_5}-1}
\varphi_{_{\{1,2,3\}}}^{(3)}({\boldsymbol{\beta}},\;{p^2\over m_{_1}^2},\;{m_{_2}^2\over m_{_1}^2})
\nonumber\\
&&\hspace{0.0cm}
+C_{_{\{1,2,3\}}}^{(7)}({\boldsymbol{\beta}})(m_{_1}^2)^{-\beta_{_1}}(p^2)^{\beta_{_1}+\beta_{_5}-1}
\varphi_{_{\{1,2,3\}}}^{(7)}({\boldsymbol{\beta}},\;{m_{_2}^2\over p^2},\;{p^2\over m_{_1}^2})
\nonumber\\
&&\hspace{0.0cm}
+C_{_{\{1,2,3\}}}^{(8)}({\boldsymbol{\beta}})(m_{_1}^2)^{-\beta_{_1}}(m_{_2}^2)^{1-\beta_{_2}-\beta_{_4}}(p^2)^{-\beta_{_3}}
\varphi_{_{\{1,2,3\}}}^{(8)}({\boldsymbol{\beta}},\;{m_{_2}^2\over p^2},\;{m_{_2}^2\over m_{_1}^2})\;.
\label{self-energy3}
\end{eqnarray}

In a proper subset of the parameter space $m_{_2}^2<m_{_1}^2<|p^2|$, the Feynman integral can be written as
\begin{eqnarray}
&&\Big(\Lambda_{_{\rm RE}}^2\Big)^{D/2-2}A_{_{1SE}}(p^2,m_{_1}^2,m_{_2}^2)
\nonumber\\
&&\hspace{-0.5cm}=
C_{_{\{1,2,3\}}}^{(8)}({\boldsymbol{\beta}})(m_{_1}^2)^{-\beta_{_1}}(m_{_2}^2)^{1-\beta_{_2}-\beta_{_4}}(p^2)^{-\beta_{_3}}
\varphi_{_{\{1,2,3\}}}^{(8)}({\boldsymbol{\beta}},\;{m_{_2}^2\over p^2},\;{m_{_2}^2\over m_{_1}^2})
\nonumber\\
&&\hspace{0.0cm}
+C_{_{\{1,2,3\}}}^{(9)}({\boldsymbol{\beta}})(m_{_1}^2)^{\beta_{_3}+\beta_{_5}-1}(p^2)^{-\beta_{_3}}
\varphi_{_{\{1,2,3\}}}^{(9)}({\boldsymbol{\beta}},\;{m_{_1}^2\over p^2},\;{m_{_2}^2\over m_{_1}^2})
\nonumber\\
&&\hspace{0.0cm}
+C_{_{\{1,2,3\}}}^{(10)}({\boldsymbol{\beta}})(p^2)^{\beta_{_5}-1}
\varphi_{_{\{1,2,3\}}}^{(10)}({\boldsymbol{\beta}},\;{m_{_2}^2\over p^2},\;{m_{_1}^2\over p^2})\;.
\label{self-energy4}
\end{eqnarray}

In a proper subset of the parameter space $m_{_1}^2<m_{_2}^2<|p^2|$, the Feynman integral is generally written as
\begin{eqnarray}
&&\Big(\Lambda_{_{\rm RE}}^2\Big)^{D/2-2}A_{_{1SE}}(p^2,m_{_1}^2,m_{_2}^2)
\nonumber\\
&&\hspace{-0.5cm}=
C_{_{\{1,2,3\}}}^{(10)}({\boldsymbol{\beta}})(p^2)^{\beta_{_5}-1}
\varphi_{_{\{1,2,3\}}}^{(10)}({\boldsymbol{\beta}},\;{m_{_2}^2\over p^2},\;{m_{_1}^2\over p^2})
\nonumber\\
&&\hspace{0.0cm}
+C_{_{\{1,2,3\}}}^{(11)}({\boldsymbol{\beta}})(m_{_2}^2)^{\beta_{_3}+\beta_{_5}-1}(p^2)^{-\beta_{_3}}
\varphi_{_{\{1,2,3\}}}^{(11)}({\boldsymbol{\beta}},\;{m_{_2}^2\over p^2},\;{m_{_1}^2\over m_{_2}^2})
\nonumber\\
&&\hspace{0.0cm}
+C_{_{\{1,2,3\}}}^{(12)}({\boldsymbol{\beta}})(m_{_1}^2)^{1-\beta_{_1}-\beta_{_4}}(m_{_2}^2)^{-\beta_{_2}}(p^2)^{-\beta_{_3}}
\varphi_{_{\{1,2,3\}}}^{(12)}({\boldsymbol{\beta}},\;{m_{_1}^2\over p^2},\;{m_{_1}^2\over m_{_2}^2})\;.
\label{self-energy5}
\end{eqnarray}

In a proper subset of the parameter space $m_{_1}^2<|p^2|<m_{_2}^2$, the Feynman integral is generally written as
\begin{eqnarray}
&&\Big(\Lambda_{_{\rm RE}}^2\Big)^{D/2-2}A_{_{1SE}}(p^2,m_{_1}^2,m_{_2}^2)
\nonumber\\
&&\hspace{-0.5cm}=
C_{_{\{1,2,3\}}}^{(4)}({\boldsymbol{\beta}})(m_{_2}^2)^{-\beta_{_2}}(p^2)^{\beta_{_2}+\beta_{_5}-1}
\varphi_{_{\{1,2,3\}}}^{(4)}({\boldsymbol{\beta}},\;{p^2\over m_{_2}^2},\;{m_{_1}^2\over p^2})
\nonumber\\
&&\hspace{0.0cm}
+C_{_{\{1,2,3\}}}^{(5)}({\boldsymbol{\beta}})(m_{_2}^2)^{\beta_{_5}-1}
\varphi_{_{\{1,2,3\}}}^{(5)}({\boldsymbol{\beta}},\;{p^2\over m_{_2}^2},\;{m_{_1}^2\over m_{_2}^2})
\nonumber\\
&&\hspace{0.0cm}
+C_{_{\{1,2,3\}}}^{(12)}({\boldsymbol{\beta}})(m_{_1}^2)^{1-\beta_{_1}-\beta_{_4}}(m_{_2}^2)^{-\beta_{_2}}(p^2)^{-\beta_{_3}}
\varphi_{_{\{1,2,3\}}}^{(12)}({\boldsymbol{\beta}},\;{m_{_1}^2\over p^2},\;{m_{_1}^2\over m_{_2}^2})\;.
\label{self-energy6}
\end{eqnarray}

Using the boundary conditions in Eq.(\ref{boundary-conditions}), we have
\begin{eqnarray}
&&C_{_{\{1,2,3\}}}^{(3)}({\boldsymbol{\beta}})
=C_{_{\{1,2,3\}}}^{(5)}({\boldsymbol{\beta}})
={\Gamma({D\over2}-1)\Gamma(2-{D\over2})\over(4\pi)^{D/2}\Gamma({D\over2})}
\;,\nonumber\\
&&C_{_{\{1,2,3\}}}^{(10)}({\boldsymbol{\beta}})
={(-1)^{D/2-2}\Gamma^2({D\over2}-1)\Gamma(2-{D\over2})\over(4\pi)^{D/2}\Gamma(D-2)}\;.
\label{self-energy7}
\end{eqnarray}
Other coefficients are some linear combinations of the above coefficients through the inverse Gauss relations.
In order to derive other combinatorial coefficients, we apply the analytic continuations presented in the section~\ref{sec2}.
Performing the inverse transformation of suitable variables in Eq.(\ref{self-energy1}) and Eq.(\ref{self-energy2}), we have
\begin{eqnarray}
&&C_{_{\{1,2,3\}}}^{(6)}({\boldsymbol{\beta}})=(-1)^{\beta_{_2}}{\Gamma(1-\beta_{_3}-\beta_{_4})\Gamma(\beta_{_5})
\over\Gamma(\beta_{_1}+\beta_{_5}-1)\Gamma(\beta_{_2}+\beta_{_5})}C_{_{\{1,2,3\}}}^{(3)}({\boldsymbol{\beta}})
\nonumber\\
&&\hspace{2.3cm}
+(-1)^{\beta_{_2}+\beta_{_5}}{\Gamma(2-\beta_{_2}-\beta_{_5})
\Gamma(1-\beta_{_3}-\beta_{_4})\Gamma(\beta_{_5})\over\Gamma(\beta_{_1})\Gamma(1-\beta_{_2})\Gamma(\beta_{_2}+\beta_{_5})}
C_{_{\{1,2,3\}}}^{(5)}({\boldsymbol{\beta}})
\;,\nonumber\\
&&C_{_{\{1,2,3\}}}^{(2)}({\boldsymbol{\beta}})=
(-1)^{\beta_{_1}+\beta_{_5}}{\Gamma(2-\beta_{_1}-\beta_{_5})\Gamma(1-\beta_{_3}-\beta_{_4})\Gamma(\beta_{_5})\over
\Gamma(1-\beta_{_1})\Gamma(\beta_{_2})\Gamma(\beta_{_1}+\beta_{_5})}C_{_{\{1,2,3\}}}^{(3)}({\boldsymbol{\beta}})
\nonumber\\
&&\hspace{2.3cm}
+(-1)^{\beta_{_1}}{\Gamma(1-\beta_{_3}-\beta_{_4})\Gamma(\beta_{_5})\over\Gamma(\beta_{_1}+\beta_{_5})\Gamma(\beta_{_2}+\beta_{_5}-1)}
C_{_{\{1,2,3\}}}^{(5)}({\boldsymbol{\beta}})\;.
\label{self-energy7-3}
\end{eqnarray}
In a similar way, we formulate other coefficients as some linear combinations on $C_{_{\{1,2,3\}}}^{(3)}({\boldsymbol{\beta}})$,
$C_{_{\{1,2,3\}}}^{(5)}({\boldsymbol{\beta}})$, and $C_{_{\{1,2,3\}}}^{(10)}({\boldsymbol{\beta}})$. To shorten the length
of text, we put those expressions in the appendix~\ref{app5}. Taking the space-time dimension $D=4-2\varepsilon$ in dimensional regularization,
we find that the ultraviolet divergence in Eq(\ref{self-energy1})$\sim$Eq.(\ref{self-energy6}) is $1/\varepsilon$.

Taking the Mellin-Barnes transformation, Ref.~\cite{Davydychev3} presents the Feynman integral
of the 1-loop self energy in terms of the fourth type Appell function $F_{_4}$. The representation
in terms of $F_{_4}$ functions of variables $p^2/m_{_2}^2$, $m_{_1}^2/m_{_2}^2$ is given in Eq.~(20)
of Ref.~\cite{Davydychev3}, and that in terms of $F_{_4}$ functions of variables $m_{_2}^2/p^2$, $m_{_1}^2/p^2$
is given in Eq.~(21), respectively. Furthermore, Eq.~(20) of Ref.~\cite{Davydychev3}
can be changed into Eq.~(21) of Ref.~\cite{Davydychev3} through the well-known inverse transformation~\cite{Slater1966}
\begin{eqnarray}
&&F_{_4}\left(\left.\begin{array}{c}a,\;b\\ c,\;d\end{array}\right|\;x,\;y\right)
\nonumber\\
&&\hspace{-0.5cm}=
{\Gamma(d)\Gamma(b-a)\over\Gamma(b)\Gamma(d-a)}(-y)^{-a}
F_{_4}\left(\left.\begin{array}{c}a,\;1+a-d\\ c,\;1-b+a\end{array}\right|\;{x\over y},\;{1\over y}\right)
\nonumber\\
&&\hspace{0.0cm}
+{\Gamma(d)\Gamma(a-b)\over\Gamma(a)\Gamma(d-b)}(-y)^{-b}
F_{_4}\left(\left.\begin{array}{c}b,\;1+b-d\\ c,\;1-a+b\end{array}\right|\;{x\over y},\;{1\over y}\right)\;.
\label{PE-1-2}
\end{eqnarray}

To compare our expressions with that of Ref.~\cite{Davydychev3}, we assume that all powers of propagators are unit.
When the powers of the propagators are some other positive integers, the corresponding result is obtained
by differentiating the propagators with respect to the virtual squared masses.

Under the assumption above, the $F_{_4}$ functions in Ref.~\cite{Davydychev3} are reduced to the Gauss functions through the
transformations~\cite{Bailey1964}
\begin{eqnarray}
&&F_{_4}\left(\left.\begin{array}{c}a,\;b\\ b,\;b\end{array}
\right|\;-{s\over(1-s)(1-t)},\;-{t\over(1-s)(1-t)}\right)
\nonumber\\
&&\hspace{-0.5cm}=
(1-s)^a(1-t)^a\;_{_2}F_{_1}\left(\left.\begin{array}{c}a,\;1+a-b\\ b\end{array}\right|\;st\right)
\;,\nonumber\\
&&F_{_4}\left(\left.\begin{array}{c}a,\;b\\ 1+a-b,\;b\end{array}
\right|\;-{s\over(1-s)(1-t)},-{t\over(1-s)(1-t)}\right)
\nonumber\\
&&\hspace{-0.5cm}=
(1-t)^a\;_{_2}F_{_1}\left(\left.\begin{array}{c}a,\;b\\ 1+a-b\end{array}\right|\;-{s(1-t)\over1-s}\right)\;.
\label{PE-1-1}
\end{eqnarray}

As ${\boldsymbol{\beta}}={\boldsymbol{\beta}}_{_{(1S)}}$, the hypergeometric series in our
results are reduced to the Gauss functions through Eq.(\ref{sec4-2b}) and Eq.(\ref{sec4-4b}) also.
As an example, the hypergeometric series $\varphi_{_{\{1,2,3\}}}^{(10)}$ in Eq.~(\ref{self-energy4}) and Eq.~(\ref{self-energy5})
is reduced to
\begin{eqnarray}
&&\varphi_{_{\{1,2,3\}}}^{(10)}({\boldsymbol{\beta}},\;x,\;y)
=F_{_1}\left(\left.\begin{array}{c}2-{D\over2},\;2-{D\over2},\;2-{D\over2}\\ 3-D\end{array}\right|\;x,\;y\right)
\nonumber\\
&&\hspace{-0.5cm}=
(1-y)^{D/2-2}F_{_1}\left(\left.\begin{array}{c}2-{D\over2},\;{D\over2},\;-1,\\ 3-D\end{array}\right|\;{x-y\over1-y},{y\over y-1}\right)
\nonumber\\
&&\hspace{-0.5cm}=
(1-y)^{D/2-2}\Big\{\;_{_2}F_{_1}\left(\left.\begin{array}{c}2-{D\over2},\;2-{D\over2}\\ 3-D\end{array}\right|\;
{x-y\over1-y}\right)
\nonumber\\
&&\hspace{0.0cm}
-{4-D\over6-2D}{y\over y-1}\;_{_2}F_{_1}\left(\left.\begin{array}{c}3-{D\over2},\;2-{D\over2}\\ 4-D\end{array}\right|\;
{x-y\over1-y}\right)\Big\}
\label{PE-2-9}
\end{eqnarray}
through the second identity in Eq.(\ref{sec4-2b}). The hypergeometric series $\varphi_{_{\{1,2,3\}}}^{(8)}$,
$\varphi_{_{\{1,2,3\}}}^{(9)}$, $\varphi_{_{\{1,2,3\}}}^{(11)}$, and $\varphi_{_{\{1,2,3\}}}^{(12)}$ can also be expressed
in terms of the Gauss functions. Using those equations and some transformations of the Gauss function,
we find that Eq.~(\ref{self-energy4}) and Eq.~(\ref{self-energy5}) respectively coincide with Eq.~(21) of Ref.~\cite{Davydychev3}.
In order to shorten the length of text, we present the relevant expressions in the appendix~\ref{app5}.
Similarly we find that Eq.~(\ref{self-energy2}) and Eq.~(\ref{self-energy6}) coincide with
Eq.~(20) of Ref.~\cite{Davydychev3}, respectively.

A connection between Feynman parametrization of a 1-loop N-point function and the simplex in N-dimensional Euclidean space
is presented in Refs.~\cite{Davydychev1998,Davydychev2017} because there is a one-to-one correspondence between mass sides
and vertices of the 1-loop Feynman diagrams, and the Feynman integral of  a 1-loop N-point diagram
is proportional to the ratio of an N-dimensional solid angle at the meeting point of the mass sides to the content of
the N-dimensional basic simplex. Using the inverse transformation in Eq.~(\ref{Intro2}) and the quadratic transformation~\cite{Davydychev1993NPB}
\begin{eqnarray}
&&\;_2F_{_1}\left(\left.\begin{array}{c}a,\;b\\ a-b+1\end{array}\right|\;\zeta\right)
=(1-\zeta)^{-a}\;_2F_{_1}\left(\left.\begin{array}{c}{a\over2},\;{a+1\over2}-b\\ a-b+1\end{array}\right|\;-{4\zeta\over(1-\zeta)^2}\right)\;,
\label{PE-2-9a}
\end{eqnarray}
one transforms Eq.~(\ref{PE-2-18a}) to Eq.~(4.7) in Ref.~\cite{Davydychev1998}
(to Eq.~(1) in Ref.~\cite{Davydychev2017}) exactly.

Definitely the geometrical approach is a fascinating idea to evaluate the Feynman integrals of 1-loop N-point diagrams.
Since there is not a one-to-one correspondence between mass sides and vertices of higher loop Feynman diagrams,
how to generalize this approach to higher loops is a challenge so far.
The approach based on GKZ-system can provide the analytic expressions and singular loci for Feynman integrals of high loops
simultaneously in a consistent way.

\section{The expressions of a 2-loop massless triangle diagram\label{sec7}}
\indent\indent
\begin{figure}[ht]
\setlength{\unitlength}{1cm}
\centering
\vspace{0.0cm}\hspace{-1.5cm}
\includegraphics[height=5cm,width=9.0cm]{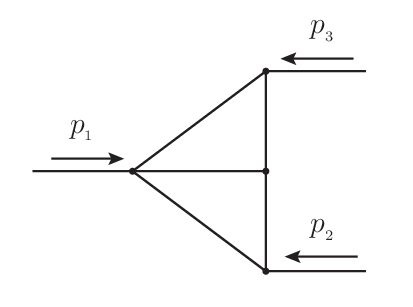}
\vspace{0cm}
\caption[]{A 2-loop triangle diagram with zero virtual masses.}
\label{fig2}
\end{figure}

In $\alpha$-parametrization, the Feynman integral of the 2-loop massless triangle diagram drawn in Fig.\ref{fig2}
is written as
\begin{eqnarray}
&&iA_{_{WE}}(p_{_1}^2,p_{_2}^2,p_{_3}^2)=
-i\Big(\Lambda_{_{\rm RE}}^2\Big)^{5-D}\int{d^Dq_{_1}\over(2\pi)^D}{d^Dq_{_2}\over(2\pi)^D}\int_{_0}^\infty\prod\limits_{i=1}^5d\alpha_{_i}
\exp\Big\{i\Big[\alpha_{_1}q_{_1}^2+\alpha_{_2}q_{_2}^2
\nonumber\\
&&\hspace{3.2cm}
+\alpha_{_3}(q_{_1}+p_{_2})^2+\alpha_{_4}(q_{_2}+p_{_3})^2+\alpha_{_5}(q_{_1}+q_{_2}-p_{_1})^2\Big]\Big\}
\nonumber\\
&&\hspace{2.7cm}=
{\Big(\Lambda_{_{\rm RE}}^2\Big)^{5-D}\Gamma(5-D)\exp\{i\pi(1-{D\over2})\}\over(4\pi)^D(-i)^{4-D}}
\int_{_S}\omega_{_5}(t)
\nonumber\\
&&\hspace{3.2cm}\times
[(t_{_1}+t_{_3})(t_{_2}+t_{_4})+(t_{_1}+t_{_2}+t_{_3}+t_{_4})t_{_5}]^{5-3D/2}
\nonumber\\
&&\hspace{3.2cm}\times
\Big[t_{_1}t_{_2}t_{_5}p_{_1}^2+t_{_1}(t_{_2}t_{_3}+t_{_3}t_{_4}+t_{_3}t_{_5}+t_{_4}t_{_5})p_{_2}^2
\nonumber\\
&&\hspace{3.2cm}
+t_{_2}(t_{_1}t_{_4}+t_{_3}t_{_4}+t_{_3}t_{_5}+t_{_4}t_{_5})p_{_3}^2\Big]^{D-5}\;,
\label{2WE-2-1}
\end{eqnarray}
where the hyperplane $S$ in the 5-dimensional affine space is defined by $t_{_5}=1$,
and $\omega_{_5}(t)$ is the volume element of $4$-dimensional projective space, respectively. The integral can be
embedded in the subvariety of the Grassmannian $G_{_{5,10}}$ which is reduced further as the matroid
$\boldsymbol{{\xi}}^{WE}$ of size $3\times 5$
\begin{eqnarray}
&&\boldsymbol{{\xi}}^{WE}=\left(\begin{array}{ccccc}1&0&0&1&p_{_1}^2\\0&1&0&1&p_{_2}^2\\
0&0&1&1&p_{_3}^2\end{array}\right)\;,
\label{2WE-2-2}
\end{eqnarray}
where the splitting coordinates $r_{_1}=p_{_1}^2$, $r_{_2}=p_{_2}^2$, and $r_{_3}=p_{_3}^2$, respectively.
In other words, the Feynman integral can be also written as some linear combinations of the generalized hypergeometric
functions $\varphi_{_{\cal B}}^{(i)}$, $i=1,\cdots,12$ presented in the sections~$\ref{sec1}-\ref{sec5}$,
where the exponent vector and boundary conditions depend on the concrete topology of the diagram.

For convenience we take the exponent vector ${\boldsymbol{\beta}}={\boldsymbol{\beta}}_{_{(WE)}}
+2\Delta\;{\bf e}_{_1}-\Delta\;{\bf e}_{_2}-\Delta\;{\bf e}_{_3}$ in Eq.(\ref{sec2-2}),
where ${\boldsymbol{\beta}}_{_{(WE)}}=(1,\;1,\;1,\;3-D,\;D-4)\in{\bf C}^5$ denotes the exponent vector,
and $\Delta$ is a nonzero c-number. Certainly at the end of calculation, we take the limit $\Delta\rightarrow0$. The boundary
values of the Feynman integral of the 2-loop massless triangle are
\begin{eqnarray}
&&A_{_{WE}}(p^2,p^2,0)=A_{_{WE}}(p^2,0,p^2)=
\Big({\Lambda_{_{\rm RE}}^2\over-p^2}\Big)^{5-D}{\cal B}_1
\;,\nonumber\\
&&A_{_{WE}}(0,p^2,p^2)=
\Big({\Lambda_{_{\rm RE}}^2\over-p^2}\Big)^{5-D}{\cal B}_2\;,
\label{2WE-2-10}
\end{eqnarray}
where ${\cal B}_1$ and ${\cal B}_2$ are derived through the Gegenbauer polynomial technique~\cite{Chetyrkin1980},
\begin{eqnarray}
&&{\cal B}_1={\Gamma^3({D\over2}-1)\over(4\pi)^{D}\Gamma({D\over2})}
\Big\{{\Gamma(2-{D\over2})\Gamma(4-D)\over\Gamma({D\over2}-1)}\Big({m^2\over-p^2}\Big)^{D-4}
\nonumber\\
&&\hspace{1.2cm}
+{(D-2)\Gamma({D\over2}-2)\Gamma^2(3-{D\over2})\over(4-D)\Gamma({D\over2}-1)\Gamma(D-3)}\Big({m^2\over-p^2}\Big)^{{D\over2}-2}
\nonumber\\
&&\hspace{1.2cm}
+{(2-D)\Gamma({D\over2}-2)\Gamma(4-D)\over(6-D)\Gamma({3D\over2}-5)}\Big\}
\;,\nonumber\\
&&{\cal B}_2={2\Gamma^3({D\over2}-1)\Gamma(5-D)\over(4\pi)^D\Gamma(D-2)\Gamma({3D\over2}-5)}
\sum\limits_{n=0}^\infty{\Gamma(D-2+n)\over n!({D\over2}-1+n)^2}
\nonumber\\
&&\hspace{1.2cm}\times
\Big\{{1\over(1+n)(3-{D\over2}+n)}+{1\over(1+n)(D-3+n)}
\nonumber\\
&&\hspace{1.2cm}
+{1\over(D-3+n)({3D\over2}-5+n)}\Big\}\;.
\label{2WE-2-11}
\end{eqnarray}
The boundary values of the Feynman integral are used to obtain the combinatorial coefficients
where the infrared (IR) divergence is regularized by $\ln(m^2/(-p^2))$ in the Laurent series around $D=4$.

In a proper subset of the parameter space $|p_{_3}^2|<|p_{_2}^2|<|p_{_1}^2|$, the Feynman integral is generally written as
\begin{eqnarray}
&&\Big(\Lambda_{_{\rm RE}}^2\Big)^{5-D}A_{_{WE}}(p_{_1}^2,p_{_2}^2,p_{_3}^2)
\nonumber\\
&&\hspace{-0.5cm}=
C_{_{\{1,2,3\}}}^{(1)}({\boldsymbol{\beta}})(p_{_1}^2)^{-\beta_{_1}}(p_{_2}^2)^{-\beta_{_2}}(p_{_3}^2)^{1-\beta_{_3}-\beta_{_4}}
\varphi_{_{\{1,2,3\}}}^{(1)}({\boldsymbol{\beta}},\;{p_{_3}^2\over p_{_2}^2},\;{p_{_3}^2\over p_{_1}^2})
\nonumber\\
&&\hspace{0.0cm}
+C_{_{\{1,2,3\}}}^{(2)}({\boldsymbol{\beta}})(p_{_1}^2)^{-\beta_{_1}}(p_{_2}^2)^{\beta_{_1}+\beta_{_5}-1}
\varphi_{_{\{1,2,3\}}}^{(2)}({\boldsymbol{\beta}},\;{p_{_3}^2\over p_{_2}^2},\;{p_{_2}^2\over p_{_1}^2})
\nonumber\\
&&\hspace{0.0cm}
+C_{_{\{1,2,3\}}}^{(3)}({\boldsymbol{\beta}})(p_{_1}^2)^{\beta_{_5}-1}
\varphi_{_{\{1,2,3\}}}^{(3)}({\boldsymbol{\beta}},\;{p_{_3}^2\over p_{_1}^2},\;{p_{_2}^2\over p_{_1}^2})\;.
\label{2WE-2-12}
\end{eqnarray}

In a proper subset of the parameter space $|p_{_3}^2|<|p_{_1}^2|<|p_{_2}^2|$, the Feynman integral is similarly written as
\begin{eqnarray}
&&\Big(\Lambda_{_{\rm RE}}^2\Big)^{5-D}A_{_{WE}}(p_{_1}^2,p_{_2}^2,p_{_3}^2)
\nonumber\\
&&\hspace{-0.5cm}=
C_{_{\{1,2,3\}}}^{(1)}({\boldsymbol{\beta}})(p_{_1}^2)^{-\beta_{_1}}(p_{_2}^2)^{-\beta_{_2}}(p_{_3}^2)^{1-\beta_{_3}-\beta_{_4}}
\varphi_{_{\{1,2,3\}}}^{(1)}({\boldsymbol{\beta}},\;{p_{_3}^2\over p_{_2}^2},\;{p_{_3}^2\over p_{_1}^2})
\nonumber\\
&&\hspace{0.0cm}
+C_{_{\{1,2,3\}}}^{(5)}({\boldsymbol{\beta}})(p_{_2}^2)^{\beta_{_5}-1}
\varphi_{_{\{1,2,3\}}}^{(5)}({\boldsymbol{\beta}},\;{p_{_3}^2\over p_{_2}^2},\;{p_{_1}^2\over p_{_2}^2})
\nonumber\\
&&\hspace{0.0cm}
+C_{_{\{1,2,3\}}}^{(6)}({\boldsymbol{\beta}})(p_{_1}^2)^{\beta_{_2}+\beta_{_5}-1}(p_{_2}^2)^{-\beta_{_2}}
\varphi_{_{\{1,2,3\}}}^{(6)}({\boldsymbol{\beta}},\;{p_{_3}^2\over p_{_1}^2},\;{p_{_1}^2\over p_{_2}^2})\;.
\label{2WE-2-13}
\end{eqnarray}

In a proper subset of the parameter space $|p_{_2}^2|<|p_{_3}^2|<|p_{_1}^2|$, the Feynman integral can be written as
\begin{eqnarray}
&&\Big(\Lambda_{_{\rm RE}}^2\Big)^{5-D}A_{_{WE}}(p_{_1}^2,p_{_2}^2,p_{_3}^2)
\nonumber\\
&&\hspace{-0.5cm}=
C_{_{\{1,2,3\}}}^{(3)}({\boldsymbol{\beta}})(p_{_1}^2)^{\beta_{_5}-1}
\varphi_{_{\{1,2,3\}}}^{(3)}({\boldsymbol{\beta}},\;{p_{_3}^2\over p_{_1}^2},\;{p_{_2}^2\over p_{_1}^2})
\nonumber\\
&&\hspace{0.0cm}
+C_{_{\{1,2,3\}}}^{(7)}({\boldsymbol{\beta}})(p_{_1}^2)^{-\beta_{_1}}(p_{_3}^2)^{\beta_{_1}+\beta_{_5}-1}
\varphi_{_{\{1,2,3\}}}^{(7)}({\boldsymbol{\beta}},\;{p_{_2}^2\over p_{_3}^2},\;{p_{_3}^2\over p_{_1}^2})
\nonumber\\
&&\hspace{0.0cm}
+C_{_{\{1,2,3\}}}^{(8)}({\boldsymbol{\beta}})(p_{_1}^2)^{-\beta_{_1}}(p_{_2}^2)^{1-\beta_{_2}-\beta_{_4}}(p_{_3}^2)^{-\beta_{_3}}
\varphi_{_{\{1,2,3\}}}^{(8)}({\boldsymbol{\beta}},\;{p_{_2}^2\over p_{_3}^2},\;{p_{_2}^2\over p_{_1}^2})\;.
\label{2WE-2-14}
\end{eqnarray}

In a proper subset of the parameter space $|p_{_2}^2|<|p_{_1}^2|<|p_{_3}^2|$, the Feynman integral is accordingly written as
\begin{eqnarray}
&&\Big(\Lambda_{_{\rm RE}}^2\Big)^{5-D}A_{_{WE}}(p_{_1}^2,p_{_2}^2,p_{_3}^2)
\nonumber\\
&&\hspace{-0.5cm}=
C_{_{\{1,2,3\}}}^{(8)}({\boldsymbol{\beta}})(p_{_1}^2)^{-\beta_{_1}}(p_{_2}^2)^{1-\beta_{_2}-\beta_{_4}}(p_{_3}^2)^{-\beta_{_3}}
\varphi_{_{\{1,2,3\}}}^{(8)}({\boldsymbol{\beta}},\;{p_{_2}^2\over p_{_3}^2},\;{p_{_2}^2\over p_{_1}^2})
\nonumber\\
&&\hspace{0.0cm}
+C_{_{\{1,2,3\}}}^{(9)}({\boldsymbol{\beta}})(p_{_1}^2)^{\beta_{_3}+\beta_{_5}-1}(p_{_3}^2)^{-\beta_{_3}}
\varphi_{_{\{1,2,3\}}}^{(9)}({\boldsymbol{\beta}},\;{p_{_1}^2\over p_{_3}^2},\;{p_{_2}^2\over p_{_1}^2})
\nonumber\\
&&\hspace{0.0cm}
+C_{_{\{1,2,3\}}}^{(10)}({\boldsymbol{\beta}})(p_{_3}^2)^{\beta_{_5}-1}
\varphi_{_{\{1,2,3\}}}^{(10)}({\boldsymbol{\beta}},\;{p_{_2}^2\over p_{_3}^2},\;{p_{_1}^2\over p_{_3}^2})\;.
\label{2WE-2-15}
\end{eqnarray}

In a proper subset of the parameter space $|p_{_1}^2|<|p_{_2}^2|<|p_{_3}^2|$, the Feynman integral is generally written as
\begin{eqnarray}
&&\Big(\Lambda_{_{\rm RE}}^2\Big)^{5-D}A_{_{WE}}(p_{_1}^2,p_{_2}^2,p_{_3}^2)
\nonumber\\
&&\hspace{-0.5cm}=
C_{_{\{1,2,3\}}}^{(10)}({\boldsymbol{\beta}})(p_{_3}^2)^{\beta_{_5}-1}
\varphi_{_{\{1,2,3\}}}^{(10)}({\boldsymbol{\beta}},\;{p_{_2}^2\over p_{_3}^2},\;{p_{_1}^2\over p_{_3}^2})
\nonumber\\
&&\hspace{0.0cm}
+C_{_{\{1,2,3\}}}^{(11)}({\boldsymbol{\beta}})(p_{_2}^2)^{\beta_{_3}+\beta_{_5}-1}(p_{_3}^2)^{-\beta_{_3}}
\varphi_{_{\{1,2,3\}}}^{(11)}({\boldsymbol{\beta}},\;{p_{_2}^2\over p_{_3}^2},\;{p_{_1}^2\over p_{_2}^2})
\nonumber\\
&&\hspace{0.0cm}
+C_{_{\{1,2,3\}}}^{(12)}({\boldsymbol{\beta}})(p_{_1}^2)^{1-\beta_{_1}-\beta_{_4}}(p_{_2}^2)^{-\beta_{_2}}(p_{_3}^2)^{-\beta_{_3}}
\varphi_{_{\{1,2,3\}}}^{(12)}({\boldsymbol{\beta}},\;{p_{_1}^2\over p_{_3}^2},\;{p_{_1}^2\over p_{_2}^2})\;.
\label{2WE-2-16}
\end{eqnarray}

In a proper subset of the parameter space $|p_{_1}^2|<|p_{_3}^2|<|p_{_2}^2|$, the Feynman integral is generally written as
\begin{eqnarray}
&&\Big(\Lambda_{_{\rm RE}}^2\Big)^{5-D}A_{_{WE}}(p_{_1}^2,p_{_2}^2,p_{_3}^2)
\nonumber\\
&&\hspace{-0.5cm}=
C_{_{\{1,2,3\}}}^{(4)}({\boldsymbol{\beta}})(p_{_2}^2)^{-\beta_{_2}}(p_{_3}^2)^{\beta_{_2}+\beta_{_5}-1}
\varphi_{_{\{1,2,3\}}}^{(4)}({\boldsymbol{\beta}},\;{p_{_3}^2\over p_{_2}^2},\;{p_{_1}^2\over p_{_3}^2})
\nonumber\\
&&\hspace{0.0cm}
+C_{_{\{1,2,3\}}}^{(5)}({\boldsymbol{\beta}})(p_{_2}^2)^{\beta_{_5}-1}
\varphi_{_{\{1,2,3\}}}^{(5)}({\boldsymbol{\beta}},\;{p_{_3}^2\over p_{_2}^2},\;{p_{_1}^2\over p_{_2}^2})
\nonumber\\
&&\hspace{0.0cm}
+C_{_{\{1,2,3\}}}^{(12)}({\boldsymbol{\beta}})(p_{_1}^2)^{1-\beta_{_1}-\beta_{_4}}(p_{_2}^2)^{-\beta_{_2}}(p_{_3}^2)^{-\beta_{_3}}
\varphi_{_{\{1,2,3\}}}^{(12)}({\boldsymbol{\beta}},\;{p_{_1}^2\over p_{_3}^2},\;{p_{_1}^2\over p_{_2}^2})\;.
\label{2WE-2-17}
\end{eqnarray}

Using Eq.(\ref{self-energy7-3}), Eq.(\ref{self-energy2-2}), Eq.(\ref{self-energy5-3}), we write the boundary
conditions presented in Eq.(\ref{2WE-2-10}) as
\begin{eqnarray}
&&-2C_{_{\{1,2,3\}}}^{(3)}({\boldsymbol{\beta}})
+C_{_{\{1,2,3\}}}^{(5)}({\boldsymbol{\beta}})={{\cal B}_1\over4-D}
\;,\nonumber\\
&&-2C_{_{\{1,2,3\}}}^{(3)}({\boldsymbol{\beta}})
+C_{_{\{1,2,3\}}}^{(10)}({\boldsymbol{\beta}})={{\cal B}_1\over4-D}
\;,\nonumber\\
&&C_{_{\{1,2,3\}}}^{(5)}({\boldsymbol{\beta}})
+C_{_{\{1,2,3\}}}^{(10)}({\boldsymbol{\beta}})=-{2{\cal B}_2\over4-D}\;.
\label{2WE-2-23}
\end{eqnarray}
The system implies
\begin{eqnarray}
&&C_{_{\{1,2,3\}}}^{(3)}({\boldsymbol{\beta}})=-{{\cal B}_1+{\cal B}_2\over2(4-D)}
\;,\nonumber\\
&&C_{_{\{1,2,3\}}}^{(5)}=C_{_{\{1,2,3\}}}^{(10)}=-{{\cal B}_2\over4-D}\;.
\label{2WE-2-24}
\end{eqnarray}
The other coefficients $C_{_{\{1,2,3\}}}^{(i)}({\boldsymbol{\beta}})$, $i\in \{1,\cdots,12\}\setminus\{3,5,10\}$ are obtained from
Eq.(\ref{self-energy7-3}), Eq.(\ref{self-energy2-2}), and Eq.(\ref{self-energy5-3}), respectively.

The Feynman integral of the 2-loop massless triangle diagram can also be represented
in terms of polylogarithms as the time-space dimension $D=4$~\cite{Ussyukina1994}.
In order to compare our result with that in terms of polylogarithms, we define the dimensionless variables
\begin{eqnarray}
&&x={p_{_2}^2\over p_{_1}^2},\;\;y={p_{_3}^2\over p_{_1}^2}\;,
\label{2WE-2-25}
\end{eqnarray}
thus
\begin{eqnarray}
&&\Big(\Lambda_{_{\rm RE}}^2\Big)A_{_{WE}}(p_{_1}^2,p_{_2}^2,p_{_3}^2)
={\Phi^{(2)}(x,y)\over(4\pi)^4}
\label{2WE-2-26}
\end{eqnarray}
as $|p_{_1}^2|\ge\max(|p_{_2}^2|,\;|p_{_3}^2|)$, where $\Phi^{(2)}$ is defined in Eq.(16) and Eq.(17) of~Ref.~\cite{Ussyukina1994}.
Obviously the Feynman integral contains the IR divergence in the scenario $p_{_1}^2=p_{_2}^2$, $p_{_3}^2=0$.
In the expression in terms of polylogarithms,
\begin{eqnarray}
&&{\Phi^{(2)}(x,y)\over(4\pi)^4}\Big|_{\left.x=1,y\rightarrow0\right.}=
{1\over(4\pi)^4}\Big\{6-3\ln y+{1\over2}\ln^2y\Big\}\;.
\label{2WE-2-27}
\end{eqnarray}
In other words, the IR divergence is regularized by $\ln y$ in the approach adopted in Ref.~\cite{Ussyukina1994}.
As $D=4$, one similarly derives
\begin{eqnarray}
&&\Big\{{\rm The\;right-hand\;side\;of\;Eq.~(\ref{2WE-2-12})}\Big\}\Big|_{x=1,y=0}
\nonumber\\
&&\hspace{-0.5cm}=
{1\over(4\pi)^4}\Big\{1+{\pi^2\over6}-\ln\Big({m^2\over-p_{_1}^2}\Big)
+{1\over2}\ln^2\Big({m^2\over-p_{_1}^2}\Big)\Big\}\;,
\label{2WE-2-28}
\end{eqnarray}
where the IR divergence is regularized by $\ln(m^2/(-p_{_1}^2))$. Note that
Eq.~(\ref{2WE-2-28}) can also be gotten through the Mellin-Barnes contour. The matching program
between the results of two different approaches indicates
\begin{eqnarray}
&&\ln\Big({m^2\over-p_{_1}^2}\Big)+{1\over2}\ln^2\Big({m^2\over-p_{_1}^2}\Big)
\rightarrow{\pi^2\over6}-5+3\ln y-{1\over2}\ln^2y\;.
\label{2WE-2-29}
\end{eqnarray}
Inserting the replacement above in the right-hand side of Eq.~(\ref{2WE-2-12}),
one derives
\begin{eqnarray}
&&{\partial\over\partial x}\Big\{{\rm The\;right-hand\;side\;of\;Eq.~(\ref{2WE-2-12})}\Big\}\Big|_{x=1,y=0}
\nonumber\\
&&\hspace{-0.5cm}=
{1\over(4\pi)^4}\Big\{-{21\over8}+{13\over8}\ln y-{3\over8}\ln^2y\Big\}
\;,\nonumber\\
&&{\partial\over\partial y}\Big\{{\rm The\;right-hand\;side\;of\;Eq.~(\ref{2WE-2-12})}\Big\}\Big|_{x=1,y=0}
\nonumber\\
&&\hspace{-0.5cm}=
{1\over(4\pi)^4}\Big\{{65\over216}-{19\over72}\ln y+{5\over72}\ln^2y+{\ln y-3\over y}\Big\}\;,
\label{2WE-2-30}
\end{eqnarray}
which are coincide with the result of Ref.~\cite{Ussyukina1994}. In a similar way, we can verify
Eq.~(\ref{2WE-2-14}) coinciding with the result in terms of polylogarithms.

\section{Summary\label{sec8}}
\indent\indent
The important properties of generalized hypergeometric functions are the various Gauss relations among them.
After embedding Feynman integrals in the subvarieties of Grassmannians through homogenization
of the integrands, we formulate Feynman integrals as some finite linear combinations
of generalized hypergeometric functions in some regions of the domain of definition. Taking the generalized hypergeometric functions
on the Grassmannian $G_{_{3,5}}$ as an example, we present the method to obtain Gauss relations among those generalized hypergeometric functions
in this work. The hypergeometric expression of a
Feynman integral is continued from a proper subset of certain connected component to the corresponding
proper subset of another connected component by the inverse Gauss relations, and then continued to the whole domain
of definition of the Feynman integral by the Gauss-Kummer relations accordingly. Furthermore the Laurent series of the analytic expression around the
dimension $D=4$ can be obtained by the Gauss adjacent relations where the coefficient of term with power of $D-4$ is
given by a linear combination of hypergeometric functions with integer parameters.
As examples, we illustrate how to obtain the expressions of the Feynman integrals of the 1-loop self energy and
a 2-loop massless triangle diagram in their domains of definition, respectively.

Another important issue is that the singular locus of certain Feynman integral
is a complement to the domain of definition of Feynman integral in the whole parameter space. It is widely believed that the physical
thresholds obtained from Cutkosky cuts are proper subsets of the singular locus of the Feynman integral~\cite{Eden1966,Britto}.
The dominant contributions of the Feynman integral at the threshold are derived by the expansion of its analytic expressions
in the limit of heavy masses and large momenta~\cite{Smirnov1990}. Actually the singular locus of a Feynman integral depends on
determinant of the resultant complex in an affine algebra~\cite{Gelfand1994}, and we will release our relevant calculations elsewhere~\cite{Feng2025}.

\begin{acknowledgments}
\indent\indent
The work has been supported partly by the National Natural Science Foundation
of China (NNSFC) with Grant No. 12075074, No. 12235008, Natural Science
Foundation of Guangxi Autonomous Region with Grant No. 2022GXNSFDA035068,
Natural Science Foundation of Hebei Province with Grant No. A2022201017, No. A2023201041,
and the youth top-notch talent support program of Hebei province.
\end{acknowledgments}

\appendix

\section{The Hypergeometric functions for all possible affine spanning\label{app1}}
\subsection{${\bf{\cal B}}=\{1,2,3\}$\label{sec6-1}}
\indent\indent
Corresponding to the integer lattice $n_{_1}E_{_3}^{(2)}+ n_{_2}E_{_3}^{(3)}$,
the generalized hypergeometric function is formulated as
\begin{eqnarray}
&&\Phi_{_{\{1,2,3\}}}^{(3)}({\boldsymbol{\beta}},\;\boldsymbol{{\xi}})=A_{_{\{1,2,3\}}}^{(3)}({\boldsymbol{\beta}})(r_{_1})^{\beta_{_5}-1}
\varphi_{_{\{1,2,3\}}}^{(3)}({\boldsymbol{\beta}},\;{r_{_3}\over r_{_1}},\;{r_{_2}\over r_{_1}})\;,\nonumber\\
&&\varphi_{_{\{1,2,3\}}}^{(3)}({\boldsymbol{\beta}},\;x_{_1},\;x_{_2})=\sum\limits_{n_{_1},n_{_2}}c_{_{\{1,2,3\}}}^{(3)}({\boldsymbol{\beta}},\;n_{_1},n_{_2})
x_{_1}^{n_{_1}}x_{_2}^{n_{_2}}\;,
\label{Interger-lattice3-2}
\end{eqnarray}
where
\begin{eqnarray}
&&A_{_{\{1,2,3\}}}^{(3)}({\boldsymbol{\beta}})={\Gamma(\beta_{_4})\over\Gamma(1-\beta_{_2})\Gamma(1-\beta_{_3})\Gamma(2-\beta_{_1}-\beta_{_5})}\;,\nonumber\\
&&c_{_{\{1,2,3\}}}^{(3)}({\boldsymbol{\beta}},\;n_{_1},n_{_2})={(\beta_{_3})_{_{n_{_1}}}(\beta_{_2})_{_{n_{_2}}}(1-\beta_{_5})_{_{n_{_1}+n_{_2}}}\over n_{_1}!n_{_2}!
(2-\beta_{_1}-\beta_{_5})_{_{n_{_1}+n_{_2}}}}\;.
\label{Interger-lattice3-3}
\end{eqnarray}

As the integer lattice is $n_{_1}E_{_3}^{(1)}- n_{_2}E_{_3}^{(2)}$,
the generalized hypergeometric function is formulated as
\begin{eqnarray}
&&\Phi_{_{\{1,2,3\}}}^{(4)}({\boldsymbol{\beta}},\;\boldsymbol{{\xi}})=A_{_{\{1,2,3\}}}^{(4)}({\boldsymbol{\beta}})(r_{_2})^{-\beta_{_2}}(r_{_3})^{\beta_{_2}+\beta_{_5}-1}
\varphi_{_{\{1,2,3\}}}^{(4)}({\boldsymbol{\beta}},\;{r_{_3}\over r_{_2}},\;{r_{_1}\over r_{_3}})\;,\nonumber\\
&&\varphi_{_{\{1,2,3\}}}^{(4)}({\boldsymbol{\beta}},\;x_{_1},\;x_{_2})=\sum\limits_{n_{_1},n_{_2}}c_{_{\{1,2,3\}}}^{(4)}({\boldsymbol{\beta}},\;n_{_1},n_{_2})
x_{_1}^{n_{_1}}x_{_2}^{n_{_2}}\;,
\label{Interger-lattice4-2}
\end{eqnarray}
where
\begin{eqnarray}
&&A_{_{\{1,2,3\}}}^{(4)}({\boldsymbol{\beta}})={\Gamma(\beta_{_4})\Gamma(\beta_{_5})\over\Gamma(1-\beta_{_1})\Gamma(1-\beta_{_2})\Gamma(\beta_{_1}+\beta_{_4})\Gamma(\beta_{_2}+\beta_{_5})}\;,\nonumber\\
&&c_{_{\{1,2,3\}}}^{(4)}({\boldsymbol{\beta}},\;n_{_1},n_{_2})={(-1)^{n_{_1}+n_{_2}}(\beta_{_2})_{_{n_{_1}}}(\beta_{_1})_{_{n_{_2}}}\over n_{_1}!n_{_2}!
(\beta_{_1}+\beta_{_4})_{_{-n_{_1}+n_{_2}}}(\beta_{_2}+\beta_{_5})_{_{n_{_1}-n_{_2}}}}\;.
\label{Interger-lattice4-3}
\end{eqnarray}

As the integer lattice is chosen as $n_{_1}E_{_3}^{(1)}- n_{_2}E_{_3}^{(3)}$, the generalized hypergeometric function is formulated as
\begin{eqnarray}
&&\Phi_{_{\{1,2,3\}}}^{(5)}({\boldsymbol{\beta}},\;\boldsymbol{{\xi}})=A_{_{\{1,2,3\}}}^{(5)}({\boldsymbol{\beta}})(r_{_2})^{\beta_{_5}-1}
\varphi_{_{\{1,2,3\}}}^{(5)}({\boldsymbol{\beta}},\;{r_{_3}\over r_{_2}},\;{r_{_1}\over r_{_2}})\;,\nonumber\\
&&\varphi_{_{\{1,2,3\}}}^{(5)}({\boldsymbol{\beta}},\;x_{_1},\;x_{_2})=\sum\limits_{n_{_1},n_{_2}}c_{_{\{1,2,3\}}}^{(5)}({\boldsymbol{\beta}},\;n_{_1},n_{_2})
x_{_1}^{n_{_1}}x_{_2}^{n_{_2}}\;,
\label{Interger-lattice5-2}
\end{eqnarray}
where
\begin{eqnarray}
&&A_{_{\{1,2,3\}}}^{(5)}({\boldsymbol{\beta}})={\Gamma(\beta_{_4})\over\Gamma(1-\beta_{_1})\Gamma(1-\beta_{_3})\Gamma(2-\beta_{_2}-\beta_{_5})}\;,\nonumber\\
&&c_{_{\{1,2,3\}}}^{(5)}({\boldsymbol{\beta}},\;n_{_1},n_{_2})={(\beta_{_3})_{_{n_{_1}}}(\beta_{_1})_{_{n_{_2}}}(1-\beta_{_5})_{_{n_{_1}+n_{_2}}}\over n_{_1}!n_{_2}!
(2-\beta_{_2}-\beta_{_5})_{_{n_{_1}+n_{_2}}}}\;.
\label{Interger-lattice5-3}
\end{eqnarray}

Corresponding to the integer lattice $n_{_1}E_{_3}^{(2)}- n_{_2}E_{_3}^{(3)}$,
the generalized hypergeometric function is formulated as
\begin{eqnarray}
&&\Phi_{_{\{1,2,3\}}}^{(6)}({\boldsymbol{\beta}},\;\boldsymbol{{\xi}})=A_{_{\{1,2,3\}}}^{(6)}({\boldsymbol{\beta}})(r_{_1})^{\beta_{_2}+\beta_{_5}-1}(r_{_2})^{-\beta_{_2}}
\varphi_{_{\{1,2,3\}}}^{(6)}({\boldsymbol{\beta}},\;{r_{_3}\over r_{_1}},\;{r_{_1}\over r_{_2}})\;,\nonumber\\
&&\varphi_{_{\{1,2,3\}}}^{(6)}({\boldsymbol{\beta}},\;x_{_1},\;x_{_2})=\sum\limits_{n_{_1},n_{_2}}c_{_{\{1,2,3\}}}^{(6)}({\boldsymbol{\beta}},\;n_{_1},n_{_2})
x_{_1}^{n_{_1}}x_{_2}^{n_{_2}}\;,
\label{Interger-lattice6-2}
\end{eqnarray}
where
\begin{eqnarray}
&&A_{_{\{1,2,3\}}}^{(6)}({\boldsymbol{\beta}})={\Gamma(\beta_{_4})\Gamma(\beta_{_5})\over\Gamma(1-\beta_{_2})\Gamma(1-\beta_{_3})\Gamma(\beta_{_2}+\beta_{_5})\Gamma(\beta_{_3}+\beta_{_4})}\;,\nonumber\\
&&c_{_{\{1,2,3\}}}^{(6)}({\boldsymbol{\beta}},\;n_{_1},n_{_2})={(-1)^{n_{_1}+n_{_2}}(\beta_{_3})_{_{n_{_1}}}(\beta_{_2})_{_{n_{_2}}}\over n_{_1}!n_{_2}!
(\beta_{_2}+\beta_{_5})_{_{-n_{_1}+n_{_2}}}(\beta_{_3}+\beta_{_4})_{_{n_{_1}-n_{_2}}}}\;.
\label{Interger-lattice6-3}
\end{eqnarray}

When the integer lattice is formulated as $-n_{_1}E_{_3}^{(1)}+ n_{_2}E_{_3}^{(2)}$,
the generalized hypergeometric function is formulated as
\begin{eqnarray}
&&\Phi_{_{\{1,2,3\}}}^{(7)}({\boldsymbol{\beta}},\;\boldsymbol{{\xi}})=A_{_{\{1,2,3\}}}^{(7)}({\boldsymbol{\beta}})(r_{_1})^{-\beta_{_1}}(r_{_3})^{\beta_{_1}+\beta_{_5}-1}
\varphi_{_{\{1,2,3\}}}^{(7)}({\boldsymbol{\beta}},\;{r_{_2}\over r_{_3}},\;{r_{_3}\over r_{_1}})\;,\nonumber\\
&&\varphi_{_{\{1,2,3\}}}^{(7)}({\boldsymbol{\beta}},\;x_{_1},\;x_{_2})=\sum\limits_{n_{_1},n_{_2}}c_{_{\{1,2,3\}}}^{(7)}({\boldsymbol{\beta}},\;n_{_1},n_{_2})
x_{_1}^{n_{_1}}x_{_2}^{n_{_2}}\;,
\label{Interger-lattice7-2}
\end{eqnarray}
where
\begin{eqnarray}
&&A_{_{\{1,2,3\}}}^{(7)}({\boldsymbol{\beta}})={\Gamma(\beta_{_4})\Gamma(\beta_{_5})\over\Gamma(1-\beta_{_1})\Gamma(1-\beta_{_2})\Gamma(\beta_{_1}+\beta_{_5})\Gamma(\beta_{_2}+\beta_{_4})}\;,\nonumber\\
&&c_{_{\{1,2,3\}}}^{(7)}({\boldsymbol{\beta}},\;n_{_1},n_{_2})={(-1)^{n_{_1}+n_{_2}}(\beta_{_2})_{_{n_{_1}}}(\beta_{_1})_{_{n_{_2}}}\over n_{_1}!n_{_2}!
(\beta_{_1}+\beta_{_5})_{_{-n_{_1}+n_{_2}}}(\beta_{_2}+\beta_{_4})_{_{n_{_1}-n_{_2}}}}\;.
\label{Interger-lattice7-3}
\end{eqnarray}

Corresponding to the integer lattice $-n_{_1}E_{_3}^{(1)}+ n_{_2}E_{_3}^{(3)}$, the generalized hypergeometric function is formulated as
\begin{eqnarray}
&&\Phi_{_{\{1,2,3\}}}^{(8)}({\boldsymbol{\beta}},\;\boldsymbol{{\xi}})=A_{_{\{1,2,3\}}}^{(8)}({\boldsymbol{\beta}})(r_{_1})^{-\beta_{_1}}(r_{_2})^{1-\beta_{_2}-\beta_{_4}}(r_{_3})^{-\beta_{_3}}
\varphi_{_{\{1,2,3\}}}^{(8)}({\boldsymbol{\beta}},\;{r_{_2}\over r_{_3}},\;{r_{_2}\over r_{_1}})\;,\nonumber\\
&&\varphi_{_{\{1,2,3\}}}^{(8)}({\boldsymbol{\beta}},\;x_{_1},\;x_{_2})=\sum\limits_{n_{_1},n_{_2}}c_{_{\{1,2,3\}}}^{(8)}({\boldsymbol{\beta}},\;n_{_1},n_{_2})
x_{_1}^{n_{_1}}x_{_2}^{n_{_2}}\;,
\label{Interger-lattice8-2}
\end{eqnarray}
where
\begin{eqnarray}
&&A_{_{\{1,2,3\}}}^{(8)}({\boldsymbol{\beta}})={\Gamma(\beta_{_5})\over\Gamma(1-\beta_{_1})\Gamma(1-\beta_{_3})\Gamma(2-\beta_{_2}-\beta_{_4})}\;,\nonumber\\
&&c_{_{\{1,2,3\}}}^{(8)}({\boldsymbol{\beta}},\;n_{_1},n_{_2})={(\beta_{_3})_{_{n_{_1}}}(\beta_{_1})_{_{n_{_2}}}(1-\beta_{_4})_{_{n_{_1}+n_{_2}}}\over n_{_1}!n_{_2}!
(2-\beta_{_2}-\beta_{_4})_{_{n_{_1}+n_{_2}}}}\;.
\label{Interger-lattice8-3}
\end{eqnarray}

For the integer lattice $-n_{_1}E_{_3}^{(2)}+ n_{_2}E_{_3}^{(3)}$, the generalized hypergeometric function is formulated as
\begin{eqnarray}
&&\Phi_{_{\{1,2,3\}}}^{(9)}({\boldsymbol{\beta}},\;\boldsymbol{{\xi}})=A_{_{\{1,2,3\}}}^{(9)}({\boldsymbol{\beta}})(r_{_1})^{\beta_{_3}+\beta_{_5}-1}(r_{_3})^{-\beta_{_3}}
\varphi_{_{\{1,2,3\}}}^{(9)}({\boldsymbol{\beta}},\;{r_{_1}\over r_{_3}},\;{r_{_2}\over r_{_1}})\;,\nonumber\\
&&\varphi_{_{\{1,2,3\}}}^{(9)}({\boldsymbol{\beta}},\;x_{_1},\;x_{_2})=\sum\limits_{n_{_1},n_{_2}}c_{_{\{1,2,3\}}}^{(9)}({\boldsymbol{\beta}},\;n_{_1},n_{_2})
x_{_1}^{n_{_1}}x_{_2}^{n_{_2}}\;,
\label{Interger-lattice9-2}
\end{eqnarray}
where
\begin{eqnarray}
&&A_{_{\{1,2,3\}}}^{(9)}({\boldsymbol{\beta}})={\Gamma(\beta_{_4})\Gamma(\beta_{_5})\over\Gamma(1-\beta_{_2})\Gamma(1-\beta_{_3})\Gamma(\beta_{_2}+\beta_{_4})\Gamma(\beta_{_3}+\beta_{_5})}\;,\nonumber\\
&&c_{_{\{1,2,3\}}}^{(9)}({\boldsymbol{\beta}},\;n_{_1},n_{_2})={(-1)^{n_{_1}+n_{_2}}(\beta_{_3})_{_{n_{_1}}}(\beta_{_2})_{_{n_{_2}}}\over n_{_1}!n_{_2}!
(\beta_{_3}+\beta_{_5})_{_{n_{_1}-n_{_2}}}(\beta_{_2}+\beta_{_4})_{_{-n_{_1}+n_{_2}}}}\;.
\label{Interger-lattice9-3}
\end{eqnarray}

As the integer lattice is given by $-n_{_1}E_{_3}^{(1)}- n_{_2}E_{_3}^{(2)}$,
the generalized hypergeometric function is formulated as
\begin{eqnarray}
&&\Phi_{_{\{1,2,3\}}}^{(10)}({\boldsymbol{\beta}},\;\boldsymbol{{\xi}})=A_{_{\{1,2,3\}}}^{(10)}({\boldsymbol{\beta}})(r_{_3})^{\beta_{_5}-1}
\varphi_{_{\{1,2,3\}}}^{(10)}({\boldsymbol{\beta}},\;{r_{_2}\over r_{_3}},\;{r_{_1}\over r_{_3}})\;,\nonumber\\
&&\varphi_{_{\{1,2,3\}}}^{(10)}({\boldsymbol{\beta}},\;x_{_1},\;x_{_2})=\sum\limits_{n_{_1},n_{_2}}c_{_{\{1,2,3\}}}^{(10)}({\boldsymbol{\beta}},\;n_{_1},n_{_2})
x_{_1}^{n_{_1}}x_{_2}^{n_{_2}}\;,
\label{Interger-lattice10-2}
\end{eqnarray}
where
\begin{eqnarray}
&&A_{_{\{1,2,3\}}}^{(10)}({\boldsymbol{\beta}})={\Gamma(\beta_{_4})\over\Gamma(1-\beta_{_1})\Gamma(1-\beta_{_2})\Gamma(2-\beta_{_3}-\beta_{_5})}\;,\nonumber\\
&&c_{_{\{1,2,3\}}}^{(10)}({\boldsymbol{\beta}},\;n_{_1},n_{_2})={(\beta_{_2})_{_{n_{_1}}}(\beta_{_1})_{_{n_{_2}}}(1-\beta_{_5})_{_{n_{_1}+n_{_2}}}\over n_{_1}!n_{_2}!
(2-\beta_{_3}-\beta_{_5})_{_{n_{_1}+n_{_2}}}}\;.
\label{Interger-lattice10-3}
\end{eqnarray}

Corresponding to the integer lattice $-n_{_1}E_{_3}^{(1)}- n_{_2}E_{_3}^{(3)}$, the generalized hypergeometric function is formulated as
\begin{eqnarray}
&&\Phi_{_{\{1,2,3\}}}^{(11)}({\boldsymbol{\beta}},\;\boldsymbol{{\xi}})=A_{_{\{1,2,3\}}}^{(11)}({\boldsymbol{\beta}})(r_{_2})^{\beta_{_3}+\beta_{_5}-1}(r_{_3})^{-\beta_{_3}}
\varphi_{_{\{1,2,3\}}}^{(11)}({\boldsymbol{\beta}},\;{r_{_2}\over r_{_3}},\;{r_{_1}\over r_{_2}})\;,\nonumber\\
&&\varphi_{_{\{1,2,3\}}}^{(11)}({\boldsymbol{\beta}},\;x_{_1},\;x_{_2})=\sum\limits_{n_{_1},n_{_2}}c_{_{\{1,2,3\}}}^{(11)}({\boldsymbol{\beta}},\;n_{_1},n_{_2})
x_{_1}^{n_{_1}}x_{_2}^{n_{_2}}\;,
\label{Interger-lattice11-2}
\end{eqnarray}
where
\begin{eqnarray}
&&A_{_{\{1,2,3\}}}^{(11)}({\boldsymbol{\beta}})={\Gamma(\beta_{_4})\Gamma(\beta_{_5})\over\Gamma(1-\beta_{_1})\Gamma(1-\beta_{_3})\Gamma(\beta_{_1}+\beta_{_4})\Gamma(\beta_{_3}+\beta_{_5})}\;,\nonumber\\
&&c_{_{\{1,2,3\}}}^{(11)}({\boldsymbol{\beta}},\;n_{_1},n_{_2})={(-1)^{n_{_1}+n_{_2}}(\beta_{_3})_{_{n_{_1}}}(\beta_{_1})_{_{n_{_2}}}\over n_{_1}!n_{_2}!
(\beta_{_3}+\beta_{_5})_{_{n_{_1}-n_{_2}}}(\beta_{_1}+\beta_{_4})_{_{-n_{_1}+n_{_2}}}}\;.
\label{Interger-lattice11-3}
\end{eqnarray}

For the integer lattice $-n_{_1}E_{_3}^{(2)}- n_{_2}E_{_3}^{(3)}$, the generalized hypergeometric function is formulated as
\begin{eqnarray}
&&\Phi_{_{\{1,2,3\}}}^{(12)}({\boldsymbol{\beta}},\;\boldsymbol{{\xi}})=A_{_{\{1,2,3\}}}^{(12)}({\boldsymbol{\beta}})(r_{_1})^{1-\beta_{_1}-\beta_{_4}}(r_{_2})^{-\beta_{_2}}(r_{_3})^{-\beta_{_3}}
\varphi_{_{\{1,2,3\}}}^{(12)}({\boldsymbol{\beta}},\;{r_{_1}\over r_{_3}},\;{r_{_1}\over r_{_2}})\;,\nonumber\\
&&\varphi_{_{\{1,2,3\}}}^{(12)}({\boldsymbol{\beta}},\;x_{_1},\;x_{_2})=\sum\limits_{n_{_1},n_{_2}}c_{_{\{1,2,3\}}}^{(12)}({\boldsymbol{\beta}},\;n_{_1},n_{_2})
x_{_1}^{n_{_1}}x_{_2}^{n_{_2}}\;,
\label{Interger-lattice12-2}
\end{eqnarray}
where
\begin{eqnarray}
&&A_{_{\{1,2,3\}}}^{(12)}({\boldsymbol{\beta}})={\Gamma(\beta_{_5})\over\Gamma(1-\beta_{_2})\Gamma(1-\beta_{_3})\Gamma(2-\beta_{_1}-\beta_{_4})}\;,\nonumber\\
&&c_{_{\{1,2,3\}}}^{(12)}({\boldsymbol{\beta}},\;n_{_1},n_{_2})={(\beta_{_3})_{_{n_{_1}}}(\beta_{_2})_{_{n_{_2}}}(1-\beta_{_4})_{_{n_{_1}+n_{_2}}}\over n_{_1}!n_{_2}!
(2-\beta_{_1}-\beta_{_4})_{_{n_{_1}+n_{_2}}}}\;.
\label{Interger-lattice12-3}
\end{eqnarray}

\subsection{${\bf{\cal B}}=\{1,2,4\}$\label{sec6-2}}
\indent\indent
\begin{eqnarray}
&&{\rm Det}({\boldsymbol{\xi}}_{_{\{1,2,4\}}})=1
\;,\nonumber\\
&&{\boldsymbol{\xi}}_{_{\{1,2,4\}}}^{-1}\cdot {\boldsymbol{\xi}}=
\left(\begin{array}{ccccc}\;1\;&\;0\;&\;-1\;&\;0\;&\;r_{_1}-r_{_3}\;\\\;0\;&\;1\;&\;-1\;&\;0\;&\;r_{_2}-r_{_3}\;\\
\;0\;&\;0\;&\;1\;&\;1\;&\;r_{_3}\;\end{array}\right)\;.
\label{Gauss-Kummer6-2-1}
\end{eqnarray}

Corresponding to the integer lattice $n_{_1}E_{_3}^{(1)}+ n_{_2}E_{_3}^{(2)}$, the generalized hypergeometric function is formulated as
\begin{eqnarray}
&&\Phi_{_{\{1,2,4\}}}^{(1)}({\boldsymbol{\beta}},\;\boldsymbol{{\xi}})=A_{_{\{1,2,4\}}}^{(1)}({\boldsymbol{\beta}})(r_{_1}-r_{_3})^{-\beta_{_1}}
(r_{_2}-r_{_3})^{-\beta_{_2}}(r_{_3})^{1-\beta_{_3}-\beta_{_4}}
\nonumber\\
&&\hspace{3.2cm}\times
\varphi_{_{\{1,2,4\}}}^{(1)}({\boldsymbol{\beta}},\;{r_{_3}\over r_{_3}-r_{_2}},\;{r_{_3}\over r_{_3}-r_{_1}})\;,\nonumber\\
&&\varphi_{_{\{1,2,4\}}}^{(1)}({\boldsymbol{\beta}},\;x_{_1},\;x_{_2})=\varphi_{_{\{1,2,3\}}}^{(1)}(\widehat{(34)}{\boldsymbol{\beta}},\;x_{_1},\;x_{_2})\;.
\label{Gauss-Kummer6-2-2b}
\end{eqnarray}

As the integer lattice is $n_{_1}E_{_3}^{(1)}+ n_{_2}E_{_3}^{(3)}$, the generalized hypergeometric function is formulated as
\begin{eqnarray}
&&\Phi_{_{\{1,2,4\}}}^{(2)}({\boldsymbol{\beta}},\;\boldsymbol{{\xi}})=A_{_{\{1,2,4\}}}^{(2)}({\boldsymbol{\beta}})(r_{_1}-r_{_3})^{-\beta_{_1}}(r_{_2}-r_{_3})^{\beta_{_1}+\beta_{_5}-1}
\nonumber\\
&&\hspace{3.2cm}\times
\varphi_{_{\{1,2,4\}}}^{(2)}({\boldsymbol{\beta}},\;{r_{_3}\over r_{_3}-r_{_2}},\;{r_{_3}-r_{_2}\over r_{_3}-r_{_1}})\;,\nonumber\\
&&\varphi_{_{\{1,2,4\}}}^{(2)}({\boldsymbol{\beta}},\;x_{_1},\;x_{_2})=\varphi_{_{\{1,2,3\}}}^{(2)}(\widehat{(34)}{\boldsymbol{\beta}},\;x_{_1},\;x_{_2})\;.
\label{Gauss-Kummer6-2-3b}
\end{eqnarray}

As the integer lattice is $n_{_1}E_{_3}^{(2)}+ n_{_2}E_{_3}^{(3)}$, the generalized hypergeometric function is written as
\begin{eqnarray}
&&\Phi_{_{\{1,2,4\}}}^{(3)}({\boldsymbol{\beta}},\;\boldsymbol{{\xi}})=A_{_{\{1,2,4\}}}^{(3)}({\boldsymbol{\beta}})(r_{_1}-r_{_3})^{\beta_{_5}-1}
\nonumber\\
&&\hspace{3.2cm}\times
\varphi_{_{\{1,2,4\}}}^{(3)}({\boldsymbol{\beta}},\;{r_{_3}\over r_{_3}-r_{_1}},\;{r_{_3}-r_{_2}\over r_{_3}-r_{_1}})\;,\nonumber\\
&&\varphi_{_{\{1,2,4\}}}^{(3)}({\boldsymbol{\beta}},\;x_{_1},\;x_{_2})=\varphi_{_{\{1,2,3\}}}^{(3)}(\widehat{(34)}{\boldsymbol{\beta}},\;x_{_1},\;x_{_2})\;.
\label{Gauss-Kummer6-2-4b}
\end{eqnarray}

When the integer lattice is $n_{_1}E_{_3}^{(1)}- n_{_2}E_{_3}^{(2)}$, the generalized hypergeometric function is formulated as
\begin{eqnarray}
&&\Phi_{_{\{1,2,4\}}}^{(4)}({\boldsymbol{\beta}},\;\boldsymbol{{\xi}})=A_{_{\{1,2,3\}}}^{(4)}({\boldsymbol{\beta}})(r_{_2}-r_{_3})^{-\beta_{_2}}(r_{_3})^{\beta_{_2}+\beta_{_5}-1}
\nonumber\\
&&\hspace{3.2cm}\times
\varphi_{_{\{1,2,4\}}}^{(4)}({\boldsymbol{\beta}},\;{r_{_3}\over r_{_3}-r_{_2}},\;{r_{_3}-r_{_1}\over r_{_3}})\;,\nonumber\\
&&\varphi_{_{\{1,2,4\}}}^{(4)}({\boldsymbol{\beta}},\;x_{_1},\;x_{_2})=\varphi_{_{\{1,2,3\}}}^{(4)}(\widehat{(34)}{\boldsymbol{\beta}},\;x_{_1},\;x_{_2})\;.
\label{Gauss-Kummer6-2-5b}
\end{eqnarray}

Corresponding to the integer lattice $n_{_1}E_{_3}^{(1)}- n_{_2}E_{_3}^{(3)}$, the generalized hypergeometric function is formulated as
\begin{eqnarray}
&&\Phi_{_{\{1,2,4\}}}^{(5)}({\boldsymbol{\beta}},\;\boldsymbol{{\xi}})=A_{_{\{1,2,4\}}}^{(5)}({\boldsymbol{\beta}})(r_{_2}-r_{_3})^{\beta_{_5}-1}
\nonumber\\
&&\hspace{3.2cm}\times
\varphi_{_{\{1,2,4\}}}^{(5)}({\boldsymbol{\beta}},\;{r_{_3}\over r_{_3}-r_{_2}},\;{r_{_3}-r_{_1}\over r_{_3}-r_{_2}})\;,\nonumber\\
&&\varphi_{_{\{1,2,4\}}}^{(5)}({\boldsymbol{\beta}},\;x_{_1},\;x_{_2})=\varphi_{_{\{1,2,3\}}}^{(5)}(\widehat{(34)}{\boldsymbol{\beta}},\;x_{_1},\;x_{_2})\;.
\label{Gauss-Kummer6-2-6b}
\end{eqnarray}

Corresponding to the integer lattice $n_{_1}E_{_3}^{(2)}- n_{_2}E_{_3}^{(3)}$, the generalized hypergeometric function is formulated as
\begin{eqnarray}
&&\Phi_{_{\{1,2,4\}}}^{(6)}({\boldsymbol{\beta}},\;\boldsymbol{{\xi}})=A_{_{\{1,2,3\}}}^{(6)}({\boldsymbol{\beta}})(r_{_1}-r_{_3})^{\beta_{_2}+\beta_{_5}-1}(r_{_2}-r_{_3})^{-\beta_{_2}}
\nonumber\\
&&\hspace{3.2cm}\times
\varphi_{_{\{1,2,4\}}}^{(6)}({\boldsymbol{\beta}},\;{r_{_3}\over r_{_3}-r_{_1}},\;{r_{_3}-r_{_1}\over r_{_3}-r_{_2}})\;,\nonumber\\
&&\varphi_{_{\{1,2,4\}}}^{(6)}({\boldsymbol{\beta}},\;x_{_1},\;x_{_2})=\varphi_{_{\{1,2,3\}}}^{(6)}(\widehat{(34)}{\boldsymbol{\beta}},\;x_{_1},\;x_{_2})\;.
\label{Gauss-Kummer6-2-7b}
\end{eqnarray}

When the integer lattice is $-n_{_1}E_{_3}^{(1)}+ n_{_2}E_{_3}^{(2)}$, the generalized hypergeometric function is formulated as
\begin{eqnarray}
&&\Phi_{_{\{1,2,4\}}}^{(7)}({\boldsymbol{\beta}},\;\boldsymbol{{\xi}})=A_{_{\{1,2,3\}}}^{(7)}({\boldsymbol{\beta}})(r_{_1}-r_{_3})^{-\beta_{_1}}(r_{_3})^{\beta_{_1}+\beta_{_5}-1}
\nonumber\\
&&\hspace{3.2cm}\times
\varphi_{_{\{1,2,4\}}}^{(7)}({\boldsymbol{\beta}},\;{r_{_3}-r_{_2}\over r_{_3}},\;{r_{_3}\over r_{_3}-r_{_1}})\;,\nonumber\\
&&\varphi_{_{\{1,2,4\}}}^{(7)}({\boldsymbol{\beta}},\;x_{_1},\;x_{_2})=\varphi_{_{\{1,2,3\}}}^{(7)}(\widehat{(34)}{\boldsymbol{\beta}},\;x_{_1},\;x_{_2})\;.
\label{Gauss-Kummer6-2-8b}
\end{eqnarray}

When the integer lattice is $-n_{_1}E_{_3}^{(1)}+ n_{_2}E_{_3}^{(3)}$, the generalized hypergeometric function is formulated as
\begin{eqnarray}
&&\Phi_{_{\{1,2,4\}}}^{(8)}({\boldsymbol{\beta}},\;\boldsymbol{{\xi}})=A_{_{\{1,2,4\}}}^{(8)}({\boldsymbol{\beta}})(r_{_1}-r_{_3})^{-\beta_{_1}}(r_{_2}-r_{_3})^{1-\beta_{_2}-\beta_{_3}}(r_{_3})^{-\beta_{_4}}
\nonumber\\
&&\hspace{3.2cm}\times
\varphi_{_{\{1,2,4\}}}^{(8)}({\boldsymbol{\beta}},\;{r_{_3}-r_{_2}\over r_{_3}},\;{r_{_3}-r_{_2}\over r_{_3}-r_{_1}})\;,\nonumber\\
&&\varphi_{_{\{1,2,4\}}}^{(8)}({\boldsymbol{\beta}},\;x_{_1},\;x_{_2})=\varphi_{_{\{1,2,3\}}}^{(8)}(\widehat{(34)}{\boldsymbol{\beta}},\;x_{_1},\;x_{_2})\;.
\label{Gauss-Kummer6-2-9b}
\end{eqnarray}

As the integer lattice is $-n_{_1}E_{_3}^{(2)}+ n_{_2}E_{_3}^{(3)}$, the generalized hypergeometric function is formulated as
\begin{eqnarray}
&&\Phi_{_{\{1,2,4\}}}^{(9)}({\boldsymbol{\beta}},\;\boldsymbol{{\xi}})=A_{_{\{1,2,4\}}}^{(9)}({\boldsymbol{\beta}})(r_{_1}-r_{_3})^{\beta_{_4}+\beta_{_5}-1}(r_{_3})^{-\beta_{_4}}
\nonumber\\
&&\hspace{3.2cm}\times
\varphi_{_{\{1,2,4\}}}^{(9)}({\boldsymbol{\beta}},\;{r_{_3}-r_{_1}\over r_{_3}},\;{r_{_3}-r_{_2}\over r_{_3}-r_{_1}})\;,\nonumber\\
&&\varphi_{_{\{1,2,4\}}}^{(9)}({\boldsymbol{\beta}},\;x_{_1},\;x_{_2})=\varphi_{_{\{1,2,3\}}}^{(9)}(\widehat{(34)}{\boldsymbol{\beta}},\;x_{_1},\;x_{_2})\;.
\label{Gauss-Kummer6-2-10b}
\end{eqnarray}

Corresponding to the integer lattice $-n_{_1}E_{_3}^{(1)}- n_{_2}E_{_3}^{(2)}$, the generalized hypergeometric function is formulated as
\begin{eqnarray}
&&\Phi_{_{\{1,2,4\}}}^{(10)}({\boldsymbol{\beta}},\;\boldsymbol{{\xi}})=A_{_{\{1,2,4\}}}^{(10)}({\boldsymbol{\beta}})(r_{_3})^{\beta_{_5}-1}
\nonumber\\
&&\hspace{3.2cm}\times
\varphi_{_{\{1,2,4\}}}^{(10)}({\boldsymbol{\beta}},\;{r_{_3}-r_{_2}\over r_{_3}},\;{r_{_3}-r_{_1}\over r_{_3}})\;,\nonumber\\
&&\varphi_{_{\{1,2,4\}}}^{(10)}({\boldsymbol{\beta}},\;x_{_1},\;x_{_2})=\varphi_{_{\{1,2,3\}}}^{(10)}(\widehat{(34)}{\boldsymbol{\beta}},\;x_{_1},\;x_{_2})\;.
\label{Gauss-Kummer6-2-11b}
\end{eqnarray}

When the integer lattice is $-n_{_1}E_{_3}^{(1)}- n_{_2}E_{_3}^{(3)}$, the generalized hypergeometric function is formulated as
\begin{eqnarray}
&&\Phi_{_{\{1,2,4\}}}^{(11)}({\boldsymbol{\beta}},\;\boldsymbol{{\xi}})=A_{_{\{1,2,4\}}}^{(11)}({\boldsymbol{\beta}})(r_{_2}-r_{_3})^{\beta_{_4}+\beta_{_5}-1}(r_{_3})^{-\beta_{_4}}
\nonumber\\
&&\hspace{3.2cm}\times
\varphi_{_{\{1,2,4\}}}^{(11)}({\boldsymbol{\beta}},\;{r_{_3}-r_{_2}\over r_{_3}},\;{r_{_3}-r_{_1}\over r_{_3}-r_{_2}})\;,\nonumber\\
&&\varphi_{_{\{1,2,4\}}}^{(11)}({\boldsymbol{\beta}},\;x_{_1},\;x_{_2})=\varphi_{_{\{1,2,3\}}}^{(11)}(\widehat{(34)}{\boldsymbol{\beta}},\;x_{_1},\;x_{_2})\;.
\label{Gauss-Kummer6-2-12b}
\end{eqnarray}

Corresponding to the integer lattice $-n_{_1}E_{_3}^{(2)}- n_{_2}E_{_3}^{(3)}$, the generalized hypergeometric function is formulated as
\begin{eqnarray}
&&\Phi_{_{\{1,2,4\}}}^{(12)}({\boldsymbol{\beta}},\;\boldsymbol{{\xi}})=A_{_{\{1,2,4\}}}^{(12)}({\boldsymbol{\beta}})(r_{_1}-r_{_3})^{1-\beta_{_1}-\beta_{_3}}(r_{_2}-r_{_3})^{-\beta_{_2}}(r_{_3})^{-\beta_{_4}}
\nonumber\\
&&\hspace{3.2cm}\times
\varphi_{_{\{1,2,4\}}}^{(12)}({\boldsymbol{\beta}},\;{r_{_3}-r_{_1}\over r_{_3}},\;{r_{_3}-r_{_1}\over r_{_3}-r_{_2}})\;,\nonumber\\
&&\varphi_{_{\{1,2,4\}}}^{(12)}({\boldsymbol{\beta}},\;x_{_1},\;x_{_2})=\varphi_{_{\{1,2,3\}}}^{(12)}(\widehat{(34)}{\boldsymbol{\beta}},\;x_{_1},\;x_{_2})\;.
\label{Gauss-Kummer6-2-13b}
\end{eqnarray}

\subsection{${\bf{\cal B}}=\{1,2,5\}$\label{sec6-3}}
\indent\indent
\begin{eqnarray}
&&{\rm Det}({\boldsymbol{\xi}}_{_{\{1,2,5\}}})=r_{_3}
\;,\nonumber\\
&&{\boldsymbol{\xi}}_{_{\{1,2,5\}}}^{-1}\cdot {\boldsymbol{\xi}}=
\left(\begin{array}{ccccc}\;1\;&\;0\;&\;-{r_{_1}\over r_{_3}}\;&\;1-{r_{_1}\over r_{_3}}\;&\;0\;\\\;0\;&\;1\;&\;-{r_{_2}\over r_{_3}}\;&\;1-{r_{_2}\over r_{_3}}\;&\;0\;\\
\;0\;&\;0\;&\;{1\over r_{_3}}\;&\;{1\over r_{_3}}\;&\;1\;\end{array}\right)\;.
\label{Gauss-Kummer6-3-1}
\end{eqnarray}

Corresponding to the integer lattice $n_{_1}E_{_3}^{(1)}+ n_{_2}E_{_3}^{(2)}$, the generalized hypergeometric function is formulated as
\begin{eqnarray}
&&\Phi_{_{\{1,2,5\}}}^{(1)}({\boldsymbol{\beta}},\;\boldsymbol{{\xi}})=A_{_{\{1,2,5\}}}^{(1)}({\boldsymbol{\beta}})(r_{_3}-r_{_1})^{-\beta_{_1}}(r_{_3}-r_{_2})^{-\beta_{_2}}(r_{_3})^{1-\beta_{_3}-\beta_{_4}}
\nonumber\\
&&\hspace{3.2cm}\times
\varphi_{_{\{1,2,5\}}}^{(1)}({\boldsymbol{\beta}},\;{r_{_2}\over r_{_2}-r_{_3}},\;{r_{_1}\over r_{_1}-r_{_3}})\;,\nonumber\\
&&\varphi_{_{\{1,2,5\}}}^{(1)}({\boldsymbol{\beta}},\;x_{_1},\;x_{_2})=\varphi_{_{\{1,2,3\}}}^{(1)}(\widehat{(354)}{\boldsymbol{\beta}},\;x_{_1},\;x_{_2})\;.
\label{Gauss-Kummer6-3-2b}
\end{eqnarray}

As the integer lattice is $n_{_1}E_{_3}^{(1)}+ n_{_2}E_{_3}^{(3)}$, the generalized hypergeometric function is formulated as
\begin{eqnarray}
&&\Phi_{_{\{1,2,5\}}}^{(2)}({\boldsymbol{\beta}},\;\boldsymbol{{\xi}})=A_{_{\{1,2,5\}}}^{(2)}({\boldsymbol{\beta}})
(r_{_3})^{1-\beta_{_3}-\beta_{_4}}(-r_{_2})^{\beta_{_3}+\beta_{_5}-1}(r_{_3}-r_{_1})^{-\beta_{_1}}
\nonumber\\
&&\hspace{3.2cm}\times
(r_{_3}-r_{_2})^{\beta_{_1}+\beta_{_4}-1}\varphi_{_{\{1,2,5\}}}^{(2)}({\boldsymbol{\beta}},\;{r_{_2}\over r_{_2}-r_{_3}},\;{r_{_1}(r_{_2}-r_{_3})\over r_{_2}(r_{_1}-r_{_3})})\;,\nonumber\\
&&\varphi_{_{\{1,2,5\}}}^{(2)}({\boldsymbol{\beta}},\;x_{_1},\;x_{_2})=\varphi_{_{\{1,2,3\}}}^{(2)}(\widehat{(354)}{\boldsymbol{\beta}},\;x_{_1},\;x_{_2})\;.
\label{Gauss-Kummer6-3-3b}
\end{eqnarray}

When the integer lattice is $n_{_1}E_{_3}^{(2)}+ n_{_2}E_{_3}^{(3)}$, the generalized hypergeometric function is formulated as
\begin{eqnarray}
&&\Phi_{_{\{1,2,5\}}}^{(3)}({\boldsymbol{\beta}},\;\boldsymbol{{\xi}})=A_{_{\{1,2,5\}}}^{(3)}({\boldsymbol{\beta}})(r_{_3})^{1-\beta_{_3}-\beta_{_4}}
(-r_{_1})^{1-\beta_{_1}-\beta_{_4}}(-r_{_2})^{-\beta_{_2}}(r_{_3}-r_{_1})^{\beta_{_4}-1}
\nonumber\\
&&\hspace{3.2cm}\times
\varphi_{_{\{1,2,5\}}}^{(3)}({\boldsymbol{\beta}},\;{r_{_1}\over r_{_1}-r_{_3}},\;{r_{_1}(r_{_2}-r_{_3})\over r_{_2}(r_{_1}-r_{_3})})\;,\nonumber\\
&&\varphi_{_{\{1,2,5\}}}^{(3)}({\boldsymbol{\beta}},\;x_{_1},\;x_{_2})=\varphi_{_{\{1,2,3\}}}^{(3)}(\widehat{(354)}{\boldsymbol{\beta}},\;x_{_1},\;x_{_2})\;.
\label{Gauss-Kummer6-3-4b}
\end{eqnarray}

Corresponding to the integer lattice $n_{_1}E_{_3}^{(1)}- n_{_2}E_{_3}^{(2)}$, the generalized hypergeometric function is formulated as
\begin{eqnarray}
&&\Phi_{_{\{1,2,5\}}}^{(4)}({\boldsymbol{\beta}},\;\boldsymbol{{\xi}})=A_{_{\{1,2,5\}}}^{(4)}({\boldsymbol{\beta}})(r_{_3})^{1-\beta_{_3}-\beta_{_4}}
(-r_{_1})^{-\beta_{_1}}(r_{_3}-r_{_2})^{-\beta_{_2}}
\nonumber\\
&&\hspace{3.2cm}\times
\varphi_{_{\{1,2,5\}}}^{(4)}({\boldsymbol{\beta}},\;{r_{_2}\over r_{_2}-r_{_3}},\;{r_{_1}-r_{_3}\over r_{_1}})\;,\nonumber\\
&&\varphi_{_{\{1,2,5\}}}^{(4)}({\boldsymbol{\beta}},\;x_{_1},\;x_{_2})=\varphi_{_{\{1,2,3\}}}^{(4)}(\widehat{(354)}{\boldsymbol{\beta}},\;x_{_1},\;x_{_2})\;.
\label{Gauss-Kummer6-3-5b}
\end{eqnarray}

Corresponding to the integer lattice $n_{_1}E_{_3}^{(1)}- n_{_2}E_{_3}^{(3)}$, the generalized hypergeometric function is formulated as
\begin{eqnarray}
&&\Phi_{_{\{1,2,5\}}}^{(5)}({\boldsymbol{\beta}},\;\boldsymbol{{\xi}})=A_{_{\{1,2,5\}}}^{(5)}({\boldsymbol{\beta}})
(r_{_3})^{1-\beta_{_3}-\beta_{_4}}(-r_{_1})^{-\beta_{_1}}(-r_{_2})^{1-\beta_{_2}-\beta_{_4}}(r_{_3}-r_{_2})^{\beta_{_4}-1}
\nonumber\\
&&\hspace{3.2cm}\times
\varphi_{_{\{1,2,5\}}}^{(5)}({\boldsymbol{\beta}},\;{r_{_2}\over r_{_2}-r_{_3}},\;{r_{_2}(r_{_1}-r_{_3})\over r_{_1}(r_{_2}-r_{_3})})\;,\nonumber\\
&&\varphi_{_{\{1,2,5\}}}^{(5)}({\boldsymbol{\beta}},\;x_{_1},\;x_{_2})=\varphi_{_{\{1,2,3\}}}^{(5)}(\widehat{(354)}{\boldsymbol{\beta}},\;x_{_1},\;x_{_2})\;.
\label{Gauss-Kummer6-3-6b}
\end{eqnarray}

When the integer lattice is $n_{_1}E_{_3}^{(2)}- n_{_2}E_{_3}^{(3)}$, the generalized hypergeometric function is formulated as
\begin{eqnarray}
&&\Phi_{_{\{1,2,5\}}}^{(6)}({\boldsymbol{\beta}},\;\boldsymbol{{\xi}})=A_{_{\{1,2,5\}}}^{(6)}({\boldsymbol{\beta}})
(r_{_3})^{1-\beta_{_3}-\beta_{_4}}(-r_{_1})^{\beta_{_3}+\beta_{_5}-1}(r_{_3}-r_{_1})^{\beta_{_2}+\beta_{_4}-1}
\nonumber\\
&&\hspace{3.2cm}\times
(r_{_3}-r_{_2})^{-\beta_{_2}}
\varphi_{_{\{1,2,5\}}}^{(6)}({\boldsymbol{\beta}},\;{r_{_1}\over r_{_1}-r_{_3}},\;{r_{_2}(r_{_1}-r_{_3})\over r_{_1}(r_{_2}-r_{_3})})\;,\nonumber\\
&&\varphi_{_{\{1,2,5\}}}^{(6)}({\boldsymbol{\beta}},\;x_{_1},\;x_{_2})=\varphi_{_{\{1,2,3\}}}^{(6)}(\widehat{(354)}{\boldsymbol{\beta}},\;x_{_1},\;x_{_2})\;.
\label{Gauss-Kummer6-3-7b}
\end{eqnarray}

As the integer lattice is $-n_{_1}E_{_3}^{(1)}+ n_{_2}E_{_3}^{(2)}$, the generalized hypergeometric function is formulated as
\begin{eqnarray}
&&\Phi_{_{\{1,2,5\}}}^{(7)}({\boldsymbol{\beta}},\;\boldsymbol{{\xi}})=A_{_{\{1,2,3\}}}^{(7)}({\boldsymbol{\beta}})
(r_{_3})^{1-\beta_{_3}-\beta_{_4}}(-r_{_2})^{-\beta_{_2}}(r_{_3}-r_{_1})^{-\beta_{_1}}
\nonumber\\
&&\hspace{3.2cm}\times
\varphi_{_{\{1,2,5\}}}^{(7)}({\boldsymbol{\beta}},\;{r_{_2}-r_{_3}\over r_{_2}},\;{r_{_1}\over r_{_1}-r_{_3}})\;,\nonumber\\
&&\varphi_{_{\{1,2,5\}}}^{(7)}({\boldsymbol{\beta}},\;x_{_1},\;x_{_2})=\varphi_{_{\{1,2,3\}}}^{(7)}(\widehat{(354)}{\boldsymbol{\beta}},\;x_{_1},\;x_{_2})\;.
\label{Gauss-Kummer6-3-8b}
\end{eqnarray}

When the integer lattice is $-n_{_1}E_{_3}^{(1)}+ n_{_2}E_{_3}^{(3)}$, the generalized hypergeometric function is formulated as
\begin{eqnarray}
&&\Phi_{_{\{1,2,5\}}}^{(8)}({\boldsymbol{\beta}},\;\boldsymbol{{\xi}})=A_{_{\{1,2,5\}}}^{(8)}({\boldsymbol{\beta}})
(r_{_3})^{1-\beta_{_3}-\beta_{_4}}(-r_{_2})^{\beta_{_3}-1}(r_{_3}-r_{_1})^{-\beta_{_1}}
\nonumber\\
&&\hspace{3.2cm}\times
(r_{_3}-r_{_2})^{1-\beta_{_2}-\beta_{_3}}
\varphi_{_{\{1,2,5\}}}^{(8)}({\boldsymbol{\beta}},\;{r_{_2}-r_{_3}\over r_{_2}},\;{r_{_1}(r_{_2}-r_{_3})\over r_{_2}(r_{_1}-r_{_3})})\;,\nonumber\\
&&\varphi_{_{\{1,2,5\}}}^{(8)}({\boldsymbol{\beta}},\;x_{_1},\;x_{_2})=\varphi_{_{\{1,2,3\}}}^{(8)}(\widehat{(354)}{\boldsymbol{\beta}},\;x_{_1},\;x_{_2})\;.
\label{Gauss-Kummer6-3-9b}
\end{eqnarray}

Corresponding to the integer lattice $-n_{_1}E_{_3}^{(2)}+ n_{_2}E_{_3}^{(3)}$, the generalized hypergeometric function is formulated as
\begin{eqnarray}
&&\Phi_{_{\{1,2,5\}}}^{(9)}({\boldsymbol{\beta}},\;\boldsymbol{{\xi}})=A_{_{\{1,2,5\}}}^{(9)}({\boldsymbol{\beta}})
(r_{_3})^{1-\beta_{_3}-\beta_{_4}}(-r_{_1})^{\beta_{_2}+\beta_{_3}-1}(-r_{_2})^{-\beta_{_2}}
\nonumber\\
&&\hspace{3.2cm}\times
(r_{_3}-r_{_1})^{\beta_{_4}+\beta_{_5}-1}
\varphi_{_{\{1,2,5\}}}^{(9)}({\boldsymbol{\beta}},\;{r_{_1}-r_{_3}\over r_{_1}},\;{r_{_1}(r_{_2}-r_{_3})\over r_{_2}(r_{_1}-r_{_3})})\;,\nonumber\\
&&\varphi_{_{\{1,2,5\}}}^{(9)}({\boldsymbol{\beta}},\;x_{_1},\;x_{_2})=\varphi_{_{\{1,2,3\}}}^{(9)}(\widehat{(354)}{\boldsymbol{\beta}},\;x_{_1},\;x_{_2})\;.
\label{Gauss-Kummer6-3-10b}
\end{eqnarray}

Corresponding to the integer lattice $-n_{_1}E_{_3}^{(1)}- n_{_2}E_{_3}^{(2)}$, the generalized hypergeometric function is formulated as
\begin{eqnarray}
&&\Phi_{_{\{1,2,5\}}}^{(10)}({\boldsymbol{\beta}},\;\boldsymbol{{\xi}})=A_{_{\{1,2,3\}}}^{(10)}({\boldsymbol{\beta}})(r_{_3})^{1-\beta_{_3}-\beta_{_4}}
(-r_{_1})^{-\beta_{_1}}(-r_{_2})^{-\beta_{_2}}
\nonumber\\
&&\hspace{3.2cm}\times
\varphi_{_{\{1,2,5\}}}^{(10)}({\boldsymbol{\beta}},\;{r_{_2}-r_{_3}\over r_{_2}},\;{r_{_1}-r_{_3}\over r_{_1}})\;,\nonumber\\
&&\varphi_{_{\{1,2,5\}}}^{(10)}({\boldsymbol{\beta}},\;x_{_1},\;x_{_2})=\varphi_{_{\{1,2,3\}}}^{(10)}(\widehat{(354)}{\boldsymbol{\beta}},\;x_{_1},\;x_{_2})\;.
\label{Gauss-Kummer6-3-11b}
\end{eqnarray}

As the integer lattice is $-n_{_1}E_{_3}^{(1)}- n_{_2}E_{_3}^{(3)}$, the generalized hypergeometric function is formulated as
\begin{eqnarray}
&&\Phi_{_{\{1,2,5\}}}^{(11)}({\boldsymbol{\beta}},\;\boldsymbol{{\xi}})=A_{_{\{1,2,5\}}}^{(11)}({\boldsymbol{\beta}})
(r_{_3})^{1-\beta_{_3}-\beta_{_4}}(-r_{_1})^{-\beta_{_1}}(-r_{_2})^{\beta_{_1}+\beta_{_3}-1}
\nonumber\\
&&\hspace{3.2cm}\times
(r_{_3}-r_{_2})^{\beta_{_4}+\beta_{_5}-1}
\varphi_{_{\{1,2,5\}}}^{(11)}({\boldsymbol{\beta}},\;{r_{_2}-r_{_3}\over r_{_2}},\;{r_{_2}(r_{_1}-r_{_3})\over r_{_1}(r_{_2}-r_{_3})})\;,\nonumber\\
&&\varphi_{_{\{1,2,5\}}}^{(11)}({\boldsymbol{\beta}},\;x_{_1},\;x_{_2})=\varphi_{_{\{1,2,3\}}}^{(11)}(\widehat{(354)}{\boldsymbol{\beta}},\;x_{_1},\;x_{_2})\;.
\label{Gauss-Kummer6-3-12b}
\end{eqnarray}

Corresponding to the integer lattice $-n_{_1}E_{_3}^{(2)}- n_{_2}E_{_3}^{(3)}$, the generalized hypergeometric function is formulated as
\begin{eqnarray}
&&\Phi_{_{\{1,2,5\}}}^{(12)}({\boldsymbol{\beta}},\;\boldsymbol{{\xi}})=A_{_{\{1,2,5\}}}^{(12)}({\boldsymbol{\beta}})
(r_{_3})^{1-\beta_{_3}-\beta_{_4}}(-r_{_1})^{\beta_{_3}-1}(r_{_3}-r_{_1})^{1-\beta_{_1}-\beta_{_3}}
\nonumber\\
&&\hspace{3.2cm}\times
(r_{_3}-r_{_2})^{-\beta_{_2}}
\varphi_{_{\{1,2,5\}}}^{(12)}({\boldsymbol{\beta}},\;{r_{_1}-r_{_3}\over r_{_1}},\;{r_{_2}(r_{_1}-r_{_3})\over r_{_1}(r_{_2}-r_{_3})})\;,\nonumber\\
&&\varphi_{_{\{1,2,5\}}}^{(12)}({\boldsymbol{\beta}},\;x_{_1},\;x_{_2})=\varphi_{_{\{1,2,3\}}}^{(12)}(\widehat{(354)}{\boldsymbol{\beta}},\;x_{_1},\;x_{_2})\;.
\label{Gauss-Kummer6-3-13b}
\end{eqnarray}

\subsection{${\bf{\cal B}}=\{1,3,4\}$\label{sec6-4}}
\indent\indent
\begin{eqnarray}
&&{\rm Det}({\boldsymbol{\xi}}_{_{\{1,3,4\}}})=-1
\;,\nonumber\\
&&{\boldsymbol{\xi}}_{_{\{1,3,4\}}}^{-1}\cdot {\boldsymbol{\xi}}=
\left(\begin{array}{ccccc}\;1\;&\;-1\;&\;0\;&\;0\;&\;r_{_1}-r_{_2}\;\\\;0\;&\;-1\;&\;1\;&\;0\;&\;-r_{_2}+r_{_3}\;\\
\;0\;&\;1\;&\;0\;&\;1\;&\;r_{_2}\;\end{array}\right)\;.
\label{Gauss-Kummer6-4-1}
\end{eqnarray}

Corresponding to the integer lattice $n_{_1}E_{_3}^{(1)}+ n_{_2}E_{_3}^{(2)}$, the generalized hypergeometric function is formulated as
\begin{eqnarray}
&&\Phi_{_{\{1,3,4\}}}^{(1)}({\boldsymbol{\beta}},\;\boldsymbol{{\xi}})=A_{_{\{1,3,4\}}}^{(1)}({\boldsymbol{\beta}})(r_{_1}-r_{_2})^{-\beta_{_1}}(r_{_3}-r_{_2})^{-\beta_{_3}}(r_{_2})^{1-\beta_{_2}-\beta_{_4}}
\nonumber\\
&&\hspace{3.2cm}\times
\varphi_{_{\{1,3,4\}}}^{(1)}({\boldsymbol{\beta}},\;{r_{_2}\over r_{_2}-r_{_3}},\;{r_{_2}\over r_{_2}-r_{_1}})\;,\nonumber\\
&&\varphi_{_{\{1,3,4\}}}^{(1)}({\boldsymbol{\beta}},\;x_{_1},\;x_{_2})=\varphi_{_{\{1,2,3\}}}^{(1)}(\widehat{(234)}{\boldsymbol{\beta}},\;x_{_1},\;x_{_2})\;.
\label{Gauss-Kummer6-4-1-2}
\end{eqnarray}

As the integer lattice is $n_{_1}E_{_3}^{(1)}+ n_{_2}E_{_3}^{(3)}$, the generalized hypergeometric function is formulated as
\begin{eqnarray}
&&\Phi_{_{\{1,3,4\}}}^{(2)}({\boldsymbol{\beta}},\;\boldsymbol{{\xi}})=A_{_{\{1,3,4\}}}^{(2)}({\boldsymbol{\beta}})(r_{_1}-r_{_2})^{-\beta_{_1}}(r_{_3}-r_{_2})^{\beta_{_1}+\beta_{_5}-1}
\nonumber\\
&&\hspace{3.2cm}\times
\varphi_{_{\{1,3,4\}}}^{(2)}({\boldsymbol{\beta}},\;{r_{_2}\over r_{_2}-r_{_3}},\;{r_{_2}-r_{_3}\over r_{_2}-r_{_1}})\;,\nonumber\\
&&\varphi_{_{\{1,3,4\}}}^{(2)}({\boldsymbol{\beta}},\;x_{_1},\;x_{_2})=\varphi_{_{\{1,2,3\}}}^{(2)}(\widehat{(234)}{\boldsymbol{\beta}},\;x_{_1},\;x_{_2})\;.
\label{Gauss-Kummer6-4-2-2}
\end{eqnarray}

As the integer lattice is $n_{_1}E_{_3}^{(2)}+ n_{_2}E_{_3}^{(3)}$, the generalized hypergeometric function is formulated as
\begin{eqnarray}
&&\Phi_{_{\{1,3,4\}}}^{(3)}({\boldsymbol{\beta}},\;\boldsymbol{{\xi}})=A_{_{\{1,3,4\}}}^{(3)}({\boldsymbol{\beta}})(r_{_1}-m_{_2})^{\beta_{_5}-1}
\varphi_{_{\{1,3,4\}}}^{(3)}({\boldsymbol{\beta}},\;{r_{_2}\over r_{_2}-r_{_1}},\;{r_{_2}-r_{_3}\over r_{_2}-r_{_1}})\;,\nonumber\\
&&\varphi_{_{\{1,3,4\}}}^{(3)}({\boldsymbol{\beta}},\;x_{_1},\;x_{_2})=\varphi_{_{\{1,2,3\}}}^{(3)}(\widehat{(234)}{\boldsymbol{\beta}},\;x_{_1},\;x_{_2})\;.
\label{Gauss-Kummer6-4-3-2}
\end{eqnarray}

When the integer lattice is $n_{_1}E_{_3}^{(1)}- n_{_2}E_{_3}^{(2)}$, the generalized hypergeometric function is formulated as
\begin{eqnarray}
&&\Phi_{_{\{1,3,4\}}}^{(4)}({\boldsymbol{\beta}},\;\boldsymbol{{\xi}})=A_{_{\{1,3,4\}}}^{(4)}({\boldsymbol{\beta}})(r_{_3}-r_{_2})^{-\beta_{_3}}(r_{_2})^{\beta_{_3}+\beta_{_5}-1}
\nonumber\\
&&\hspace{3.2cm}\times
\varphi_{_{\{1,3,4\}}}^{(4)}({\boldsymbol{\beta}},\;{r_{_2}\over r_{_2}-r_{_3}},\;{r_{_2}-r_{_1}\over r_{_2}})\;,\nonumber\\
&&\varphi_{_{\{1,3,4\}}}^{(4)}({\boldsymbol{\beta}},\;x_{_1},\;x_{_2})=\varphi_{_{\{1,2,3\}}}^{(4)}(\widehat{(234)}{\boldsymbol{\beta}},\;x_{_1},\;x_{_2})\;.
\label{Gauss-Kummer6-4-4-2}
\end{eqnarray}

Corresponding to the integer lattice $n_{_1}E_{_3}^{(1)}- n_{_2}E_{_3}^{(3)}$, the generalized hypergeometric function is formulated as
\begin{eqnarray}
&&\Phi_{_{\{1,3,4\}}}^{(5)}({\boldsymbol{\beta}},\;\boldsymbol{{\xi}})=A_{_{\{1,3,4\}}}^{(5)}({\boldsymbol{\beta}})(r_{_3}-r_{_2})^{\beta_{_5}-1}
\nonumber\\
&&\hspace{3.2cm}\times
\varphi_{_{\{1,3,4\}}}^{(5)}({\boldsymbol{\beta}},\;{r_{_2}\over r_{_2}-r_{_3}},\;{r_{_2}-r_{_1}\over r_{_2}-r_{_3}})\;,\nonumber\\
&&\varphi_{_{\{1,3,4\}}}^{(5)}({\boldsymbol{\beta}},\;x_{_1},\;x_{_2})=\varphi_{_{\{1,2,3\}}}^{(5)}(\widehat{(234)}{\boldsymbol{\beta}},\;x_{_1},\;x_{_2})\;.
\label{Gauss-Kummer6-4-5-2}
\end{eqnarray}

Corresponding to the integer lattice $n_{_1}E_{_3}^{(2)}- n_{_2}E_{_3}^{(3)}$, the generalized hypergeometric function is formulated as
\begin{eqnarray}
&&\Phi_{_{\{1,3,4\}}}^{(6)}({\boldsymbol{\beta}},\;\boldsymbol{{\xi}})=A_{_{\{1,3,4\}}}^{(6)}({\boldsymbol{\beta}})(r_{_1}-r_{_2})^{\beta_{_3}+\beta_{_5}-1}(r_{_3}-r_{_2})^{-\beta_{_3}}
\nonumber\\
&&\hspace{3.2cm}\times
\varphi_{_{\{1,3,4\}}}^{(6)}({\boldsymbol{\beta}},\;{r_{_2}\over r_{_2}-r_{_1}},\;{r_{_2}-r_{_1}\over r_{_2}-r_{_3}})\;,\nonumber\\
&&\varphi_{_{\{1,3,4\}}}^{(6)}({\boldsymbol{\beta}},\;x_{_1},\;x_{_2})=\varphi_{_{\{1,2,3\}}}^{(6)}(\widehat{(234)}{\boldsymbol{\beta}},\;x_{_1},\;x_{_2})\;.
\label{Gauss-Kummer6-4-6-2}
\end{eqnarray}

As the integer lattice is $-n_{_1}E_{_3}^{(1)}+ n_{_2}E_{_3}^{(2)}$, the generalized hypergeometric function is formulated as
\begin{eqnarray}
&&\Phi_{_{\{1,3,4\}}}^{(7)}({\boldsymbol{\beta}},\;\boldsymbol{{\xi}})=A_{_{\{1,3,4\}}}^{(7)}({\boldsymbol{\beta}})(r_{_1}-r_{_2})^{-\beta_{_1}}(r_{_2})^{\beta_{_1}+\beta_{_5}-1}
\nonumber\\
&&\hspace{3.2cm}\times
\varphi_{_{\{1,3,4\}}}^{(7)}({\boldsymbol{\beta}},\;{r_{_2}-r_{_3}\over r_{_2}},\;{r_{_2}\over r_{_2}-r_{_1}})\;,\nonumber\\
&&\varphi_{_{\{1,3,4\}}}^{(7)}({\boldsymbol{\beta}},\;x_{_1},\;x_{_2})=\varphi_{_{\{1,2,3\}}}^{(7)}(\widehat{(234)}{\boldsymbol{\beta}},\;x_{_1},\;x_{_2})\;.
\label{Gauss-Kummer6-4-7-2}
\end{eqnarray}

When the integer lattice is $-n_{_1}E_{_3}^{(1)}+ n_{_2}E_{_3}^{(3)}$, the generalized hypergeometric function is formulated as
\begin{eqnarray}
&&\Phi_{_{\{1,3,4\}}}^{(8)}({\boldsymbol{\beta}},\;\boldsymbol{{\xi}})=A_{_{\{1,3,4\}}}^{(8)}({\boldsymbol{\beta}})(r_{_1}-r_{_2})^{-\beta_{_1}}(r_{_3}-r_{_2})^{1-\beta_{_2}-\beta_{_3}}
(r_{_2})^{-\beta_{_4}}
\nonumber\\
&&\hspace{3.2cm}\times
\varphi_{_{\{1,3,4\}}}^{(8)}({\boldsymbol{\beta}},\;{r_{_2}-r_{_3}\over r_{_2}},\;{r_{_2}-r_{_3}\over r_{_2}-r_{_1}})\;,\nonumber\\
&&\varphi_{_{\{1,3,4\}}}^{(8)}({\boldsymbol{\beta}},\;x_{_1},\;x_{_2})=\varphi_{_{\{1,2,3\}}}^{(8)}(\widehat{(234)}{\boldsymbol{\beta}},\;x_{_1},\;x_{_2})\;.
\label{Gauss-Kummer6-4-8-2}
\end{eqnarray}

Corresponding to the integer lattice $-n_{_1}E_{_3}^{(2)}+ n_{_2}E_{_3}^{(3)}$, the generalized hypergeometric function is formulated as
\begin{eqnarray}
&&\Phi_{_{\{1,3,4\}}}^{(9)}({\boldsymbol{\beta}},\;\boldsymbol{{\xi}})=A_{_{\{1,3,4\}}}^{(9)}({\boldsymbol{\beta}})(r_{_1}-r_{_2})^{\beta_{_4}+\beta_{_5}-1}(r_{_2})^{-\beta_{_4}}
\nonumber\\
&&\hspace{3.2cm}\times
\varphi_{_{\{1,3,4\}}}^{(9)}({\boldsymbol{\beta}},\;{r_{_2}-r_{_1}\over r_{_2}},\;{r_{_2}-r_{_3}\over r_{_2}-r_{_1}})\;,\nonumber\\
&&\varphi_{_{\{1,3,4\}}}^{(9)}({\boldsymbol{\beta}},\;x_{_1},\;x_{_2})=\varphi_{_{\{1,2,3\}}}^{(9)}(\widehat{(234)}{\boldsymbol{\beta}},\;x_{_1},\;x_{_2})\;.
\label{Gauss-Kummer6-4-9-2}
\end{eqnarray}

Corresponding to the integer lattice $-n_{_1}E_{_3}^{(1)}- n_{_2}E_{_3}^{(2)}$, the generalized hypergeometric function is formulated as
\begin{eqnarray}
&&\Phi_{_{\{1,3,4\}}}^{(10)}({\boldsymbol{\beta}},\;\boldsymbol{{\xi}})=A_{_{\{1,3,4\}}}^{(10)}({\boldsymbol{\beta}})(r_{_2})^{\beta_{_5}-1}
\nonumber\\
&&\hspace{3.2cm}\times
\varphi_{_{\{1,3,4\}}}^{(10)}({\boldsymbol{\beta}},\;{r_{_2}-r_{_3}\over r_{_2}},\;{r_{_2}-r_{_1}\over r_{_2}})\;,\nonumber\\
&&\varphi_{_{\{1,3,4\}}}^{(10)}({\boldsymbol{\beta}},\;x_{_1},\;x_{_2})=\varphi_{_{\{1,2,3\}}}^{(10)}(\widehat{(234)}{\boldsymbol{\beta}},\;x_{_1},\;x_{_2})\;.
\label{Gauss-Kummer6-4-10-2}
\end{eqnarray}

Corresponding to the integer lattice $-n_{_1}E_{_3}^{(1)}- n_{_2}E_{_3}^{(3)}$, the generalized hypergeometric function is formulated as
\begin{eqnarray}
&&\Phi_{_{\{1,3,4\}}}^{(11)}({\boldsymbol{\beta}},\;\boldsymbol{{\xi}})=A_{_{\{1,3,4\}}}^{(11)}({\boldsymbol{\beta}})(r_{_3}-r_{_2})^{\beta_{_4}+\beta_{_5}-1}(r_{_2})^{-\beta_{_4}}
\nonumber\\
&&\hspace{3.2cm}\times
\varphi_{_{\{1,3,4\}}}^{(11)}({\boldsymbol{\beta}},\;{r_{_2}-r_{_3}\over r_{_2}},\;{r_{_2}-r_{_1}\over r_{_2}-r_{_3}})\;,\nonumber\\
&&\varphi_{_{\{1,3,4\}}}^{(11)}({\boldsymbol{\beta}},\;x_{_1},\;x_{_2})=\varphi_{_{\{1,2,3\}}}^{(11)}(\widehat{(234)}{\boldsymbol{\beta}},\;x_{_1},\;x_{_2})\;.
\label{Gauss-Kummer6-4-11-2}
\end{eqnarray}

As the integer lattice is $-n_{_1}E_{_3}^{(2)}- n_{_2}E_{_3}^{(3)}$, the generalized hypergeometric function is formulated as
\begin{eqnarray}
&&\Phi_{_{\{1,3,4\}}}^{(12)}({\boldsymbol{\beta}},\;\boldsymbol{{\xi}})=A_{_{\{1,3,4\}}}^{(12)}({\boldsymbol{\beta}})(r_{_1}-r_{_2})^{1-\beta_{_1}-\beta_{_2}}(r_{_3}-r_{_2})^{-\beta_{_3}}(r_{_2})^{-\beta_{_4}}
\nonumber\\
&&\hspace{3.2cm}\times
\varphi_{_{\{1,3,4\}}}^{(12)}({\boldsymbol{\beta}},\;{r_{_2}-r_{_1}\over r_{_2}},\;{r_{_2}-r_{_1}\over r_{_2}-r_{_3}})\;,\nonumber\\
&&\varphi_{_{\{1,3,4\}}}^{(12)}({\boldsymbol{\beta}},\;x_{_1},\;x_{_2})=\varphi_{_{\{1,2,3\}}}^{(12)}(\widehat{(234)}{\boldsymbol{\beta}},\;x_{_1},\;x_{_2})\;.
\label{Gauss-Kummer6-4-12-2}
\end{eqnarray}

\subsection{${\bf{\cal B}}=\{1,3,5\}$\label{sec6-5}}
\indent\indent
\begin{eqnarray}
&&{\rm Det}({\boldsymbol{\xi}}_{_{\{1,3,5\}}})=-r_{_2}
\;,\nonumber\\
&&{\boldsymbol{\xi}}_{_{\{1,3,5\}}}^{-1}\cdot {\boldsymbol{\xi}}=
\left(\begin{array}{ccccc}\;1\;&\;-{r_{_1}\over r_{_2}}\;&\;0\;&\;1-{r_{_1}\over r_{_2}}\;&\;0\;\\\;0\;&\;-{r_{_3}\over r_{_2}}\;&\;1\;&\;1-{r_{_3}\over r_{_2}}\;&\;0\;\\
\;0\;&\;{1\over r_{_2}}\;&\;0\;&\;{1\over r_{_2}}\;&\;1\;\end{array}\right)\;.
\label{Gauss-Kummer6-5-1}
\end{eqnarray}

Corresponding to the integer lattice $n_{_1}E_{_3}^{(1)}+ n_{_2}E_{_3}^{(2)}$, the generalized hypergeometric function is formulated as
\begin{eqnarray}
&&\Phi_{_{\{1,3,5\}}}^{(1)}({\boldsymbol{\beta}},\;\boldsymbol{{\xi}})=A_{_{\{1,3,5\}}}^{(1)}({\boldsymbol{\beta}})(r_{_2})^{1-\beta_{_2}-\beta_{_4}}(r_{_2}-r_{_1})^{-\beta_{_1}}
\nonumber\\
&&\hspace{3.2cm}\times
(r_{_2}-r_{_3})^{-\beta_{_3}}
\varphi_{_{\{1,3,5\}}}^{(1)}({\boldsymbol{\beta}},\;{r_{_3}\over r_{_3}-r_{_2}},\;{r_{_1}\over r_{_1}-r_{_2}})\;,\nonumber\\
&&\varphi_{_{\{1,3,5\}}}^{(1)}({\boldsymbol{\beta}},\;x_{_1},\;x_{_2})=\varphi_{_{\{1,2,3\}}}^{(1)}(\widehat{(2354)}{\boldsymbol{\beta}},\;x_{_1},\;x_{_2})\;.
\label{Gauss-Kummer6-5-1-2}
\end{eqnarray}

As the integer lattice is $n_{_1}E_{_3}^{(1)}+ n_{_2}E_{_3}^{(3)}$, the generalized hypergeometric function is formulated as
\begin{eqnarray}
&&\Phi_{_{\{1,3,5\}}}^{(2)}({\boldsymbol{\beta}},\;\boldsymbol{{\xi}})=A_{_{\{1,3,5\}}}^{(2)}({\boldsymbol{\beta}})
(r_{_2})^{1-\beta_{_2}-\beta_{_4}}(-r_{_3})^{\beta_{_2}+\beta_{_5}-1}(r_{_2}-r_{_1})^{-\beta_{_1}}
\nonumber\\
&&\hspace{3.2cm}\times
(r_{_2}-r_{_3})^{\beta_{_1}+\beta_{_4}-1}
\varphi_{_{\{1,3,5\}}}^{(2)}({\boldsymbol{\beta}},\;{r_{_3}\over r_{_3}-r_{_2}},\;{r_{_1}(r_{_3}-r_{_2})\over r_{_3}(r_{_1}-r_{_2})})\;,\nonumber\\
&&\varphi_{_{\{1,3,5\}}}^{(2)}({\boldsymbol{\beta}},\;x_{_1},\;x_{_2})=\varphi_{_{\{1,2,3\}}}^{(2)}(\widehat{(2354)}{\boldsymbol{\beta}},\;x_{_1},\;x_{_2})\;.
\label{Gauss-Kummer6-5-2-2}
\end{eqnarray}

Corresponding to the integer lattice $n_{_1}E_{_3}^{(2)}+ n_{_2}E_{_3}^{(3)}$, the generalized hypergeometric function is formulated as
\begin{eqnarray}
&&\Phi_{_{\{1,3,5\}}}^{(3)}({\boldsymbol{\beta}},\;\boldsymbol{{\xi}})=A_{_{\{1,3,5\}}}^{(3)}({\boldsymbol{\beta}})
(r_{_2})^{1-\beta_{_2}-\beta_{_4}}(-r_{_1})^{1-\beta_{_1}-\beta_{_4}}(-r_{_3})^{-\beta_{_3}}
\nonumber\\
&&\hspace{3.2cm}\times
(r_{_2}-r_{_1})^{\beta_{_4}-1}
\varphi_{_{\{1,3,5\}}}^{(3)}({\boldsymbol{\beta}},\;{r_{_1}\over r_{_1}-r_{_2}},\;{r_{_1}(r_{_3}-r_{_2})\over r_{_3}(r_{_1}-r_{_2})})\;,\nonumber\\
&&\varphi_{_{\{1,3,5\}}}^{(3)}({\boldsymbol{\beta}},\;x_{_1},\;x_{_2})=\varphi_{_{\{1,2,3\}}}^{(3)}(\widehat{(2354)}{\boldsymbol{\beta}},\;x_{_1},\;x_{_2})\;.
\label{Gauss-Kummer6-5-3-2}
\end{eqnarray}

Corresponding to the integer lattice $n_{_1}E_{_3}^{(1)}- n_{_2}E_{_3}^{(2)}$, the generalized hypergeometric function is formulated as
\begin{eqnarray}
&&\Phi_{_{\{1,3,5\}}}^{(4)}({\boldsymbol{\beta}},\;\boldsymbol{{\xi}})=A_{_{\{1,3,5\}}}^{(4)}({\boldsymbol{\beta}})(r_{_2})^{1-\beta_{_2}-\beta_{_4}}(-r_{_1})^{-\beta_{_1}}
\nonumber\\
&&\hspace{3.2cm}\times
(r_{_2}-r_{_3})^{-\beta_{_3}}
\varphi_{_{\{1,3,5\}}}^{(4)}({\boldsymbol{\beta}},\;{r_{_3}\over r_{_3}-r_{_2}},\;{r_{_1}-r_{_2}\over r_{_1}})\;,\nonumber\\
&&\varphi_{_{\{1,3,5\}}}^{(4)}({\boldsymbol{\beta}},\;x_{_1},\;x_{_2})=\varphi_{_{\{1,2,3\}}}^{(4)}(\widehat{(2354)}{\boldsymbol{\beta}},\;x_{_1},\;x_{_2})\;.
\label{Gauss-Kummer6-5-4-2}
\end{eqnarray}

When the integer lattice is $n_{_1}E_{_3}^{(1)}- n_{_2}E_{_3}^{(3)}$, the generalized hypergeometric function is formulated as
\begin{eqnarray}
&&\Phi_{_{\{1,3,5\}}}^{(5)}({\boldsymbol{\beta}},\;\boldsymbol{{\xi}})=A_{_{\{1,3,5\}}}^{(5)}({\boldsymbol{\beta}})
(r_{_2})^{1-\beta_{_2}-\beta_{_4}}(-r_{_1})^{-\beta_{_1}}(-r_{_3})^{1-\beta_{_3}-\beta_{_4}}
\nonumber\\
&&\hspace{3.2cm}\times
(r_{_2}-r_{_3})^{\beta_{_4}-1}
\varphi_{_{\{1,3,5\}}}^{(5)}({\boldsymbol{\beta}},\;{r_{_3}\over r_{_3}-r_{_2}},\;{r_{_3}(r_{_1}-r_{_2})\over r_{_1}(r_{_3}-r_{_2})})\;,\nonumber\\
&&\varphi_{_{\{1,3,5\}}}^{(5)}({\boldsymbol{\beta}},\;x_{_1},\;x_{_2})=\varphi_{_{\{1,2,3\}}}^{(5)}(\widehat{(2354)}{\boldsymbol{\beta}},\;x_{_1},\;x_{_2})\;.
\label{Gauss-Kummer6-5-5-2}
\end{eqnarray}

When the integer lattice is $n_{_1}E_{_3}^{(2)}- n_{_2}E_{_3}^{(3)}$, the generalized hypergeometric function is formulated as
\begin{eqnarray}
&&\Phi_{_{\{1,3,5\}}}^{(6)}({\boldsymbol{\beta}},\;\boldsymbol{{\xi}})=A_{_{\{1,3,5\}}}^{(6)}({\boldsymbol{\beta}})
(r_{_2})^{1-\beta_{_2}-\beta_{_4}}(-r_{_1})^{\beta_{_2}+\beta_{_5}-1}(r_{_2}-r_{_1})^{\beta_{_3}+\beta_{_4}-1}
\nonumber\\
&&\hspace{3.2cm}\times
(r_{_2}-r_{_3})^{-\beta_{_3}}
\varphi_{_{\{1,3,5\}}}^{(6)}({\boldsymbol{\beta}},\;{r_{_1}\over r_{_1}-r_{_2}},\;{r_{_3}(r_{_1}-r_{_2})\over r_{_1}(r_{_3}-r_{_2})})\;,\nonumber\\
&&\varphi_{_{\{1,3,5\}}}^{(6)}({\boldsymbol{\beta}},\;x_{_1},\;x_{_2})=\varphi_{_{\{1,2,3\}}}^{(6)}(\widehat{(2354)}{\boldsymbol{\beta}},\;x_{_1},\;x_{_2})\;.
\label{Gauss-Kummer6-5-6-2}
\end{eqnarray}

Corresponding to the integer lattice $-n_{_1}E_{_3}^{(1)}+ n_{_2}E_{_3}^{(2)}$, the generalized hypergeometric function is formulated as
\begin{eqnarray}
&&\Phi_{_{\{1,3,5\}}}^{(7)}({\boldsymbol{\beta}},\;\boldsymbol{{\xi}})=A_{_{\{1,3,5\}}}^{(7)}({\boldsymbol{\beta}})
(r_{_2})^{1-\beta_{_2}-\beta_{_4}}(-r_{_3})^{-\beta_{_3}}(r_{_2}-r_{_1})^{-\beta_{_1}}
\nonumber\\
&&\hspace{3.2cm}\times
\varphi_{_{\{1,3,5\}}}^{(7)}({\boldsymbol{\beta}},\;{r_{_3}-r_{_2}\over r_{_3}},\;{r_{_1}\over r_{_1}-r_{_2}})\;,\nonumber\\
&&\varphi_{_{\{1,3,5\}}}^{(7)}({\boldsymbol{\beta}},\;x_{_1},\;x_{_2})=\varphi_{_{\{1,2,3\}}}^{(7)}(\widehat{(2354)}{\boldsymbol{\beta}},\;x_{_1},\;x_{_2})\;.
\label{Gauss-Kummer6-5-7-2}
\end{eqnarray}

As the integer lattice is $-n_{_1}E_{_3}^{(1)}+ n_{_2}E_{_3}^{(3)}$, the generalized hypergeometric function is formulated as
\begin{eqnarray}
&&\Phi_{_{\{1,3,5\}}}^{(8)}({\boldsymbol{\beta}},\;\boldsymbol{{\xi}})=A_{_{\{1,3,5\}}}^{(8)}({\boldsymbol{\beta}})(r_{_2})^{1-\beta_{_2}-\beta_{_4}}
(-r_{_3})^{\beta_{_2}-1}(r_{_2}-r_{_1})^{-\beta_{_1}}
\nonumber\\
&&\hspace{3.2cm}\times
(r_{_2}-r_{_3})^{1-\beta_{_2}-\beta_{_3}}
\varphi_{_{\{1,3,5\}}}^{(8)}({\boldsymbol{\beta}},\;{r_{_3}-r_{_2}\over r_{_3}},\;{r_{_1}(r_{_3}-r_{_2})\over r_{_3}(r_{_1}-r_{_2})})\;,\nonumber\\
&&\varphi_{_{\{1,3,5\}}}^{(8)}({\boldsymbol{\beta}},\;x_{_1},\;x_{_2})=\varphi_{_{\{1,2,3\}}}^{(8)}(\widehat{(2354)}{\boldsymbol{\beta}},\;x_{_1},\;x_{_2})\;.
\label{Gauss-Kummer6-5-8-2}
\end{eqnarray}

As the integer lattice is $-n_{_1}E_{_3}^{(2)}+ n_{_2}E_{_3}^{(3)}$, the generalized hypergeometric function is formulated as
\begin{eqnarray}
&&\Phi_{_{\{1,3,5\}}}^{(9)}({\boldsymbol{\beta}},\;\boldsymbol{{\xi}})=A_{_{\{1,3,5\}}}^{(9)}({\boldsymbol{\beta}})
(r_{_2})^{1-\beta_{_2}-\beta_{_4}}(-r_{_1})^{\beta_{_2}+\beta_{_3}-1}(-r_{_3})^{-\beta_{_3}}
\nonumber\\
&&\hspace{3.2cm}\times
(r_{_2}-r_{_1})^{\beta_{_4}+\beta_{_5}-1}
\varphi_{_{\{1,3,5\}}}^{(9)}({\boldsymbol{\beta}},\;{r_{_1}-r_{_2}\over r_{_1}},\;{r_{_1}(r_{_3}-r_{_2})\over r_{_3}(r_{_1}-r_{_2})})\;,\nonumber\\
&&\varphi_{_{\{1,3,5\}}}^{(9)}({\boldsymbol{\beta}},\;x_{_1},\;x_{_2})=\varphi_{_{\{1,2,3\}}}^{(9)}(\widehat{(2354)}{\boldsymbol{\beta}},\;x_{_1},\;x_{_2})\;.
\label{Gauss-Kummer6-5-9-2}
\end{eqnarray}

Corresponding to the integer lattice $-n_{_1}E_{_3}^{(1)}- n_{_2}E_{_3}^{(2)}$, the generalized hypergeometric function is formulated as
\begin{eqnarray}
&&\Phi_{_{\{1,3,5\}}}^{(10)}({\boldsymbol{\beta}},\;\boldsymbol{{\xi}})=A_{_{\{1,3,5\}}}^{(10)}({\boldsymbol{\beta}})
(r_{_2})^{1-\beta_{_2}-\beta_{_4}}(-r_{_1})^{-\beta_{_1}}(-r_{_3})^{-\beta_{_3}}
\nonumber\\
&&\hspace{3.2cm}\times
\varphi_{_{\{1,3,5\}}}^{(10)}({\boldsymbol{\beta}},\;{r_{_3}-r_{_2}\over r_{_3}},\;{r_{_1}-r_{_2}\over r_{_1}})\;,\nonumber\\
&&\varphi_{_{\{1,3,5\}}}^{(10)}({\boldsymbol{\beta}},\;x_{_1},\;x_{_2})=\varphi_{_{\{1,2,3\}}}^{(10)}(\widehat{(2354)}{\boldsymbol{\beta}},\;x_{_1},\;x_{_2})\;.
\label{Gauss-Kummer6-5-10-2}
\end{eqnarray}

When the integer lattice is $-n_{_1}E_{_3}^{(1)}- n_{_2}E_{_3}^{(3)}$, the generalized hypergeometric function is formulated as
\begin{eqnarray}
&&\Phi_{_{\{1,3,5\}}}^{(11)}({\boldsymbol{\beta}},\;\boldsymbol{{\xi}})=A_{_{\{1,3,5\}}}^{(11)}({\boldsymbol{\beta}})
(r_{_2})^{1-\beta_{_2}-\beta_{_4}}(-r_{_1})^{-\beta_{_1}}(-r_{_3})^{\beta_{_1}+\beta_{_2}-1}
\nonumber\\
&&\hspace{3.2cm}\times
(r_{_2}-r_{_3})^{\beta_{_4}+\beta_{_5}-1}
\varphi_{_{\{1,3,5\}}}^{(11)}({\boldsymbol{\beta}},\;{r_{_3}-r_{_2}\over r_{_3}},\;{r_{_3}(r_{_1}-r_{_2})\over r_{_1}(r_{_3}-r_{_2})})\;,\nonumber\\
&&\varphi_{_{\{1,3,5\}}}^{(11)}({\boldsymbol{\beta}},\;x_{_1},\;x_{_2})=\varphi_{_{\{1,2,3\}}}^{(11)}(\widehat{(2354)}{\boldsymbol{\beta}},\;x_{_1},\;x_{_2})\;.
\label{Gauss-Kummer6-5-11-2}
\end{eqnarray}

Corresponding to the integer lattice $-n_{_1}E_{_3}^{(2)}- n_{_2}E_{_3}^{(3)}$, the generalized hypergeometric function is formulated as
\begin{eqnarray}
&&\Phi_{_{\{1,3,5\}}}^{(12)}({\boldsymbol{\beta}},\;\boldsymbol{{\xi}})=A_{_{\{1,3,5\}}}^{(12)}({\boldsymbol{\beta}})
(r_{_2})^{1-\beta_{_2}-\beta_{_4}}(-r_{_1})^{\beta_{_2}-1}(r_{_2}-r_{_1})^{1-\beta_{_1}-\beta_{_2}}
\nonumber\\
&&\hspace{3.2cm}\times
(r_{_2}-r_{_3})^{-\beta_{_3}}
\varphi_{_{\{1,3,5\}}}^{(12)}({\boldsymbol{\beta}},\;{r_{_1}-r_{_2}\over r_{_1}},\;{r_{_3}(r_{_1}-r_{_2})\over r_{_1}(r_{_3}-r_{_2})})\;,\nonumber\\
&&\varphi_{_{\{1,3,5\}}}^{(12)}({\boldsymbol{\beta}},\;x_{_1},\;x_{_2})=\varphi_{_{\{1,2,3\}}}^{(12)}(\widehat{(2354)}{\boldsymbol{\beta}},\;x_{_1},\;x_{_2})\;.
\label{Gauss-Kummer6-5-12-2}
\end{eqnarray}

\subsection{${\bf{\cal B}}=\{2,3,4\}$\label{sec6-6}}
\indent\indent
\begin{eqnarray}
&&{\rm Det}({\boldsymbol{\xi}}_{_{\{2,3,4\}}})=1
\;,\nonumber\\
&&{\boldsymbol{\xi}}_{_{\{2,3,4\}}}^{-1}\cdot {\boldsymbol{\xi}}=
\left(\begin{array}{ccccc}\;-1\;&\;1\;&\;0\;&\;0\;&\;-r_{_1}+r_{_2}\;\\\;-1\;&\;0\;&\;1\;&\;0\;&\;-r_{_1}+r_{_3}\;\\
\;1\;&\;0\;&\;0\;&\;1\;&\;r_{_1}\;\end{array}\right)\;.
\label{Gauss-Kummer6-6-1}
\end{eqnarray}

As the integer lattice is $n_{_1}E_{_3}^{(1)}+ n_{_2}E_{_3}^{(2)}$, the generalized hypergeometric function is formulated as
\begin{eqnarray}
&&\Phi_{_{\{2,3,4\}}}^{(1)}({\boldsymbol{\beta}},\;\boldsymbol{{\xi}})=A_{_{\{2,3,4\}}}^{(1)}({\boldsymbol{\beta}})
(r_{_2}-r_{_1})^{-\beta_{_2}}(r_{_3}-r_{_1})^{-\beta_{_3}}
\nonumber\\
&&\hspace{3.2cm}\times
(r_{_1})^{1-\beta_{_1}-\beta_{_4}}
\varphi_{_{\{2,3,4\}}}^{(1)}({\boldsymbol{\beta}},\;{r_{_1}\over r_{_1}-r_{_3}},\;{r_{_1}\over r_{_1}-r_{_2}})\;,\nonumber\\
&&\varphi_{_{\{2,3,4\}}}^{(1)}({\boldsymbol{\beta}},\;x_{_1},\;x_{_2})=\varphi_{_{\{1,2,3\}}}^{(1)}(\widehat{(1234)}{\boldsymbol{\beta}},\;x_{_1},\;x_{_2})\;.
\label{Gauss-Kummer6-6-1-2}
\end{eqnarray}

As the integer lattice is $n_{_1}E_{_3}^{(1)}+ n_{_2}E_{_3}^{(3)}$, the generalized hypergeometric function is formulated as
\begin{eqnarray}
&&\Phi_{_{\{2,3,4\}}}^{(2)}({\boldsymbol{\beta}},\;\boldsymbol{{\xi}})=A_{_{\{2,3,4\}}}^{(2)}({\boldsymbol{\beta}})(r_{_2}-r_{_1})^{-\beta_{_2}}
(r_{_3}-r_{_1})^{\beta_{_2}+\beta_{_5}-1}
\nonumber\\
&&\hspace{3.2cm}\times
\varphi_{_{\{2,3,4\}}}^{(2)}({\boldsymbol{\beta}},\;{r_{_1}\over r_{_1}-r_{_3}},\;{r_{_3}-r_{_1}\over r_{_2}-r_{_1}})\;,\nonumber\\
&&\varphi_{_{\{2,3,4\}}}^{(2)}({\boldsymbol{\beta}},\;x_{_1},\;x_{_2})=\varphi_{_{\{1,2,3\}}}^{(2)}(\widehat{(1234)}{\boldsymbol{\beta}},\;x_{_1},\;x_{_2})\;.
\label{Gauss-Kummer6-6-2-2}
\end{eqnarray}

Corresponding to the integer lattice $n_{_1}E_{_3}^{(2)}+ n_{_2}E_{_3}^{(3)}$, the generalized hypergeometric function is formulated as
\begin{eqnarray}
&&\Phi_{_{\{2,3,4\}}}^{(3)}({\boldsymbol{\beta}},\;\boldsymbol{{\xi}})=A_{_{\{2,3,4\}}}^{(3)}({\boldsymbol{\beta}})(r_{_2}-r_{_1})^{\beta_{_5}-1}
\varphi_{_{\{2,3,4\}}}^{(3)}({\boldsymbol{\beta}},\;{r_{_1}\over r_{_1}-r_{_2}},\;{r_{_3}-r_{_1}\over r_{_2}-r_{_1}})\;,\nonumber\\
&&\varphi_{_{\{2,3,4\}}}^{(3)}({\boldsymbol{\beta}},\;x_{_1},\;x_{_2})=\varphi_{_{\{1,2,3\}}}^{(3)}(\widehat{(1234)}{\boldsymbol{\beta}},\;x_{_1},\;x_{_2})\;.
\label{Gauss-Kummer6-6-3-2}
\end{eqnarray}

Corresponding to the integer lattice $n_{_1}E_{_3}^{(1)}- n_{_2}E_{_3}^{(2)}$, the generalized hypergeometric function is formulated as
\begin{eqnarray}
&&\Phi_{_{\{2,3,4\}}}^{(4)}({\boldsymbol{\beta}},\;\boldsymbol{{\xi}})=A_{_{\{2,3,4\}}}^{(4)}({\boldsymbol{\beta}})
(r_{_1})^{\beta_{_3}+\beta_{_5}-1}(r_{_3}-r_{_1})^{-\beta_{_3}}
\nonumber\\
&&\hspace{3.2cm}\times
\varphi_{_{\{2,3,4\}}}^{(4)}({\boldsymbol{\beta}},\;{r_{_1}\over r_{_1}-r_{_3}},\;{r_{_1}-r_{_2}\over r_{_1}})\;,\nonumber\\
&&\varphi_{_{\{2,3,4\}}}^{(4)}({\boldsymbol{\beta}},\;x_{_1},\;x_{_2})=\varphi_{_{\{1,2,3\}}}^{(4)}(\widehat{(1234)}{\boldsymbol{\beta}},\;x_{_1},\;x_{_2})\;.
\label{Gauss-Kummer6-6-4-2}
\end{eqnarray}

When the integer lattice is $n_{_1}E_{_3}^{(1)}- n_{_2}E_{_3}^{(3)}$, the generalized hypergeometric function is formulated as
\begin{eqnarray}
&&\Phi_{_{\{2,3,4\}}}^{(5)}({\boldsymbol{\beta}},\;\boldsymbol{{\xi}})=A_{_{\{2,3,4\}}}^{(5)}({\boldsymbol{\beta}})(r_{_3}-r_{_1})^{\beta_{_5}-1}
\nonumber\\
&&\hspace{3.2cm}\times
\varphi_{_{\{2,3,4\}}}^{(5)}({\boldsymbol{\beta}},\;{r_{_1}\over r_{_1}-r_{_3}},\;{r_{_1}-r_{_2}\over r_{_1}-r_{_3}})\;,\nonumber\\
&&\varphi_{_{\{2,3,4\}}}^{(5)}({\boldsymbol{\beta}},\;x_{_1},\;x_{_2})=\varphi_{_{\{1,2,3\}}}^{(5)}(\widehat{(1234)}{\boldsymbol{\beta}},\;x_{_1},\;x_{_2})\;.
\label{Gauss-Kummer6-6-5-2}
\end{eqnarray}

When the integer lattice is $n_{_1}E_{_3}^{(2)}- n_{_2}E_{_3}^{(3)}$, the generalized hypergeometric function is formulated as
\begin{eqnarray}
&&\Phi_{_{\{2,3,4\}}}^{(6)}({\boldsymbol{\beta}},\;\boldsymbol{{\xi}})=A_{_{\{2,3,4\}}}^{(6)}({\boldsymbol{\beta}})(r_{_2}-r_{_1})^{\beta_{_3}+\beta_{_5}-1}(r_{_3}-r_{_1})^{-\beta_{_3}}
\nonumber\\
&&\hspace{3.2cm}\times
\varphi_{_{\{2,3,4\}}}^{(6)}({\boldsymbol{\beta}},\;{r_{_1}\over r_{_1}-r_{_2}},\;{r_{_2}-r_{_1}\over r_{_3}-r_{_1}})\;,\nonumber\\
&&\varphi_{_{\{2,3,4\}}}^{(6)}({\boldsymbol{\beta}},\;x_{_1},\;x_{_2})=\varphi_{_{\{1,2,3\}}}^{(6)}(\widehat{(1234)}{\boldsymbol{\beta}},\;x_{_1},\;x_{_2})\;.
\label{Gauss-Kummer6-6-6-2}
\end{eqnarray}

As the integer lattice is $-n_{_1}E_{_3}^{(1)}+ n_{_2}E_{_3}^{(2)}$, the generalized hypergeometric function is formulated as
\begin{eqnarray}
&&\Phi_{_{\{2,3,4\}}}^{(7)}({\boldsymbol{\beta}},\;\boldsymbol{{\xi}})=A_{_{\{2,3,4\}}}^{(7)}({\boldsymbol{\beta}})(r_{_2}-r_{_1})^{-\beta_{_2}}(r_{_1})^{\beta_{_2}+\beta_{_5}-1}
\nonumber\\
&&\hspace{3.2cm}\times
\varphi_{_{\{2,3,4\}}}^{(7)}({\boldsymbol{\beta}},\;{r_{_1}-r_{_3}\over r_{_1}},\;{r_{_1}\over r_{_1}-r_{_2}})\;,\nonumber\\
&&\varphi_{_{\{2,3,4\}}}^{(7)}({\boldsymbol{\beta}},\;x_{_1},\;x_{_2})=\varphi_{_{\{1,2,3\}}}^{(7)}(\widehat{(1234)}{\boldsymbol{\beta}},\;x_{_1},\;x_{_2})\;.
\label{Gauss-Kummer6-6-7-2}
\end{eqnarray}

As the integer lattice is $-n_{_1}E_{_3}^{(1)}+ n_{_2}E_{_3}^{(3)}$, the generalized hypergeometric function is formulated as
\begin{eqnarray}
&&\Phi_{_{\{2,3,4\}}}^{(8)}({\boldsymbol{\beta}},\;\boldsymbol{{\xi}})=A_{_{\{2,3,4\}}}^{(8)}({\boldsymbol{\beta}})(r_{_1})^{-\beta_{_4}}(r_{_3}-r_{_1})^{1-\beta_{_1}-\beta_{_3}}
(r_{_2}-r_{_1})^{-\beta_{_2}}
\nonumber\\
&&\hspace{3.2cm}\times
\varphi_{_{\{2,3,4\}}}^{(8)}({\boldsymbol{\beta}},\;{r_{_1}-r_{_3}\over r_{_1}},\;{r_{_3}-r_{_1}\over r_{_2}-r_{_1}})\;,\nonumber\\
&&\varphi_{_{\{2,3,4\}}}^{(8)}({\boldsymbol{\beta}},\;x_{_1},\;x_{_2})=\varphi_{_{\{1,2,3\}}}^{(8)}(\widehat{(1234)}{\boldsymbol{\beta}},\;x_{_1},\;x_{_2})\;.
\label{Gauss-Kummer6-6-8-2}
\end{eqnarray}

Corresponding to the integer lattice $-n_{_1}E_{_3}^{(2)}+ n_{_2}E_{_3}^{(3)}$, the generalized hypergeometric function is formulated as
\begin{eqnarray}
&&\Phi_{_{\{2,3,4\}}}^{(9)}({\boldsymbol{\beta}},\;\boldsymbol{{\xi}})=A_{_{\{2,3,4\}}}^{(9)}({\boldsymbol{\beta}})(r_{_2}-r_{_1})^{\beta_{_4}+\beta_{_5}-1}(r_{_1})^{-\beta_{_4}}
\nonumber\\
&&\hspace{3.2cm}\times
\varphi_{_{\{2,3,4\}}}^{(9)}({\boldsymbol{\beta}},\;{r_{_1}-r_{_2}\over r_{_1}},\;{r_{_3}-r_{_1}\over r_{_2}-r_{_1}})\;,\nonumber\\
&&\varphi_{_{\{2,3,4\}}}^{(9)}({\boldsymbol{\beta}},\;x_{_1},\;x_{_2})=\varphi_{_{\{1,2,3\}}}^{(9)}(\widehat{(1234)}{\boldsymbol{\beta}},\;x_{_1},\;x_{_2})\;.
\label{Gauss-Kummer6-6-9-2}
\end{eqnarray}

Corresponding to the integer lattice $-n_{_1}E_{_3}^{(1)}- n_{_2}E_{_3}^{(2)}$, the generalized hypergeometric function is formulated as
\begin{eqnarray}
&&\Phi_{_{\{2,3,4\}}}^{(10)}({\boldsymbol{\beta}},\;\boldsymbol{{\xi}})=A_{_{\{2,3,4\}}}^{(10)}({\boldsymbol{\beta}})(r_{_1})^{\beta_{_5}-1}
\nonumber\\
&&\hspace{3.2cm}\times
\varphi_{_{\{2,3,4\}}}^{(10)}({\boldsymbol{\beta}},\;{r_{_1}-r_{_3}\over r_{_1}},\;{r_{_1}-r_{_2}\over r_{_1}})\;,\nonumber\\
&&\varphi_{_{\{2,3,4\}}}^{(10)}({\boldsymbol{\beta}},\;x_{_1},\;x_{_2})=\varphi_{_{\{1,2,3\}}}^{(10)}(\widehat{(1234)}{\boldsymbol{\beta}},\;x_{_1},\;x_{_2})\;.
\label{Gauss-Kummer6-6-10-2}
\end{eqnarray}

Corresponding to the integer lattice $-n_{_1}E_{_3}^{(1)}- n_{_2}E_{_3}^{(3)}$, the generalized hypergeometric function is formulated as
\begin{eqnarray}
&&\Phi_{_{\{2,3,4\}}}^{(11)}({\boldsymbol{\beta}},\;\boldsymbol{{\xi}})=A_{_{\{2,3,4\}}}^{(11)}({\boldsymbol{\beta}})(r_{_3}-r_{_1})^{\beta_{_4}+\beta_{_5}-1}(r_{_1})^{-\beta_{_4}}
\nonumber\\
&&\hspace{3.2cm}\times
\varphi_{_{\{2,3,4\}}}^{(11)}({\boldsymbol{\beta}},\;{r_{_1}-r_{_3}\over r_{_1}},\;{r_{_2}-r_{_1}\over r_{_3}-r_{_1}})\;,\nonumber\\
&&\varphi_{_{\{2,3,4\}}}^{(11)}({\boldsymbol{\beta}},\;x_{_1},\;x_{_2})=\varphi_{_{\{1,2,3\}}}^{(11)}(\widehat{(1234)}{\boldsymbol{\beta}},\;x_{_1},\;x_{_2})\;.
\label{Gauss-Kummer6-6-11-2}
\end{eqnarray}

Corresponding to the integer lattice $-n_{_1}E_{_3}^{(2)}- n_{_2}E_{_3}^{(3)}$, the generalized hypergeometric function is formulated as
\begin{eqnarray}
&&\Phi_{_{\{2,3,4\}}}^{(12)}({\boldsymbol{\beta}},\;\boldsymbol{{\xi}})=A_{_{\{2,3,4\}}}^{(12)}({\boldsymbol{\beta}})
(r_{_2}-r_{_1})^{1-\beta_{_1}-\beta_{_2}}(r_{_3}-r_{_1})^{-\beta_{_3}}(r_{_1})^{-\beta_{_4}}
\nonumber\\
&&\hspace{3.2cm}\times
\varphi_{_{\{2,3,4\}}}^{(12)}({\boldsymbol{\beta}},\;{r_{_1}-r_{_2}\over r_{_1}},\;{r_{_2}-r_{_1}\over r_{_3}-r_{_1}})\;,\nonumber\\
&&\varphi_{_{\{2,3,4\}}}^{(12)}({\boldsymbol{\beta}},\;x_{_1},\;x_{_2})=\varphi_{_{\{1,2,3\}}}^{(12)}(\widehat{(1234)}{\boldsymbol{\beta}},\;x_{_1},\;x_{_2})\;.
\label{Gauss-Kummer6-6-12-2}
\end{eqnarray}

\subsection{${\bf{\cal B}}=\{2,3,5\}$\label{sec6-7}}
\indent\indent
\begin{eqnarray}
&&{\rm Det}({\boldsymbol{\xi}}_{_{\{2,3,5\}}})=r_{_1}
\;,\nonumber\\
&&{\boldsymbol{\xi}}_{_{\{2,3,5\}}}^{-1}\cdot {\boldsymbol{\xi}}=
\left(\begin{array}{ccccc}\;-{r_{_2}\over r_{_1}}\;&\;1\;&\;0\;&\;1-{r_{_2}\over r_{_1}}\;&\;0\;\\\;-{r_{_3}\over r_{_1}}\;&\;0\;&\;1\;&\;1-{r_{_3}\over r_{_1}}\;&\;0\;\\
\;{1\over r_{_1}}\;&\;0\;&\;0\;&\;{1\over r_{_1}}\;&\;1\;\end{array}\right)\;.
\label{Gauss-Kummer6-7-1}
\end{eqnarray}

Corresponding to the integer lattice $n_{_1}E_{_3}^{(1)}+ n_{_2}E_{_3}^{(2)}$, the generalized hypergeometric function is formulated as
\begin{eqnarray}
&&\Phi_{_{\{2,3,5\}}}^{(1)}({\boldsymbol{\beta}},\;\boldsymbol{{\xi}})=A_{_{\{2,3,5\}}}^{(1)}({\boldsymbol{\beta}})
(r_{_1})^{1-\beta_{_1}-\beta_{_4}}(r_{_1}-r_{_2})^{-\beta_{_2}}(r_{_1}-r_{_3})^{-\beta_{_3}}
\nonumber\\
&&\hspace{3.2cm}\times
\varphi_{_{\{2,3,5\}}}^{(1)}({\boldsymbol{\beta}},\;{r_{_3}\over r_{_3}-r_{_1}},\;{r_{_2}\over r_{_2}-r_{_1}})\;,\nonumber\\
&&\varphi_{_{\{2,3,5\}}}^{(1)}({\boldsymbol{\beta}},\;x_{_1},\;x_{_2})=\varphi_{_{\{1,2,3\}}}^{(1)}(\widehat{(12354)}{\boldsymbol{\beta}},\;x_{_1},\;x_{_2})\;.
\label{Gauss-Kummer6-7-1-2}
\end{eqnarray}

Corresponding to the integer lattice $n_{_1}E_{_3}^{(1)}+ n_{_2}E_{_3}^{(3)}$, the generalized hypergeometric function is formulated as
\begin{eqnarray}
&&\Phi_{_{\{2,3,5\}}}^{(2)}({\boldsymbol{\beta}},\;\boldsymbol{{\xi}})=A_{_{\{2,3,5\}}}^{(2)}({\boldsymbol{\beta}})
(r_{_1})^{1-\beta_{_1}-\beta_{_4}}(-r_{_3})^{\beta_{_1}+\beta_{_5}-1}(r_{_1}-r_{_2})^{-\beta_{_2}}
\nonumber\\
&&\hspace{3.2cm}\times
(r_{_1}-r_{_3})^{\beta_{_2}+\beta_{_4}-1}
\varphi_{_{\{2,3,5\}}}^{(2)}({\boldsymbol{\beta}},\;{r_{_3}\over r_{_3}-r_{_1}},\;{r_{_2}(r_{_3}-r_{_1})\over r_{_3}(r_{_2}-r_{_1})})\;,\nonumber\\
&&\varphi_{_{\{2,3,5\}}}^{(2)}({\boldsymbol{\beta}},\;x_{_1},\;x_{_2})=\varphi_{_{\{1,2,3\}}}^{(2)}(\widehat{(12354)}{\boldsymbol{\beta}},\;x_{_1},\;x_{_2})\;.
\label{Gauss-Kummer6-7-2-2}
\end{eqnarray}

Corresponding to the integer lattice $n_{_1}E_{_3}^{(2)}+ n_{_2}E_{_3}^{(3)}$, the generalized hypergeometric function is formulated as
\begin{eqnarray}
&&\Phi_{_{\{2,3,5\}}}^{(3)}({\boldsymbol{\beta}},\;\boldsymbol{{\xi}})=A_{_{\{2,3,5\}}}^{(3)}({\boldsymbol{\beta}})
(r_{_1})^{1-\beta_{_1}-\beta_{_4}}(-r_{_2})^{1-\beta_{_2}-\beta_{_4}}(-r_{_3})^{-\beta_{_3}}
\nonumber\\
&&\hspace{3.2cm}\times
(r_{_1}-r_{_2})^{\beta_{_4}-1}
\varphi_{_{\{2,3,5\}}}^{(3)}({\boldsymbol{\beta}},\;{r_{_2}\over r_{_2}-r_{_1}},\;{r_{_2}(r_{_3}-r_{_1})\over r_{_3}(r_{_2}-r_{_1})})\;,\nonumber\\
&&\varphi_{_{\{2,3,5\}}}^{(3)}({\boldsymbol{\beta}},\;x_{_1},\;x_{_2})=\varphi_{_{\{1,2,3\}}}^{(3)}(\widehat{(12354)}{\boldsymbol{\beta}},\;x_{_1},\;x_{_2})\;.
\label{Gauss-Kummer6-7-3-2}
\end{eqnarray}

As the integer lattice is $n_{_1}E_{_3}^{(1)}- n_{_2}E_{_3}^{(2)}$, the generalized hypergeometric function is formulated as
\begin{eqnarray}
&&\Phi_{_{\{2,3,5\}}}^{(4)}({\boldsymbol{\beta}},\;\boldsymbol{{\xi}})=A_{_{\{2,3,5\}}}^{(4)}({\boldsymbol{\beta}})
(r_{_1})^{1-\beta_{_1}-\beta_{_4}}(-r_{_2})^{-\beta_{_2}}(r_{_1}-r_{_3})^{-\beta_{_3}}
\nonumber\\
&&\hspace{3.2cm}\times
\varphi_{_{\{2,3,5\}}}^{(4)}({\boldsymbol{\beta}},\;{r_{_3}\over r_{_3}-r_{_1}},\;{r_{_2}-r_{_1}\over r_{_2}})\;,\nonumber\\
&&\varphi_{_{\{2,3,5\}}}^{(4)}({\boldsymbol{\beta}},\;x_{_1},\;x_{_2})=\varphi_{_{\{1,2,3\}}}^{(4)}(\widehat{(12354)}{\boldsymbol{\beta}},\;x_{_1},\;x_{_2})\;.
\label{Gauss-Kummer6-7-4-2}
\end{eqnarray}

As the integer lattice is $n_{_1}E_{_3}^{(1)}- n_{_2}E_{_3}^{(3)}$, the generalized hypergeometric function is formulated as
\begin{eqnarray}
&&\Phi_{_{\{2,3,5\}}}^{(5)}({\boldsymbol{\beta}},\;\boldsymbol{{\xi}})=A_{_{\{2,3,5\}}}^{(5)}({\boldsymbol{\beta}})
(r_{_1})^{1-\beta_{_1}-\beta_{_4}}(-r_{_2})^{-\beta_{_2}}(-r_{_3})^{1-\beta_{_3}-\beta_{_4}}
\nonumber\\
&&\hspace{3.2cm}\times
(r_{_1}-r_{_3})^{\beta_{_4}-1}
\varphi_{_{\{2,3,5\}}}^{(5)}({\boldsymbol{\beta}},\;{r_{_3}\over r_{_3}-r_{_1}},\;{r_{_3}(r_{_2}-r_{_1})\over r_{_2}(r_{_3}-r_{_1})})\;,\nonumber\\
&&\varphi_{_{\{2,3,5\}}}^{(5)}({\boldsymbol{\beta}},\;x_{_1},\;x_{_2})=\varphi_{_{\{1,2,3\}}}^{(5)}(\widehat{(12354)}{\boldsymbol{\beta}},\;x_{_1},\;x_{_2})\;.
\label{Gauss-Kummer6-7-5-2}
\end{eqnarray}

When the integer lattice is $n_{_1}E_{_3}^{(2)}- n_{_2}E_{_3}^{(3)}$, the generalized hypergeometric function is formulated as
\begin{eqnarray}
&&\Phi_{_{\{2,3,5\}}}^{(6)}({\boldsymbol{\beta}},\;\boldsymbol{{\xi}})=A_{_{\{2,3,5\}}}^{(6)}({\boldsymbol{\beta}})
(r_{_1})^{1-\beta_{_1}-\beta_{_4}}(-r_{_2})^{\beta_{_1}+\beta_{_5}-1}(r_{_1}-r_{_2})^{\beta_{_3}+\beta_{_4}-1}
\nonumber\\
&&\hspace{3.2cm}\times
(r_{_1}-r_{_3})^{-\beta_{_3}}
\varphi_{_{\{2,3,5\}}}^{(6)}({\boldsymbol{\beta}},\;{r_{_2}\over r_{_2}-r_{_1}},\;{r_{_3}(r_{_2}-r_{_1})\over r_{_2}(r_{_3}-r_{_1})})\;,\nonumber\\
&&\varphi_{_{\{2,3,5\}}}^{(6)}({\boldsymbol{\beta}},\;x_{_1},\;x_{_2})=\varphi_{_{\{1,2,3\}}}^{(6)}(\widehat{(12354)}{\boldsymbol{\beta}},\;x_{_1},\;x_{_2})\;.
\label{Gauss-Kummer6-7-6-2}
\end{eqnarray}

Corresponding to the integer lattice $-n_{_1}E_{_3}^{(1)}+ n_{_2}E_{_3}^{(2)}$, the generalized hypergeometric function is formulated as
\begin{eqnarray}
&&\Phi_{_{\{2,3,5\}}}^{(7)}({\boldsymbol{\beta}},\;\boldsymbol{{\xi}})=A_{_{\{2,3,5\}}}^{(7)}({\boldsymbol{\beta}})
(r_{_1})^{1-\beta_{_1}-\beta_{_4}}(-r_{_3})^{-\beta_{_3}}(r_{_1}-r_{_2})^{-\beta_{_2}}
\nonumber\\
&&\hspace{3.2cm}\times
\varphi_{_{\{2,3,5\}}}^{(7)}({\boldsymbol{\beta}},\;{r_{_3}-r_{_1}\over r_{_3}},\;{r_{_2}\over r_{_2}-r_{_1}})\;,\nonumber\\
&&\varphi_{_{\{2,3,5\}}}^{(7)}({\boldsymbol{\beta}},\;x_{_1},\;x_{_2})=\varphi_{_{\{1,2,3\}}}^{(7)}(\widehat{(12354)}{\boldsymbol{\beta}},\;x_{_1},\;x_{_2})\;.
\label{Gauss-Kummer6-7-7-2}
\end{eqnarray}

Corresponding to the integer lattice $-n_{_1}E_{_3}^{(1)}+ n_{_2}E_{_3}^{(3)}$, the generalized hypergeometric function is formulated as
\begin{eqnarray}
&&\Phi_{_{\{2,3,5\}}}^{(8)}({\boldsymbol{\beta}},\;\boldsymbol{{\xi}})=A_{_{\{2,3,5\}}}^{(8)}({\boldsymbol{\beta}})
(r_{_1})^{1-\beta_{_1}-\beta_{_4}}(-r_{_3})^{\beta_{_1}-1}(r_{_1}-r_{_2})^{-\beta_{_2}}
\nonumber\\
&&\hspace{3.2cm}\times
(r_{_1}-r_{_3})^{1-\beta_{_1}-\beta_{_3}}
\varphi_{_{\{2,3,5\}}}^{(8)}({\boldsymbol{\beta}},\;{r_{_3}-r_{_1}\over r_{_3}},\;{r_{_2}(r_{_3}-r_{_1})\over r_{_3}(r_{_2}-r_{_1})})\;,\nonumber\\
&&\varphi_{_{\{2,3,5\}}}^{(8)}({\boldsymbol{\beta}},\;x_{_1},\;x_{_2})=\varphi_{_{\{1,2,3\}}}^{(8)}(\widehat{(12354)}{\boldsymbol{\beta}},\;x_{_1},\;x_{_2})\;.
\label{Gauss-Kummer6-7-8-2}
\end{eqnarray}

As the integer lattice is $-n_{_1}E_{_3}^{(2)}+ n_{_2}E_{_3}^{(3)}$, the generalized hypergeometric function is formulated as
\begin{eqnarray}
&&\Phi_{_{\{2,3,5\}}}^{(9)}({\boldsymbol{\beta}},\;\boldsymbol{{\xi}})=A_{_{\{2,3,5\}}}^{(9)}({\boldsymbol{\beta}})
(r_{_1})^{1-\beta_{_1}-\beta_{_4}}(-r_{_2})^{\beta_{_1}+\beta_{_3}-1}(-r_{_3})^{-\beta_{_3}}
\nonumber\\
&&\hspace{3.2cm}\times
(r_{_1}-r_{_2})^{\beta_{_4}+\beta_{_5}-1}
\varphi_{_{\{2,3,5\}}}^{(9)}({\boldsymbol{\beta}},\;{r_{_2}-r_{_1}\over r_{_2}},\;{r_{_2}(r_{_3}-r_{_1})\over r_{_3}(r_{_2}-r_{_1})})\;,\nonumber\\
&&\varphi_{_{\{2,3,5\}}}^{(9)}({\boldsymbol{\beta}},\;x_{_1},\;x_{_2})=\varphi_{_{\{1,2,3\}}}^{(9)}(\widehat{(12354)}{\boldsymbol{\beta}},\;x_{_1},\;x_{_2})\;.
\label{Gauss-Kummer6-7-9-2}
\end{eqnarray}

When the integer lattice is $-n_{_1}E_{_3}^{(1)}- n_{_2}E_{_3}^{(2)}$, the generalized hypergeometric function is formulated as
\begin{eqnarray}
&&\Phi_{_{\{2,3,5\}}}^{(10)}({\boldsymbol{\beta}},\;\boldsymbol{{\xi}})=A_{_{\{2,3,5\}}}^{(10)}({\boldsymbol{\beta}})
(r_{_1})^{1-\beta_{_1}-\beta_{_4}}(-r_{_2})^{-\beta_{_2}}(-r_{_3})^{-\beta_{_3}}
\nonumber\\
&&\hspace{3.2cm}\times
\varphi_{_{\{2,3,5\}}}^{(10)}({\boldsymbol{\beta}},\;{r_{_3}-r_{_1}\over r_{_3}},\;{r_{_2}-r_{_1}\over r_{_2}})\;,\nonumber\\
&&\varphi_{_{\{2,3,5\}}}^{(10)}({\boldsymbol{\beta}},\;x_{_1},\;x_{_2})=\varphi_{_{\{1,2,3\}}}^{(10)}(\widehat{(12354)}{\boldsymbol{\beta}},\;x_{_1},\;x_{_2})\;.
\label{Gauss-Kummer6-7-10-2}
\end{eqnarray}

Corresponding to the integer lattice $-n_{_1}E_{_3}^{(1)}- n_{_2}E_{_3}^{(3)}$, the generalized hypergeometric function is formulated as
\begin{eqnarray}
&&\Phi_{_{\{2,3,5\}}}^{(11)}({\boldsymbol{\beta}},\;\boldsymbol{{\xi}})=A_{_{\{2,3,5\}}}^{(11)}({\boldsymbol{\beta}})
(r_{_1})^{1-\beta_{_1}-\beta_{_4}}(-r_{_2})^{-\beta_{_2}}(-r_{_3})^{\beta_{_1}+\beta_{_2}-1}
\nonumber\\
&&\hspace{3.2cm}\times
(r_{_1}-r_{_3})^{\beta_{_4}+\beta_{_5}-1}
\varphi_{_{\{2,3,5\}}}^{(11)}({\boldsymbol{\beta}},\;{r_{_3}-r_{_1}\over r_{_3}},\;{r_{_3}(r_{_2}-r_{_1})\over r_{_2}(r_{_3}-r_{_1})})\;,\nonumber\\
&&\varphi_{_{\{2,3,5\}}}^{(11)}({\boldsymbol{\beta}},\;x_{_1},\;x_{_2})=\varphi_{_{\{1,2,3\}}}^{(11)}(\widehat{(12354)}{\boldsymbol{\beta}},\;x_{_1},\;x_{_2})\;.
\label{Gauss-Kummer6-7-11-2}
\end{eqnarray}

Corresponding to the integer lattice $-n_{_1}E_{_3}^{(2)}- n_{_2}E_{_3}^{(3)}$, the generalized hypergeometric function is formulated as
\begin{eqnarray}
&&\Phi_{_{\{2,3,5\}}}^{(12)}({\boldsymbol{\beta}},\;\boldsymbol{{\xi}})=A_{_{\{2,3,5\}}}^{(12)}({\boldsymbol{\beta}})
(r_{_1})^{1-\beta_{_1}-\beta_{_4}}(-r_{_2})^{\beta_{_1}-1}(r_{_1}-r_{_2})^{1-\beta_{_1}-\beta_{_2}}
\nonumber\\
&&\hspace{3.2cm}\times
(r_{_1}-r_{_3})^{-\beta_{_3}}
\varphi_{_{\{2,3,5\}}}^{(12)}({\boldsymbol{\beta}},\;{r_{_2}-r_{_1}\over r_{_2}},\;{r_{_3}(r_{_2}-r_{_1})\over r_{_2}(r_{_3}-r_{_1})})\;,\nonumber\\
&&\varphi_{_{\{2,3,5\}}}^{(12)}({\boldsymbol{\beta}},\;x_{_1},\;x_{_2})=\varphi_{_{\{1,2,3\}}}^{(12)}(\widehat{(12354)}{\boldsymbol{\beta}},\;x_{_1},\;x_{_2})\;.
\label{Gauss-Kummer6-7-12-2}
\end{eqnarray}

\subsection{${\bf{\cal B}}=\{1,4,5\}$\label{sec6-8}}
\indent\indent
\begin{eqnarray}
&&{\rm Det}({\boldsymbol{\xi}}_{_{\{1,4,5\}}})=r_{_3}-r_{_2}
\;,\nonumber\\
&&{\boldsymbol{\xi}}_{_{\{1,4,5\}}}^{-1}\cdot {\boldsymbol{\xi}}=
\left(\begin{array}{ccccc}\;1\;&\;-{r_{_3}-r_{_1}\over r_{_3}-r_{_2}}\;&\;-{r_{_1}-r_{_2}\over r_{_3}-r_{_2}}\;&\;0\;&\;0\;\\\;0\;&\;{r_{_3}\over r_{_3}-r_{_2}}\;&\;-{r_{_2}\over r_{_3}-r_{_2}}\;&\;1\;&\;0\;\\
\;0\;&\;-{1\over r_{_3}-r_{_2}}\;&\;{1\over r_{_3}-r_{_2}}\;&\;0\;&\;1\;\end{array}\right)\;.
\label{Gauss-Kummer6-8-1}
\end{eqnarray}

Corresponding to the integer lattice $n_{_1}E_{_3}^{(1)}+ n_{_2}E_{_3}^{(2)}$, the generalized hypergeometric function is formulated as
\begin{eqnarray}
&&\Phi_{_{\{1,4,5\}}}^{(1)}({\boldsymbol{\beta}},\;\boldsymbol{{\xi}})=A_{_{\{1,4,5\}}}^{(1)}({\boldsymbol{\beta}})
(r_{_3}-r_{_2})^{1-\beta_{_2}-\beta_{_3}}(r_{_2}-r_{_1})^{-\beta_{_1}}
\nonumber\\
&&\hspace{3.2cm}\times
(-r_{_2})^{-\beta_{_4}}
\varphi_{_{\{1,4,5\}}}^{(1)}({\boldsymbol{\beta}},\;{r_{_3}\over r_{_2}},\;{r_{_3}-r_{_1}\over r_{_2}-r_{_1}})\;,\nonumber\\
&&\varphi_{_{\{1,4,5\}}}^{(1)}({\boldsymbol{\beta}},\;x_{_1},\;x_{_2})=\varphi_{_{\{1,2,3\}}}^{(1)}(\widehat{(24)}\widehat{(35)}{\boldsymbol{\beta}},\;x_{_1},\;x_{_2})\;.
\label{Gauss-Kummer6-8-1-2}
\end{eqnarray}

Corresponding to the integer lattice $n_{_1}E_{_3}^{(1)}+ n_{_2}E_{_3}^{(3)}$, the generalized hypergeometric function is formulated as
\begin{eqnarray}
&&\Phi_{_{\{1,4,5\}}}^{(2)}({\boldsymbol{\beta}},\;\boldsymbol{{\xi}})=A_{_{\{1,4,5\}}}^{(2)}({\boldsymbol{\beta}})
(r_{_3}-r_{_2})^{1-\beta_{_2}-\beta_{_3}}(r_{_3})^{\beta_{_2}+\beta_{_5}-1}(r_{_2}-r_{_1})^{-\beta_{_1}}
\nonumber\\
&&\hspace{3.2cm}\times
(-r_{_2})^{\beta_{_1}+\beta_{_3}-1}
\varphi_{_{\{1,4,5\}}}^{(2)}({\boldsymbol{\beta}},\;{r_{_3}\over r_{_2}},\;{r_{_2}(r_{_3}-r_{_1})\over r_{_3}(r_{_2}-r_{_1})})\;,\nonumber\\
&&\varphi_{_{\{1,4,5\}}}^{(2)}({\boldsymbol{\beta}},\;x_{_1},\;x_{_2})=\varphi_{_{\{1,2,3\}}}^{(2)}(\widehat{(24)}\widehat{(35)}{\boldsymbol{\beta}},\;x_{_1},\;x_{_2})\;.
\label{Gauss-Kummer6-8-2-2}
\end{eqnarray}

As the integer lattice is $n_{_1}E_{_3}^{(2)}+ n_{_2}E_{_3}^{(3)}$, the generalized hypergeometric function is formulated as
\begin{eqnarray}
&&\Phi_{_{\{1,4,5\}}}^{(3)}({\boldsymbol{\beta}},\;\boldsymbol{{\xi}})=A_{_{\{1,4,5\}}}^{(3)}({\boldsymbol{\beta}})
(r_{_3}-r_{_2})^{1-\beta_{_2}-\beta_{_3}}(r_{_1}-r_{_3})^{1-\beta_{_1}-\beta_{_3}}(r_{_3})^{-\beta_{_4}}
\nonumber\\
&&\hspace{3.2cm}\times
(r_{_2}-r_{_1})^{\beta_{_3}-1}
\varphi_{_{\{1,4,5\}}}^{(3)}({\boldsymbol{\beta}},\;{r_{_3}-r_{_1}\over r_{_2}-r_{_1}},\;{r_{_2}(r_{_3}-r_{_1})\over r_{_3}(r_{_2}-r_{_1})})\;,\nonumber\\
&&\varphi_{_{\{1,4,5\}}}^{(3)}({\boldsymbol{\beta}},\;x_{_1},\;x_{_2})=\varphi_{_{\{1,2,3\}}}^{(3)}(\widehat{(24)}\widehat{(35)}{\boldsymbol{\beta}},\;x_{_1},\;x_{_2})\;.
\label{Gauss-Kummer6-8-3-2}
\end{eqnarray}

As the integer lattice is $n_{_1}E_{_3}^{(1)}- n_{_2}E_{_3}^{(2)}$, the generalized hypergeometric function is formulated as
\begin{eqnarray}
&&\Phi_{_{\{1,4,5\}}}^{(4)}({\boldsymbol{\beta}},\;\boldsymbol{{\xi}})=A_{_{\{1,4,5\}}}^{(4)}({\boldsymbol{\beta}})
(r_{_3}-r_{_2})^{1-\beta_{_2}-\beta_{_3}}(r_{_1}-r_{_3})^{-\beta_{_1}}(-r_{_2})^{-\beta_{_4}}
\nonumber\\
&&\hspace{3.2cm}\times
\varphi_{_{\{1,4,5\}}}^{(4)}({\boldsymbol{\beta}},\;{r_{_3}\over r_{_2}},\;{r_{_2}-r_{_1}\over r_{_3}-r_{_1}})\;,\nonumber\\
&&\varphi_{_{\{1,4,5\}}}^{(4)}({\boldsymbol{\beta}},\;x_{_1},\;x_{_2})=\varphi_{_{\{1,2,3\}}}^{(4)}(\widehat{(24)}\widehat{(35)}{\boldsymbol{\beta}},\;x_{_1},\;x_{_2})\;.
\label{Gauss-Kummer6-8-4-2}
\end{eqnarray}

Corresponding to the integer lattice $n_{_1}E_{_3}^{(1)}- n_{_2}E_{_3}^{(3)}$, the generalized hypergeometric function is formulated as
\begin{eqnarray}
&&\Phi_{_{\{1,4,5\}}}^{(5)}({\boldsymbol{\beta}},\;\boldsymbol{{\xi}})=A_{_{\{1,4,5\}}}^{(5)}({\boldsymbol{\beta}})
(r_{_3}-r_{_2})^{1-\beta_{_2}-\beta_{_3}}(r_{_1}-r_{_3})^{-\beta_{_1}}(r_{_3})^{1-\beta_{_3}-\beta_{_4}}
\nonumber\\
&&\hspace{3.2cm}\times
(-r_{_2})^{\beta_{_3}-1}
\varphi_{_{\{1,4,5\}}}^{(5)}({\boldsymbol{\beta}},\;{r_{_3}\over r_{_2}},\;{r_{_3}(r_{_2}-r_{_1})\over r_{_2}(r_{_3}-r_{_1})})\;,\nonumber\\
&&\varphi_{_{\{1,4,5\}}}^{(5)}({\boldsymbol{\beta}},\;x_{_1},\;x_{_2})=\varphi_{_{\{1,2,3\}}}^{(5)}(\widehat{(24)}\widehat{(35)}{\boldsymbol{\beta}},\;x_{_1},\;x_{_2})\;.
\label{Gauss-Kummer6-8-5-2}
\end{eqnarray}

Corresponding to the integer lattice $n_{_1}E_{_3}^{(2)}- n_{_2}E_{_3}^{(3)}$, the generalized hypergeometric function is formulated as
\begin{eqnarray}
&&\Phi_{_{\{1,4,5\}}}^{(6)}({\boldsymbol{\beta}},\;\boldsymbol{{\xi}})=A_{_{\{1,4,5\}}}^{(6)}({\boldsymbol{\beta}})
(r_{_3}-r_{_2})^{1-\beta_{_2}-\beta_{_3}}(r_{_1}-r_{_3})^{\beta_{_2}+\beta_{_5}-1}(-r_{_2})^{-\beta_{_4}}
\nonumber\\
&&\hspace{3.2cm}\times
(r_{_2}-r_{_1})^{\beta_{_3}+\beta_{_4}-1}
\varphi_{_{\{1,4,5\}}}^{(6)}({\boldsymbol{\beta}},\;{r_{_3}-r_{_1}\over r_{_2}-r_{_1}},\;{r_{_3}(r_{_2}-r_{_1})\over r_{_2}(r_{_3}-r_{_1})})\;,\nonumber\\
&&\varphi_{_{\{1,4,5\}}}^{(6)}({\boldsymbol{\beta}},\;x_{_1},\;x_{_2})=\varphi_{_{\{1,2,3\}}}^{(6)}(\widehat{(24)}\widehat{(35)}{\boldsymbol{\beta}},\;x_{_1},\;x_{_2})\;.
\label{Gauss-Kummer6-8-6-2}
\end{eqnarray}

When the integer lattice is $-n_{_1}E_{_3}^{(1)}+ n_{_2}E_{_3}^{(2)}$, the generalized hypergeometric function is formulated as
\begin{eqnarray}
&&\Phi_{_{\{1,4,5\}}}^{(7)}({\boldsymbol{\beta}},\;\boldsymbol{{\xi}})=A_{_{\{1,4,5\}}}^{(7)}({\boldsymbol{\beta}})
(r_{_3}-r_{_2})^{1-\beta_{_2}-\beta_{_3}}(r_{_3})^{-\beta_{_4}}
\nonumber\\
&&\hspace{3.2cm}\times
(r_{_2}-r_{_1})^{-\beta_{_1}}
\varphi_{_{\{1,4,5\}}}^{(7)}({\boldsymbol{\beta}},\;{r_{_2}\over r_{_3}},\;{r_{_3}-r_{_1}\over r_{_2}-r_{_1}})\;,\nonumber\\
&&\varphi_{_{\{1,4,5\}}}^{(7)}({\boldsymbol{\beta}},\;x_{_1},\;x_{_2})=\varphi_{_{\{1,2,3\}}}^{(7)}(\widehat{(24)}\widehat{(35)}{\boldsymbol{\beta}},\;x_{_1},\;x_{_2})\;.
\label{Gauss-Kummer6-8-7-2}
\end{eqnarray}

When the integer lattice is $-n_{_1}E_{_3}^{(1)}+ n_{_2}E_{_3}^{(3)}$, the generalized hypergeometric function is formulated as
\begin{eqnarray}
&&\Phi_{_{\{1,4,5\}}}^{(8)}({\boldsymbol{\beta}},\;\boldsymbol{{\xi}})=A_{_{\{1,4,5\}}}^{(8)}({\boldsymbol{\beta}})
(r_{_3}-r_{_2})^{1-\beta_{_2}-\beta_{_3}}(r_{_3})^{\beta_{_2}-1}(-r_{_2})^{1-\beta_{_2}-\beta_{_4}}
\nonumber\\
&&\hspace{3.2cm}\times
(r_{_2}-r_{_1})^{-\beta_{_1}}
\varphi_{_{\{1,4,5\}}}^{(8)}({\boldsymbol{\beta}},\;{r_{_2}\over r_{_3}},\;{r_{_2}(r_{_3}-r_{_1})\over r_{_3}(r_{_2}-r_{_1})})\;,\nonumber\\
&&\varphi_{_{\{1,4,5\}}}^{(8)}({\boldsymbol{\beta}},\;x_{_1},\;x_{_2})=\varphi_{_{\{1,2,3\}}}^{(8)}(\widehat{(24)}\widehat{(35)}{\boldsymbol{\beta}},\;x_{_1},\;x_{_2})\;.
\label{Gauss-Kummer6-8-8-2}
\end{eqnarray}

As the integer lattice is $-n_{_1}E_{_3}^{(2)}+ n_{_2}E_{_3}^{(3)}$, the generalized hypergeometric function is formulated as
\begin{eqnarray}
&&\Phi_{_{\{1,4,5\}}}^{(9)}({\boldsymbol{\beta}},\;\boldsymbol{{\xi}})=A_{_{\{1,4,5\}}}^{(9)}({\boldsymbol{\beta}})
(r_{_3}-r_{_2})^{1-\beta_{_2}-\beta_{_3}}(r_{_1}-r_{_3})^{\beta_{_2}+\beta_{_4}-1}(r_{_3})^{-\beta_{_4}}
\nonumber\\
&&\hspace{3.2cm}\times
(r_{_2}-r_{_1})^{\beta_{_3}+\beta_{_5}-1}
\varphi_{_{\{1,4,5\}}}^{(9)}({\boldsymbol{\beta}},\;{r_{_2}-r_{_1}\over r_{_3}-r_{_1}},\;{r_{_2}(r_{_3}-r_{_1})\over r_{_3}(r_{_2}-r_{_1})})\;,\nonumber\\
&&\varphi_{_{\{1,4,5\}}}^{(9)}({\boldsymbol{\beta}},\;x_{_1},\;x_{_2})=\varphi_{_{\{1,2,3\}}}^{(9)}(\widehat{(24)}\widehat{(35)}{\boldsymbol{\beta}},\;x_{_1},\;x_{_2})\;.
\label{Gauss-Kummer6-8-9-2}
\end{eqnarray}

As the integer lattice is $-n_{_1}E_{_3}^{(1)}- n_{_2}E_{_3}^{(2)}$, the generalized hypergeometric function is formulated as
\begin{eqnarray}
&&\Phi_{_{\{1,4,5\}}}^{(10)}({\boldsymbol{\beta}},\;\boldsymbol{{\xi}})=A_{_{\{1,4,5\}}}^{(10)}({\boldsymbol{\beta}})
(r_{_3}-r_{_2})^{1-\beta_{_2}-\beta_{_3}}(r_{_1}-r_{_3})^{-\beta_{_1}}(r_{_3})^{-\beta_{_4}}
\nonumber\\
&&\hspace{3.2cm}\times
\varphi_{_{\{1,4,5\}}}^{(10)}({\boldsymbol{\beta}},\;{r_{_2}\over r_{_3}},\;{r_{_2}-r_{_1}\over r_{_3}-r_{_1}})\;,\nonumber\\
&&\varphi_{_{\{1,4,5\}}}^{(10)}({\boldsymbol{\beta}},\;x_{_1},\;x_{_2})=\varphi_{_{\{1,2,3\}}}^{(10)}(\widehat{(24)}\widehat{(35)}{\boldsymbol{\beta}},\;x_{_1},\;x_{_2})\;.
\label{Gauss-Kummer6-8-10-2}
\end{eqnarray}

Corresponding to the integer lattice $-n_{_1}E_{_3}^{(1)}- n_{_2}E_{_3}^{(3)}$, the generalized hypergeometric function is formulated as
\begin{eqnarray}
&&\Phi_{_{\{1,4,5\}}}^{(11)}({\boldsymbol{\beta}},\;\boldsymbol{{\xi}})=A_{_{\{1,4,5\}}}^{(11)}({\boldsymbol{\beta}})
(r_{_3}-r_{_2})^{1-\beta_{_2}-\beta_{_3}}(r_{_1}-r_{_3})^{-\beta_{_1}}(r_{_3})^{\beta_{_1}+\beta_{_2}-1}
\nonumber\\
&&\hspace{3.2cm}\times
(-r_{_2})^{\beta_{_3}+\beta_{_5}-1}
\varphi_{_{\{1,4,5\}}}^{(11)}({\boldsymbol{\beta}},\;{r_{_2}\over r_{_3}},\;{r_{_3}(r_{_2}-r_{_1})\over r_{_2}(r_{_3}-r_{_1})})\;,\nonumber\\
&&\varphi_{_{\{1,4,5\}}}^{(11)}({\boldsymbol{\beta}},\;x_{_1},\;x_{_2})=\varphi_{_{\{1,2,3\}}}^{(11)}(\widehat{(24)}\widehat{(35)}{\boldsymbol{\beta}},\;x_{_1},\;x_{_2})\;.
\label{Gauss-Kummer6-8-11-2}
\end{eqnarray}

As the integer lattice is $-n_{_1}E_{_3}^{(2)}- n_{_2}E_{_3}^{(3)}$, the generalized hypergeometric function is formulated as
\begin{eqnarray}
&&\Phi_{_{\{1,4,5\}}}^{(12)}({\boldsymbol{\beta}},\;\boldsymbol{{\xi}})=A_{_{\{1,4,5\}}}^{(12)}({\boldsymbol{\beta}})
(r_{_3}-r_{_2})^{1-\beta_{_2}-\beta_{_3}}(r_{_1}-r_{_3})^{\beta_{_2}-1}(-r_{_2})^{-\beta_{_4}}
\nonumber\\
&&\hspace{3.2cm}\times
(r_{_2}-r_{_1})^{1-\beta_{_1}-\beta_{_2}}
\varphi_{_{\{1,4,5\}}}^{(12)}({\boldsymbol{\beta}},\;{r_{_2}-r_{_1}\over r_{_3}-r_{_1}},\;{r_{_3}(r_{_2}-r_{_1})\over r_{_2}(r_{_3}-r_{_1})})\;,\nonumber\\
&&\varphi_{_{\{1,4,5\}}}^{(12)}({\boldsymbol{\beta}},\;x_{_1},\;x_{_2})=\varphi_{_{\{1,2,3\}}}^{(12)}(\widehat{(24)}\widehat{(35)}{\boldsymbol{\beta}},\;x_{_1},\;x_{_2})\;.
\label{Gauss-Kummer6-8-12-2}
\end{eqnarray}

\subsection{${\bf{\cal B}}=\{2,4,5\}$\label{sec6-9}}
\indent\indent
\begin{eqnarray}
&&{\rm Det}({\boldsymbol{\xi}}_{_{\{2,4,5\}}})=r_{_1}-r_{_3}
\;,\nonumber\\
&&{\boldsymbol{\xi}}_{_{\{2,4,5\}}}^{-1}\cdot {\boldsymbol{\xi}}=
\left(\begin{array}{ccccc}\;-{r_{_3}-r_{_2}\over r_{_3}-r_{_1}}\;&\;1\;&\;-{r_{_2}-r_{_1}\over r_{_3}-r_{_1}}\;&\;0\;&\;0\;\\\;{r_{_3}\over r_{_3}-r_{_1}}\;&\;0\;&\;-{r_{_1}\over r_{_3}-r_{_1}}\;&\;1\;&\;0\;\\
\;-{1\over r_{_3}-r_{_1}}\;&\;0\;&\;{1\over r_{_3}-r_{_1}}\;&\;0\;&\;1\;\end{array}\right)\;.
\label{Gauss-Kummer6-9-1}
\end{eqnarray}

Corresponding to the integer lattice $n_{_1}E_{_3}^{(1)}+ n_{_2}E_{_3}^{(2)}$, the generalized hypergeometric function is formulated as
\begin{eqnarray}
&&\Phi_{_{\{2,4,5\}}}^{(1)}({\boldsymbol{\beta}},\;\boldsymbol{{\xi}})=A_{_{\{2,4,5\}}}^{(1)}({\boldsymbol{\beta}})
(r_{_3}-r_{_1})^{1-\beta_{_1}-\beta_{_3}}(r_{_1}-r_{_2})^{-\beta_{_2}}
\nonumber\\
&&\hspace{3.2cm}\times
(-r_{_1})^{-\beta_{_4}}
\varphi_{_{\{2,4,5\}}}^{(1)}({\boldsymbol{\beta}},\;{r_{_3}\over r_{_1}},\;{r_{_3}-r_{_2}\over r_{_1}-r_{_2}})\;,\nonumber\\
&&\varphi_{_{\{2,4,5\}}}^{(1)}({\boldsymbol{\beta}},\;x_{_1},\;x_{_2})=\varphi_{_{\{1,2,3\}}}^{(1)}(\widehat{(124)}\widehat{(35)}{\boldsymbol{\beta}},\;x_{_1},\;x_{_2})\;.
\label{Gauss-Kummer6-9-1-2}
\end{eqnarray}

Corresponding to the integer lattice $n_{_1}E_{_3}^{(1)}+ n_{_2}E_{_3}^{(3)}$, the generalized hypergeometric function is formulated as
\begin{eqnarray}
&&\Phi_{_{\{2,4,5\}}}^{(2)}({\boldsymbol{\beta}},\;\boldsymbol{{\xi}})=A_{_{\{2,4,5\}}}^{(2)}({\boldsymbol{\beta}})
(r_{_3}-r_{_1})^{1-\beta_{_1}-\beta_{_3}}(r_{_3})^{\beta_{_1}+\beta_{_5}-1}(r_{_1}-r_{_2})^{-\beta_{_2}}
\nonumber\\
&&\hspace{3.2cm}\times
(-r_{_1})^{\beta_{_2}+\beta_{_3}-1}
\varphi_{_{\{2,4,5\}}}^{(2)}({\boldsymbol{\beta}},\;{r_{_3}\over r_{_1}},\;{r_{_1}(r_{_3}-r_{_2})\over r_{_3}(r_{_1}-r_{_2})})\;,\nonumber\\
&&\varphi_{_{\{2,4,5\}}}^{(2)}({\boldsymbol{\beta}},\;x_{_1},\;x_{_2})=\varphi_{_{\{1,2,3\}}}^{(2)}(\widehat{(124)}\widehat{(35)}{\boldsymbol{\beta}},\;x_{_1},\;x_{_2})\;.
\label{Gauss-Kummer6-9-2-2}
\end{eqnarray}

When the integer lattice is $n_{_1}E_{_3}^{(2)}+ n_{_2}E_{_3}^{(3)}$, the generalized hypergeometric function is formulated as
\begin{eqnarray}
&&\Phi_{_{\{2,4,5\}}}^{(3)}({\boldsymbol{\beta}},\;\boldsymbol{{\xi}})=A_{_{\{2,4,5\}}}^{(3)}({\boldsymbol{\beta}})
(r_{_3}-r_{_1})^{1-\beta_{_1}-\beta_{_3}}(r_{_2}-r_{_3})^{1-\beta_{_2}-\beta_{_3}}(r_{_3})^{-\beta_{_4}}
\nonumber\\
&&\hspace{3.2cm}\times
(r_{_1}-r_{_2})^{\beta_{_3}-1}
\varphi_{_{\{2,4,5\}}}^{(3)}({\boldsymbol{\beta}},\;{r_{_3}-r_{_2}\over r_{_1}-r_{_2}},\;{r_{_1}(r_{_3}-r_{_2})\over r_{_3}(r_{_1}-r_{_2})})\;,\nonumber\\
&&\varphi_{_{\{2,4,5\}}}^{(3)}({\boldsymbol{\beta}},\;x_{_1},\;x_{_2})=\varphi_{_{\{1,2,3\}}}^{(3)}(\widehat{(124)}\widehat{(35)}{\boldsymbol{\beta}},\;x_{_1},\;x_{_2})\;.
\label{Gauss-Kummer6-9-3-2}
\end{eqnarray}

As the integer lattice is $n_{_1}E_{_3}^{(1)}- n_{_2}E_{_3}^{(2)}$, the generalized hypergeometric function is formulated as
\begin{eqnarray}
&&\Phi_{_{\{2,4,5\}}}^{(4)}({\boldsymbol{\beta}},\;\boldsymbol{{\xi}})=A_{_{\{2,4,5\}}}^{(4)}({\boldsymbol{\beta}})
(r_{_3}-r_{_1})^{1-\beta_{_1}-\beta_{_3}}(r_{_2}-r_{_3})^{-\beta_{_2}}
\nonumber\\
&&\hspace{3.2cm}\times
(-r_{_1})^{-\beta_{_4}}
\varphi_{_{\{2,4,5\}}}^{(4)}({\boldsymbol{\beta}},\;{r_{_3}\over r_{_1}},\;{r_{_1}-r_{_2}\over r_{_3}-r_{_2}})\;,\nonumber\\
&&\varphi_{_{\{2,4,5\}}}^{(4)}({\boldsymbol{\beta}},\;x_{_1},\;x_{_2})=\varphi_{_{\{1,2,3\}}}^{(4)}(\widehat{(124)}\widehat{(35)}{\boldsymbol{\beta}},\;x_{_1},\;x_{_2})\;.
\label{Gauss-Kummer6-9-4-2}
\end{eqnarray}

Corresponding to the integer lattice $n_{_1}E_{_3}^{(1)}- n_{_2}E_{_3}^{(3)}$, the generalized hypergeometric function is formulated as
\begin{eqnarray}
&&\Phi_{_{\{2,4,5\}}}^{(5)}({\boldsymbol{\beta}},\;\boldsymbol{{\xi}})=A_{_{\{2,4,5\}}}^{(5)}({\boldsymbol{\beta}})
(r_{_3}-r_{_1})^{1-\beta_{_1}-\beta_{_3}}(r_{_2}-r_{_3})^{-\beta_{_2}}(r_{_3})^{1-\beta_{_3}-\beta_{_4}}
\nonumber\\
&&\hspace{3.2cm}\times
(-r_{_1})^{\beta_{_3}-1}
\varphi_{_{\{2,4,5\}}}^{(5)}({\boldsymbol{\beta}},\;{r_{_3}\over r_{_1}},\;{r_{_3}(r_{_1}-r_{_2})\over r_{_1}(r_{_3}-r_{_2})})\;,\nonumber\\
&&\varphi_{_{\{2,4,5\}}}^{(5)}({\boldsymbol{\beta}},\;x_{_1},\;x_{_2})=\varphi_{_{\{1,2,3\}}}^{(5)}(\widehat{(124)}\widehat{(35)}{\boldsymbol{\beta}},\;x_{_1},\;x_{_2})\;.
\label{Gauss-Kummer6-9-5-2}
\end{eqnarray}

As the integer lattice is $n_{_1}E_{_3}^{(2)}- n_{_2}E_{_3}^{(3)}$, the generalized hypergeometric function is formulated as
\begin{eqnarray}
&&\Phi_{_{\{2,4,5\}}}^{(6)}({\boldsymbol{\beta}},\;\boldsymbol{{\xi}})=A_{_{\{2,4,5\}}}^{(6)}({\boldsymbol{\beta}})
(r_{_3}-r_{_1})^{1-\beta_{_1}-\beta_{_3}}(r_{_2}-r_{_3})^{\beta_{_1}+\beta_{_5}-1}(-r_{_1})^{-\beta_{_4}}
\nonumber\\
&&\hspace{3.2cm}\times
(r_{_1}-r_{_2})^{\beta_{_3}+\beta_{_4}-1}
\varphi_{_{\{2,4,5\}}}^{(6)}({\boldsymbol{\beta}},\;{r_{_3}-r_{_2}\over r_{_1}-r_{_2}},\;{r_{_3}(r_{_1}-r_{_2})\over r_{_1}(r_{_3}-r_{_2})})\;,\nonumber\\
&&\varphi_{_{\{2,4,5\}}}^{(6)}({\boldsymbol{\beta}},\;x_{_1},\;x_{_2})=\varphi_{_{\{1,2,3\}}}^{(6)}(\widehat{(124)}\widehat{(35)}{\boldsymbol{\beta}},\;x_{_1},\;x_{_2})\;.
\label{Gauss-Kummer6-9-6-2}
\end{eqnarray}

When the integer lattice is $-n_{_1}E_{_3}^{(1)}+ n_{_2}E_{_3}^{(2)}$, the generalized hypergeometric function is formulated as
\begin{eqnarray}
&&\Phi_{_{\{2,4,5\}}}^{(7)}({\boldsymbol{\beta}},\;\boldsymbol{{\xi}})=A_{_{\{2,4,5\}}}^{(7)}({\boldsymbol{\beta}})
(r_{_3}-r_{_1})^{1-\beta_{_1}-\beta_{_3}}(r_{_3})^{-\beta_{_4}}(r_{_1}-r_{_2})^{-\beta_{_2}}
\nonumber\\
&&\hspace{3.2cm}\times
\varphi_{_{\{2,4,5\}}}^{(7)}({\boldsymbol{\beta}},\;{r_{_1}\over r_{_3}},\;{r_{_3}-r_{_2}\over r_{_1}-r_{_2}})\;,\nonumber\\
&&\varphi_{_{\{2,4,5\}}}^{(7)}({\boldsymbol{\beta}},\;x_{_1},\;x_{_2})=\varphi_{_{\{1,2,3\}}}^{(7)}(\widehat{(124)}\widehat{(35)}{\boldsymbol{\beta}},\;x_{_1},\;x_{_2})\;.
\label{Gauss-Kummer6-9-7-2}
\end{eqnarray}

When the integer lattice is $-n_{_1}E_{_3}^{(1)}+ n_{_2}E_{_3}^{(3)}$, the generalized hypergeometric function is formulated as
\begin{eqnarray}
&&\Phi_{_{\{2,4,5\}}}^{(8)}({\boldsymbol{\beta}},\;\boldsymbol{{\xi}})=A_{_{\{2,4,5\}}}^{(8)}({\boldsymbol{\beta}})
(r_{_3}-r_{_1})^{1-\beta_{_1}-\beta_{_3}}(r_{_3})^{\beta_{_1}-1}(-r_{_1})^{1-\beta_{_1}-\beta_{_4}}
\nonumber\\
&&\hspace{3.2cm}\times
(r_{_1}-r_{_2})^{-\beta_{_2}}
\varphi_{_{\{2,4,5\}}}^{(8)}({\boldsymbol{\beta}},\;{r_{_1}\over r_{_3}},\;{r_{_1}(r_{_3}-r_{_2})\over r_{_3}(r_{_1}-r_{_2})})\;,\nonumber\\
&&\varphi_{_{\{2,4,5\}}}^{(8)}({\boldsymbol{\beta}},\;x_{_1},\;x_{_2})=\varphi_{_{\{1,2,3\}}}^{(8)}(\widehat{(124)}\widehat{(35)}{\boldsymbol{\beta}},\;x_{_1},\;x_{_2})\;.
\label{Gauss-Kummer6-9-8-2}
\end{eqnarray}

Corresponding to the integer lattice $-n_{_1}E_{_3}^{(2)}+ n_{_2}E_{_3}^{(3)}$, the generalized hypergeometric function is formulated as
\begin{eqnarray}
&&\Phi_{_{\{2,4,5\}}}^{(9)}({\boldsymbol{\beta}},\;\boldsymbol{{\xi}})=A_{_{\{2,4,5\}}}^{(9)}({\boldsymbol{\beta}})
(r_{_3}-r_{_1})^{1-\beta_{_1}-\beta_{_3}}(r_{_2}-r_{_3})^{\beta_{_1}+\beta_{_4}-1}(r_{_3})^{-\beta_{_4}}
\nonumber\\
&&\hspace{3.2cm}\times
(r_{_1}-r_{_2})^{\beta_{_3}+\beta_{_5}-1}
\varphi_{_{\{2,4,5\}}}^{(9)}({\boldsymbol{\beta}},\;{r_{_1}-r_{_2}\over r_{_3}-r_{_2}},\;{r_{_1}(r_{_3}-r_{_2})\over r_{_3}(r_{_1}-r_{_2})})\;,\nonumber\\
&&\varphi_{_{\{2,4,5\}}}^{(9)}({\boldsymbol{\beta}},\;x_{_1},\;x_{_2})=\varphi_{_{\{1,2,3\}}}^{(9)}(\widehat{(124)}\widehat{(35)}{\boldsymbol{\beta}},\;x_{_1},\;x_{_2})\;.
\label{Gauss-Kummer6-9-9-2}
\end{eqnarray}

Corresponding to the integer lattice $-n_{_1}E_{_3}^{(1)}- n_{_2}E_{_3}^{(2)}$, the generalized hypergeometric function is formulated as
\begin{eqnarray}
&&\Phi_{_{\{2,4,5\}}}^{(10)}({\boldsymbol{\beta}},\;\boldsymbol{{\xi}})=A_{_{\{2,4,5\}}}^{(10)}({\boldsymbol{\beta}})
(r_{_3}-r_{_1})^{1-\beta_{_1}-\beta_{_3}}(r_{_2}-r_{_3})^{-\beta_{_2}}(r_{_3})^{-\beta_{_4}}
\nonumber\\
&&\hspace{3.2cm}\times
\varphi_{_{\{2,4,5\}}}^{(10)}({\boldsymbol{\beta}},\;{r_{_1}\over r_{_3}},\;{r_{_1}-r_{_2}\over r_{_3}-r_{_2}})\;,\nonumber\\
&&\varphi_{_{\{2,4,5\}}}^{(10)}({\boldsymbol{\beta}},\;x_{_1},\;x_{_2})=\varphi_{_{\{1,2,3\}}}^{(10)}(\widehat{(124)}\widehat{(35)}{\boldsymbol{\beta}},\;x_{_1},\;x_{_2})\;.
\label{Gauss-Kummer6-9-10-2}
\end{eqnarray}

Corresponding to the integer lattice $-n_{_1}E_{_3}^{(1)}- n_{_2}E_{_3}^{(3)}$, the generalized hypergeometric function is formulated as
\begin{eqnarray}
&&\Phi_{_{\{2,4,5\}}}^{(11)}({\boldsymbol{\beta}},\;\boldsymbol{{\xi}})=A_{_{\{2,4,5\}}}^{(11)}({\boldsymbol{\beta}})
(r_{_3}-r_{_1})^{1-\beta_{_1}-\beta_{_3}}(r_{_2}-r_{_3})^{-\beta_{_2}}(r_{_3})^{\beta_{_1}+\beta_{_2}-1}
\nonumber\\
&&\hspace{3.2cm}\times
(-r_{_1})^{\beta_{_3}+\beta_{_5}-1}
\varphi_{_{\{2,4,5\}}}^{(11)}({\boldsymbol{\beta}},\;{r_{_1}\over r_{_3}},\;{r_{_3}(r_{_1}-r_{_2})\over r_{_1}(r_{_3}-r_{_2})})\;,\nonumber\\
&&\varphi_{_{\{2,4,5\}}}^{(11)}({\boldsymbol{\beta}},\;x_{_1},\;x_{_2})=\varphi_{_{\{1,2,3\}}}^{(11)}(\widehat{(124)}\widehat{(35)}{\boldsymbol{\beta}},\;x_{_1},\;x_{_2})\;.
\label{Gauss-Kummer6-9-11-2}
\end{eqnarray}

Corresponding to the integer lattice $-n_{_1}E_{_3}^{(2)}- n_{_2}E_{_3}^{(3)}$, the generalized hypergeometric function is formulated as
\begin{eqnarray}
&&\Phi_{_{\{2,4,5\}}}^{(12)}({\boldsymbol{\beta}},\;\boldsymbol{{\xi}})=A_{_{\{2,4,5\}}}^{(12)}({\boldsymbol{\beta}})
(r_{_3}-r_{_1})^{1-\beta_{_1}-\beta_{_3}}(r_{_2}-r_{_3})^{\beta_{_1}-1}(-r_{_1})^{-\beta_{_4}}
\nonumber\\
&&\hspace{3.2cm}\times
(r_{_1}-r_{_2})^{1-\beta_{_1}-\beta_{_2}}
\varphi_{_{\{2,4,5\}}}^{(12)}({\boldsymbol{\beta}},\;{r_{_1}-r_{_2}\over r_{_3}-r_{_2}},\;{r_{_3}(r_{_1}-r_{_2})\over r_{_1}(r_{_3}-r_{_2})})\;,\nonumber\\
&&\varphi_{_{\{2,4,5\}}}^{(12)}({\boldsymbol{\beta}},\;x_{_1},\;x_{_2})=\varphi_{_{\{1,2,3\}}}^{(12)}(\widehat{(124)}\widehat{(35)}{\boldsymbol{\beta}},\;x_{_1},\;x_{_2})\;.
\label{Gauss-Kummer6-9-12-2}
\end{eqnarray}

\subsection{${\bf{\cal B}}=\{3,4,5\}$\label{sec6-10}}
\indent\indent
\begin{eqnarray}
&&{\rm Det}({\boldsymbol{\xi}}_{_{\{3,4,5\}}})=r_{_2}-r_{_1}
\;,\nonumber\\
&&{\boldsymbol{\xi}}_{_{\{3,4,5\}}}^{-1}\cdot {\boldsymbol{\xi}}=
\left(\begin{array}{ccccc}\;{r_{_3}-r_{_2}\over r_{_2}-r_{_1}}\;&\;-{r_{_3}-r_{_1}\over r_{_2}-r_{_1}}\;&\;1\;&\;0\;&\;0\;\\\;{r_{_2}\over r_{_2}-r_{_1}}\;&\;-{r_{_1}\over r_{_2}-r_{_1}}\;&\;0\;&\;1\;&\;0\;\\
\;-{1\over r_{_2}-r_{_1}}\;&\;{1\over r_{_2}-r_{_1}}\;&\;0\;&\;0\;&\;1\;\end{array}\right)\;.
\label{Gauss-Kummer6-10-1}
\end{eqnarray}

As the integer lattice is $n_{_1}E_{_3}^{(1)}+ n_{_2}E_{_3}^{(2)}$, the generalized hypergeometric function is formulated as
\begin{eqnarray}
&&\Phi_{_{\{3,4,5\}}}^{(1)}({\boldsymbol{\beta}},\;\boldsymbol{{\xi}})=A_{_{\{3,4,5\}}}^{(1)}({\boldsymbol{\beta}})
(r_{_2}-r_{_1})^{1-\beta_{_1}-\beta_{_2}}(r_{_1}-r_{_3})^{-\beta_{_3}}
\nonumber\\
&&\hspace{3.2cm}\times
(-r_{_1})^{-\beta_{_4}}
\varphi_{_{\{3,4,5\}}}^{(1)}({\boldsymbol{\beta}},\;{r_{_2}\over r_{_1}},\;{r_{_2}-r_{_3}\over r_{_1}-r_{_3}})\;,\nonumber\\
&&\varphi_{_{\{3,4,5\}}}^{(1)}({\boldsymbol{\beta}},\;x_{_1},\;x_{_2})=\varphi_{_{\{1,2,3\}}}^{(1)}(\widehat{(13524)}{\boldsymbol{\beta}},\;x_{_1},\;x_{_2})\;.
\label{Gauss-Kummer6-10-1-2}
\end{eqnarray}

As the integer lattice is $n_{_1}E_{_3}^{(1)}+ n_{_2}E_{_3}^{(3)}$, the generalized hypergeometric function is formulated as
\begin{eqnarray}
&&\Phi_{_{\{3,4,5\}}}^{(2)}({\boldsymbol{\beta}},\;\boldsymbol{{\xi}})=A_{_{\{3,4,5\}}}^{(2)}({\boldsymbol{\beta}})
(r_{_2}-r_{_1})^{1-\beta_{_1}-\beta_{_2}}(r_{_2})^{\beta_{_1}+\beta_{_5}-1}(r_{_1}-r_{_3})^{-\beta_{_3}}
\nonumber\\
&&\hspace{3.2cm}\times
(-r_{_1})^{\beta_{_2}+\beta_{_3}-1}
\varphi_{_{\{3,4,5\}}}^{(2)}({\boldsymbol{\beta}},\;{r_{_2}\over r_{_1}},\;{r_{_1}(r_{_2}-r_{_3})\over r_{_2}(r_{_1}-r_{_3})})\;,\nonumber\\
&&\varphi_{_{\{3,4,5\}}}^{(2)}({\boldsymbol{\beta}},\;x_{_1},\;x_{_2})=\varphi_{_{\{1,2,3\}}}^{(2)}(\widehat{(13524)}{\boldsymbol{\beta}},\;x_{_1},\;x_{_2})\;.
\label{Gauss-Kummer6-10-2-2}
\end{eqnarray}

Corresponding to the integer lattice $n_{_1}E_{_3}^{(2)}+ n_{_2}E_{_3}^{(3)}$, the generalized hypergeometric function is formulated as
\begin{eqnarray}
&&\Phi_{_{\{3,4,5\}}}^{(3)}({\boldsymbol{\beta}},\;\boldsymbol{{\xi}})=A_{_{\{3,4,5\}}}^{(3)}({\boldsymbol{\beta}})
(r_{_2}-r_{_1})^{1-\beta_{_1}-\beta_{_2}}(r_{_3}-r_{_2})^{1-\beta_{_2}-\beta_{_3}}(r_{_2})^{-\beta_{_4}}
\nonumber\\
&&\hspace{3.2cm}\times
(r_{_1}-r_{_3})^{\beta_{_2}-1}
\varphi_{_{\{3,4,5\}}}^{(3)}({\boldsymbol{\beta}},\;{r_{_2}-r_{_3}\over r_{_1}-r_{_3}},\;{r_{_1}(r_{_2}-r_{_3})\over r_{_2}(r_{_1}-r_{_3})})\;,\nonumber\\
&&\varphi_{_{\{3,4,5\}}}^{(3)}({\boldsymbol{\beta}},\;x_{_1},\;x_{_2})=\varphi_{_{\{1,2,3\}}}^{(3)}(\widehat{(13524)}{\boldsymbol{\beta}},\;x_{_1},\;x_{_2})\;.
\label{Gauss-Kummer6-10-3-2}
\end{eqnarray}

Corresponding to the integer lattice $n_{_1}E_{_3}^{(1)}- n_{_2}E_{_3}^{(2)}$, the generalized hypergeometric function is formulated as
\begin{eqnarray}
&&\Phi_{_{\{3,4,5\}}}^{(4)}({\boldsymbol{\beta}},\;\boldsymbol{{\xi}})=A_{_{\{3,4,5\}}}^{(4)}({\boldsymbol{\beta}})
(r_{_2}-r_{_1})^{1-\beta_{_1}-\beta_{_2}}(r_{_3}-r_{_2})^{-\beta_{_3}}
\nonumber\\
&&\hspace{3.2cm}\times
(-r_{_1})^{-\beta_{_4}}
\varphi_{_{\{3,4,5\}}}^{(4)}({\boldsymbol{\beta}},\;{r_{_2}\over r_{_1}},\;{r_{_1}-r_{_3}\over r_{_2}-r_{_3}})\;,\nonumber\\
&&\varphi_{_{\{3,4,5\}}}^{(4)}({\boldsymbol{\beta}},\;x_{_1},\;x_{_2})=\varphi_{_{\{1,2,3\}}}^{(4)}(\widehat{(13524)}{\boldsymbol{\beta}},\;x_{_1},\;x_{_2})\;.
\label{Gauss-Kummer6-10-4-2}
\end{eqnarray}

Corresponding to the integer lattice $n_{_1}E_{_3}^{(1)}- n_{_2}E_{_3}^{(3)}$, the generalized hypergeometric function is formulated as
\begin{eqnarray}
&&\Phi_{_{\{3,4,5\}}}^{(5)}({\boldsymbol{\beta}},\;\boldsymbol{{\xi}})=A_{_{\{3,4,5\}}}^{(5)}({\boldsymbol{\beta}})
(r_{_2}-r_{_1})^{1-\beta_{_1}-\beta_{_2}}(r_{_3}-r_{_2})^{-\beta_{_3}}(r_{_2})^{1-\beta_{_2}-\beta_{_4}}
\nonumber\\
&&\hspace{3.2cm}\times
(-r_{_1})^{\beta_{_2}-1}
\varphi_{_{\{3,4,5\}}}^{(5)}({\boldsymbol{\beta}},\;{r_{_2}\over r_{_1}},\;{r_{_2}(r_{_1}-r_{_3})\over r_{_1}(r_{_2}-r_{_3})})\;,\nonumber\\
&&\varphi_{_{\{3,4,5\}}}^{(5)}({\boldsymbol{\beta}},\;x_{_1},\;x_{_2})=\varphi_{_{\{1,2,3\}}}^{(5)}(\widehat{(13524)}{\boldsymbol{\beta}},\;x_{_1},\;x_{_2})\;.
\label{Gauss-Kummer6-10-5-2}
\end{eqnarray}

When the integer lattice is $n_{_1}E_{_3}^{(2)}- n_{_2}E_{_3}^{(3)}$, the generalized hypergeometric function is formulated as
\begin{eqnarray}
&&\Phi_{_{\{3,4,5\}}}^{(6)}({\boldsymbol{\beta}},\;\boldsymbol{{\xi}})=A_{_{\{3,4,5\}}}^{(6)}({\boldsymbol{\beta}})
(r_{_2}-r_{_1})^{1-\beta_{_1}-\beta_{_2}}(r_{_3}-r_{_2})^{\beta_{_1}+\beta_{_5}-1}(-r_{_1})^{-\beta_{_4}}
\nonumber\\
&&\hspace{3.2cm}\times
(r_{_1}-r_{_3})^{\beta_{_2}+\beta_{_4}-1}
\varphi_{_{\{3,4,5\}}}^{(6)}({\boldsymbol{\beta}},\;{r_{_2}-r_{_3}\over r_{_1}-r_{_3}},\;{r_{_2}(r_{_1}-r_{_3})\over r_{_1}(r_{_2}-r_{_3})})\;,\nonumber\\
&&\varphi_{_{\{3,4,5\}}}^{(6)}({\boldsymbol{\beta}},\;x_{_1},\;x_{_2})=\varphi_{_{\{1,2,3\}}}^{(6)}(\widehat{(13524)}{\boldsymbol{\beta}},\;x_{_1},\;x_{_2})\;.
\label{Gauss-Kummer6-10-6-2}
\end{eqnarray}

When the integer lattice is $-n_{_1}E_{_3}^{(1)}+ n_{_2}E_{_3}^{(2)}$, the generalized hypergeometric function is formulated as
\begin{eqnarray}
&&\Phi_{_{\{3,4,5\}}}^{(7)}({\boldsymbol{\beta}},\;\boldsymbol{{\xi}})=A_{_{\{3,4,5\}}}^{(7)}({\boldsymbol{\beta}})
(r_{_2}-r_{_1})^{1-\beta_{_1}-\beta_{_2}}(r_{_2})^{-\beta_{_4}}
\nonumber\\
&&\hspace{3.2cm}\times
(r_{_1}-r_{_3})^{-\beta_{_3}}
\varphi_{_{\{3,4,5\}}}^{(7)}({\boldsymbol{\beta}},\;{r_{_1}\over r_{_2}},\;{r_{_2}-r_{_3}\over r_{_1}-r_{_3}})\;,\nonumber\\
&&\varphi_{_{\{3,4,5\}}}^{(7)}({\boldsymbol{\beta}},\;x_{_1},\;x_{_2})=\varphi_{_{\{1,2,3\}}}^{(7)}(\widehat{(13524)}{\boldsymbol{\beta}},\;x_{_1},\;x_{_2})\;.
\label{Gauss-Kummer6-10-7-2}
\end{eqnarray}

Corresponding to the integer lattice $-n_{_1}E_{_3}^{(1)}+ n_{_2}E_{_3}^{(3)}$, the generalized hypergeometric function is formulated as
\begin{eqnarray}
&&\Phi_{_{\{3,4,5\}}}^{(8)}({\boldsymbol{\beta}},\;\boldsymbol{{\xi}})=A_{_{\{3,4,5\}}}^{(8)}({\boldsymbol{\beta}})
(r_{_2}-r_{_1})^{1-\beta_{_1}-\beta_{_2}}(r_{_2})^{\beta_{_1}-1}(r_{_1}-r_{_3})^{-\beta_{_3}}
\nonumber\\
&&\hspace{3.2cm}\times
(-r_{_1})^{1-\beta_{_1}-\beta_{_4}}
\varphi_{_{\{3,4,5\}}}^{(8)}({\boldsymbol{\beta}},\;{r_{_1}\over r_{_2}},\;{r_{_1}(r_{_2}-r_{_3})\over r_{_2}(r_{_1}-r_{_3})})\;,\nonumber\\
&&\varphi_{_{\{3,4,5\}}}^{(8)}({\boldsymbol{\beta}},\;x_{_1},\;x_{_2})=\varphi_{_{\{1,2,3\}}}^{(8)}(\widehat{(13524)}{\boldsymbol{\beta}},\;x_{_1},\;x_{_2})\;.
\label{Gauss-Kummer6-10-8-2}
\end{eqnarray}

Corresponding to the integer lattice $-n_{_1}E_{_3}^{(2)}+ n_{_2}E_{_3}^{(3)}$, the generalized hypergeometric function is formulated as
\begin{eqnarray}
&&\Phi_{_{\{3,4,5\}}}^{(9)}({\boldsymbol{\beta}},\;\boldsymbol{{\xi}})=A_{_{\{3,4,5\}}}^{(9)}({\boldsymbol{\beta}})
(r_{_2}-r_{_1})^{1-\beta_{_1}-\beta_{_2}}(r_{_3}-r_{_2})^{\beta_{_1}+\beta_{_4}-1}(r_{_2})^{-\beta_{_4}}
\nonumber\\
&&\hspace{3.2cm}\times
(r_{_1}-r_{_3})^{\beta_{_2}+\beta_{_5}-1}
\varphi_{_{\{3,4,5\}}}^{(9)}({\boldsymbol{\beta}},\;{r_{_1}-r_{_3}\over r_{_2}-r_{_3}},\;{r_{_1}(r_{_2}-r_{_3})\over r_{_2}(r_{_1}-r_{_3})})\;,\nonumber\\
&&\varphi_{_{\{3,4,5\}}}^{(9)}({\boldsymbol{\beta}},\;x_{_1},\;x_{_2})=\varphi_{_{\{1,2,3\}}}^{(9)}(\widehat{(13524)}{\boldsymbol{\beta}},\;x_{_1},\;x_{_2})\;.
\label{Gauss-Kummer6-10-9-2}
\end{eqnarray}

As the integer lattice is $-n_{_1}E_{_3}^{(1)}- n_{_2}E_{_3}^{(2)}$, the generalized hypergeometric function is formulated as
\begin{eqnarray}
&&\Phi_{_{\{3,4,5\}}}^{(10)}({\boldsymbol{\beta}},\;\boldsymbol{{\xi}})=A_{_{\{3,4,5\}}}^{(10)}({\boldsymbol{\beta}})
(r_{_2}-r_{_1})^{1-\beta_{_1}-\beta_{_2}}(r_{_3}-r_{_2})^{-\beta_{_3}}
\nonumber\\
&&\hspace{3.2cm}\times
(r_{_2})^{-\beta_{_4}}
\varphi_{_{\{3,4,5\}}}^{(10)}({\boldsymbol{\beta}},\;{r_{_1}\over r_{_2}},\;{r_{_1}-r_{_3}\over r_{_2}-r_{_3}})\;,\nonumber\\
&&\varphi_{_{\{3,4,5\}}}^{(10)}({\boldsymbol{\beta}},\;x_{_1},\;x_{_2})=\varphi_{_{\{1,2,3\}}}^{(10)}(\widehat{(13524)}{\boldsymbol{\beta}},\;x_{_1},\;x_{_2})\;.
\label{Gauss-Kummer6-10-10-2}
\end{eqnarray}

Corresponding to the integer lattice $-n_{_1}E_{_3}^{(1)}- n_{_2}E_{_3}^{(3)}$, the generalized hypergeometric function is formulated as
\begin{eqnarray}
&&\Phi_{_{\{3,4,5\}}}^{(11)}({\boldsymbol{\beta}},\;\boldsymbol{{\xi}})=A_{_{\{3,4,5\}}}^{(11)}({\boldsymbol{\beta}})
(r_{_2}-r_{_1})^{1-\beta_{_1}-\beta_{_2}}(r_{_3}-r_{_2})^{-\beta_{_3}}(r_{_2})^{\beta_{_1}+\beta_{_3}-1}
\nonumber\\
&&\hspace{3.2cm}\times
(-r_{_1})^{\beta_{_2}+\beta_{_5}-1}
\varphi_{_{\{3,4,5\}}}^{(11)}({\boldsymbol{\beta}},\;{r_{_1}\over r_{_2}},\;{r_{_2}(r_{_1}-r_{_3})\over r_{_1}(r_{_2}-r_{_3})})\;,\nonumber\\
&&\varphi_{_{\{3,4,5\}}}^{(11)}({\boldsymbol{\beta}},\;x_{_1},\;x_{_2})=\varphi_{_{\{1,2,3\}}}^{(11)}(\widehat{(13524)}{\boldsymbol{\beta}},\;x_{_1},\;x_{_2})\;.
\label{Gauss-Kummer6-10-11-2}
\end{eqnarray}

As the integer lattice is $-n_{_1}E_{_3}^{(2)}- n_{_2}E_{_3}^{(3)}$, the generalized hypergeometric function is formulated as
\begin{eqnarray}
&&\Phi_{_{\{3,4,5\}}}^{(12)}({\boldsymbol{\beta}},\;\boldsymbol{{\xi}})=A_{_{\{3,4,5\}}}^{(12)}({\boldsymbol{\beta}})
(r_{_2}-r_{_1})^{1-\beta_{_1}-\beta_{_2}}(r_{_3}-r_{_2})^{\beta_{_1}-1}(r_{_1}-r_{_3})^{1-\beta_{_1}-\beta_{_3}}
\nonumber\\
&&\hspace{3.2cm}\times
(-r_{_1})^{-\beta_{_4}}
\varphi_{_{\{3,4,5\}}}^{(12)}({\boldsymbol{\beta}},\;{r_{_1}-r_{_3}\over r_{_2}-r_{_3}},\;{r_{_2}(r_{_1}-r_{_3})\over r_{_1}(r_{_2}-r_{_3})})\;,\nonumber\\
&&\varphi_{_{\{3,4,5\}}}^{(12)}({\boldsymbol{\beta}},\;x_{_1},\;x_{_2})=\varphi_{_{\{1,2,3\}}}^{(12)}(\widehat{(13524)}{\boldsymbol{\beta}},\;x_{_1},\;x_{_2})\;.
\label{Gauss-Kummer6-10-12-2}
\end{eqnarray}

\section{The inverse Gauss relations\label{app2}}
\indent\indent
Under the inverse transformations of the variables in the function $\varphi_{_{\{1,2,3\}}}^{(2)}$, the inverse Gauss relations are
\begin{eqnarray}
&&\varphi_{_{\{1,2,3\}}}^{(2)}({\boldsymbol{\beta}},\;x_{_1},\;x_{_2})
\nonumber\\
&&\hspace{-0.5cm}=
{\Gamma(\beta_{_1}+\beta_{_5})\Gamma(\beta_{_2}+\beta_{_5}-1)\over\Gamma(1-\beta_{_3}-\beta_{_4})\Gamma(\beta_{_5})}
(-x_{_2})^{-\beta_{_1}}\varphi_{_{\{1,2,3\}}}^{(5)}({\boldsymbol{\beta}},\;x_{_1},\;{1\over x_{_2}})
\nonumber\\
&&\hspace{0.0cm}
+{\Gamma(\beta_{_1}+\beta_{_5})\Gamma(1-\beta_{_2}-\beta_{_5})\over\Gamma(\beta_{_1})\Gamma(1-\beta_{_2})}(-x_{_2})^{\beta_{_3}+\beta_{_4}-1}
\varphi_{_{\{1,2,3\}}}^{(6)}({\boldsymbol{\beta}},\;x_{_1}x_{_2},\;{1\over x_{_2}})
\nonumber\\
&&\hspace{-0.5cm}=
{\Gamma(\beta_{_2}+\beta_{_4}-1)\Gamma(\beta_{_3}+\beta_{_4})\over\Gamma(1-\beta_{_1}-\beta_{_5})\Gamma(\beta_{_4})}
(-x_{_1})^{-\beta_{_3}}\varphi_{_{\{1,2,3\}}}^{(8)}({\boldsymbol{\beta}},\;{1\over x_{_1}},\;x_{_2})
\nonumber\\
&&\hspace{0.0cm}
+{\Gamma(\beta_{_3}+\beta_{_4})\Gamma(1-\beta_{_2}-\beta_{_4})\over\Gamma(\beta_{_3})\Gamma(1-\beta_{_2})}(-x_{_1})^{\beta_{_1}+\beta_{_5}-1}
\varphi_{_{\{1,2,3\}}}^{(7)}({\boldsymbol{\beta}},\;{1\over x_{_1}},\;x_{_1}x_{_2})\;.
\label{Inverse-lattice2-6}
\end{eqnarray}

The inverse Gauss relations corresponding to the inverse transformations of the variables in the function $\varphi_{_{\{1,2,3\}}}^{(3)}$ are
\begin{eqnarray}
&&\varphi_{_{\{1,2,3\}}}^{(3)}({\boldsymbol{\beta}},\;x_{_1},\;x_{_2})
\nonumber\\
&&\hspace{-0.5cm}=
{\Gamma(2-\beta_{_1}-\beta_{_5})\Gamma(1-\beta_{_2}-\beta_{_5})\over\Gamma(\beta_{_3}+\beta_{_4})\Gamma(1-\beta_{_5})}
(-x_{_2})^{-\beta_{_2}}\varphi_{_{\{1,2,3\}}}^{(6)}({\boldsymbol{\beta}},\;x_{_1},\;{1\over x_{_2}})
\nonumber\\
&&\hspace{0.0cm}
+{\Gamma(\beta_{_2}+\beta_{_5}-1)\Gamma(2-\beta_{_1}-\beta_{_5})\over\Gamma(\beta_{_2})\Gamma(1-\beta_{_1})}(-x_{_2})^{\beta_{_5}-1}
\varphi_{_{\{1,2,3\}}}^{(5)}({\boldsymbol{\beta}},\;{x_{_1}\over x_{_2}}\;,{1\over x_{_2}})
\nonumber\\
&&\hspace{-0.5cm}=
{\Gamma(2-\beta_{_1}-\beta_{_5})\Gamma(1-\beta_{_3}-\beta_{_5})\over\Gamma(\beta_{_2}+\beta_{_4})\Gamma(1-\beta_{_5})}
(-x_{_1})^{-\beta_{_3}}\varphi_{_{\{1,2,3\}}}^{(9)}({\boldsymbol{\beta}},\;{1\over x_{_1}},\;x_{_2})
\nonumber\\
&&\hspace{0.0cm}
+{\Gamma(\beta_{_3}+\beta_{_5}-1)\Gamma(2-\beta_{_1}-\beta_{_5})\over\Gamma(\beta_{_3})\Gamma(1-\beta_{_1})}(-x_{_1})^{\beta_{_5}-1}
\varphi_{_{\{1,2,3\}}}^{(10)}({\boldsymbol{\beta}},\;{x_{_2}\over x_{_1}},\;{1\over x_{_1}})\;.
\label{Inverse-lattice3-6}
\end{eqnarray}

The inverse Gauss relations corresponding to the inverse transformations of the variables in the function $\varphi_{_{\{1,2,3\}}}^{(4)}$ are
\begin{eqnarray}
&&\varphi_{_{\{1,2,3\}}}^{(4)}({\boldsymbol{\beta}},\;x_{_1},\;x_{_2})
\nonumber\\
&&\hspace{-0.5cm}=
{\Gamma(\beta_{_3}+\beta_{_4}-1)\Gamma(\beta_{_1}+\beta_{_4})\over\Gamma(1-\beta_{_2}-\beta_{_5})\Gamma(\beta_{_4})}
(-x_{_2})^{-\beta_{_1}}\varphi_{_{\{1,2,3\}}}^{(1)}({\boldsymbol{\beta}},\;x_{_1},\;{1\over x_{_2}})
\nonumber\\
&&\hspace{0.0cm}
+{\Gamma(\beta_{_1}+\beta_{_4})\Gamma(1-\beta_{_3}-\beta_{_4})\over\Gamma(\beta_{_1})\Gamma(1-\beta_{_3})}(-x_{_2})^{\beta_{_2}+\beta_{_5}-1}
\varphi_{_{\{1,2,3\}}}^{(6)}({\boldsymbol{\beta}},\;{1\over x_{_2}},\;x_{_1}x_{_2})
\nonumber\\
&&\hspace{-0.5cm}=
{\Gamma(\beta_{_2}+\beta_{_5})\Gamma(\beta_{_3}+\beta_{_5}-1)\over\Gamma(1-\beta_{_1}-\beta_{_4})\Gamma(\beta_{_5})}
(-x_{_1})^{-\beta_{_2}}\varphi_{_{\{1,2,3\}}}^{(10)}({\boldsymbol{\beta}},\;{1\over x_{_1}},\;x_{_2})
\nonumber\\
&&\hspace{0.0cm}
+{\Gamma(\beta_{_2}+\beta_{_5})\Gamma(1-\beta_{_3}-\beta_{_5})\over\Gamma(\beta_{_2})\Gamma(1-\beta_{_3})}(-x_{_1})^{\beta_{_1}+\beta_{_4}-1}
\varphi_{_{\{1,2,3\}}}^{(11)}({\boldsymbol{\beta}},\;{1\over x_{_1}},\;x_{_1}x_{_2})\;.
\label{Inverse-lattice4-6}
\end{eqnarray}

Under the inverse transformations of the variables in the function $\varphi_{_{\{1,2,3\}}}^{(5)}$, the inverse Gauss relations are
\begin{eqnarray}
&&\varphi_{_{\{1,2,3\}}}^{(5)}({\boldsymbol{\beta}},\;x_{_1},\;x_{_2})
\nonumber\\
&&\hspace{-0.5cm}=
{\Gamma(1-\beta_{_1}-\beta_{_5})\Gamma(2-\beta_{_2}-\beta_{_5})\over\Gamma(\beta_{_3}+\beta_{_4})\Gamma(1-\beta_{_5})}
(-x_{_2})^{-\beta_{_1}}\varphi_{_{\{1,2,3\}}}^{(2)}({\boldsymbol{\beta}},\;x_{_1},\;{1\over x_{_2}})
\nonumber\\
&&\hspace{0.0cm}
+{\Gamma(\beta_{_1}+\beta_{_5}-1)\Gamma(2-\beta_{_2}-\beta_{_5})\over\Gamma(\beta_{_1})\Gamma(1-\beta_{_2})}(-x_{_2})^{\beta_{_5}-1}
\varphi_{_{\{1,2,3\}}}^{(3)}({\boldsymbol{\beta}},\;{x_{_1}\over x_{_2}}\;,{1\over x_{_2}})
\nonumber\\
&&\hspace{-0.5cm}=
{\Gamma(1-\beta_{_3}-\beta_{_5})\Gamma(2-\beta_{_2}-\beta_{_5})\over\Gamma(\beta_{_1}+\beta_{_4})\Gamma(1-\beta_{_5})}
(-x_{_1})^{-\beta_{_3}}\varphi_{_{\{1,2,3\}}}^{(11)}({\boldsymbol{\beta}},\;{1\over x_{_1}},\;x_{_2})
\nonumber\\
&&\hspace{0.0cm}
+{\Gamma(\beta_{_3}+\beta_{_5}-1)\Gamma(2-\beta_{_2}-\beta_{_5})\over\Gamma(\beta_{_3})\Gamma(1-\beta_{_2})}(-x_{_1})^{\beta_{_5}-1}
\varphi_{_{\{1,2,3\}}}^{(10)}({\boldsymbol{\beta}},\;{1\over x_{_1}},\;{x_{_2}\over x_{_1}})\;.
\label{Inverse-lattice5-6}
\end{eqnarray}

Under the inverse transformations of the variables in the function $\varphi_{_{\{1,2,3\}}}^{(6)}$, the inverse Gauss relations are
\begin{eqnarray}
&&\varphi_{_{\{1,2,3\}}}^{(6)}({\boldsymbol{\beta}},\;x_{_1},\;x_{_2})
\nonumber\\
&&\hspace{-0.5cm}=
{\Gamma(\beta_{_1}+\beta_{_5}-1)\Gamma(\beta_{_2}+\beta_{_5})\over\Gamma(1-\beta_{_3}-\beta_{_4})\Gamma(\beta_{_5})}
(-x_{_2})^{-\beta_{_2}}\varphi_{_{\{1,2,3\}}}^{(3)}({\boldsymbol{\beta}},\;x_{_1},\;{1\over x_{_2}})
\nonumber\\
&&\hspace{0.0cm}
+{\Gamma(\beta_{_2}+\beta_{_5})\Gamma(1-\beta_{_1}-\beta_{_5})\over\Gamma(\beta_{_2})\Gamma(1-\beta_{_1})}(-x_{_2})^{\beta_{_3}+\beta_{_4}-1}
\varphi_{_{\{1,2,3\}}}^{(2)}({\boldsymbol{\beta}},\;x_{_1}x_{_2},\;{1\over x_{_2}})
\nonumber\\
&&\hspace{-0.5cm}=
{\Gamma(\beta_{_1}+\beta_{_4}-1)\Gamma(\beta_{_3}+\beta_{_4})\over\Gamma(1-\beta_{_2}-\beta_{_5})\Gamma(\beta_{_4})}
(-x_{_1})^{-\beta_{_3}}\varphi_{_{\{1,2,3\}}}^{(12)}({\boldsymbol{\beta}},\;{1\over x_{_1}},\;x_{_2})
\nonumber\\
&&\hspace{0.0cm}
+{\Gamma(\beta_{_3}+\beta_{_4})\Gamma(1-\beta_{_1}-\beta_{_4})\over\Gamma(\beta_{_3})\Gamma(1-\beta_{_1})}(-x_{_1})^{\beta_{_2}+\beta_{_5}-1}
\varphi_{_{\{1,2,3\}}}^{(4)}({\boldsymbol{\beta}},\;x_{_1}x_{_2},\;{1\over x_{_1}})\;.
\label{Inverse-lattice6-6}
\end{eqnarray}

Similarly the inverse Gauss relations originating from the inverse transformations of the variables in the function $\varphi_{_{\{1,2,3\}}}^{(7)}$ are
\begin{eqnarray}
&&\varphi_{_{\{1,2,3\}}}^{(7)}({\boldsymbol{\beta}},\;x_{_1},\;x_{_2})
\nonumber\\
&&\hspace{-0.5cm}=
{\Gamma(\beta_{_3}+\beta_{_5}-1)\Gamma(\beta_{_1}+\beta_{_5})\over\Gamma(1-\beta_{_2}-\beta_{_4})\Gamma(\beta_{_5})}
(-x_{_2})^{-\beta_{_1}}\varphi_{_{\{1,2,3\}}}^{(10)}({\boldsymbol{\beta}},\;x_{_1},\;{1\over x_{_2}})
\nonumber\\
&&\hspace{0.0cm}
+{\Gamma(\beta_{_1}+\beta_{_5})\Gamma(1-\beta_{_3}-\beta_{_5})\over\Gamma(\beta_{_1})\Gamma(1-\beta_{_3})}(-x_{_2})^{\beta_{_2}+\beta_{_4}-1}
\varphi_{_{\{1,2,3\}}}^{(9)}({\boldsymbol{\beta}},\;{1\over x_{_2}},\;x_{_1}x_{_2})
\nonumber\\
&&\hspace{-0.5cm}=
{\Gamma(\beta_{_3}+\beta_{_4}-1)\Gamma(\beta_{_2}+\beta_{_4})\over\Gamma(1-\beta_{_1}-\beta_{_5})\Gamma(\beta_{_4})}
(-x_{_1})^{-\beta_{_2}}\varphi_{_{\{1,2,3\}}}^{(1)}({\boldsymbol{\beta}},\;{1\over x_{_1}},\;x_{_2})
\nonumber\\
&&\hspace{0.0cm}
+{\Gamma(\beta_{_2}+\beta_{_4})\Gamma(1-\beta_{_3}-\beta_{_4})\over\Gamma(\beta_{_2})\Gamma(1-\beta_{_3})}(-x_{_1})^{\beta_{_1}+\beta_{_5}-1}
\varphi_{_{\{1,2,3\}}}^{(2)}({\boldsymbol{\beta}},\;{1\over x_{_1}},\;x_{_1}x_{_2})\;.
\label{Inverse-lattice7-6}
\end{eqnarray}

The inverse Gauss relations corresponding to the inverse transformations of the variables in the function $\varphi_{_{\{1,2,3\}}}^{(8)}$ are
\begin{eqnarray}
&&\varphi_{_{\{1,2,3\}}}^{(8)}({\boldsymbol{\beta}},\;x_{_1},\;x_{_2})
\nonumber\\
&&\hspace{-0.5cm}=
{\Gamma(1-\beta_{_1}-\beta_{_4})\Gamma(2-\beta_{_2}-\beta_{_4})\over\Gamma(\beta_{_3}+\beta_{_5})\Gamma(1-\beta_{_4})}
(-x_{_2})^{-\beta_{_1}}\varphi_{_{\{1,2,3\}}}^{(11)}({\boldsymbol{\beta}},\;x_{_1},\;{1\over x_{_2}})
\nonumber\\
&&\hspace{0.0cm}
+{\Gamma(\beta_{_1}+\beta_{_4}-1)\Gamma(2-\beta_{_2}-\beta_{_4})\over\Gamma(\beta_{_1})\Gamma(1-\beta_{_2})}(-x_{_2})^{\beta_{_4}-1}
\varphi_{_{\{1,2,3\}}}^{(12)}({\boldsymbol{\beta}},\;{x_{_1}\over x_{_2}},\;{1\over x_{_2}})
\nonumber\\
&&\hspace{-0.5cm}=
{\Gamma(1-\beta_{_3}-\beta_{_4})\Gamma(2-\beta_{_2}-\beta_{_4})\over\Gamma(\beta_{_1}+\beta_{_5})\Gamma(1-\beta_{_4})}
(-x_{_1})^{-\beta_{_3}}\varphi_{_{\{1,2,3\}}}^{(2)}({\boldsymbol{\beta}},\;{1\over x_{_1}},\;x_{_2})
\nonumber\\
&&\hspace{0.0cm}
+{\Gamma(\beta_{_3}+\beta_{_4}-1)\Gamma(2-\beta_{_2}-\beta_{_4})\over\Gamma(\beta_{_3})\Gamma(1-\beta_{_2})}(-x_{_1})^{\beta_{_4}-1}
\varphi_{_{\{1,2,3\}}}^{(1)}({\boldsymbol{\beta}},\;{1\over x_{_1}},\;{x_{_2}\over x_{_1}})\;.
\label{Inverse-lattice8-6}
\end{eqnarray}

The inverse Gauss relations originating from the inverse transformations of the variables in the function $\varphi_{_{\{1,2,3\}}}^{(9)}$ are
\begin{eqnarray}
&&\varphi_{_{\{1,2,3\}}}^{(9)}({\boldsymbol{\beta}},\;x_{_1},\;x_{_2})
\nonumber\\
&&\hspace{-0.5cm}=
{\Gamma(\beta_{_1}+\beta_{_4}-1)\Gamma(\beta_{_2}+\beta_{_4})\over\Gamma(1-\beta_{_3}-\beta_{_5})\Gamma(\beta_{_4})}
(-x_{_2})^{-\beta_{_2}}\varphi_{_{\{1,2,3\}}}^{(12)}({\boldsymbol{\beta}},\;x_{_1},\;{1\over x_{_2}})
\nonumber\\
&&\hspace{0.0cm}
+{\Gamma(\beta_{_2}+\beta_{_4})\Gamma(1-\beta_{_1}-\beta_{_4})\over\Gamma(\beta_{_2})\Gamma(1-\beta_{_1})}(-x_{_2})^{\beta_{_3}+\beta_{_5}-1}
\varphi_{_{\{1,2,3\}}}^{(11)}({\boldsymbol{\beta}},\;x_{_1}x_{_2},\;{1\over x_{_2}})
\nonumber\\
&&\hspace{-0.5cm}=
{\Gamma(\beta_{_1}+\beta_{_5}-1)\Gamma(\beta_{_3}+\beta_{_5})\over\Gamma(1-\beta_{_2}-\beta_{_4})\Gamma(\beta_{_5})}
(-x_{_1})^{-\beta_{_3}}\varphi_{_{\{1,2,3\}}}^{(3)}({\boldsymbol{\beta}},\;{1\over x_{_1}},\;x_{_2})
\nonumber\\
&&\hspace{0.0cm}
+{\Gamma(\beta_{_3}+\beta_{_5})\Gamma(1-\beta_{_1}-\beta_{_5})\over\Gamma(\beta_{_3})\Gamma(1-\beta_{_1})}(-x_{_1})^{\beta_{_2}+\beta_{_4}-1}
\varphi_{_{\{1,2,3\}}}^{(7)}({\boldsymbol{\beta}},\;x_{_1}x_{_2},\;{1\over x_{_1}})\;.
\label{Inverse-lattice9-6}
\end{eqnarray}

The inverse Gauss relations corresponding to the inverse transformations of the variables in the function $\varphi_{_{\{1,2,3\}}}^{(10)}$ are
\begin{eqnarray}
&&\varphi_{_{\{1,2,3\}}}^{(10)}({\boldsymbol{\beta}},\;x_{_1},\;x_{_2})
\nonumber\\
&&\hspace{-0.5cm}=
{\Gamma(1-\beta_{_1}-\beta_{_5})\Gamma(2-\beta_{_3}-\beta_{_5})\over\Gamma(\beta_{_2}+\beta_{_4})\Gamma(1-\beta_{_5})}
(-x_{_2})^{-\beta_{_1}}\varphi_{_{\{1,2,3\}}}^{(7)}({\boldsymbol{\beta}},\;x_{_1},\;{1\over x_{_2}})
\nonumber\\
&&\hspace{0.0cm}
+{\Gamma(\beta_{_1}+\beta_{_5}-1)\Gamma(2-\beta_{_3}-\beta_{_5})\over\Gamma(\beta_{_1})\Gamma(1-\beta_{_3})}(-x_{_2})^{\beta_{_5}-1}
\varphi_{_{\{1,2,3\}}}^{(3)}({\boldsymbol{\beta}},\;{1\over x_{_2}},\;{x_{_1}\over x_{_2}})
\nonumber\\
&&\hspace{-0.5cm}=
{\Gamma(1-\beta_{_2}-\beta_{_5})\Gamma(2-\beta_{_3}-\beta_{_5})\over\Gamma(\beta_{_1}+\beta_{_4})\Gamma(1-\beta_{_5})}
(-x_{_1})^{-\beta_{_2}}\varphi_{_{\{1,2,3\}}}^{(4)}({\boldsymbol{\beta}},\;{1\over x_{_1}},\;x_{_2})
\nonumber\\
&&\hspace{0.0cm}
+{\Gamma(\beta_{_2}+\beta_{_5}-1)\Gamma(2-\beta_{_3}-\beta_{_5})\over\Gamma(\beta_{_2})\Gamma(1-\beta_{_3})}(-x_{_1})^{\beta_{_5}-1}
\varphi_{_{\{1,2,3\}}}^{(5)}({\boldsymbol{\beta}},\;{1\over x_{_1}},\;{x_{_2}\over x_{_1}})\;.
\label{Inverse-lattice10-6}
\end{eqnarray}

The inverse Gauss relations corresponding to the inverse transformations of the variables in the function $\varphi_{_{\{1,2,3\}}}^{(11)}$ are
\begin{eqnarray}
&&\varphi_{_{\{1,2,3\}}}^{(11)}({\boldsymbol{\beta}},\;x_{_1},\;x_{_2})
\nonumber\\
&&\hspace{-0.5cm}=
{\Gamma(\beta_{_2}+\beta_{_4}-1)\Gamma(\beta_{_1}+\beta_{_4})\over\Gamma(1-\beta_{_3}-\beta_{_5})\Gamma(\beta_{_4})}
(-x_{_2})^{-\beta_{_1}}\varphi_{_{\{1,2,3\}}}^{(8)}({\boldsymbol{\beta}},\;x_{_1},\;{1\over x_{_2}})
\nonumber\\
&&\hspace{0.0cm}
+{\Gamma(\beta_{_1}+\beta_{_4})\Gamma(1-\beta_{_2}-\beta_{_4})\over\Gamma(\beta_{_1})\Gamma(1-\beta_{_2})}(-x_{_2})^{\beta_{_3}+\beta_{_5}-1}
\varphi_{_{\{1,2,3\}}}^{(9)}({\boldsymbol{\beta}},\;x_{_1}x_{_2},\;{1\over x_{_2}})
\nonumber\\
&&\hspace{-0.5cm}=
{\Gamma(\beta_{_2}+\beta_{_5}-1)\Gamma(\beta_{_3}+\beta_{_5})\over\Gamma(1-\beta_{_1}-\beta_{_4})\Gamma(\beta_{_5})}
(-x_{_1})^{-\beta_{_3}}\varphi_{_{\{1,2,3\}}}^{(5)}({\boldsymbol{\beta}},\;{1\over x_{_1}},\;x_{_2})
\nonumber\\
&&\hspace{0.0cm}
+{\Gamma(\beta_{_3}+\beta_{_5})\Gamma(1-\beta_{_2}-\beta_{_5})\over\Gamma(\beta_{_3})\Gamma(1-\beta_{_2})}(-x_{_1})^{\beta_{_1}+\beta_{_4}-1}
\varphi_{_{\{1,2,3\}}}^{(4)}({\boldsymbol{\beta}},\;{1\over x_{_1}},\;x_{_1}x_{_2})\;.
\label{Inverse-lattice11-6}
\end{eqnarray}

The inverse Gauss relations originating from the inverse transformations of the variables in the function $\varphi_{_{\{1,2,3\}}}^{(12)}$ are
\begin{eqnarray}
&&\varphi_{_{\{1,2,3\}}}^{(12)}({\boldsymbol{\beta}},\;x_{_1},\;x_{_2})
\nonumber\\
&&\hspace{-0.5cm}=
{\Gamma(2-\beta_{_1}-\beta_{_4})\Gamma(1-\beta_{_2}-\beta_{_4})\over\Gamma(\beta_{_3}+\beta_{_5})\Gamma(1-\beta_{_4})}
(-x_{_2})^{-\beta_{_2}}\varphi_{_{\{1,2,3\}}}^{(9)}({\boldsymbol{\beta}},\;x_{_1},\;{1\over x_{_2}})
\nonumber\\
&&\hspace{0.0cm}
+{\Gamma(\beta_{_2}+\beta_{_4}-1)\Gamma(2-\beta_{_1}-\beta_{_4})\over\Gamma(\beta_{_2})\Gamma(1-\beta_{_1})}(-x_{_2})^{\beta_{_4}-1}
\varphi_{_{\{1,2,3\}}}^{(8)}({\boldsymbol{\beta}},\;{x_{_1}\over x_{_2}}\;,{1\over x_{_2}})
\nonumber\\
&&\hspace{-0.5cm}=
{\Gamma(2-\beta_{_1}-\beta_{_4})\Gamma(1-\beta_{_3}-\beta_{_4})\over\Gamma(\beta_{_2}+\beta_{_5})\Gamma(1-\beta_{_4})}
(-x_{_1})^{-\beta_{_3}}\varphi_{_{\{1,2,3\}}}^{(6)}({\boldsymbol{\beta}},\;{1\over x_{_1}},\;x_{_2})
\nonumber\\
&&\hspace{0.0cm}
+{\Gamma(\beta_{_3}+\beta_{_4}-1)\Gamma(2-\beta_{_1}-\beta_{_4})\over\Gamma(\beta_{_3})\Gamma(1-\beta_{_1})}(-x_{_1})^{\beta_{_4}-1}
\varphi_{_{\{1,2,3\}}}^{(1)}({\boldsymbol{\beta}},\;{x_{_2}\over x_{_1}},\;{1\over x_{_1}})\;.
\label{Inverse-lattice12-6}
\end{eqnarray}

\section{The Gauss adjacent relations\label{app3}}
\indent\indent
For $\Phi_{_{\{1,2,3\}}}^{(2)}$, the equations in Eq.(\ref{Gauss-d},\ref{Gauss-e}) induce the adjacent relations of the
generalized hypergeometric function $\varphi_{_{\{1,2,3\}}}^{(2)}$ as
\begin{eqnarray}
&&(\beta_{_1}+\beta_{_5})\Big[\varphi_{_{\{1,2,3\}}}^{(2)}({\boldsymbol{\beta}})
-\varphi_{_{\{1,2,3\}}}^{(2)}({\boldsymbol{\beta}}+{\bf e}_{_1}-{\bf e}_{_5})\Big]
\nonumber\\
&&+(1-\beta_{_3}-\beta_{_4})x_{_2}\varphi_{_{\{1,2,3\}}}^{(2)}({\boldsymbol{\beta}}+{\bf e}_{_1}-{\bf e}_{_4})
\equiv0
\;,\nonumber\\
&&\beta_{_2}\varphi_{_{\{1,2,3\}}}^{(2)}({\boldsymbol{\beta}})
+(\beta_{_3}+\beta_{_4}-1)\varphi_{_{\{1,2,3\}}}^{(2)}({\boldsymbol{\beta}}+{\bf e}_{_2}-{\bf e}_{_4})
\nonumber\\
&&+(\beta_{_1}+\beta_{_5}-1)\varphi_{_{\{1,2,3\}}}^{(2)}({\boldsymbol{\beta}}+{\bf e}_{_2}-{\bf e}_{_5})\equiv0
\;,\nonumber\\
&&(\beta_{_3}+\beta_{_4})\Big[\varphi_{_{\{1,2,3\}}}^{(2)}({\boldsymbol{\beta}})
-\varphi_{_{\{1,2,3\}}}^{(2)}({\boldsymbol{\beta}}+{\bf e}_{_3}-{\bf e}_{_4})\Big]
\nonumber\\
&&+(1-\beta_{_1}-\beta_{_5})x_{_1}\varphi_{_{\{1,2,3\}}}^{(2)}({\boldsymbol{\beta}}+{\bf e}_{_3}-{\bf e}_{_5})\equiv0
\;,\nonumber\\
&&(\beta_{_3}+\beta_{_4})\Big[(\beta_{_1}-1)x_{_2}\varphi_{_{\{1,2,3\}}}^{(2)}({\boldsymbol{\beta}})
+(1-\beta_{_1}-\beta_{_5})\varphi_{_{\{1,2,3\}}}^{(2)}({\boldsymbol{\beta}}-{\bf e}_{_1}+{\bf e}_{_2})\Big]
\nonumber\\
&&+(\beta_{_1}+\beta_{_5}-1)\Big[\beta_{_3}\varphi_{_{\{1,2,3\}}}^{(2)}({\boldsymbol{\beta}}-{\bf e}_{_1}+{\bf e}_{_3})
+\beta_{_4}\varphi_{_{\{1,2,3\}}}^{(2)}({\boldsymbol{\beta}}-{\bf e}_{_1}+{\bf e}_{_4})\Big]\equiv0
\;,\nonumber\\
&&(\beta_{_3}+\beta_{_4})\Big[(\beta_{_1}-1)\varphi_{_{\{1,2,3\}}}^{(2)}({\boldsymbol{\beta}})
+\beta_{_5}\varphi_{_{\{1,2,3\}}}^{(2)}({\boldsymbol{\beta}}-{\bf e}_{_1}+{\bf e}_{_5})\Big]
\nonumber\\
&&+(1-\beta_{_1}-\beta_{_5})\Big[(\beta_{_3}+\beta_{_4})\varphi_{_{\{1,2,3\}}}^{(2)}({\boldsymbol{\beta}}-{\bf e}_{_1}+{\bf e}_{_2})
-\beta_{_3}x_{_1}\varphi_{_{\{1,2,3\}}}^{(2)}({\boldsymbol{\beta}}-{\bf e}_{_1}+{\bf e}_{_3})\Big]\equiv0\;.
\label{Gauss-adjacent2-1}
\end{eqnarray}

For $\Phi_{_{\{1,2,3\}}}^{(3)}$, the equations in Eq.(\ref{Gauss-d},\ref{Gauss-e}) induce the adjacent relations of the
generalized hypergeometric function $\varphi_{_{\{1,2,3\}}}^{(3)}$ as
\begin{eqnarray}
&&\beta_{_1}\varphi_{_{\{1,2,3\}}}^{(3)}({\boldsymbol{\beta}})
++(1-\beta_{_1}-\beta_{_5})\varphi_{_{\{1,2,3\}}}^{(3)}({\boldsymbol{\beta}}+{\bf e}_{_1}-{\bf e}_{_4})
\nonumber\\
&&+(\beta_{_5}-1)\varphi_{_{\{1,2,3\}}}^{(3)}({\boldsymbol{\beta}}+{\bf e}_{_1}-{\bf e}_{_5},\;x_{_1},\;x_{_2})\equiv0
\;,\nonumber\\
&&(2-\beta_{_1}-\beta_{_5})\varphi_{_{\{1,2,3\}}}^{(3)}({\boldsymbol{\beta}})
-\varphi_{_{\{1,2,3\}}}^{(3)}({\boldsymbol{\beta}}+{\bf e}_{_2}-{\bf e}_{_4})\Big]
\nonumber\\
&&+(\beta_{_5}-1)x_{_2}\varphi_{_{\{1,2,3\}}}^{(3)}({\boldsymbol{\beta}}+{\bf e}_{_2}-{\bf e}_{_5})\equiv0
\;,\nonumber\\
&&(2-\beta_{_1}-\beta_{_5})\Big[\varphi_{_{\{1,2,3\}}}^{(3)}({\boldsymbol{\beta}})
-\varphi_{_{\{1,2,3\}}}^{(3)}({\boldsymbol{\beta}}+{\bf e}_{_3}-{\bf e}_{_4})\Big]
\nonumber\\
&&+(\beta_{_5}-1)x_{_1}\varphi_{_{\{1,2,3\}}}^{(3)}({\boldsymbol{\beta}}+{\bf e}_{_3}-{\bf e}_{_5})\equiv0
\;,\nonumber\\
&&(\beta_{_1}+\beta_{_5}-2)\varphi_{_{\{1,2,3\}}}^{(3)}({\boldsymbol{\beta}})
+\beta_{_2}\varphi_{_{\{1,2,3\}}}^{(3)}({\boldsymbol{\beta}}-{\bf e}_{_1}+{\bf e}_{_2})
\nonumber\\
&&+\beta_{_3}\varphi_{_{\{1,2,3\}}}^{(3)}({\boldsymbol{\beta}}-{\bf e}_{_1}+{\bf e}_{_3})
+\beta_{_4}\varphi_{_{\{1,2,3\}}}^{(3)}({\boldsymbol{\beta}}-{\bf e}_{_1}+{\bf e}_{_4})\equiv0
\;,\nonumber\\
&&(\beta_{_1}+\beta_{_5}-2)\Big[\varphi_{_{\{1,2,3\}}}^{(3)}({\boldsymbol{\beta}})
-\varphi_{_{\{1,2,3\}}}^{(3)}({\boldsymbol{\beta}}-{\bf e}_{_1}+{\bf e}_{_5})\Big]
\nonumber\\
&&\hspace{0.0cm}
+\beta_{_2}x_{_2}\varphi_{_{\{1,2,3\}}}^{(3)}({\boldsymbol{\beta}}-{\bf e}_{_1}+{\bf e}_{_2})
+\beta_{_3}x_{_1}\varphi_{_{\{1,2,3\}}}^{(3)}({\boldsymbol{\beta}}-{\bf e}_{_1}+{\bf e}_{_3})\equiv0\;.
\label{Gauss-adjacent3-1}
\end{eqnarray}

For $\Phi_{_{\{1,2,3\}}}^{(4)}$, the equations in Eq.(\ref{Gauss-d},\ref{Gauss-e}) induce the adjacent relations of the
generalized hypergeometric function $\varphi_{_{\{1,2,3\}}}^{(4)}$ as
\begin{eqnarray}
&&(\beta_{_1}+\beta_{_4})\Big[\varphi_{_{\{1,2,3\}}}^{(4)}({\boldsymbol{\beta}})
-\varphi_{_{\{1,2,3\}}}^{(4)}({\boldsymbol{\beta}}+{\bf e}_{_1}-{\bf e}_{_4})\Big]
\nonumber\\
&&\hspace{0.0cm}
+(1-\beta_{_2}-\beta_{_5})x_{_2}\varphi_{_{\{1,2,3\}}}^{(4)}({\boldsymbol{\beta}}+{\bf e}_{_1}-{\bf e}_{_5})\equiv0
\;,\nonumber\\
&&(\beta_{_2}+\beta_{_5})\Big[\varphi_{_{\{1,2,3\}}}^{(4)}({\boldsymbol{\beta}})
-\varphi_{_{\{1,2,3\}}}^{(4)}({\boldsymbol{\beta}}+{\bf e}_{_2}-{\bf e}_{_5})\Big]
\nonumber\\
&&\hspace{0.0cm}
+(1-\beta_{_1}-\beta_{_4})x_{_1}\varphi_{_{\{1,2,3\}}}^{(4)}({\boldsymbol{\beta}}+{\bf e}_{_2}-{\bf e}_{_4})\equiv0
\;,\nonumber\\
&&\beta_{_3}\varphi_{_{\{1,2,3\}}}^{(4)}({\boldsymbol{\beta}})
+(\beta_{_1}+\beta_{_4}-1)\varphi_{_{\{1,2,3\}}}^{(4)}({\boldsymbol{\beta}}+{\bf e}_{_3}-{\bf e}_{_4})
\nonumber\\
&&\hspace{0.0cm}
+(\beta_{_2}+\beta_{_5}-1)\varphi_{_{\{1,2,3\}}}^{(4)}({\boldsymbol{\beta}}+{\bf e}_{_3}-{\bf e}_{_5})\equiv0
\;,\nonumber\\
&&(\beta_{_2}+\beta_{_5})\Big[(\beta_{_1}-1)\varphi_{_{\{1,2,3\}}}^{(4)}({\boldsymbol{\beta}})
+\beta_{_4}\varphi_{_{\{1,2,3\}}}^{(4)}({\boldsymbol{\beta}}-{\bf e}_{_1}+{\bf e}_{_4})\Big]
\nonumber\\
&&\hspace{0.0cm}
+(\beta_{_1}+\beta_{_4}-1)\Big[\beta_{_2}x_{_1}\varphi_{_{\{1,2,3\}}}^{(4)}({\boldsymbol{\beta}}-{\bf e}_{_1}+{\bf e}_{_2})
-(\beta_{_2}+\beta_{_5})\varphi_{_{\{1,2,3\}}}^{(4)}({\boldsymbol{\beta}}-{\bf e}_{_1}+{\bf e}_{_3})\Big]
\equiv0
\;,\nonumber\\
&&(\beta_{_2}+\beta_{_5})\Big[(\beta_{_1}-1)x_{_2}\varphi_{_{\{1,2,3\}}}^{(4)}({\boldsymbol{\beta}})
+(1-\beta_{_1}-\beta_{_4})\varphi_{_{\{1,2,3\}}}^{(4)}({\boldsymbol{\beta}}-{\bf e}_{_1}+{\bf e}_{_3})\Big]
\nonumber\\
&&\hspace{0.0cm}
+(\beta_{_1}+\beta_{_4}-1)\Big[\beta_{_2}\varphi_{_{\{1,2,3\}}}^{(4)}({\boldsymbol{\beta}}-{\bf e}_{_1}+{\bf e}_{_2})
+\beta_{_5}\varphi_{_{\{1,2,3\}}}^{(4)}({\boldsymbol{\beta}}-{\bf e}_{_1}+{\bf e}_{_5})\Big]\equiv0\;.
\label{Gauss-adjacent4-1}
\end{eqnarray}

For $\Phi_{_{\{1,2,3\}}}^{(5)}$, the equations in Eq.(\ref{Gauss-d},\ref{Gauss-e}) induce the adjacent relations of the
generalized hypergeometric function $\varphi_{_{\{1,2,3\}}}^{(5)}$ as
\begin{eqnarray}
&&(2-\beta_{_2}-\beta_{_5})\Big[\varphi_{_{\{1,2,3\}}}^{(5)}({\boldsymbol{\beta}})
-\varphi_{_{\{1,2,3\}}}^{(5)}({\boldsymbol{\beta}}+{\bf e}_{_1}-{\bf e}_{_4})\Big]
\nonumber\\
&&\hspace{0.0cm}
+(1-\beta_{_5})x_{_2}\varphi_{_{\{1,2,3\}}}^{(5)}({\boldsymbol{\beta}}+{\bf e}_{_1}-{\bf e}_{_5})\equiv0
\;,\nonumber\\
&&\beta_{_2}\varphi_{_{\{1,2,3\}}}^{(5)}({\boldsymbol{\beta}})
+(1-\beta_{_2}-\beta_{_5})\varphi_{_{\{1,2,3\}}}^{(5)}({\boldsymbol{\beta}}+{\bf e}_{_2}-{\bf e}_{_4})
\nonumber\\
&&\hspace{0.0cm}
+(\beta_{_5}-1)\varphi_{_{\{1,2,3\}}}^{(5)}({\boldsymbol{\beta}}+{\bf e}_{_2}-{\bf e}_{_5})\equiv0
\;,\nonumber\\
&&(2-\beta_{_2}-\beta_{_5})\Big[\varphi_{_{\{1,2,3\}}}^{(5)}({\boldsymbol{\beta}})
-\varphi_{_{\{1,2,3\}}}^{(5)}({\boldsymbol{\beta}}+{\bf e}_{_3}-{\bf e}_{_4})\Big]
\nonumber\\
&&\hspace{0.0cm}
+(1-\beta_{_5})x_{_1}\varphi_{_{\{1,2,3\}}}^{(5)}({\boldsymbol{\beta}}+{\bf e}_{_3}-{\bf e}_{_5})\equiv0
\;,\nonumber\\
&&(\beta_{_1}-1)\varphi_{_{\{1,2,3\}}}^{(5)}({\boldsymbol{\beta}})
+(\beta_{_2}+\beta_{_5}-1)\varphi_{_{\{1,2,3\}}}^{(5)}({\boldsymbol{\beta}}-{\bf e}_{_1}+{\bf e}_{_2})
\nonumber\\
&&\hspace{0.0cm}
+\beta_{_3}\varphi_{_{\{1,2,3\}}}^{(5)}({\boldsymbol{\beta}}-{\bf e}_{_1}+{\bf e}_{_3})
+\beta_{_4}\varphi_{_{\{1,2,3\}}}^{(5)}({\boldsymbol{\beta}}-{\bf e}_{_1}+{\bf e}_{_4})\equiv0
\;,\nonumber\\
&&(\beta_{_3}-1)x_{_2}\varphi_{_{\{1,2,3\}}}^{(5)}({\boldsymbol{\beta}})
+\beta_{_3}x_{_1}\varphi_{_{\{1,2,3\}}}^{(5)}({\boldsymbol{\beta}}-{\bf e}_{_1}+{\bf e}_{_3})
\nonumber\\
&&\hspace{0.0cm}
+(\beta_{_2}+\beta_{_5}-1)\Big[\varphi_{_{\{1,2,3\}}}^{(5)}({\boldsymbol{\beta}}-{\bf e}_{_1}+{\bf e}_{_2})
-\varphi_{_{\{1,2,3\}}}^{(5)}({\boldsymbol{\beta}}-{\bf e}_{_1}+{\bf e}_{_5})\Big]\equiv0\;.
\label{Gauss-adjacent5-1}
\end{eqnarray}

For $\Phi_{_{\{1,2,3\}}}^{(6)}$, the equations in Eq.(\ref{Gauss-d},\ref{Gauss-e}) induce the adjacent relations of the
generalized hypergeometric function $\varphi_{_{\{1,2,3\}}}^{(6)}$ as
\begin{eqnarray}
&&\beta_{_1}\varphi_{_{\{1,2,3\}}}^{(6)}({\boldsymbol{\beta}})
+(\beta_{_3}+\beta_{_4}-1)\varphi_{_{\{1,2,3\}}}^{(6)}({\boldsymbol{\beta}}+{\bf e}_{_1}-{\bf e}_{_4})
\nonumber\\
&&\hspace{0.0cm}
+(\beta_{_2}+\beta_{_5}-1)\varphi_{_{\{1,2,3\}}}^{(6)}({\boldsymbol{\beta}}+{\bf e}_{_1}-{\bf e}_{_5})\equiv0
\;,\nonumber\\
&&(\beta_{_2}+\beta_{_5})\Big[\varphi_{_{\{1,2,3\}}}^{(6)}({\boldsymbol{\beta}})
-\varphi_{_{\{1,2,3\}}}^{(6)}({\boldsymbol{\beta}}+{\bf e}_{_2}-{\bf e}_{_5})\Big]
\nonumber\\
&&\hspace{0.0cm}
+(1-\beta_{_3}-\beta_{_4})x_{_2}\varphi_{_{\{1,2,3\}}}^{(6)}({\boldsymbol{\beta}}+{\bf e}_{_2}-{\bf e}_{_4})\equiv0
\;,\nonumber\\
&&(\beta_{_3}+\beta_{_4})\Big[\varphi_{_{\{1,2,3\}}}^{(6)}({\boldsymbol{\beta}})
-\varphi_{_{\{1,2,3\}}}^{(6)}({\boldsymbol{\beta}}+{\bf e}_{_3}-{\bf e}_{_4})\Big]
\nonumber\\
&&\hspace{0.0cm}
+(1-\beta_{_2}-\beta_{_5})x_{_1}\varphi_{_{\{1,2,3\}}}^{(6)}({\boldsymbol{\beta}}+{\bf e}_{_3}-{\bf e}_{_5})\equiv0
\;,\nonumber\\
&&(\beta_{_3}+\beta_{_4})\Big[(\beta_{_2}+\beta_{_5})\varphi_{_{\{1,2,3\}}}^{(6)}({\boldsymbol{\beta}})
-\beta_{_2}x_{_2}\varphi_{_{\{1,2,3\}}}^{(6)}({\boldsymbol{\beta}}-{\bf e}_{_1}+{\bf e}_{_2})\Big]
\nonumber\\
&&\hspace{0.0cm}
-(\beta_{_2}+\beta_{_5})\Big[\beta_{_3}\varphi_{_{\{1,2,3\}}}^{(6)}({\boldsymbol{\beta}}-{\bf e}_{_1}+{\bf e}_{_3})
+\beta_{_4}\varphi_{_{\{1,2,3\}}}^{(6)}({\boldsymbol{\beta}}-{\bf e}_{_1}+{\bf e}_{_4})\Big]\equiv0
\;,\nonumber\\
&&(\beta_{_3}+\beta_{_4})\Big[(\beta_{_2}+\beta_{_5})\varphi_{_{\{1,2,3\}}}^{(6)}({\boldsymbol{\beta}})
-\beta_{_3}x_{_1}\varphi_{_{\{1,2,3\}}}^{(6)}({\boldsymbol{\beta}}-{\bf e}_{_1}+{\bf e}_{_3})\Big]
\nonumber\\
&&\hspace{0.0cm}
-(\beta_{_3}+\beta_{_4})\Big[\beta_{_2}\varphi_{_{\{1,2,3\}}}^{(6)}({\boldsymbol{\beta}}-{\bf e}_{_1}+{\bf e}_{_2})
+\beta_{_5}\varphi_{_{\{1,2,3\}}}^{(6)}({\boldsymbol{\beta}}-{\bf e}_{_1}+{\bf e}_{_5})\Big]\equiv0\;.
\label{Gauss-adjacent6-1}
\end{eqnarray}

For $\Phi_{_{\{1,2,3\}}}^{(7)}$, the equations in Eq.(\ref{Gauss-d},\ref{Gauss-e}) induce the adjacent relations of the
generalized hypergeometric function $\varphi_{_{\{1,2,3\}}}^{(7)}$ as
\begin{eqnarray}
&&(\beta_{_1}+\beta_{_5})\Big[\varphi_{_{\{1,2,3\}}}^{(7)}({\boldsymbol{\beta}})
-\varphi_{_{\{1,2,3\}}}^{(7)}({\boldsymbol{\beta}}+{\bf e}_{_1}-{\bf e}_{_5})\Big]
\nonumber\\
&&\hspace{0.0cm}
+(1-\beta_{_2}-\beta_{_4})x_{_2}\varphi_{_{\{1,2,3\}}}^{(7)}({\boldsymbol{\beta}}+{\bf e}_{_1}-{\bf e}_{_4})\equiv0
\;,\nonumber\\
&&(\beta_{_2}+\beta_{_4})\Big[\varphi_{_{\{1,2,3\}}}^{(7)}({\boldsymbol{\beta}})
-\varphi_{_{\{1,2,3\}}}^{(7)}({\boldsymbol{\beta}}+{\bf e}_{_2}-{\bf e}_{_4})\Big]
\nonumber\\
&&\hspace{0.0cm}
+(1-\beta_{_1}-\beta_{_5})x_{_1}\varphi_{_{\{1,2,3\}}}^{(7)}({\boldsymbol{\beta}}+{\bf e}_{_2}-{\bf e}_{_5})\equiv0
\;,\nonumber\\
&&\beta_{_3}\varphi_{_{\{1,2,3\}}}^{(7)}({\boldsymbol{\beta}})
+(\beta_{_2}+\beta_{_4}-1)\varphi_{_{\{1,2,3\}}}^{(7)}({\boldsymbol{\beta}}+{\bf e}_{_3}-{\bf e}_{_4})
\nonumber\\
&&\hspace{0.0cm}
+(\beta_{_1}+\beta_{_5}-1)\varphi_{_{\{1,2,3\}}}^{(7)}({\boldsymbol{\beta}}+{\bf e}_{_3}-{\bf e}_{_5})\equiv0
\;,\nonumber\\
&&(\beta_{_2}+\beta_{_4})\Big[(\beta_{_1}-1)x_{_2}\varphi_{_{\{1,2,3\}}}^{(7)}({\boldsymbol{\beta}})
+(1-\beta_{_1}-\beta_{_5})\varphi_{_{\{1,2,3\}}}^{(7)}({\boldsymbol{\beta}}-{\bf e}_{_1}+{\bf e}_{_3})\Big]
\nonumber\\
&&\hspace{0.0cm}
+(\beta_{_1}+\beta_{_5}-1)\Big[\beta_{_2}\varphi_{_{\{1,2,3\}}}^{(7)}({\boldsymbol{\beta}}-{\bf e}_{_1}+{\bf e}_{_2})
+\beta_{_4}\varphi_{_{\{1,2,3\}}}^{(7)}({\boldsymbol{\beta}}-{\bf e}_{_1}+{\bf e}_{_4})\Big]\equiv0
\;,\nonumber\\
&&(\beta_{_2}+\beta_{_4})\Big[(\beta_{_1}-1)\varphi_{_{\{1,2,3\}}}^{(7)}({\boldsymbol{\beta}})
+\beta_{_5}\varphi_{_{\{1,2,3\}}}^{(7)}({\boldsymbol{\beta}}-{\bf e}_{_1}+{\bf e}_{_5})\Big]
\nonumber\\
&&\hspace{0.0cm}
+(\beta_{_1}+\beta_{_5}-1)\Big[\beta_{_2}x_{_1}\varphi_{_{\{1,2,3\}}}^{(7)}({\boldsymbol{\beta}}-{\bf e}_{_1}+{\bf e}_{_2})
-(\beta_{_2}+\beta_{_4})\varphi_{_{\{1,2,3\}}}^{(7)}({\boldsymbol{\beta}}-{\bf e}_{_1}+{\bf e}_{_3})\Big]\equiv0\;.
\label{Gauss-adjacent7-1}
\end{eqnarray}

For $\Phi_{_{\{1,2,3\}}}^{(8)}$, the equations in Eq.(\ref{Gauss-d},\ref{Gauss-e}) induce the adjacent relations of the
generalized hypergeometric function $\varphi_{_{\{1,2,3\}}}^{(8)}$ as
\begin{eqnarray}
&&(2-\beta_{_2}-\beta_{_4})\Big[\varphi_{_{\{1,2,3\}}}^{(8)}({\boldsymbol{\beta}})
-\varphi_{_{\{1,2,3\}}}^{(8)}({\boldsymbol{\beta}}+{\bf e}_{_1}-{\bf e}_{_5})\Big]
\nonumber\\
&&\hspace{0.0cm}
+(1-\beta_{_4})x_{_2}\varphi_{_{\{1,2,3\}}}^{(8)}({\boldsymbol{\beta}}+{\bf e}_{_1}-{\bf e}_{_4})\equiv0
\;,\nonumber\\
&&\beta_{_2}\varphi_{_{\{1,2,3\}}}^{(8)}({\boldsymbol{\beta}})
+(\beta_{_4}-1)\varphi_{_{\{1,2,3\}}}^{(8)}({\boldsymbol{\beta}}+{\bf e}_{_2}-{\bf e}_{_4})
\nonumber\\
&&\hspace{0.0cm}
+(1-\beta_{_2}-\beta_{_4})\varphi_{_{\{1,2,3\}}}^{(8)}({\boldsymbol{\beta}}+{\bf e}_{_2}-{\bf e}_{_5})\equiv0
\;,\nonumber\\
&&(2-\beta_{_2}-\beta_{_4})\Big[\varphi_{_{\{1,2,3\}}}^{(8)}({\boldsymbol{\beta}})
-\varphi_{_{\{1,2,3\}}}^{(8)}({\boldsymbol{\beta}}+{\bf e}_{_3}-{\bf e}_{_5})\Big]
\nonumber\\
&&\hspace{0.0cm}
+(1-\beta_{_4})x_{_1}\varphi_{_{\{1,2,3\}}}^{(8)}({\boldsymbol{\beta}}+{\bf e}_{_3}-{\bf e}_{_4})\equiv0
\;,\nonumber\\
&&(\beta_{_1}-1)x_{_2}\varphi_{_{\{1,2,3\}}}^{(8)}({\boldsymbol{\beta}})
+\beta_{_3}x_{_1}\varphi_{_{\{1,2,3\}}}^{(8)}({\boldsymbol{\beta}}-{\bf e}_{_1}+{\bf e}_{_3})
\nonumber\\
&&\hspace{0.0cm}
+(\beta_{_2}+\beta_{_4}-1)\Big[\varphi_{_{\{1,2,3\}}}^{(8)}({\boldsymbol{\beta}}-{\bf e}_{_1}+{\bf e}_{_2})
-\varphi_{_{\{1,2,3\}}}^{(8)}({\boldsymbol{\beta}}-{\bf e}_{_1}+{\bf e}_{_4})\Big]\equiv0
\;,\nonumber\\
&&(\beta_{_1}-1)\varphi_{_{\{1,2,3\}}}^{(8)}({\boldsymbol{\beta}})
+(\beta_{_2}+\beta_{_4}-1)\varphi_{_{\{1,2,3\}}}^{(8)}({\boldsymbol{\beta}}-{\bf e}_{_1}+{\bf e}_{_2})
\nonumber\\
&&\hspace{0.0cm}
+\beta_{_3}\varphi_{_{\{1,2,3\}}}^{(8)}({\boldsymbol{\beta}}-{\bf e}_{_1}+{\bf e}_{_3})
+\beta_{_5}\varphi_{_{\{1,2,3\}}}^{(8)}({\boldsymbol{\beta}}-{\bf e}_{_1}+{\bf e}_{_5})\equiv0\;.
\label{Gauss-adjacent8-1}
\end{eqnarray}

For $\Phi_{_{\{1,2,3\}}}^{(9)}$, the equations in Eq.(\ref{Gauss-d},\ref{Gauss-e}) induce the adjacent relations of the
generalized hypergeometric function $\varphi_{_{\{1,2,3\}}}^{(9)}$ as
\begin{eqnarray}
&&\beta_{_1}\varphi_{_{\{1,2,3\}}}^{(9)}({\boldsymbol{\beta}})
+(\beta_{_2}+\beta_{_4}-1)\varphi_{_{\{1,2,3\}}}^{(9)}({\boldsymbol{\beta}}+{\bf e}_{_1}-{\bf e}_{_4})
\nonumber\\
&&\hspace{0.0cm}
+(\beta_{_3}+\beta_{_5}-1)\varphi_{_{\{1,2,3\}}}^{(9)}({\boldsymbol{\beta}}+{\bf e}_{_1}-{\bf e}_{_5})\equiv0
\;,\nonumber\\
&&(\beta_{_2}+\beta_{_4})\Big[\varphi_{_{\{1,2,3\}}}^{(9)}({\boldsymbol{\beta}})
-\varphi_{_{\{1,2,3\}}}^{(9)}({\boldsymbol{\beta}}+{\bf e}_{_2}-{\bf e}_{_4})\Big]
\nonumber\\
&&\hspace{0.0cm}
+(1-\beta_{_3}-\beta_{_5})x_{_2}\varphi_{_{\{1,2,3\}}}^{(9)}({\boldsymbol{\beta}}+{\bf e}_{_2}-{\bf e}_{_5})\equiv0
\;,\nonumber\\
&&(\beta_{_3}+\beta_{_5})\Big[\varphi_{_{\{1,2,3\}}}^{(9)}({\boldsymbol{\beta}})
-\varphi_{_{\{1,2,3\}}}^{(9)}({\boldsymbol{\beta}}+{\bf e}_{_3}-{\bf e}_{_5})\Big]
\nonumber\\
&&\hspace{0.0cm}
+(1-\beta_{_2}-\beta_{_4})x_{_1}\varphi_{_{\{1,2,3\}}}^{(9)}({\boldsymbol{\beta}}+{\bf e}_{_3}-{\bf e}_{_4})\equiv0
\;,\nonumber\\
&&(\beta_{_2}+\beta_{_4})\Big[(\beta_{_3}+\beta_{_5})\varphi_{_{\{1,2,3\}}}^{(9)}({\boldsymbol{\beta}})
-\beta_{_3}x_{_1}\varphi_{_{\{1,2,3\}}}^{(9)}({\boldsymbol{\beta}}-{\bf e}_{_1}+{\bf e}_{_3})\Big]
\nonumber\\
&&\hspace{0.0cm}
-(\beta_{_3}+\beta_{_5})\Big[\beta_{_2}\varphi_{_{\{1,2,3\}}}^{(9)}({\boldsymbol{\beta}}-{\bf e}_{_1}+{\bf e}_{_2})
+\beta_{_4}\varphi_{_{\{1,2,3\}}}^{(9)}({\boldsymbol{\beta}}-{\bf e}_{_1}+{\bf e}_{_4})\Big]\equiv0
\;,\nonumber\\
&&(\beta_{_3}+\beta_{_5})\Big[(\beta_{_2}+\beta_{_4})\varphi_{_{\{1,2,3\}}}^{(9)}({\boldsymbol{\beta}})
-\beta_{_2}x_{_2}\varphi_{_{\{1,2,3\}}}^{(9)}({\boldsymbol{\beta}}-{\bf e}_{_1}+{\bf e}_{_2})\Big]
\nonumber\\
&&\hspace{0.0cm}
-(\beta_{_2}+\beta_{_4})\Big[\beta_{_3}\varphi_{_{\{1,2,3\}}}^{(9)}({\boldsymbol{\beta}}-{\bf e}_{_1}+{\bf e}_{_3})
+\beta_{_5}\varphi_{_{\{1,2,3\}}}^{(9)}({\boldsymbol{\beta}}-{\bf e}_{_1}+{\bf e}_{_5})\Big]\equiv0\;.
\label{Gauss-adjacent9-1}
\end{eqnarray}

For $\Phi_{_{\{1,2,3\}}}^{(10)}$, the equations in Eq.(\ref{Gauss-d},\ref{Gauss-e}) induce the adjacent relations of the
generalized hypergeometric function $\varphi_{_{\{1,2,3\}}}^{(10)}$ as
\begin{eqnarray}
&&(2-\beta_{_3}-\beta_{_5})\Big[\varphi_{_{\{1,2,3\}}}^{(10)}({\boldsymbol{\beta}})
-\varphi_{_{\{1,2,3\}}}^{(10)}({\boldsymbol{\beta}}+{\bf e}_{_1}-{\bf e}_{_4})\Big]
\nonumber\\
&&\hspace{0.0cm}
+(1-\beta_{_5})x_{_2}\varphi_{_{\{1,2,3\}}}^{(10)}({\boldsymbol{\beta}}+{\bf e}_{_1}-{\bf e}_{_5})\equiv0
\;,\nonumber\\
&&(2-\beta_{_3}-\beta_{_5})\Big[\varphi_{_{\{1,2,3\}}}^{(10)}({\boldsymbol{\beta}})
-\varphi_{_{\{1,2,3\}}}^{(10)}({\boldsymbol{\beta}}+{\bf e}_{_2}-{\bf e}_{_4})\Big]
\nonumber\\
&&\hspace{0.0cm}
+(1-\beta_{_5})x_{_1}\varphi_{_{\{1,2,3\}}}^{(10)}({\boldsymbol{\beta}}+{\bf e}_{_2}-{\bf e}_{_5})\equiv0
\;,\nonumber\\
&&\beta_{_3}\varphi_{_{\{1,2,3\}}}^{(10)}({\boldsymbol{\beta}})
+(1-\beta_{_3}-\beta_{_5})\varphi_{_{\{1,2,3\}}}^{(10)}({\boldsymbol{\beta}}+{\bf e}_{_3}-{\bf e}_{_4})
\nonumber\\
&&\hspace{0.0cm}
+(\beta_{_5}-1)\varphi_{_{\{1,2,3\}}}^{(10)}({\boldsymbol{\beta}}+{\bf e}_{_3}-{\bf e}_{_5})\equiv0
\;,\nonumber\\
&&(\beta_{_1}-1)\varphi_{_{\{1,2,3\}}}^{(10)}({\boldsymbol{\beta}})
+\beta_{_2}\varphi_{_{\{1,2,3\}}}^{(10)}({\boldsymbol{\beta}}-{\bf e}_{_1}+{\bf e}_{_2})
\nonumber\\
&&\hspace{0.0cm}
+(\beta_{_3}+\beta_{_5}-1)\varphi_{_{\{1,2,3\}}}^{(10)}({\boldsymbol{\beta}}-{\bf e}_{_1}+{\bf e}_{_3})
+\beta_{_4}\varphi_{_{\{1,2,3\}}}^{(10)}({\boldsymbol{\beta}}-{\bf e}_{_1}+{\bf e}_{_4})\equiv0
\;,\nonumber\\
&&(\beta_{_1}-1)x_{_2}\varphi_{_{\{1,2,3\}}}^{(10)}({\boldsymbol{\beta}})
+\beta_{_2}x_{_1}\varphi_{_{\{1,2,3\}}}^{(10)}({\boldsymbol{\beta}}-{\bf e}_{_1}+{\bf e}_{_2})
\nonumber\\
&&\hspace{0.0cm}
+(\beta_{_3}+\beta_{_5}-1)\Big[\varphi_{_{\{1,2,3\}}}^{(10)}({\boldsymbol{\beta}}-{\bf e}_{_1}+{\bf e}_{_3})
-\varphi_{_{\{1,2,3\}}}^{(10)}({\boldsymbol{\beta}}-{\bf e}_{_1}+{\bf e}_{_5})\Big]\equiv0\;.
\label{Gauss-adjacent10-1}
\end{eqnarray}

For $\Phi_{_{\{1,2,3\}}}^{(11)}$, the equations in Eq.(\ref{Gauss-d},\ref{Gauss-e}) induce the adjacent relations of the
generalized hypergeometric function $\varphi_{_{\{1,2,3\}}}^{(11)}$ as
\begin{eqnarray}
&&(\beta_{_1}+\beta_{_4})\Big[\varphi_{_{\{1,2,3\}}}^{(11)}({\boldsymbol{\beta}})
-\varphi_{_{\{1,2,3\}}}^{(11)}({\boldsymbol{\beta}}+{\bf e}_{_1}-{\bf e}_{_4})\Big]
\nonumber\\
&&\hspace{0.0cm}
+(1-\beta_{_3}-\beta_{_5})x_{_2}\varphi_{_{\{1,2,3\}}}^{(11)}({\boldsymbol{\beta}}+{\bf e}_{_1}-{\bf e}_{_5})\equiv0
\;,\nonumber\\
&&\beta_{_2}\varphi_{_{\{1,2,3\}}}^{(11)}({\boldsymbol{\beta}})
+(\beta_{_1}+\beta_{_4}-1)\varphi_{_{\{1,2,3\}}}^{(11)}({\boldsymbol{\beta}}+{\bf e}_{_2}-{\bf e}_{_4})
\nonumber\\
&&\hspace{0.0cm}
+(\beta_{_3}+\beta_{_5}-1)\varphi_{_{\{1,2,3\}}}^{(11)}({\boldsymbol{\beta}}+{\bf e}_{_2}-{\bf e}_{_5})\equiv0
\;,\nonumber\\
&&(\beta_{_3}+\beta_{_5})\Big[\varphi_{_{\{1,2,3\}}}^{(11)}({\boldsymbol{\beta}})
-\varphi_{_{\{1,2,3\}}}^{(11)}({\boldsymbol{\beta}}+{\bf e}_{_3}-{\bf e}_{_5})\Big]
\nonumber\\
&&\hspace{0.0cm}
+(1-\beta_{_1}-\beta_{_4})x_{_1}\varphi_{_{\{1,2,3\}}}^{(11)}({\boldsymbol{\beta}}+{\bf e}_{_3}-{\bf e}_{_4})\equiv0
\;,\nonumber\\
&&(\beta_{_3}+\beta_{_5})\Big[(\beta_{_1}-1)\varphi_{_{\{1,2,3\}}}^{(11)}({\boldsymbol{\beta}})
+\beta_{_4}\varphi_{_{\{1,2,3\}}}^{(11)}({\boldsymbol{\beta}}-{\bf e}_{_1}+{\bf e}_{_4})\Big]
\nonumber\\
&&\hspace{0.0cm}
+(1-\beta_{_1}-\beta_{_4})\Big[(\beta_{_3}+\beta_{_5})\varphi_{_{\{1,2,3\}}}^{(11)}({\boldsymbol{\beta}}-{\bf e}_{_1}+{\bf e}_{_2})
-\beta_{_3}x_{_1}\varphi_{_{\{1,2,3\}}}^{(11)}({\boldsymbol{\beta}}-{\bf e}_{_1}+{\bf e}_{_3})\Big]\equiv0
\;,\nonumber\\
&&(\beta_{_3}+\beta_{_5})\Big[(\beta_{_1}-1)x_{_2}\varphi_{_{\{1,2,3\}}}^{(11)}({\boldsymbol{\beta}})
+(1-\beta_{_1}-\beta_{_4})\varphi_{_{\{1,2,3\}}}^{(11)}({\boldsymbol{\beta}}-{\bf e}_{_1}+{\bf e}_{_2})\Big]
\nonumber\\
&&\hspace{0.0cm}
+(\beta_{_1}+\beta_{_4}-1)\Big[\beta_{_3}\varphi_{_{\{1,2,3\}}}^{(11)}({\boldsymbol{\beta}}-{\bf e}_{_1}+{\bf e}_{_3})
+\beta_{_5}\varphi_{_{\{1,2,3\}}}^{(11)}({\boldsymbol{\beta}}-{\bf e}_{_1}+{\bf e}_{_5})\Big]\equiv0\;.
\label{Gauss-adjacent11-1}
\end{eqnarray}

For $\Phi_{_{\{1,2,3\}}}^{(12)}$, the equations in Eq.(\ref{Gauss-d},\ref{Gauss-e}) induce the adjacent relations of the
generalized hypergeometric function $\varphi_{_{\{1,2,3\}}}^{(12)}$ as
\begin{eqnarray}
&&\beta_{_1}\varphi_{_{\{1,2,3\}}}^{(12)}({\boldsymbol{\beta}})
+(\beta_{_4}-1)\varphi_{_{\{1,2,3\}}}^{(12)}({\boldsymbol{\beta}}+{\bf e}_{_1}-{\bf e}_{_4})
\nonumber\\
&&\hspace{0.0cm}
+(1-\beta_{_1}-\beta_{_4})\varphi_{_{\{1,2,3\}}}^{(12)}({\boldsymbol{\beta}}+{\bf e}_{_1}-{\bf e}_{_5})\equiv0
\;,\nonumber\\
&&(2-\beta_{_1}-\beta_{_4})\Big[\varphi_{_{\{1,2,3\}}}^{(12)}({\boldsymbol{\beta}})
-\varphi_{_{\{1,2,3\}}}^{(12)}({\boldsymbol{\beta}}+{\bf e}_{_2}-{\bf e}_{_5})\Big]
\nonumber\\
&&\hspace{0.0cm}
+(1-\beta_{_4})x_{_2}\varphi_{_{\{1,2,3\}}}^{(12)}({\boldsymbol{\beta}}+{\bf e}_{_2}-{\bf e}_{_4},\;x_{_1},\;x_{_2})\equiv0
\;,\nonumber\\
&&(2-\beta_{_1}-\beta_{_4})\Big[\varphi_{_{\{1,2,3\}}}^{(12)}({\boldsymbol{\beta}})
-\varphi_{_{\{1,2,3\}}}^{(12)}({\boldsymbol{\beta}}+{\bf e}_{_3}-{\bf e}_{_5})\Big]
\nonumber\\
&&\hspace{0.0cm}
+(1-\beta_{_4})x_{_1}\varphi_{_{\{1,2,3\}}}^{(12)}({\boldsymbol{\beta}}+{\bf e}_{_3}-{\bf e}_{_4},\;x_{_1},\;x_{_2})\equiv0
\;,\nonumber\\
&&(\beta_{_1}+\beta_{_4}-2)\Big[\varphi_{_{\{1,2,3\}}}^{(12)}({\boldsymbol{\beta}})
-\varphi_{_{\{1,2,3\}}}^{(12)}({\boldsymbol{\beta}}-{\bf e}_{_1}+{\bf e}_{_4}\Big]
\nonumber\\
&&\hspace{0.0cm}
+\beta_{_2}x_{_2}\varphi_{_{\{1,2,3\}}}^{(12)}({\boldsymbol{\beta}}-{\bf e}_{_1}+{\bf e}_{_2})
+\beta_{_3}x_{_1}\varphi_{_{\{1,2,3\}}}^{(12)}({\boldsymbol{\beta}}-{\bf e}_{_1}+{\bf e}_{_3})
\equiv0
\;,\nonumber\\
&&(\beta_{_1}+\beta_{_4}-2)\varphi_{_{\{1,2,3\}}}^{(12)}({\boldsymbol{\beta}})
+\beta_{_2}\varphi_{_{\{1,2,3\}}}^{(12)}({\boldsymbol{\beta}}-{\bf e}_{_1}+{\bf e}_{_2})
\nonumber\\
&&\hspace{0.0cm}
+\beta_{_3}\varphi_{_{\{1,2,3\}}}^{(12)}({\boldsymbol{\beta}}-{\bf e}_{_1}+{\bf e}_{_3})
+\beta_{_5}\varphi_{_{\{1,2,3\}}}^{(12)}({\boldsymbol{\beta}}-{\bf e}_{_1}+{\bf e}_{_5})\equiv0\;.
\label{Gauss-adjacent12-1}
\end{eqnarray}

\section{The Gauss-Kummer relations\label{app4}}
\indent\indent
Corresponding to the geometric representation shown in Fig.\ref{fig1}(a) with $\{a,b\}=\{1,2\}$, $\{c,d,e\}=\{3,4,5\}$,
the following six solutions of the GKZ-system presented in Eq.(\ref{sec2-2})
only differ from each other by some constant factors,
\begin{eqnarray}
&&\Phi_{_{\{1,3,4\}}}^{(12)}\sim\Phi_{_{\{1,3,5\}}}^{(12)}\sim\Phi_{_{\{2,3,4\}}}^{(12)}
\sim\Phi_{_{\{2,3,5\}}}^{(12)}\sim\Phi_{_{\{1,4,5\}}}^{(12)}\sim\Phi_{_{\{2,4,5\}}}^{(12)}\;.
\label{Gauss-Kummer6-11-1}
\end{eqnarray}
The relations induce
\begin{eqnarray}
&&y^{-\beta_{_4}}(y-1)^{-\beta_{_3}}
\varphi_{_{\{1,3,4\}}}^{(12)}({\boldsymbol{\beta}},\;1-{x\over y},\;{y-x\over y-1})
\nonumber\\
&&\hspace{-0.5cm}=
x^{\beta_{_2}-1}y^{1-\beta_{_2}-\beta_{_4}}(y-1)^{-\beta_{_3}}
\varphi_{_{\{1,3,5\}}}^{(12)}({\boldsymbol{\beta}},\;1-{y\over x},\;{x-y\over x(1-y)})
\nonumber\\
&&\hspace{-0.5cm}=
x^{-\beta_{_4}}(x-1)^{-\beta_{_3}}
\varphi_{_{\{2,3,4\}}}^{(12)}({\boldsymbol{\beta}},\;1-{y\over x},\;{y-x\over 1-x})
\nonumber\\
&&\hspace{-0.5cm}=
x^{1-\beta_{_1}-\beta_{_4}}(x-1)^{-\beta_{_3}}y^{\beta_{_1}-1}
\varphi_{_{\{2,3,5\}}}^{(12)}({\boldsymbol{\beta}},\;1-{x\over y},\;{y-x\over y(1-x)})
\nonumber\\
&&\hspace{-0.5cm}=
(x-1)^{\beta_{_2}-1}y^{-\beta_{_4}}(y-1)^{1-\beta_{_2}-\beta_{_3}}
\varphi_{_{\{1,4,5\}}}^{(12)}({\boldsymbol{\beta}},\;{y-x\over 1-x},\;{y-x\over y(1-x)})
\nonumber\\
&&\hspace{-0.5cm}=
x^{-\beta_{_4}}(x-1)^{1-\beta_{_1}-\beta_{_3}}(y-1)^{\beta_{_1}-1}
\varphi_{_{\{2,4,5\}}}^{(12)}({\boldsymbol{\beta}},\;{x-y\over 1-y},\;{x-y\over x(1-y)})\;.
\label{Gauss-Kummer6-11-1-1}
\end{eqnarray}

To obtain the above equations in terms of the first type Appell function $F_{_1}$,
we rewrite those $\varphi$ functions explicitly as
\begin{eqnarray}
&&y^{-\beta_{_4}}(y-1)^{-\beta_{_3}}
\varphi_{_{\{1,3,4\}}}^{(12)}({\boldsymbol{\beta}},\;1-{x\over y},\;{y-x\over y-1})
\nonumber\\
&&\hspace{-0.5cm}=
y^{-\beta_{_4}}(y-1)^{-\beta_{_3}}F_{_1}\left(\left.\begin{array}{c}1-\beta_{_2},\;\beta_{_4},\;\beta_{_3}\\
2-\beta_{_1}-\beta_{_2}\end{array}\right|\;1-{x\over y},\;{y-x\over y-1}\right)
\;,\nonumber\\
&&x^{\beta_{_2}-1}y^{1-\beta_{_2}-\beta_{_4}}(y-1)^{-\beta_{_3}}
\varphi_{_{\{1,3,5\}}}^{(12)}({\boldsymbol{\beta}},\;1-{y\over x},\;{x-y\over x(1-y)})
\nonumber\\
&&\hspace{-0.5cm}=
x^{\beta_{_2}-1}y^{1-\beta_{_2}-\beta_{_4}}(y-1)^{-\beta_{_3}}
F_{_1}\left(\left.\begin{array}{c}1-\beta_{_2},\;\beta_{_5},\;\beta_{_3}\\2-\beta_{_1}-\beta_{_2}\end{array}\right|\;1-{y\over x},\;{x-y\over x(1-y)}\right)
\nonumber\\
&&x^{-\beta_{_4}}(x-1)^{-\beta_{_3}}
\varphi_{_{\{2,3,4\}}}^{(12)}({\boldsymbol{\beta}},\;1-{y\over x},\;{y-x\over 1-x})
\nonumber\\
&&\hspace{-0.5cm}=
x^{-\beta_{_4}}(x-1)^{-\beta_{_3}}F_{_1}\left(\left.\begin{array}{c}1-\beta_{_1},\;\beta_{_4},\;\beta_{_3}\\2-\beta_{_1}-\beta_{_2}
\end{array}\right|\;1-{y\over x},\;{y-x\over 1-x}\right)
\;,\nonumber\\
&&x^{1-\beta_{_1}-\beta_{_4}}(x-1)^{-\beta_{_3}}y^{\beta_{_1}-1}
\varphi_{_{\{2,3,5\}}}^{(12)}({\boldsymbol{\beta}},\;1-{x\over y},\;{y-x\over y(1-x)})
\nonumber\\
&&\hspace{-0.5cm}=
x^{1-\beta_{_1}-\beta_{_4}}(x-1)^{-\beta_{_3}}y^{\beta_{_1}-1}
F_{_1}\left(\left.\begin{array}{c}1-\beta_{_1},\;\beta_{_5},\;\beta_{_3}\\2-\beta_{_1}-\beta_{_2}
\end{array}\right|\;1-{x\over y},\;{y-x\over y(1-x)}\right)
\;,\nonumber\\
&&(x-1)^{\beta_{_2}-1}y^{-\beta_{_4}}(y-1)^{1-\beta_{_2}-\beta_{_3}}
\varphi_{_{\{1,4,5\}}}^{(12)}({\boldsymbol{\beta}},\;{y-x\over 1-x},\;{y-x\over y(1-x)})
\nonumber\\
&&\hspace{-0.5cm}=
(x-1)^{\beta_{_2}-1}y^{-\beta_{_4}}(y-1)^{1-\beta_{_2}-\beta_{_3}}
F_{_1}\left(\left.\begin{array}{c}1-\beta_{_2},\;\beta_{_5},\;\beta_{_4}\\2-\beta_{_1}-\beta_{_2}
\end{array}\right|\;{y-x\over 1-x},\;{y-x\over y(1-x)}\right)
\;,\nonumber\\
&&x^{-\beta_{_4}}(x-1)^{1-\beta_{_1}-\beta_{_3}}(y-1)^{\beta_{_1}-1}
\varphi_{_{\{2,4,5\}}}^{(12)}({\boldsymbol{\beta}},\;{x-y\over 1-y},\;{x-y\over x(1-y)})
\nonumber\\
&&\hspace{-0.5cm}=
x^{-\beta_{_4}}(x-1)^{1-\beta_{_1}-\beta_{_3}}(y-1)^{\beta_{_1}-1}
F_{_1}\left(\left.\begin{array}{c}1-\beta_{_1},\;\beta_{_5},\;\beta_{_4}\\2-\beta_{_1}-\beta_{_2}
\end{array}\right|\;{x-y\over 1-y},\;{x-y\over x(1-y)}\right)\;.
\label{Gauss-Kummer6-11-1-2}
\end{eqnarray}
Inserting the above equations into Eq.(\ref{Gauss-Kummer6-11-1-1}) and using $\sum\beta_{_i}=2$, we obtain
\begin{eqnarray}
&&y^{-\beta_{_4}}(y-1)^{-\beta_{_3}}F_{_1}\left(\left.\begin{array}{c}1-\beta_{_2},\;\beta_{_4},\;\beta_{_3}\\
2-\beta_{_1}-\beta_{_2}\end{array}\right|\;1-{x\over y},\;{y-x\over y-1}\right)
\nonumber\\
&&\hspace{-0.5cm}=
x^{\beta_{_2}-1}y^{1-\beta_{_2}-\beta_{_4}}(y-1)^{-\beta_{_3}}
F_{_1}\left(\left.\begin{array}{c}1-\beta_{_2},\;\beta_{_5},\;\beta_{_3}\\2-\beta_{_1}-\beta_{_2}\end{array}\right|\;1-{y\over x},\;{x-y\over x(1-y)}\right)
\nonumber\\
&&\hspace{-0.5cm}=
x^{-\beta_{_4}}(x-1)^{-\beta_{_3}}F_{_1}\left(\left.\begin{array}{c}1-\beta_{_1},\;\beta_{_4},\;\beta_{_3}\\2-\beta_{_1}-\beta_{_2}
\end{array}\right|\;1-{y\over x},\;{y-x\over 1-x}\right)
\nonumber\\
&&\hspace{-0.5cm}=
x^{1-\beta_{_1}-\beta_{_4}}(x-1)^{-\beta_{_3}}y^{\beta_{_1}-1}
F_{_1}\left(\left.\begin{array}{c}1-\beta_{_1},\;\beta_{_5},\;\beta_{_3}\\2-\beta_{_1}-\beta_{_2}
\end{array}\right|\;1-{x\over y},\;{y-x\over y(1-x)}\right)
\nonumber\\
&&\hspace{-0.5cm}=
(x-1)^{\beta_{_2}-1}y^{-\beta_{_4}}(y-1)^{1-\beta_{_2}-\beta_{_3}}
F_{_1}\left(\left.\begin{array}{c}1-\beta_{_2},\;\beta_{_5},\;\beta_{_4}\\2-\beta_{_1}-\beta_{_2}
\end{array}\right|\;{y-x\over 1-x},\;{y-x\over y(1-x)}\right)
\nonumber\\
&&\hspace{-0.5cm}=
x^{-\beta_{_4}}(x-1)^{1-\beta_{_1}-\beta_{_3}}(y-1)^{\beta_{_1}-1}
F_{_1}\left(\left.\begin{array}{c}1-\beta_{_1},\;\beta_{_5},\;\beta_{_4}\\2-\beta_{_1}-\beta_{_2}
\end{array}\right|\;{x-y\over 1-y},\;{x-y\over x(1-y)}\right)\;.
\label{Gauss-Kummer6-11-1-3}
\end{eqnarray}
Taking $\alpha=1-\beta_{_2}$, $\beta=\beta_{_4}$, $\beta^\prime=\beta_{_3}$, $\gamma=2-\beta_{_1}-\beta_{_2}$,
and performing the transformations of the variables
\begin{eqnarray}
&&x^\prime=1-{x\over y}, \;y^\prime={y-x\over y-1}\;,
\label{Gauss-Kummer6-11-1-4}
\end{eqnarray}
one finds that the relations are identical to the relations in Eq.(\ref{sec4-2b}) except for an extra common power factor
$(x^\prime)^{-\beta^\prime}(y^\prime)^{-\beta}(y^\prime-x^\prime)^{\beta+\beta^\prime}$.

As the geometric representation is given as Fig.\ref{fig1}(a) with $\{a,b\}=\{1,3\}$, $\{c,d,e\}=\{2,4,5\}$,
the following six solutions of the GKZ-system presented in Eq.(\ref{sec2-2})
only differ from each other by some constant factors,
\begin{eqnarray}
&&\Phi_{_{\{1,2,4\}}}^{(12)}\sim\Phi_{_{\{1,2,5\}}}^{(12)}\sim\Phi_{_{\{2,3,4\}}}^{(8)}
\sim\Phi_{_{\{2,3,5\}}}^{(8)}\sim\Phi_{_{\{1,4,5\}}}^{(3)}\sim\Phi_{_{\{3,4,5\}}}^{(12)}\;.
\label{Gauss-Kummer6-11-2}
\end{eqnarray}
The relations are explicitly written as
\begin{eqnarray}
&&(1-y)^{-\beta_{_2}}\varphi_{_{\{1,2,4\}}}^{(12)}({\boldsymbol{\beta}},\;1-x,\;{1-x\over 1-y})
\nonumber\\
&&\hspace{-0.5cm}=
x^{\beta_{_3}-1}(1-y)^{-\beta_{_2}}
\varphi_{_{\{1,2,5\}}}^{(12)}({\boldsymbol{\beta}},\;1-{1\over x},\;{y(x-1)\over x(y-1)})
\nonumber\\
&&\hspace{-0.5cm}=
x^{-\beta_{_4}}(x-y)^{-\beta_{_2}}
\varphi_{_{\{2,3,4\}}}^{(8)}({\boldsymbol{\beta}},\;1-{1\over x},\;{1-x\over y-x})
\nonumber\\
&&\hspace{-0.5cm}=
x^{1-\beta_{_1}-\beta_{_4}}(x-y)^{-\beta_{_2}}
\varphi_{_{\{2,3,5\}}}^{(8)}({\boldsymbol{\beta}},\;1-x,\;{y(1-x)\over y-x})
\nonumber\\
&&\hspace{-0.5cm}=
(1-y)^{1-\beta_{_2}-\beta_{_3}}(x-y)^{\beta_{_3}-1}
\varphi_{_{\{1,4,5\}}}^{(3)}({\boldsymbol{\beta}},\;{1-x\over y-x},\;{y(1-x)\over y-x})
\nonumber\\
&&\hspace{-0.5cm}=
x^{-\beta_{_4}}(x-y)^{1-\beta_{_1}-\beta_{_2}}(1-y)^{\beta_{_1}-1}
\varphi_{_{\{3,4,5\}}}^{(12)}({\boldsymbol{\beta}},\;{x-1\over y-1},\;{y(x-1)\over x(y-1)})\;.
\label{Gauss-Kummer6-11-2-1}
\end{eqnarray}

When the geometric representation is depicted in Fig.\ref{fig1}(a) with $\{a,b\}=\{1,4\}$, $\{c,d,e\}=\{2,3,5\}$,
the following six solutions of the GKZ-system presented in Eq.(\ref{sec2-2})
only differ from each other by some constant factors,
\begin{eqnarray}
&&\Phi_{_{\{1,2,3\}}}^{(12)}\sim\Phi_{_{\{1,2,5\}}}^{(3)}\sim\Phi_{_{\{1,3,5\}}}^{(3)}
\sim\Phi_{_{\{2,3,4\}}}^{(1)}\sim\Phi_{_{\{2,4,5\}}}^{(8)}\sim\Phi_{_{\{3,4,5\}}}^{(8)}\;.
\label{Gauss-Kummer6-11-3}
\end{eqnarray}
The relations give
\begin{eqnarray}
&&y^{-\beta_{_2}}\varphi_{_{\{1,2,3\}}}^{(12)}({\boldsymbol{\beta}},\;x,\;{x\over y})
\nonumber\\
&&\hspace{-0.5cm}=
y^{-\beta_{_2}}(1-x)^{\beta_{_4}-1}
\varphi_{_{\{1,2,5\}}}^{(3)}({\boldsymbol{\beta}},\;{x\over x-1},\;{x(y-1)\over y(x-1)})
\nonumber\\
&&\hspace{-0.5cm}=
y^{1-\beta_{_2}-\beta_{_4}}(y-x)^{\beta_{_4}-1}
\varphi_{_{\{1,3,5\}}}^{(3)}({\boldsymbol{\beta}},\;{x\over x-y},\;{x(1-y)\over x-y})
\nonumber\\
&&\hspace{-0.5cm}=
(y-x)^{-\beta_{_2}}(1-x)^{-\beta_{_3}}
\varphi_{_{\{2,3,4\}}}^{(1)}({\boldsymbol{\beta}},\;{x\over x-1},\;{x\over x-y})
\nonumber\\
&&\hspace{-0.5cm}=
(1-x)^{1-\beta_{_1}-\beta_{_3}}(y-x)^{-\beta_{_2}}
\varphi_{_{\{2,4,5\}}}^{(8)}({\boldsymbol{\beta}},\;x,\;{x(1-y)\over x-y})
\nonumber\\
&&\hspace{-0.5cm}=
y^{\beta_{_1}-1}(y-x)^{1-\beta_{_1}-\beta_{_2}}(1-x)^{-\beta_{_3}}
\varphi_{_{\{3,4,5\}}}^{(8)}({\boldsymbol{\beta}},\;{x\over y},\;{x(y-1)\over y(x-1)})\;.
\label{Gauss-Kummer6-11-3-1}
\end{eqnarray}

As the geometric representation is depicted in Fig.\ref{fig1}(a) with $\{a,b\}=\{1,5\}$, $\{c,d,e\}=\{2,3,4\}$,
the following six solutions of the GKZ-system presented in Eq.(\ref{sec2-2})
only differ from each other by some constant factors,
\begin{eqnarray}
&&\Phi_{_{\{1,2,3\}}}^{(3)}\sim\Phi_{_{\{1,2,4\}}}^{(3)}\sim\Phi_{_{\{1,3,4\}}}^{(3)}
\sim\Phi_{_{\{2,3,5\}}}^{(1)}\sim\Phi_{_{\{2,4,5\}}}^{(1)}\sim\Phi_{_{\{3,4,5\}}}^{(1)}\;.
\label{Gauss-Kummer6-11-4}
\end{eqnarray}
The relations give
\begin{eqnarray}
&&x^{\beta_{_5}-1}\varphi_{_{\{1,2,3\}}}^{(3)}({\boldsymbol{\beta}},\;{1\over x},\;{y\over x})
\nonumber\\
&&\hspace{-0.5cm}=
(x-1)^{\beta_{_5}-1}\varphi_{_{\{1,2,4\}}}^{(3)}({\boldsymbol{\beta}},\;{1\over 1-x},\;{1-y\over 1-x})
\nonumber\\
&&\hspace{-0.5cm}=
(x-y)^{\beta_{_5}-1}\varphi_{_{\{1,3,4\}}}^{(3)}({\boldsymbol{\beta}},\;{y\over y-x},\;{y-1\over y-x})
\nonumber\\
&&\hspace{-0.5cm}=
x^{1-\beta_{_1}-\beta_{_4}}(x-y)^{-\beta_{_2}}(x-1)^{-\beta_{_3}}
\varphi_{_{\{2,3,5\}}}^{(1)}({\boldsymbol{\beta}},\;{1\over 1-x},\;{y\over y-x})
\nonumber\\
&&\hspace{-0.5cm}=
x^{-\beta_{_4}}(x-1)^{1-\beta_{_1}-\beta_{_3}}(x-y)^{-\beta_{_2}}
\varphi_{_{\{2,4,5\}}}^{(1)}({\boldsymbol{\beta}},\;{1\over x},\;{1-y\over x-y})
\nonumber\\
&&\hspace{-0.5cm}=
x^{-\beta_{_4}}(x-y)^{1-\beta_{_1}-\beta_{_2}}(x-1)^{-\beta_{_3}}
\varphi_{_{\{3,4,5\}}}^{(1)}({\boldsymbol{\beta}},\;{y\over x},\;{y-1\over x-1})\;.
\label{Gauss-Kummer6-11-4-1}
\end{eqnarray}

For the geometric representation shown in Fig.\ref{fig1}(a) with $\{a,b\}=\{2,3\}$, $\{c,d,e\}=\{1,4,5\}$,
the following six solutions of the GKZ-system presented in Eq.(\ref{sec2-2})
only differ from each other by some constant factors,
\begin{eqnarray}
&&\Phi_{_{\{1,2,4\}}}^{(8)}\sim\Phi_{_{\{1,2,5\}}}^{(8)}\sim\Phi_{_{\{1,3,4\}}}^{(8)}
\sim\Phi_{_{\{1,3,5\}}}^{(8)}\sim\Phi_{_{\{2,4,5\}}}^{(3)}\sim\Phi_{_{\{3,4,5\}}}^{(3)}\;.
\label{Gauss-Kummer6-11-5}
\end{eqnarray}
The relations can be written explicitly as
\begin{eqnarray}
&&(1-x)^{-\beta_{_1}}
\varphi_{_{\{1,2,4\}}}^{(8)}({\boldsymbol{\beta}},\;1-y,\;{1-y\over 1-x})
\nonumber\\
&&\hspace{-0.5cm}=
y^{\beta_{_3}-1}(1-x)^{-\beta_{_1}}\varphi_{_{\{1,2,5\}}}^{(8)}({\boldsymbol{\beta}},\;1-{1\over y},\;{x(y-1)\over y(x-1)})
\nonumber\\
&&\hspace{-0.5cm}=
y^{-\beta_{_4}}(y-x)^{-\beta_{_1}}
\varphi_{_{\{1,3,4\}}}^{(8)}({\boldsymbol{\beta}},\;1-{1\over y},\;{y-1\over y-x})
\nonumber\\
&&\hspace{-0.5cm}=
y^{1-\beta_{_2}-\beta_{_4}}(y-x)^{-\beta_{_1}}
\varphi_{_{\{1,3,5\}}}^{(8)}({\boldsymbol{\beta}},\;1-y,\;{x(1-y)\over x-y})
\nonumber\\
&&\hspace{-0.5cm}=
(1-x)^{1-\beta_{_1}-\beta_{_3}}(y-x)^{\beta_{_3}-1}
\varphi_{_{\{2,4,5\}}}^{(3)}({\boldsymbol{\beta}},\;{1-y\over x-y},\;{x(1-y)\over x-y})
\nonumber\\
&&\hspace{-0.5cm}=
y^{-\beta_{_4}}(y-x)^{1-\beta_{_1}-\beta_{_2}}(1-x)^{\beta_{_2}-1}
\varphi_{_{\{3,4,5\}}}^{(3)}({\boldsymbol{\beta}},\;{y-1\over x-1},\;{x(y-1)\over y(x-1)})\;.
\label{Gauss-Kummer6-11-5-1}
\end{eqnarray}

As the geometric representation is shown in Fig.\ref{fig1}(a) with $\{a,b\}=\{2,4\}$, $\{c,d,e\}=\{1,3,5\}$,
the following six solutions of the GKZ-system presented in Eq.(\ref{sec2-2})
only differ from each other by some constant factors,
\begin{eqnarray}
&&\Phi_{_{\{1,2,3\}}}^{(8)}\sim\Phi_{_{\{1,2,5\}}}^{(5)}\sim\Phi_{_{\{1,3,4\}}}^{(1)}
\sim\Phi_{_{\{2,3,5\}}}^{(3)}\sim\Phi_{_{\{1,4,5\}}}^{(8)}\sim\Phi_{_{\{3,4,5\}}}^{(5)}\;.
\label{Gauss-Kummer6-11-6}
\end{eqnarray}
The relations can be written explicitly as
\begin{eqnarray}
&&x^{-\beta_{_1}}\varphi_{_{\{1,2,3\}}}^{(8)}({\boldsymbol{\beta}},\;y,\;{y\over x})
\nonumber\\
&&\hspace{-0.5cm}=
x^{-\beta_{_1}}(1-y)^{\beta_{_4}-1}
\varphi_{_{\{1,2,5\}}}^{(5)}({\boldsymbol{\beta}},\;{y\over y-1},\;{y(x-1)\over x(y-1)})
\nonumber\\
&&\hspace{-0.5cm}=
(x-y)^{-\beta_{_1}}(1-y)^{-\beta_{_3}}
\varphi_{_{\{1,3,4\}}}^{(1)}({\boldsymbol{\beta}},\;{y\over y-1},\;{y\over y-x})
\nonumber\\
&&\hspace{-0.5cm}=
x^{1-\beta_{_1}-\beta_{_4}}(x-y)^{\beta_{_4}-1}
\varphi_{_{\{2,3,5\}}}^{(3)}({\boldsymbol{\beta}},\;{y\over y-x},\;{y(1-x)\over y-x})
\nonumber\\
&&\hspace{-0.5cm}=
(x-y)^{-\beta_{_1}}(1-y)^{1-\beta_{_2}-\beta_{_3}}
\varphi_{_{\{1,4,5\}}}^{(8)}({\boldsymbol{\beta}},\;y,\;{y(1-x)\over y-x})
\nonumber\\
&&\hspace{-0.5cm}=
x^{\beta_{_2}-1}(x-y)^{1-\beta_{_1}-\beta_{_2}}(1-y)^{-\beta_{_3}}
\varphi_{_{\{3,4,5\}}}^{(5)}({\boldsymbol{\beta}},\;{y\over x},\;{y(x-1)\over x(y-1)})\;.
\label{Gauss-Kummer6-11-6-1}
\end{eqnarray}

As the geometric representation is shown in Fig.\ref{fig1}(a) with $\{a,b\}=\{2,5\}$, $\{c,d,e\}=\{1,3,4\}$,
the following six solutions of the GKZ-system presented in Eq.(\ref{sec2-2})
only differ from each other by some constant factors,
\begin{eqnarray}
&&\Phi_{_{\{1,2,3\}}}^{(5)}\sim\Phi_{_{\{1,2,4\}}}^{(5)}\sim\Phi_{_{\{1,3,5\}}}^{(1)}
\sim\Phi_{_{\{2,3,4\}}}^{(3)}\sim\Phi_{_{\{1,4,5\}}}^{(1)}\sim\Phi_{_{\{3,4,5\}}}^{(10)}\;.
\label{Gauss-Kummer6-11-7}
\end{eqnarray}
The relations can be written explicitly as
\begin{eqnarray}
&&y^{\beta_{_5}-1}\varphi_{_{\{1,2,3\}}}^{(5)}({\boldsymbol{\beta}},\;{1\over y},\;{x\over y})
\nonumber\\
&&\hspace{-0.5cm}=
(y-1)^{\beta_{_5}-1}
\varphi_{_{\{1,2,4\}}}^{(5)}({\boldsymbol{\beta}},\;{1\over 1-y},\;{1-x\over 1-y})
\nonumber\\
&&\hspace{-0.5cm}=
y^{1-\beta_{_2}-\beta_{_4}}(y-x)^{-\beta_{_1}}(y-1)^{-\beta_{_3}}
\varphi_{_{\{1,3,5\}}}^{(1)}({\boldsymbol{\beta}},\;{1\over 1-y},\;{x\over x-y})
\nonumber\\
&&\hspace{-0.5cm}=
(y-x)^{\beta_{_5}-1}
\varphi_{_{\{2,3,4\}}}^{(3)}({\boldsymbol{\beta}},\;{x\over x-y},\;{1-x\over y-x})
\nonumber\\
&&\hspace{-0.5cm}=
(y-1)^{1-\beta_{_2}-\beta_{_3}}(y-x)^{-\beta_{_1}}y^{-\beta_{_4}}
\varphi_{_{\{1,4,5\}}}^{(1)}({\boldsymbol{\beta}},\;{1\over y},\;{1-x\over y-x})
\nonumber\\
&&\hspace{-0.5cm}=
(y-x)^{1-\beta_{_1}-\beta_{_2}}(y-1)^{-\beta_{_3}}y^{-\beta_{_4}}
\varphi_{_{\{3,4,5\}}}^{(10)}({\boldsymbol{\beta}},\;{x\over y},\;{x-1\over y-1})\;.
\label{Gauss-Kummer6-11-7-1}
\end{eqnarray}

When the geometric representation is shown in Fig.\ref{fig1}(a) with $\{a,b\}=\{3,4\}$, $\{c,d,e\}=\{1,2,5\}$,
the following six solutions of the GKZ-system presented in Eq.(\ref{sec2-2})
only differ from each other by some constant factors,
\begin{eqnarray}
&&\Phi_{_{\{1,2,3\}}}^{(1)}\sim\Phi_{_{\{1,2,4\}}}^{(1)}\sim\Phi_{_{\{1,3,5\}}}^{(5)}
\sim\Phi_{_{\{2,3,5\}}}^{(5)}\sim\Phi_{_{\{1,4,5\}}}^{(5)}\sim\Phi_{_{\{2,4,5\}}}^{(5)}\;.
\label{Gauss-Kummer6-11-8}
\end{eqnarray}
The relations are further written as
\begin{eqnarray}
&&x^{-\beta_{_1}}y^{-\beta_{_2}}
\varphi_{_{\{1,2,3\}}}^{(1)}({\boldsymbol{\beta}},\;{1\over y},\;{1\over x})
\nonumber\\
&&\hspace{-0.5cm}=
(x-1)^{-\beta_{_1}}(y-1)^{-\beta_{_2}}
\varphi_{_{\{1,2,4\}}}^{(1)}({\boldsymbol{\beta}},\;{1\over 1-y},\;{1\over 1-x})
\nonumber\\
&&\hspace{-0.5cm}=
x^{-\beta_{_1}}y^{1-\beta_{_2}-\beta_{_4}}(y-1)^{\beta_{_4}-1}
\varphi_{_{\{1,3,5\}}}^{(5)}({\boldsymbol{\beta}},\;{1\over 1-y},\;{x-y\over x(1-y)})
\nonumber\\
&&\hspace{-0.5cm}=
x^{1-\beta_{_1}-\beta_{_4}}y^{-\beta_{_2}}(x-1)^{\beta_{_4}-1}
\varphi_{_{\{2,3,5\}}}^{(5)}({\boldsymbol{\beta}},\;{1\over 1-x},\;{y-x\over y(1-x)})
\nonumber\\
&&\hspace{-0.5cm}=
y^{\beta_{_3}-1}(y-1)^{1-\beta_{_2}-\beta_{_3}}(x-1)^{-\beta_{_1}}
\varphi_{_{\{1,4,5\}}}^{(5)}({\boldsymbol{\beta}},\;{1\over y},\;{y-x\over y(1-x)})
\nonumber\\
&&\hspace{-0.5cm}=
x^{\beta_{_3}-1}(x-1)^{1-\beta_{_1}-\beta_{_3}}(y-1)^{-\beta_{_2}}
\varphi_{_{\{2,4,5\}}}^{(5)}({\boldsymbol{\beta}},\;{1\over x},\;{y-x\over x(1-y)})\;.
\label{Gauss-Kummer6-11-8-1}
\end{eqnarray}

When the geometric representation is shown in Fig.\ref{fig1}(a) with $\{a,b\}=\{4,5\}$, $\{c,d,e\}=\{1,2,3\}$,
the following six solutions of the GKZ-system presented in Eq.(\ref{sec2-2})
only differ from each other by some constant factors,
\begin{eqnarray}
&&\Phi_{_{\{1,2,4\}}}^{(10)}\sim\Phi_{_{\{1,2,5\}}}^{(10)}\sim\Phi_{_{\{1,3,4\}}}^{(10)}
\sim\Phi_{_{\{1,3,5\}}}^{(10)}\sim\Phi_{_{\{2,3,4\}}}^{(10)}\sim\Phi_{_{\{2,3,5\}}}^{(10)}\;.
\label{Gauss-Kummer6-11-10}
\end{eqnarray}
The relations are further written as
\begin{eqnarray}
&&\varphi_{_{\{1,2,4\}}}^{(10)}({\boldsymbol{\beta}},\;1-y,\;1-x)
\nonumber\\
&&\hspace{-0.5cm}=
x^{-\beta_{_1}}y^{-\beta_{_2}}
\varphi_{_{\{1,2,5\}}}^{(10)}({\boldsymbol{\beta}},\;1-{1\over y},\;1-{1\over x})
\nonumber\\
&&\hspace{-0.5cm}=
y^{\beta_{_5}-1}
\varphi_{_{\{1,3,4\}}}^{(10)}({\boldsymbol{\beta}},\;1-{1\over y},\;1-{x\over y})
\nonumber\\
&&\hspace{-0.5cm}=
y^{1-\beta_{_2}-\beta_{_4}}x^{-\beta_{_1}}
\varphi_{_{\{1,3,5\}}}^{(10)}({\boldsymbol{\beta}},\;1-y,\;1-{y\over x})
\nonumber\\
&&\hspace{-0.5cm}=
x^{\beta_{_5}-1}
\varphi_{_{\{2,3,4\}}}^{(10)}({\boldsymbol{\beta}},\;1-{1\over x},\;1-{y\over x})
\nonumber\\
&&\hspace{-0.5cm}=
x^{1-\beta_{_1}-\beta_{_4}}y^{-\beta_{_2}}
\varphi_{_{\{2,3,5\}}}^{(10)}({\boldsymbol{\beta}},\;1-x,\;1-{x\over y})\;.
\label{Gauss-Kummer6-11-10-1}
\end{eqnarray}

As the geometric representation is shown in Fig.\ref{fig1}(b) with $a=1$, $\{b,c\}=\{2,3\}$, and $\{d,e\}=\{4,5\}$,
the following four solutions of the GKZ-system presented in Eq.(\ref{sec2-2})
only differ from each other by some constant factors,
\begin{eqnarray}
&&\Phi_{_{\{1,2,4\}}}^{(9)}\sim\Phi_{_{\{1,2,5\}}}^{(9)}\sim\Phi_{_{\{1,3,4\}}}^{(9)}
\sim\Phi_{_{\{1,3,5\}}}^{(9)}\;.
\label{Gauss-Kummer6-11-11}
\end{eqnarray}
The relations are simplified further as
\begin{eqnarray}
&&(1-x)^{\beta_{_4}+\beta_{_5}-1}\varphi_{_{\{1,2,4\}}}^{(9)}({\boldsymbol{\beta}},\;1-x,\;{1-y\over 1-x})
\nonumber\\
&&\hspace{-0.5cm}=
x^{\beta_{_2}+\beta_{_3}-1}(1-x)^{\beta_{_4}+\beta_{_5}-1}y^{-\beta_{_2}}
\varphi_{_{\{1,2,5\}}}^{(9)}({\boldsymbol{\beta}},\;1-{1\over x},\;{x(y-1)\over y(x-1)})
\nonumber\\
&&\hspace{-0.5cm}=
y^{-\beta_{_4}}(y-x)^{\beta_{_4}+\beta_{_5}-1}
\varphi_{_{\{1,3,4\}}}^{(9)}({\boldsymbol{\beta}},\;1-{x\over y},\;{y-1\over y-x})
\nonumber\\
&&\hspace{-0.5cm}=
x^{\beta_{_2}+\beta_{_3}-1}y^{1-\beta_{_2}-\beta_{_4}}(y-x)^{\beta_{_4}+\beta_{_5}-1}
\varphi_{_{\{1,3,5\}}}^{(9)}({\boldsymbol{\beta}},\;1-{y\over x},\;{x(1-y)\over x-y})\;.
\label{Gauss-Kummer6-11-11-2}
\end{eqnarray}

In order to formulate the above equations in terms of the Horn function $H_{_1}$ ,
we rewrite the $\varphi$ functions as
\begin{eqnarray}
&&(1-x)^{\beta_{_4}+\beta_{_5}-1}\varphi_{_{\{1,2,4\}}}^{(9)}({\boldsymbol{\beta}},\;1-x,\;{1-y\over 1-x})
\nonumber\\
&&\hspace{-0.5cm}=
(1-x)^{\beta_{_4}+\beta_{_5}-1}H_{_1}\left(\left.\begin{array}{c}\beta_{_4},\;\beta_{_2}\\
\beta_{_4}+\beta_{_5},\;\beta_{_2}+\beta_{_3}\end{array}\right|\;1-x,\;{1-y\over 1-x}\right)
\;,\nonumber\\
&&x^{\beta_{_2}+\beta_{_3}-1}(1-x)^{\beta_{_4}+\beta_{_5}-1}y^{-\beta_{_2}}
\varphi_{_{\{1,2,5\}}}^{(9)}({\boldsymbol{\beta}},\;1-{1\over x},\;{x(y-1)\over y(x-1)})
\nonumber\\
&&\hspace{-0.5cm}=
x^{\beta_{_2}+\beta_{_3}-1}(1-x)^{\beta_{_4}+\beta_{_5}-1}y^{-\beta_{_2}}
H_{_1}\left(\left.\begin{array}{c}\beta_{_5},\;\beta_{_2}\\
\beta_{_4}+\beta_{_5},\;\beta_{_2}+\beta_{_3}\end{array}\right|\;1-{1\over x},\;{x(y-1)\over y(x-1)}\right)
\;,\nonumber\\
&&y^{-\beta_{_4}}(y-x)^{\beta_{_4}+\beta_{_5}-1}
\varphi_{_{\{1,3,4\}}}^{(9)}({\boldsymbol{\beta}},\;1-{x\over y},\;{y-1\over y-x})
\nonumber\\
&&\hspace{-0.5cm}=
y^{-\beta_{_4}}(y-x)^{\beta_{_4}+\beta_{_5}-1}H_{_1}\left(\left.\begin{array}{c}\beta_{_4},\;\beta_{_3}\\
\beta_{_4}+\beta_{_5},\;\beta_{_2}+\beta_{_3}\end{array}\right|\;1-{x\over y},\;{y-1\over y-x}\right)
\;,\nonumber\\
&&x^{\beta_{_2}+\beta_{_3}-1}y^{1-\beta_{_2}-\beta_{_4}}(y-x)^{\beta_{_4}+\beta_{_5}-1}
\varphi_{_{\{1,3,5\}}}^{(9)}({\boldsymbol{\beta}},\;1-{y\over x},\;{x(1-y)\over x-y})
\nonumber\\
&&\hspace{-0.5cm}=
x^{\beta_{_2}+\beta_{_3}-1}y^{1-\beta_{_2}-\beta_{_4}}(y-x)^{\beta_{_4}+\beta_{_5}-1}
H_{_1}\left(\left.\begin{array}{c}\beta_{_5},\;\beta_{_3}\\
\beta_{_4}+\beta_{_5},\;\beta_{_2}+\beta_{_3}\end{array}\right|\;{x-y\over 1-y},{x(1-y)\over x-y}\right)\;.
\label{Gauss-Kummer6-11-11-3}
\end{eqnarray}
Denoting $\alpha=\beta_{_4}$, $\beta=\beta_{_4}+\beta_{_5}$, $\alpha^\prime=\beta_{_2}$,
$\beta^\prime=\beta_{_2}+\beta_{_3}$, and making the transformations of the variables
\begin{eqnarray}
&&x^\prime=1-x, \;y^\prime={y-1\over x-1}\;,
\label{Gauss-Kummer6-11-11-4}
\end{eqnarray}
one finds that the relations are identical to the relations in Eq.(\ref{sec4-4b}) except for an extra common power factor
$(x^\prime)^{\beta-1}$.

As the geometric representation is shown in Fig.\ref{fig1}(b) with $a=1$, $\{b,c\}=\{2,4\}$, and $\{d,e\}=\{3,5\}$,
the following four solutions of the GKZ-system presented in Eq.(\ref{sec2-2})
only differ from each other by some constant factors,
\begin{eqnarray}
&&\Phi_{_{\{1,2,3\}}}^{(9)}\sim\Phi_{_{\{1,2,5\}}}^{(6)}\sim\Phi_{_{\{1,3,4\}}}^{(6)}
\sim\Phi_{_{\{1,4,5\}}}^{(9)}\;.
\label{Gauss-Kummer6-11-12}
\end{eqnarray}
The relations are further written as
\begin{eqnarray}
&&x^{\beta_{_3}+\beta_{_5}-1}\varphi_{_{\{1,2,3\}}}^{(9)}({\boldsymbol{\beta}},\;x,\;{y\over x})
\nonumber\\
&&\hspace{-0.5cm}=
(1-y)^{-\beta_{_2}}x^{\beta_{_3}+\beta_{_5}-1}(1-x)^{\beta_{_2}+\beta_{_4}-1}
\varphi_{_{\{1,2,5\}}}^{(6)}({\boldsymbol{\beta}},\;{x\over x-1},\;{y(x-1)\over x(y-1)})
\nonumber\\
&&\hspace{-0.5cm}=
(1-y)^{-\beta_{_3}}(x-y)^{\beta_{_3}+\beta_{_5}-1}
\varphi_{_{\{1,3,4\}}}^{(6)}({\boldsymbol{\beta}},\;{y\over y-x},\;{y-x\over y-1})
\nonumber\\
&&\hspace{-0.5cm}=
(1-y)^{1-\beta_{_2}-\beta_{_3}}(1-x)^{\beta_{_2}+\beta_{_4}-1}(x-y)^{\beta_{_3}+\beta_{_5}-1}
\varphi_{_{\{1,4,5\}}}^{(9)}({\boldsymbol{\beta}},\;{y-x\over 1-x},\;{y(1-x)\over y-x})\;.
\label{Gauss-Kummer6-11-12-2}
\end{eqnarray}

As the geometric representation is shown in Fig.\ref{fig1}(b) with $a=1$, $\{b,c\}=\{2,5\}$, and $\{d,e\}=\{3,4\}$,
the following four solutions of the GKZ-system presented in Eq.(\ref{sec2-2})
only differ from each other by some constant factors,
\begin{eqnarray}
&&\Phi_{_{\{1,2,3\}}}^{(6)}\sim\Phi_{_{\{1,2,4\}}}^{(6)}\sim\Phi_{_{\{1,3,5\}}}^{(6)}
\sim\Phi_{_{\{1,4,5\}}}^{(6)}\;.
\label{Gauss-Kummer6-11-13}
\end{eqnarray}
The relations are further written as
\begin{eqnarray}
&&y^{-\beta_{_2}}x^{\beta_{_2}+\beta_{_5}-1}
\varphi_{_{\{1,2,3\}}}^{(6)}({\boldsymbol{\beta}},\;{1\over x},\;{x\over y})
\nonumber\\
&&\hspace{-0.5cm}=
(y-1)^{-\beta_{_2}}(x-1)^{\beta_{_2}+\beta_{_5}-1}
\varphi_{_{\{1,2,4\}}}^{(6)}({\boldsymbol{\beta}},\;{1\over 1-x},\;{1-x\over 1-y})
\nonumber\\
&&\hspace{-0.5cm}=
y^{1-\beta_{_2}-\beta_{_4}}x^{\beta_{_2}+\beta_{_5}-1}(y-1)^{-\beta_{_3}}
(y-x)^{\beta_{_3}+\beta_{_4}-1}
\nonumber\\
&&\hspace{0.0cm}\times
\varphi_{_{\{1,3,5\}}}^{(6)}({\boldsymbol{\beta}},\;{x\over x-y},\;{x-y\over x(1-y)})
\nonumber\\
&&\hspace{-0.5cm}=
y^{-\beta_{_4}}(y-1)^{1-\beta_{_2}-\beta_{_3}}(x-1)^{\beta_{_2}+\beta_{_5}-1}
(y-x)^{\beta_{_3}+\beta_{_4}-1}
\nonumber\\
&&\hspace{0.0cm}\times
\varphi_{_{\{1,4,5\}}}^{(6)}({\boldsymbol{\beta}},\;{1-x\over y-x},\;{y-x\over y(1-x)})\;.
\label{Gauss-Kummer6-11-13-2}
\end{eqnarray}

As the geometric representation is shown in Fig.\ref{fig1}(b) with $a=2$, $\{b,c\}=\{1,3\}$, and $\{d,e\}=\{4,5\}$,
the following four solutions of the GKZ-system presented in Eq.(\ref{sec2-2})
only differ from each other by some constant factors,
\begin{eqnarray}
&&\Phi_{_{\{1,2,4\}}}^{(11)}\sim\Phi_{_{\{1,2,5\}}}^{(11)}\sim\Phi_{_{\{2,3,4\}}}^{(9)}
\sim\Phi_{_{\{2,3,5\}}}^{(9)}\;.
\label{Gauss-Kummer6-11-14}
\end{eqnarray}
The relations are simplified further as
\begin{eqnarray}
&&(y-1)^{\beta_{_4}+\beta_{_5}-1}
\varphi_{_{\{1,2,4\}}}^{(11)}({\boldsymbol{\beta}},\;1-y,\;{1-x\over 1-y})
\nonumber\\
&&\hspace{-0.5cm}=
y^{\beta_{_1}+\beta_{_3}-1}x^{-\beta_{_1}}(y-1)^{\beta_{_4}+\beta_{_5}-1}
\varphi_{_{\{1,2,5\}}}^{(11)}({\boldsymbol{\beta}},\;1-{1\over y},\;{y(x-1)\over x(y-1)})
\nonumber\\
&&\hspace{-0.5cm}=
x^{-\beta_{_4}}(y-x)^{\beta_{_4}+\beta_{_5}-1}
\varphi_{_{\{2,3,4\}}}^{(9)}({\boldsymbol{\beta}},\;1-{y\over x},\;{1-x\over y-x})
\nonumber\\
&&\hspace{-0.5cm}=
y^{\beta_{_1}+\beta_{_3}-1}x^{1-\beta_{_1}-\beta_{_4}}(y-x)^{\beta_{_4}+\beta_{_5}-1}
\varphi_{_{\{2,3,5\}}}^{(9)}({\boldsymbol{\beta}},\;1-{x\over y},\;{y(1-x)\over y-x})\;.
\label{Gauss-Kummer6-11-14-2}
\end{eqnarray}

As the geometric representation is shown in Fig.\ref{fig1}(b) with $a=2$, $\{b,c\}=\{1,5\}$, and $\{d,e\}=\{3,4\}$,
the following four solutions of the GKZ-system presented in Eq.(\ref{sec2-2})
only differ from each other by some constant factors,
\begin{eqnarray}
&&\Phi_{_{\{1,2,3\}}}^{(2)}\sim\Phi_{_{\{1,2,4\}}}^{(2)}\sim\Phi_{_{\{2,3,5\}}}^{(6)}
\sim\Phi_{_{\{2,4,5\}}}^{(6)}\;.
\label{Gauss-Kummer6-11-16}
\end{eqnarray}
The relations are further written as
\begin{eqnarray}
&&y^{\beta_{_1}+\beta_{_5}-1}x^{-\beta_{_1}}
\varphi_{_{\{1,2,3\}}}^{(2)}({\boldsymbol{\beta}},\;{1\over y},\;{y\over x})
\nonumber\\
&&\hspace{-0.5cm}=
(x-1)^{-\beta_{_1}}(y-1)^{\beta_{_1}+\beta_{_5}-1}
\varphi_{_{\{1,2,4\}}}^{(2)}({\boldsymbol{\beta}},\;{1\over 1-y},\;{1-y\over 1-x})
\nonumber\\
&&\hspace{-0.5cm}=
x^{1-\beta_{_1}-\beta_{_4}}y^{\beta_{_1}+\beta_{_5}-1}(x-y)^{\beta_{_3}+\beta_{_4}-1}
(x-1)^{-\beta_{_3}}
\varphi_{_{\{2,3,5\}}}^{(6)}({\boldsymbol{\beta}},\;{y\over y-x},\;{y-x\over y(1-x)})
\nonumber\\
&&\hspace{-0.5cm}=
(x-1)^{1-\beta_{_1}-\beta_{_3}}(y-1)^{\beta_{_1}+\beta_{_5}-1}x^{-\beta_{_4}}
(x-y)^{\beta_{_3}+\beta_{_4}-1}
\varphi_{_{\{2,4,5\}}}^{(6)}({\boldsymbol{\beta}},\;{1-y\over x-y},\;{x-y\over x(1-y)})\;.
\label{Gauss-Kummer6-11-16-2}
\end{eqnarray}

As the geometric representation is shown in Fig.\ref{fig1}(b) with $a=3$, $\{b,c\}=\{1,2\}$, and $\{d,e\}=\{4,5\}$,
the following four solutions of the GKZ-system presented in Eq.(\ref{sec2-2})
only differ from each other by some constant factors,
\begin{eqnarray}
&&\Phi_{_{\{1,3,4\}}}^{(11)}\sim\Phi_{_{\{1,3,5\}}}^{(11)}\sim\Phi_{_{\{2,3,4\}}}^{(11)}
\sim\Phi_{_{\{2,3,5\}}}^{(11)}\;.
\label{Gauss-Kummer6-11-17}
\end{eqnarray}
The relations are further written as
\begin{eqnarray}
&&(1-y)^{\beta_{_4}+\beta_{_5}-1}y^{-\beta_{_4}}
\varphi_{_{\{1,3,4\}}}^{(11)}({\boldsymbol{\beta}},\;1-{1\over y},\;{y-x\over y-1})
\nonumber\\
&&\hspace{-0.5cm}=
(1-y)^{\beta_{_4}+\beta_{_5}-1}y^{1-\beta_{_2}-\beta_{_4}}x^{-\beta_{_1}}
\varphi_{_{\{1,3,5\}}}^{(11)}({\boldsymbol{\beta}},\;1-y,\;{x-y\over x(1-y)})
\nonumber\\
&&\hspace{-0.5cm}=
(1-x)^{\beta_{_4}+\beta_{_5}-1}x^{-\beta_{_4}}
\varphi_{_{\{2,3,4\}}}^{(11)}({\boldsymbol{\beta}},\;1-{1\over x},\;{y-x\over 1-x})
\nonumber\\
&&\hspace{-0.5cm}=
(1-x)^{\beta_{_4}+\beta_{_5}-1}y^{-\beta_{_2}}x^{1-\beta_{_1}-\beta_{_4}}
\varphi_{_{\{2,3,5\}}}^{(11)}({\boldsymbol{\beta}},\;1-x,\;{y-x\over y(1-x)})\;.
\label{Gauss-Kummer6-11-17-2}
\end{eqnarray}

As the geometric representation is shown in Fig.\ref{fig1}(b) with $a=3$, $\{b,c\}=\{1,4\}$, and $\{d,e\}=\{2,5\}$,
the following four solutions of the GKZ-system presented in Eq.(\ref{sec2-2})
only differ from each other by some constant factors,
\begin{eqnarray}
&&\Phi_{_{\{1,2,3\}}}^{(4)}\sim\Phi_{_{\{1,3,5\}}}^{(2)}\sim\Phi_{_{\{2,3,4\}}}^{(2)}
\sim\Phi_{_{\{3,4,5\}}}^{(9)}\;.
\label{Gauss-Kummer6-11-18}
\end{eqnarray}
The relations are further written as
\begin{eqnarray}
&&y^{-\beta_{_2}}\varphi_{_{\{1,2,3\}}}^{(4)}({\boldsymbol{\beta}},\;{1\over y},\;x)
\nonumber\\
&&\hspace{-0.5cm}=
y^{1-\beta_{_2}-\beta_{_4}}(y-x)^{-\beta_{_1}}
(y-1)^{\beta_{_1}+\beta_{_4}-1}
\varphi_{_{\{1,3,5\}}}^{(2)}({\boldsymbol{\beta}},\;{1\over 1-y},\;{x(1-y)\over x-y})
\nonumber\\
&&\hspace{-0.5cm}=
(y-x)^{-\beta_{_2}}(1-x)^{\beta_{_2}+\beta_{_5}-1}
\varphi_{_{\{2,3,4\}}}^{(2)}({\boldsymbol{\beta}},\;{x\over x-1},\;{1-x\over y-x})
\nonumber\\
&&\hspace{-0.5cm}=
(y-x)^{1-\beta_{_1}-\beta_{_2}}(y-1)^{\beta_{_1}+\beta_{_4}-1}y^{-\beta_{_4}}
(1-x)^{\beta_{_2}+\beta_{_5}-1}
\varphi_{_{\{3,4,5\}}}^{(9)}({\boldsymbol{\beta}},\;{x-1\over y-1},\;{x(y-1)\over y(x-1)})\;.
\label{Gauss-Kummer6-11-18-2}
\end{eqnarray}

As the geometric representation is shown in Fig.\ref{fig1}(b) with $a=3$, $\{b,c\}=\{1,5\}$, and $\{d,e\}=\{2,4\}$,
the following four solutions of the GKZ-system presented in Eq.(\ref{sec2-2})
only differ from each other by some constant factors,
\begin{eqnarray}
&&\Phi_{_{\{1,2,3\}}}^{(7)}\sim\Phi_{_{\{1,3,4\}}}^{(2)}\sim\Phi_{_{\{2,3,5\}}}^{(2)}
\sim\Phi_{_{\{3,4,5\}}}^{(6)}\;.
\label{Gauss-Kummer6-11-19}
\end{eqnarray}
The relations are further modified as
\begin{eqnarray}
&&x^{-\beta_{_1}}\varphi_{_{\{1,2,3\}}}^{(7)}({\boldsymbol{\beta}},\;y,\;{1\over x})
\nonumber\\
&&\hspace{-0.5cm}=
(x-y)^{-\beta_{_1}}(1-y)^{\beta_{_1}+\beta_{_5}-1}
\varphi_{_{\{1,3,4\}}}^{(2)}({\boldsymbol{\beta}},\;{y\over y-1},\;{y-1\over y-x})
\nonumber\\
&&\hspace{-0.5cm}=
x^{1-\beta_{_1}-\beta_{_4}}(x-y)^{-\beta_{_2}}(x-1)^{\beta_{_2}+\beta_{_4}-1}
\varphi_{_{\{2,3,5\}}}^{(2)}({\boldsymbol{\beta}},\;{1\over 1-x},\;{y(1-x)\over y-x})
\nonumber\\
&&\hspace{-0.5cm}=
(x-y)^{1-\beta_{_1}-\beta_{_2}}(1-y)^{\beta_{_1}+\beta_{_5}-1}x^{-\beta_{_4}}
(x-1)^{\beta_{_2}+\beta_{_4}-1}
\varphi_{_{\{3,4,5\}}}^{(6)}({\boldsymbol{\beta}},\;{y-1\over x-1},\;{y(x-1)\over x(y-1)})\;.
\label{Gauss-Kummer6-11-19-1}
\end{eqnarray}

As the geometric representation is shown in Fig.\ref{fig1}(b) with $a=4$, $\{b,c\}=\{1,2\}$, and $\{d,e\}=\{3,5\}$,
the following four solutions of the GKZ-system presented in Eq.(\ref{sec2-2})
only differ from each other by some constant factors,
\begin{eqnarray}
&&\Phi_{_{\{1,3,4\}}}^{(4)}\sim\Phi_{_{\{1,4,5\}}}^{(11)}\sim\Phi_{_{\{2,4,5\}}}^{(11)}
\sim\Phi_{_{\{2,3,4\}}}^{(4)}\;.
\label{Gauss-Kummer6-11-20}
\end{eqnarray}
The relations are further written as
\begin{eqnarray}
&&y^{\beta_{_3}+\beta_{_5}-1}(1-y)^{-\beta_{_3}}
\varphi_{_{\{1,3,4\}}}^{(4)}({\boldsymbol{\beta}},\;{y\over y-1},\;1-{x\over y})
\nonumber\\
&&\hspace{-0.5cm}=
x^{\beta_{_3}+\beta_{_5}-1}(1-x)^{-\beta_{_3}}
\varphi_{_{\{2,3,4\}}}^{(4)}({\boldsymbol{\beta}},\;{x\over x-1},\;1-{y\over x})
\nonumber\\
&&\hspace{-0.5cm}=
y^{\beta_{_3}+\beta_{_5}-1}(1-y)^{1-\beta_{_2}-\beta_{_3}}(1-x)^{-\beta_{_1}}
\varphi_{_{\{1,4,5\}}}^{(11)}({\boldsymbol{\beta}},\;y,\;{y-x\over y(1-x)})
\nonumber\\
&&\hspace{-0.5cm}=
x^{\beta_{_3}+\beta_{_5}-1}(1-y)^{-\beta_{_2}}(1-x)^{1-\beta_{_1}-\beta_{_3}}
\varphi_{_{\{2,4,5\}}}^{(11)}({\boldsymbol{\beta}},\;x,\;{x-y\over x(1-y)})\;.
\label{Gauss-Kummer6-11-20-2}
\end{eqnarray}

As the geometric representation is shown in Fig.\ref{fig1}(b) with $a=4$, $\{b,c\}=\{1,3\}$, and $\{d,e\}=\{2,5\}$,
the following four solutions of the GKZ-system presented in Eq.(\ref{sec2-2})
only differ from each other by some constant factors,
\begin{eqnarray}
&&\Phi_{_{\{1,2,4\}}}^{(4)}\sim\Phi_{_{\{2,3,4\}}}^{(7)}\sim\Phi_{_{\{1,4,5\}}}^{(2)}
\sim\Phi_{_{\{3,4,5\}}}^{(11)}\;.
\label{Gauss-Kummer6-11-21}
\end{eqnarray}
The relations are further written as
\begin{eqnarray}
&&(y-1)^{-\beta_{_2}}\varphi_{_{\{1,2,4\}}}^{(4)}({\boldsymbol{\beta}},\;{1\over 1-y},\;1-x)
\nonumber\\
&&\hspace{-0.5cm}=
(y-x)^{-\beta_{_2}}x^{\beta_{_2}+\beta_{_5}-1}
\varphi_{_{\{2,3,4\}}}^{(7)}({\boldsymbol{\beta}},\;1-{1\over x},\;{x\over x-y})
\nonumber\\
&&\hspace{-0.5cm}=
y^{\beta_{_1}+\beta_{_3}-1}(y-1)^{1-\beta_{_2}-\beta_{_3}}(y-x)^{-\beta_{_1}}
\varphi_{_{\{1,4,5\}}}^{(2)}({\boldsymbol{\beta}},\;{1\over y},\;{y(1-x)\over y-x})
\nonumber\\
&&\hspace{-0.5cm}=
y^{\beta_{_1}+\beta_{_3}-1}(y-1)^{-\beta_{_3}}(y-x)^{1-\beta_{_1}-\beta_{_2}}
x^{\beta_{_2}+\beta_{_5}-1}
\varphi_{_{\{3,4,5\}}}^{(11)}({\boldsymbol{\beta}},\;{x\over y},\;{y(x-1)\over x(y-1)})\;.
\label{Gauss-Kummer6-11-21-2}
\end{eqnarray}

As the geometric representation is shown in Fig.\ref{fig1}(b) with $a=4$, $\{b,c\}=\{1,5\}$, and $\{d,e\}=\{2,3\}$,
the following four solutions of the GKZ-system presented in Eq.(\ref{sec2-2})
only differ from each other by some constant factors,
\begin{eqnarray}
&&\Phi_{_{\{1,2,4\}}}^{(7)}\sim\Phi_{_{\{1,3,4\}}}^{(7)}\sim\Phi_{_{\{2,4,5\}}}^{(2)}
\sim\Phi_{_{\{3,4,5\}}}^{(2)}\;.
\label{Gauss-Kummer6-11-22}
\end{eqnarray}
The relations are further written as
\begin{eqnarray}
&&(x-1)^{-\beta_{_1}}\varphi_{_{\{1,2,4\}}}^{(7)}({\boldsymbol{\beta}},\;1-y,\;{1\over 1-x})
\nonumber\\
&&\hspace{-0.5cm}=
y^{\beta_{_1}+\beta_{_5}-1}(x-y)^{-\beta_{_1}}
\varphi_{_{\{1,3,4\}}}^{(7)}({\boldsymbol{\beta}},\;1-{1\over y},\;{y\over y-x})
\nonumber\\
&&\hspace{-0.5cm}=
(x-1)^{1-\beta_{_1}-\beta_{_3}}(x-y)^{-\beta_{_2}}x^{\beta_{_2}+\beta_{_3}-1}
\varphi_{_{\{2,4,5\}}}^{(2)}({\boldsymbol{\beta}},\;{1\over x},\;{x(1-y)\over x-y})
\nonumber\\
&&\hspace{-0.5cm}=
y^{\beta_{_1}+\beta_{_5}-1}(x-y)^{1-\beta_{_1}-\beta_{_2}}(x-1)^{-\beta_{_3}}x^{\beta_{_2}+\beta_{_3}-1}
\varphi_{_{\{3,4,5\}}}^{(2)}({\boldsymbol{\beta}},\;{y\over x},\;{x(y-1)\over y(x-1)})\;.
\label{Gauss-Kummer6-11-22-2}
\end{eqnarray}

As the geometric representation is shown in Fig.\ref{fig1}(b) with $a=5$, $\{b,c\}=\{1,2\}$, and $\{d,e\}=\{3,4\}$,
the following four solutions of the GKZ-system presented in Eq.(\ref{sec2-2})
only differ from each other by some constant factors,
\begin{eqnarray}
&&\Phi_{_{\{1,3,5\}}}^{(4)}\sim\Phi_{_{\{2,3,5\}}}^{(4)}\sim\Phi_{_{\{1,4,5\}}}^{(4)}
\sim\Phi_{_{\{2,4,5\}}}^{(4)}\;.
\label{Gauss-Kummer6-11-23}
\end{eqnarray}
The relations are further written as
\begin{eqnarray}
&&y^{1-\beta_{_2}-\beta_{_4}}(1-y)^{-\beta_{_3}}x^{-\beta_{_1}}
\varphi_{_{\{1,3,5\}}}^{(4)}({\boldsymbol{\beta}},\;{1\over 1-y},\;1-{y\over x})
\nonumber\\
&&\hspace{-0.5cm}=
x^{1-\beta_{_1}-\beta_{_4}}y^{-\beta_{_2}}(1-x)^{-\beta_{_3}}
\varphi_{_{\{2,3,5\}}}^{(4)}({\boldsymbol{\beta}},\;{1\over 1-x},\;1-{x\over y})
\nonumber\\
&&\hspace{-0.5cm}=
(1-y)^{1-\beta_{_2}-\beta_{_3}}(1-x)^{-\beta_{_1}}y^{-\beta_{_4}}
\varphi_{_{\{1,4,5\}}}^{(4)}({\boldsymbol{\beta}},\;{1\over y},\;{y-x\over 1-x})
\nonumber\\
&&\hspace{-0.5cm}=
(1-x)^{1-\beta_{_1}-\beta_{_3}}(1-y)^{-\beta_{_2}}x^{-\beta_{_4}}
\varphi_{_{\{2,4,5\}}}^{(4)}({\boldsymbol{\beta}},\;{1\over x},\;{x-y\over 1-y})\;.
\label{Gauss-Kummer6-11-23-2}
\end{eqnarray}

As the geometric representation is shown in Fig.\ref{fig1}(b) with $a=5$, $\{b,c\}=\{1,3\}$, and $\{d,e\}=\{2,4\}$,
the following four solutions of the GKZ-system presented in Eq.(\ref{sec2-2})
only differ from each other by some constant factors,
\begin{eqnarray}
&&\Phi_{_{\{1,2,5\}}}^{(4)}\sim\Phi_{_{\{2,3,5\}}}^{(7)}\sim\Phi_{_{\{1,4,5\}}}^{(7)}
\sim\Phi_{_{\{3,4,5\}}}^{(4)}\;.
\label{Gauss-Kummer6-11-24}
\end{eqnarray}
The relations are further written as
\begin{eqnarray}
&&x^{-\beta_{_1}}(y-1)^{-\beta_{_2}}
\varphi_{_{\{1,2,5\}}}^{(4)}({\boldsymbol{\beta}},\;{y\over y-1},\;1-{1\over x})
\nonumber\\
&&\hspace{-0.5cm}=
x^{1-\beta_{_1}-\beta_{_4}}(y-x)^{-\beta_{_2}}
\varphi_{_{\{2,3,5\}}}^{(7)}({\boldsymbol{\beta}},\;1-x,\;{y\over y-x})
\nonumber\\
&&\hspace{-0.5cm}=
(y-1)^{1-\beta_{_2}-\beta_{_3}}(y-x)^{-\beta_{_1}}
\varphi_{_{\{1,4,5\}}}^{(7)}({\boldsymbol{\beta}},\;y,\;{1-x\over y-x})
\nonumber\\
&&\hspace{-0.5cm}=
(y-x)^{1-\beta_{_1}-\beta_{_2}}(y-1)^{-\beta_{_3}}x^{-\beta_{_4}}
\varphi_{_{\{3,4,5\}}}^{(4)}({\boldsymbol{\beta}},\;{y\over x},\;{x-1\over y-1})\;.
\label{Gauss-Kummer6-11-24-2}
\end{eqnarray}

As the geometric representation is shown in Fig.\ref{fig1}(b) with $a=5$, $\{b,c\}=\{1,4\}$, and $\{d,e\}=\{2,3\}$,
the following four solutions of the GKZ-system presented in Eq.(\ref{sec2-2})
only differ from each other by some constant factors,
\begin{eqnarray}
&&\Phi_{_{\{1,2,5\}}}^{(7)}\sim\Phi_{_{\{1,3,5\}}}^{(7)}\sim\Phi_{_{\{2,4,5\}}}^{(7)}
\sim\Phi_{_{\{3,4,5\}}}^{(7)}\;.
\label{Gauss-Kummer6-11-25}
\end{eqnarray}
The relations are further written as
\begin{eqnarray}
&&y^{-\beta_{_2}}(x-1)^{-\beta_{_1}}
\varphi_{_{\{1,2,5\}}}^{(7)}({\boldsymbol{\beta}},\;1-{1\over y},\;{x\over x-1})
\nonumber\\
&&\hspace{-0.5cm}=
y^{1-\beta_{_2}-\beta_{_4}}(x-y)^{-\beta_{_1}}
\varphi_{_{\{1,3,5\}}}^{(7)}({\boldsymbol{\beta}},\;1-y,\;{x\over x-y})
\nonumber\\
&&\hspace{-0.5cm}=
(x-1)^{1-\beta_{_1}-\beta_{_3}}(x-y)^{-\beta_{_2}}
\varphi_{_{\{2,4,5\}}}^{(7)}({\boldsymbol{\beta}},\;x,\;{1-y\over x-y})
\nonumber\\
&&\hspace{-0.5cm}=
(x-y)^{1-\beta_{_1}-\beta_{_2}}y^{-\beta_{_4}}(x-1)^{-\beta_{_3}}
\varphi_{_{\{3,4,5\}}}^{(7)}({\boldsymbol{\beta}},\;{x\over y},\;{y-1\over x-1})\;.
\label{Gauss-Kummer6-11-25-2}
\end{eqnarray}

\section{The combinatorial coefficients in one-loop self energy\label{app5}}
\indent\indent
The combinatorial coefficients of $\Phi_{_{\{1,2,3\}}}^{(i)}({\boldsymbol{\beta}}),\;i=4,7,9,11$ are respectively written as
\begin{eqnarray}
&&C_{_{\{1,2,3\}}}^{(4)}({\boldsymbol{\beta}})=
(-1)^{\beta_{_2}+\beta_{_5}}{\Gamma(1-\beta_{_1}-\beta_{_4})\Gamma(\beta_{_5})\Gamma(2-\beta_{_2}-\beta_{_5})
\over\Gamma(\beta_{_2}+\beta_{_5})\Gamma(\beta_{_3})\Gamma(1-\beta_{_2})}C_{_{\{1,2,3\}}}^{(5)}({\boldsymbol{\beta}})
\nonumber\\
&&\hspace{2.3cm}
+(-1)^{\beta_{_2}}{\Gamma(1-\beta_{_1}-\beta_{_4})\Gamma(\beta_{_5})
\over\Gamma(\beta_{_2}+\beta_{_5})\Gamma(\beta_{_3}+\beta_{_5}-1)}C_{_{\{1,2,3\}}}^{(10)}({\boldsymbol{\beta}})
\;,\nonumber\\
&&C_{_{\{1,2,3\}}}^{(7)}({\boldsymbol{\beta}})=
(-1)^{\beta_{_1}+\beta_{_5}}{\Gamma(2-\beta_{_1}-\beta_{_5})\Gamma(1-\beta_{_2}-\beta_{_4})\Gamma(\beta_{_5})\over\Gamma(\beta_{_3})\Gamma(1-\beta_{_1})\Gamma(\beta_{_1}+\beta_{_5})}
C_{_{\{1,2,3\}}}^{(3)}({\boldsymbol{\beta}})
\nonumber\\
&&\hspace{2.3cm}
+(-1)^{\beta_{_1}}{\Gamma(1-\beta_{_2}-\beta_{_4})\Gamma(\beta_{_5})\over\Gamma(\beta_{_1}+\beta_{_5})\Gamma(\beta_{_3}+\beta_{_5}-1)}
C_{_{\{1,2,3\}}}^{(10)}({\boldsymbol{\beta}})
\;,\nonumber\\
&&C_{_{\{1,2,3\}}}^{(9)}({\boldsymbol{\beta}})=(-1)^{\beta_{_3}}{\Gamma(1-\beta_{_2}-\beta_{_4})
\Gamma(\beta_{_5})\over\Gamma(\beta_{_1}+\beta_{_5}-1)\Gamma(\beta_{_3}+\beta_{_5})}
C_{_{\{1,2,3\}}}^{(3)}({\boldsymbol{\beta}})
\nonumber\\
&&\hspace{2.3cm}
+(-1)^{\beta_{_3}+\beta_{_5}}{\Gamma(1-\beta_{_2}-\beta_{_4})\Gamma(\beta_{_5})\Gamma(2-\beta_{_3}-\beta_{_5})
\over\Gamma(\beta_{_1})\Gamma(1-\beta_{_3})\Gamma(\beta_{_3}+\beta_{_5})}C_{_{\{1,2,3\}}}^{(10)}({\boldsymbol{\beta}})
\;,\nonumber\\
&&C_{_{\{1,2,3\}}}^{(11)}({\boldsymbol{\beta}})=
(-1)^{\beta_{_3}}{\Gamma(1-\beta_{_1}-\beta_{_4})\Gamma(\beta_{_5})\over\Gamma(\beta_{_2}+\beta_{_5}-1)\Gamma(\beta_{_3}+\beta_{_5})}
C_{_{\{1,2,3\}}}^{(5)}({\boldsymbol{\beta}})
\nonumber\\
&&\hspace{2.3cm}
+(-1)^{\beta_{_3}+\beta_{_5}}{\Gamma(1-\beta_{_1}-\beta_{_4})\Gamma(2-\beta_{_3}-\beta_{_5})\Gamma(\beta_{_5})
\over\Gamma(\beta_{_2})\Gamma(1-\beta_{_3})\Gamma(\beta_{_3}+\beta_{_5})}C_{_{\{1,2,3\}}}^{(10)}({\boldsymbol{\beta}})\;.
\label{self-energy2-2}
\end{eqnarray}

The combinatorial coefficients of $\Phi_{_{\{1,2,3\}}}^{(i)}({\boldsymbol{\beta}}),\;i=1,8,12$ are respectively written as
\begin{eqnarray}
&&C_{_{\{1,2,3\}}}^{(1)}({\boldsymbol{\beta}})=
(-1)^{\beta_{_1}+\beta_{_5}-1}{\Gamma(\beta_{_3}+\beta_{_4}-1)\Gamma(1-\beta_{_4})\Gamma(\beta_{_5})
\over\Gamma(1-\beta_{_1})\Gamma(\beta_{_3})\Gamma(\beta_{_1}+\beta_{_5}-1)}C_{_{\{1,2,3\}}}^{(3)}({\boldsymbol{\beta}})
\nonumber\\
&&\hspace{2.3cm}
+(-1)^{1-\beta_{_2}-\beta_{_5}}{{\Gamma(\beta_{_3}+\beta_{_4}-1)\Gamma(1-\beta_{_4})\Gamma(\beta_{_5})
\over\Gamma(1-\beta_{_2})\Gamma(\beta_{_3})\Gamma(\beta_{_2}+\beta_{_5}-1)}}C_{_{\{1,2,3\}}}^{(5)}({\boldsymbol{\beta}})
\nonumber\\
&&\hspace{2.3cm}
+(-1)^{\beta_{_1}+\beta_{_2}}{{\Gamma(1-\beta_{_4})\Gamma(\beta_{_5})
\over\Gamma(2-\beta_{_3}-\beta_{_4})\Gamma(\beta_{_3}+\beta_{_5}-1)}}C_{_{\{1,2,3\}}}^{(10)}({\boldsymbol{\beta}})
\;,\nonumber\\
&&C_{_{\{1,2,3\}}}^{(8)}({\boldsymbol{\beta}})=
(-1)^{\beta_{_1}+\beta_{_5}-1}{\Gamma(\beta_{_2}+\beta_{_4}-1)\Gamma(1-\beta_{_4})\Gamma(\beta_{_5})
\over\Gamma(1-\beta_{_1})\Gamma(\beta_{_2})\Gamma(\beta_{_1}+\beta_{_5}-1)}C_{_{\{1,2,3\}}}^{(3)}({\boldsymbol{\beta}})
\nonumber\\
&&\hspace{2.3cm}
+(-1)^{\beta_{_1}+\beta_{_3}}{{\Gamma(1-\beta_{_4})\Gamma(\beta_{_5})
\over\Gamma(2-\beta_{_2}-\beta_{_4})\Gamma(\beta_{_2}+\beta_{_5}-1)}}C_{_{\{1,2,3\}}}^{(5)}({\boldsymbol{\beta}})
\nonumber\\
&&\hspace{2.3cm}
+(-1)^{1-\beta_{_3}-\beta_{_5}}{{\Gamma(\beta_{_2}+\beta_{_4}-1)\Gamma(1-\beta_{_4})\Gamma(\beta_{_5})
\over\Gamma(\beta_{_2})\Gamma(1-\beta_{_3})\Gamma(\beta_{_3}+\beta_{_5}-1)}}C_{_{\{1,2,3\}}}^{(10)}({\boldsymbol{\beta}})
\;,\nonumber\\
&&C_{_{\{1,2,3\}}}^{(12)}({\boldsymbol{\beta}})=
(-1)^{\beta_{_2}+\beta_{_3}}{\Gamma(1-\beta_{_4})\Gamma(\beta_{_5})\over\Gamma(\beta_{_1}+\beta_{_5}-1)\Gamma(2-\beta_{_1}-\beta_{_4})}
C_{_{\{1,2,3\}}}^{(3)}({\boldsymbol{\beta}})
\nonumber\\
&&\hspace{2.3cm}
+(-1)^{\beta_{_2}+\beta_{_5}-1}{\Gamma(\beta_{_1}+\beta_{_4}-1)\Gamma(1-\beta_{_4})\Gamma(\beta_{_5})
\over\Gamma(\beta_{_1})\Gamma(1-\beta_{_2})\Gamma(\beta_{_2}+\beta_{_5}-1)}C_{_{\{1,2,3\}}}^{(5)}({\boldsymbol{\beta}})
\nonumber\\
&&\hspace{2.3cm}
+(-1)^{\beta_{_3}+\beta_{_5}-1}{\Gamma(\beta_{_1}+\beta_{_4}-1)\Gamma(1-\beta_{_4})\Gamma(\beta_{_5})
\over\Gamma(\beta_{_1})\Gamma(1-\beta_{_3})\Gamma(\beta_{_3}+\beta_{_5}-1)}C_{_{\{1,2,3\}}}^{(10)}({\boldsymbol{\beta}})\;.
\label{self-energy5-3}
\end{eqnarray}
As ${\boldsymbol{\beta}}={\boldsymbol{\beta}}_{_{(1S)}}=(2-{D\over2},\;2-{D\over2},\;{D\over2},\;-1,\;{D\over2}-1)$,
the Feynman integral of the 1-loop self energy for $m_{_2}^2<m_{_1}^2<|p^2|$ is written as
\begin{eqnarray}
&&A_{_{1SE}}(p^2,m_{_1}^2,m_{_2}^2)={\Gamma(2-{D\over2})\Gamma^2({D\over2}-1)\over(4\pi)^{D/2}\Gamma(D-2)}
\Big({-p^2\over\Lambda_{_{\rm RE}}^2}\Big)^{D/2-2}
\nonumber\\
&&\hspace{3.5cm}\times
\Big\{-{(-1)^D2\Gamma({D\over2}-1)\over D\Gamma(2-{D\over2})\Gamma(D-2)}
x^{{D\over2}-2}y^{D\over2}\varphi_{_{\{1,2,3\}}}^{(8)}({\boldsymbol{\beta}},\;y,\;{y\over x})
\nonumber\\
&&\hspace{3.5cm}
-{(-1)^D\Gamma({D\over2})\Gamma({D\over2}-1)\Gamma(3-D)\over\Gamma(2-{D\over2})\Gamma(1-{D\over2})\Gamma(D-1)}
x^{D-2}\varphi_{_{\{1,2,3\}}}^{(9)}({\boldsymbol{\beta}},\;x,\;{y\over x})
\nonumber\\
&&\hspace{3.5cm}
+\varphi_{_{\{1,2,3\}}}^{(10)}({\boldsymbol{\beta}},\;y,\;x)\Big\}\;,
\label{self-energy4-a}
\end{eqnarray}
and the Feynman integral of the 1-loop self energy for $m_{_1}^2<m_{_2}^2<|p^2|$ is written as
\begin{eqnarray}
&&A_{_{1SE}}(p^2,m_{_1}^2,m_{_2}^2)=
{\Gamma(2-{D\over2})\Gamma^2({D\over2}-1)\over(4\pi)^{D/2}\Gamma(D-2)}
\Big({-p^2\over\Lambda_{_{\rm RE}}^2}\Big)^{D/2-2}
\Big\{\varphi_{_{\{1,2,3\}}}^{(10)}({\boldsymbol{\beta}},\;y,\;x)
\nonumber\\
&&\hspace{3.5cm}
-{(-1)^D\Gamma({D\over2})\Gamma({D\over2}-1)\Gamma(3-D)\over\Gamma(2-{D\over2})\Gamma(1-{D\over2})\Gamma(D-1)}
y^{D-2}\varphi_{_{\{1,2,3\}}}^{(11)}({\boldsymbol{\beta}},\;y,\;{x\over y})
\nonumber\\
&&\hspace{3.5cm}
-{(-1)^D2\Gamma({D\over2}-1)\over D\Gamma(2-{D\over2})\Gamma(D-2)}
x^{D\over2}y^{{D\over2}-2}\varphi_{_{\{1,2,3\}}}^{(12)}({\boldsymbol{\beta}},\;x,\;{x\over y})\Big\}\;.
\label{self-energy5-a}
\end{eqnarray}

Using the relations in the Eq.~(\ref{sec4-2b}), we reduce the hypergeometric series $\varphi_{_{\{1,2,3\}}}^{(8)}$
as a Laurent polynomial
\begin{eqnarray}
&&x^{D/2-2}y^{D/2}\varphi_{_{\{1,2,3\}}}^{(8)}({\boldsymbol{\beta}},\;y,\;{y\over x})
\nonumber\\
&&\hspace{-0.5cm}=
x^{D/2-2}y^{D/2}F_{_1}\left(\left.\begin{array}{c}2,\;{D\over2},\;2-{D\over2}\\ {D\over2}-1\end{array}\right|\;y,\;{y\over x}\right)
\nonumber\\
&&\hspace{-0.5cm}=
{x^5y^{D/2}(1-y)^{4-D}\over(x-y)^{7-{D/2}}}\Big\{{D-6\over D-2}{x^2(1-y)^2\over(x-y)^2}
\nonumber\\
&&\hspace{0.0cm}
+{4\over D-2}{x(1-y)\over x-y}-{(D-6)^2\over4(D-2)}{xy(1-y)^2\over(x-y)^2}\Big\}\;.
\label{PE-2-13}
\end{eqnarray}

Similarly the hypergeometric series $\varphi_{_{\{1,2,3\}}}^{(9)}$ is reduced as
\begin{eqnarray}
&&x^{D-2}\varphi_{_{\{1,2,3\}}}^{(9)}({\boldsymbol{\beta}},\;x,\;{y\over x})
=x^{D-2}H_{_1}\left(\left.\begin{array}{c}{D\over2},\;2-{D\over2}\\ D-1,\;1-{D\over2}\end{array}\right|\;x,\;{y\over x}\right)
\nonumber\\
&&\hspace{-0.5cm}=
{x^{D-2}(1-y)^{D/2-2}\over(1-x)^{{D/2}}}\Big\{
\;_{_2}F_{_1}\left(\left.\begin{array}{c}2-{D\over2},\;3-{D\over2}\\ 2\end{array}\right|\;{y(x-1)\over x(y-1)}\right)
\nonumber\\
&&\hspace{0.0cm}
-{2\over4-D}{x\over x-1}\;_{_2}F_{_1}\left(\left.\begin{array}{c}2-{D\over2},\;2-{D\over2}\\ 1\end{array}\right|\;
{y(x-1)\over x(y-1)}\right)\Big\}\;.
\label{PE-2-14}
\end{eqnarray}
through Eq.~(\ref{sec4-4b}).

The hypergeometric series $\varphi_{_{\{1,2,3\}}}^{(10)}$ is reduced in terms of the Gauss functions
\begin{eqnarray}
&&\varphi_{_{\{1,2,3\}}}^{(10)}({\boldsymbol{\beta}},\;y,\;x)
=F_{_1}\left(\left.\begin{array}{c}2-{D\over2},\;2-{D\over2},\;2-{D\over2}\\ 3-D\end{array}\right|\;y,\;x\right)
\nonumber\\
&&\hspace{-0.5cm}=
(1-x)^{D/2-2}\Big\{{1\over1-y}\;_{_2}F_{_1}\left(\left.\begin{array}{c}1-{D\over2},\;1-{D\over2}\\ 3-D\end{array}\right|\;
{x-y\over x-1}\right)
\nonumber\\
&&\hspace{0.0cm}
+{2-D\over2(3-D)}{x\over x-1}\;_{_2}F_{_1}\left(\left.\begin{array}{c}2-{D\over2},\;2-{D\over2}\\ 4-D\end{array}\right|\;
{x-y\over x-1}\right)\Big\}
\nonumber\\
&&\hspace{-0.5cm}=
(1-y)^{D/2-2}\Big\{{1\over1-x}\;_{_2}F_{_1}\left(\left.\begin{array}{c}1-{D\over2},\;1-{D\over2}\\ 3-D\end{array}\right|\;
{x-y\over 1-y}\right)
\nonumber\\
&&\hspace{0.0cm}
+{2-D\over2(3-D)}{y\over y-1}\;_{_2}F_{_1}\left(\left.\begin{array}{c}2-{D\over2},\;2-{D\over2}\\ 4-D\end{array}\right|\;
{x-y\over 1-y}\right)\Big\}\;.
\label{PE-2-15}
\end{eqnarray}

The hypergeometric series $\varphi_{_{\{1,2,3\}}}^{(11)}$ is also reduced in terms of the Gauss functions
\begin{eqnarray}
&&y^{D-2}\varphi_{_{\{1,2,3\}}}^{(11)}({\boldsymbol{\beta}},\;y,\;{x\over y})
=y^{D-2}H_{_1}\left(\left.\begin{array}{c}{D\over2},\;2-{D\over2}\\ D-1,\;1-{D\over2}\end{array}\right|\;y,\;{x\over y}\right)
\nonumber\\
&&\hspace{-0.5cm}=
{y^{D-2}(1-x)^{D/2-2}\over(1-y)^{D/2}}\Big\{
\;_{_2}F_{_1}\left(\left.\begin{array}{c}2-{D\over2},\;3-{D\over2}\\ 2\end{array}\right|\;{x(y-1)\over y(x-1)}\right)
\nonumber\\
&&\hspace{0.0cm}
-{2\over4-D}{y\over y-1}\;_{_2}F_{_1}\left(\left.\begin{array}{c}2-{D\over2},\;2-{D\over2}\\ 1\end{array}\right|\;
{x(y-1)\over y(x-1)}\right)\Big\}\;.
\label{PE-2-16}
\end{eqnarray}

Finally the hypergeometric series $\varphi_{_{\{1,2,3\}}}^{(12)}$ is reduced
as a Laurent polynomial
\begin{eqnarray}
&&x^{D/2}y^{{D/2}-2}\varphi_{_{\{1,2,3\}}}^{(12)}({\boldsymbol{\beta}},\;x,\;{x\over y})
\nonumber\\
&&\hspace{-0.5cm}=
x^{D/2}y^{{D/2}-2}F_{_1}\left(\left.\begin{array}{c}2,\;{D\over2},\;2-{D\over2}\\ {D\over2}-1\end{array}
\right|\;x,\;{x\over y}\right)
\nonumber\\
&&\hspace{-0.5cm}=
{x^{D/2}y^5(1-x)^{4-D}\over(y-x)^{7-{D\over2}}}\Big\{{D-6\over D-2}{y^2(1-x)^2\over(y-x)^2}
\nonumber\\
&&\hspace{0.0cm}
+{4\over D-2}{y(1-x)\over y-x}-{(D-6)^2\over4(D-2)}{xy(1-x)^2\over(y-x)^2}\Big\}\;.
\label{PE-2-17}
\end{eqnarray}

Making the variable transformations
\begin{eqnarray}
&&x=-{s\over(1-s)(1-t)},\;\;\;y=-{t\over(1-s)(1-t)}\;.
\label{PE-1-5}
\end{eqnarray}
and using some well-known transformations of the Gauss function, we derive
\begin{eqnarray}
&&x^{{D\over2}-2}y^{D\over2}\varphi_{_{\{1,2,3\}}}^{(8)}({\boldsymbol{\beta}},\;y,\;{y\over x})
={D-6\over D-2}{s^7t^{D\over2}(1-s+st)^{6-D}\over(1-s)^4(1-t)^4(s-t)^{9-{D\over2}}}
\nonumber\\
&&\hspace{4.8cm}
+{4\over D-2}{s^6t^{D\over2}(1-s+st)^{5-D}\over(1-s)^3(1-t)^3(s-t)^{8-{D\over2}}}
\nonumber\\
&&\hspace{4.8cm}
-{(D-6)^2\over4(D-2)}{s^6t^{{D\over2}+1}(1-s+st)^{6-D}\over(1-s)^4(1-t)^4(s-t)^{9-{D\over2}}}
\;,\nonumber\\
&&x^{D-2}\varphi_{_{\{1,2,3\}}}^{(9)}({\boldsymbol{\beta}},\;x,\;{y\over x})=
{(D-6)\Gamma(-{D\over2})\over\Gamma({D\over2})\Gamma(3-D)}{s^7t^{D\over2}(1-s+st)^{6-D}\over(1-s)^4(1-t)^4(s-t)^{9-{D\over2}}}
\nonumber\\
&&\hspace{4.5cm}
+{4\Gamma(-{D\over2})\over\Gamma({D\over2})\Gamma(3-D)}{s^6t^{D\over2}(1-s+st)^{5-D}\over(1-s)^3(1-t)^3(s-t)^{8-{D\over2}}}
\nonumber\\
&&\hspace{4.5cm}
-{(D-6)^2\Gamma(-{D\over2})\over4\Gamma({D\over2})\Gamma(3-D)}{s^6t^{{D\over2}+1}(1-s+st)^{6-D}\over(1-s)^4(1-t)^4(s-t)^{9-{D\over2}}}
\nonumber\\
&&\hspace{4.5cm}
+{(-1)^D\Gamma(2-{D\over2})\Gamma(1-{D\over2})\over\Gamma({D\over2})\Gamma({D\over2}-1)}
\nonumber\\
&&\hspace{4.5cm}\times
{(st)^{{D\over2}-1}\over((1-s)(1-t))^{{D\over2}-2}}
\;_{_2}F_{_1}\left(\left.\begin{array}{c}1,\;2-{D\over2}\\ {D\over2}
\end{array}\right|\;st\right)
\;,\nonumber\\
&&\varphi_{_{\{1,2,3\}}}^{(10)}({\boldsymbol{\beta}},\;x,\;y)=
\Big({1-st\over(1-s)(1-t)}\Big)^{D-3}
\nonumber\\
&&\hspace{3.5cm}
+{\Gamma(3-D)(st)^{{D\over2}-1}\over\Gamma(D-1)((1-s)(1-t))^{{D\over2}-2}}
\;_{_2}F_{_1}\left(\left.\begin{array}{c}1,\;2-{D\over2}\\ {D\over2}
\end{array}\right|\;st\right)
\nonumber\\
&&\hspace{3.5cm}
-{2\Gamma(D-3)\over\Gamma^2({D\over2}-1)}{s^{{D\over2}-1}\over((1-s)(1-t))^{{D\over2}-2}}
\;_{_2}F_{_1}\left(\left.\begin{array}{c}1,\;2\\ {D\over2}\end{array}\right|\;{s(1-t)\over1-st}\right)
\nonumber\\
&&\hspace{3.5cm}
-{2\Gamma(D-3)\over\Gamma^2({D\over2}-1)}{t^{{D\over2}-1}\over((1-s)(1-t))^{{D\over2}-2}}
\;_{_2}F_{_1}\left(\left.\begin{array}{c}1,\;2\\ {D\over2}\end{array}\right|\;{t(1-s)\over1-st}\right)\;.
\label{PE-2-19}
\end{eqnarray}

Inserting the above equations into Eq.(\ref{self-energy4-a}), one derives
\begin{eqnarray}
&&A_{_{1SE}}(p^2,m_{_1}^2,m_{_2}^2)={\Gamma(2-{D\over2})\Gamma^2({D\over2}-1)\over(4\pi)^{D/2}\Gamma(D-2)}
\Big({-p^2\over\Lambda_{_{\rm RE}}^2}\Big)^{{D\over2}-2}
\Big\{\Big({1-st\over(1-s)(1-t)}\Big)^{D-3}
\nonumber\\
&&\hspace{3.5cm}
-{2\Gamma(D-3)\over\Gamma^2({D\over2}-1)}{s^{{D\over2}-1}\over((1-s)(1-t))^{{D\over2}-2}}
\;_{_2}F_{_1}\left(\left.\begin{array}{c}1,\;2\\ {D\over2}\end{array}\right|\;{s(1-t)\over1-st}\right)
\nonumber\\
&&\hspace{3.5cm}
-{2\Gamma(D-3)\over\Gamma^2({D\over2}-1)}{t^{{D\over2}-1}\over((1-s)(1-t))^{{D\over2}-2}}
\;_{_2}F_{_1}\left(\left.\begin{array}{c}1,\;2\\ {D\over2}\end{array}\right|\;{t(1-s)\over1-st}\right)\Big\}\;.
\label{PE-2-18a}
\end{eqnarray}
This result can also be obtained from Eq.(\ref{self-energy5-a}) by the similar steps above.
In our notations, Eq. (21) in Ref.\cite{Davydychev3} is written as
\begin{eqnarray}
&&A_{_{1SE}}(p^2,m_{_1}^2,m_{_2}^2)={\Gamma(2-{D\over2})\Gamma^2({D\over2}-1)\over(4\pi)^{D/2}\Gamma(D-2)}
\Big({-p^2\over\Lambda_{_{\rm RE}}^2}\Big)^{D/2-2}
\Big\{F_{_4}\left(\left.\begin{array}{c}2-{D\over2},\;3-D\\ 2-{D\over2},\;2-{D\over2}\end{array}\right|\;x,\;y\right)
\nonumber\\
&&\hspace{3.7cm}
-{2\Gamma(D-3)\over\Gamma^2({D\over2}-1)}
(-x)^{{D\over2}-1}F_{_4}\left(\left.\begin{array}{c}1,\;2-{D\over2}\\ {D\over2},\;2-{D\over2}\end{array}\right|\;x,\;y\right)
\nonumber\\
&&\hspace{3.7cm}
-{2\Gamma(D-3)\over\Gamma^2({D\over2}-1)}
(-y)^{{D\over2}-1}F_{_4}\left(\left.\begin{array}{c}1,\;2-{D\over2}\\ 2-{D\over2},\;{D\over2}\end{array}\right|\;x,\;y\right)\Big\}\;,
\label{PE-1-4}
\end{eqnarray}
which recovers the expression in Eq.(\ref{PE-2-18a}) through the formulae presented in Eq.(\ref{PE-1-1}).

\end{document}